\documentclass[10pt,oneside]{report}

\usepackage{graphicx}
\usepackage{etoolbox} 

\usepackage{titling}
\usepackage{setspace} 

\usepackage[british]{babel}
\usepackage[nodayofweek]{datetime}

\usepackage[square,numbers]{natbib}

\usepackage{notoccite}

\usepackage{scrextend}

\usepackage[a4paper,margin=20mm, bindingoffset=20mm]{geometry}
\usepackage{epstopdf}   
\usepackage{amssymb,amsmath}
\usepackage{cancel}
\usepackage{url,hyperref}
\usepackage{bbm,bm}
\usepackage{color}
\usepackage{pifont,textcomp}
\usepackage[normalem]{ulem} 
\usepackage{tabularx}
\usepackage{subfig}
\usepackage{bibentry}

\usepackage{wasysym}

\usepackage{etex}
\usepackage[all]{xy}
\usepackage{ifpdf}%
\input{Qcircuit}

\providecommand{\ket}[1]{| #1 \rangle}
\providecommand{\bra}[1]{\langle #1 |}

\providecommand{\sprod}[2]{\langle#1|#2\rangle}
\providecommand{\idop}{\mathbbm 1}

\newtheorem{prop}{Proposition}\def\PRO{\begin{prop}}\def\ORP{\end{prop}}
\newtheorem{theorem}{Theorem}\def\TH{\begin{theorem}}\def\HT{\end{theorem}}
\newtheorem{defin}[prop]{Definition}\def\DE{\begin{defin}}\def\ED{\end{defin}}

\title{Quantum computation and communication in strongly interacting systems}
\author{Robert Gerald Antonio}

\newcolumntype{C}[1]{>{\centering\arraybackslash$}p{#1}<{$}}

\newdate{thesis_date}{23}{02}{2015}

\usepackage[T1]{fontenc}
\usepackage{titlesec, blindtext, color}
\definecolor{gray75}{gray}{0.75}
\newcommand{\hsp}{\hspace{20pt}}
\titleformat{\chapter}[hang]{\Huge\bfseries}{\thechapter\hsp\textcolor{gray75}{|}\hsp}{0pt}{\Huge\bfseries}

\usepackage{fancyhdr} 

\newdateformat{monthyear}{\monthname[\THEMONTH] \THEYEAR} 
\begin{document}

\pagestyle{empty} 

\begin{center}

\vspace*{1cm}

{\Huge\textbf{\thetitle}}\\ 
\vspace*{3cm}

{\Large \theauthor}\\ 
\vspace*{2cm}
\vfill
{\Large University College London}
\vspace*{1.5cm}

\begin{figure}[h!]
\centering
\includegraphics[width=50mm]{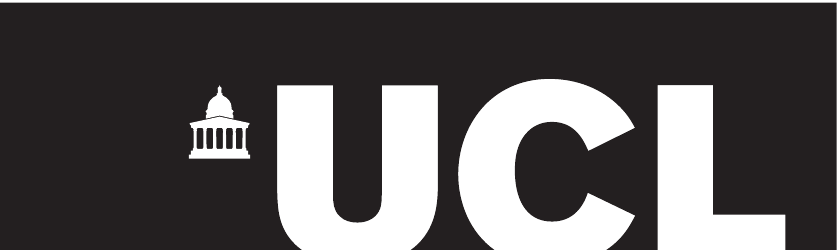}
\end{figure}

\vspace*{1.5cm}
\textsc{A dissertation submitted for the degree of}\\
\large Doctor of Philosophy \\
\monthyear\displaydate{thesis_date}
\end{center}

\newpage

\pagestyle{fancy}
\fancyfoot[C]{\thepage}
\fancyhead[LO,RE]{}

\null\thispagestyle{empty}
\newpage

\pagestyle{plain} 
\setcounter{page}{2}
\chapter*{Declaration}
I, \theauthor{}, confirm that the work presented in this thesis is my own. Where information has been derived from other sources, I confirm that this has been indicated in the thesis. Chapter~\ref{chap:DFS} is based on work done in collaboration with Sougato Bose. Chapter~\ref{chap:AGQC} is based on work done in collaboration with Damian Markham and Janet Anders. Chapter~\ref{chap:3qubit} is based on work done in collaboration with Joseph Randall, Gavin Morley Winfried Hensinger and Sougato Bose. Chapter~\ref{chap:Wigner} is based on work done in collaboration with Abofazl Bayat, Sanjeev Kumar, Michael Pepper and Sougato Bose. Section~\ref{sec:NNN} is based on work done in collaboration with Abofazl Bayat and Sougato Bose. Section~\ref{sec:EdgeLock} is based on work done in collaboration with Masudul Haque and Sougato Bose. Parts of this dissertation have been published, or submitted for publication, as follows:
\begin{itemize}
\item Bobby Antonio and Sougato Bose. Two-qubit gates for decoherence-free qubits using a ring
exchange interaction. \emph{Phys. Rev. A}, 88:042306, 2013.
\item Bobby Antonio, Damian Markham and Janet Anders, Adiabatic graph-state quantum computation. \emph{New J. Phys.} 16 113070, 2014.
\item Bobby Antonio, Abolfazl Bayat, Sanjeev Kumar, Michael Pepper and Sougato Bose. Self-Assembled Wigner Crystals as Mediators of Spin Currents and Quantum Information. \emph{Phys. Rev. Lett.}, 115 216804, 2015.
\end{itemize}

\vfill

\hfill \theauthor

\hfill \monthyear\displaydate{thesis_date}

\newpage\null\thispagestyle{empty}
\newpage
\setcounter{page}{3}
\chapter*{Abstract}
Each year, the gap between theoretical proposals and experimental endeavours to create quantum computers gets smaller, driven by the promise of fundamentally faster algorithms and quantum simulations. This occurs by the combination of experimental ingenuity and ever simpler theoretical schemes. This thesis comes from the latter perspective, aiming to find new, simpler ways in which components of a quantum computer could be built.

We first search for ways to create quantum gates, the primitive building blocks of a quantum computer. We find a novel, low-control way of performing a two-qubit gate on qubits encoded in a decoherence-free subspace, making use of many-body interactions that may already be present. This includes an analysis of the effect of control errors and magnetic field fluctuations on the gate. We then present novel ways to create three-qubit Toffoli and Fredkin gates in a single step using linear arrays of qubits, including an assessment of how well these gates could perform, for quantum or classical computation, using state-of-the-art ion trap and silicon donor technology.

We then focus on a very different model from the normal circuit model, combining ideas from measurement-based quantum computation (MBQC) and holonomic quantum computation. We generalise an earlier model to show that all MBQC patterns with a property called \emph{gflow} can be converted into a holonomic computation. The manifestation of the properties of MBQC in this adiabatically driven model is then explored.

Finally, we investigate ways in which quantum information can be communicated between distant parties, using minimally engineered spin chains. The viability of using 1D Wigner crystals as a quantum channel is analysed, as well as schemes using ideal uniform spin chains with next-neighbour interactions, and edge-locking effects.
\newpage\null\thispagestyle{empty}
\newpage
\setcounter{page}{4}
\IfFileExists{Acknowledgements.tex}{%
\chapter*{Acknowledgements}
Foremost I thank my mum and dad for creating such a loving and supportive family that has gone through much adversity but come out stronger every time. My mum in particular deserves a medal for her patience and resilience. My siblings Dave, Rache and Tommy are some of the funniest, brightest people I know, and I am constantly learning from them. Rach is the best girlfriend ever, her sense of humour and restless curiosity has kept me smiling throughout, and I feel exceptionally lucky to have her in my life. My grandparents Margeret and Gerald are also a constant source of inspiration, and I aspire to be just as active and sharp when I'm older.

Sougato Bose has been an excellent supervisor, a seemingly unending source of ideas and physics insights. I've learned a lot as his student and hope that our discussions will continue in the future. My second supervisor Janet Anders has been incredibly supportive and generous, far more than the call of duty, and through this support I have managed to make the most of my time in UCL. I've enjoyed many discussions and visits with Damian Markham, and hope that there are many more to come. I thank Abolfazl Bayat for his patience with me when I arrived as a fresh 1st-year student, and for being an excellent collaborator on some of the work presented in this thesis. I've also had the opportunity to collaborate with other world-class physicists who have all been a pleasure to work with; Gavin Morley, Winni Hensinger, Joe Randall, Masudul Haque and Tony Apollaro. 

I'm fortunate to have been part of a very dynamic and sociable group at UCL. Martin Uhrin is one of the most knowledgeable people I know, and was always on hand to answer questions on computing, politics or philosophy. Seto Balian was my partner in crime for QCumber, and could always be counted on to talk over quantum physics problems, or to hang out in Birkbeck union. I've also enjoyed talking about all kinds of things with Duncan Little, George Pender, Hugh Price and Clara Sousa-Silva, and all of the attendees of the quantum information lunchtime meetings. 

Before starting this PhD I also had a lot of help from some excellent teachers, in particular Luke Tattersall, Ian Lavender, the CNBH group in Cambridge, and my undergraduate director of studies Karl Sandeman. I thank Daniel Burgarth and Andrew Fisher for giving their time to read and examine this thesis. This PhD was funded by the Engineering and Physical Sciences Research Council.
}{}
\newpage\null\thispagestyle{empty}
\newpage

\setcounter{page}{5}

\setboolean{@twoside}{true} 
\tableofcontents
\listoffigures
\listoftables

\pagestyle{fancy}
\fancyfoot[C]{\thepage}
\fancyhead[RO,LE]{}
\fancyhead[LO]{\leftmark}
\fancyhead[RE]{\leftmark}

\chapter{Introduction}
\label{chap:Intro}

A study in 2007 estimated the total information storage capacity across all computer devices in the world to be around $295\times 10^{18}$ bytes (295 exabytes)~\cite{Hilbert2011}, roughly the same information contained by paperback novels stretching to the sun and back 200 times. This is symptomatic of our ever growing appetite for information and computing power, and how much computers have become integral to our lives over the last 50 years or so. Computers are now used in phones, coffee machines,  glasses, predicting the weather, even creating new forms of currency. 

The cause, or result, of this has been a reliable exponential growth in computational power over the years, as first noticed by Moore in his famous law~\cite{Moore1965}, which states that the number of transistors that can be put inexpensively onto a square metre approximately doubles every 2 years. The numbers are impressive; each year we gain enough computational power to perform around 60\% of the computations that could have possibly been executed by all existing general-purpose computers in the years before~\cite{Hilbert2011}. The Apollo 11 guidance computers, weighing 32kg and costing hundreds of thousands of dollars, are now outperformed by a Raspberry Pi, weighing 0.05kg and available for less than \pounds 25~\cite{NASA}. This steady trend has been possible thanks to extraordinary feats in physics and engineering, from integrated circuits to the development of complementary metal-oxide-semiconductor (CMOS) technology. However there is only so far that this technology can go, and it doesn't seem long before we reach fundamental limitations of fabrication and device reliability~\cite{Frank2002,Frank2001,Markov2014,Chien2013}.

Whilst historically our predictions about the limitations of computational power seem to be about as accurate as predictions of the end of the world, and we aren't sure of the future economics behind new computer research, it all points to a future where information is stored at the scale of molecules and atoms. With information stored at this scale, we must constantly battle the effects of quantum mechanics, or else embrace it and start to use the features of quantum mechanics to make an entirely new form of computation. This is the paradigm of \emph{quantum computation} (as opposed to the \emph{classical} computation we have today), first proposed by Benioff and Feynman~\cite{Benioff1982, Feynman1982}, and later made more concrete by Deutsch~\cite{Deutsch1989}.

Aside from satisfying our appetite for smaller computers, the idea of quantum computation raises many fundamental questions, and provides an interesting link between the seemingly different disciplines of physics and computer science. Intuitively, if we really do believe that quantum mechanics is a better description of reality than classical physics, then we expect that a quantum computer must be able to do things that a classical computer cannot. This is at odds with an enduring law of computation, the strong Church-Turing thesis, which states that any task that can be efficiently computed on a physically reasonable device can be simulated efficiently by a (probabilistic) Turing machine~\cite{Turing1936,Church1936}. Can a quantum computer be efficiently simulated by a Turing machine, or is there a violation of this rule? Whilst there are no outright proofs that this is the case, there is much evidence. The most prominent example is Shor's algorithm~\cite{Shor1994}, building on work by Simon~\cite{Simon1994}, which gives an efficient quantum algorithm for prime factorisation (a problem for which the best known classical algorithm takes close to exponential time). This is an important problem in areas such as cryptography where the hardness of prime factorisation underpins the hardness of breaking RSA~\cite{Rivest1978}, the cryptographic protocol at the heart of most secure transactions online.

Other examples of differences between quantum computers and classical computers are Grover's quadratic speed up for searching through unsorted lists~\cite{Grover1997} and the apparent difficulty (assuming that commonly held beliefs in computer science are true) of classically simulating circuits of commuting operations~\cite{Bremner2011} or linear optical networks~\cite{Aaronson2011}. Even more claims are made in the media, with the promise that quantum computers will do anything from winning elections to discovering planets~\cite{Time}.

Whether or not there is a difference between classical and quantum computing, the outcome will be positive for physics; we either unleash untapped computational power (potentially also allowing fast computer simulations of quantum systems), or we find some (perhaps fundamental) limits to our ability to control the world, as well as the numerous other offshoots that have already stemmed from quantum information theory, such as quantum cryptography, quantum sensing and random number generation. And regardless of what we do actually know about how quantum computation will change things, the history of classical computers tells us that we won't know until we try it.

But even if we accept that it is a good idea to build a quantum computer, is it \emph{practically} possible to build a quantum computer? Are there any insurmountable obstacles which prevent us from constructing a universal quantum computer? Recent work shows that in order to simulate the ground state of a molecule that is twice the size as the current largest classically simulatable molecule, a quantum computer would need on the order of $10^{18}$ gates~\cite{Wecker2014}, and to factorise a 200 digit number requires approximately 100,000 qubits~\cite{Steane1999}. Given that current state-of-the-art programmable quantum computers have reached around 20 qubits~\cite{Monz2011}, and quantum effects are often washed away in the macroscopic world, this seems like an insurmountable task. Fortunately, it is possible to encode information in ways such that the errors are correctable so, provided there are no unexpected experimental brick walls, creating a quantum computer only seems like a matter of time and effort. 

This thesis sits somewhere in between the practical and theoretical side of quantum computing, aiming to push the proposals of quantum computing just a little bit closer to the experimentalists' reach. Perhaps reflecting the diverse range of topics that make up quantum information theory, this thesis is also split into different projects. These broadly fall into three categories: work aimed at finding new ways to create quantum gates, or classical gates using quantum components, work aimed at comparing different models of computation, and work aimed at investigating how we might communicate in a quantum computer. The unifying theme behind these is the desire to make quantum computation more attainable.

In Chapter~\ref{chap:DFS}, we propose a way in which 2-qubit gates could be performed in a decoherence-free subspace. We focus on using a small number of minimal-control pulses, with minimally engineered interactions, including many-body ring exchange interactions that are predicted to be present in many systems. The gate we find has less pulses than any previous scheme, but takes roughly the same time as the best known scheme. During the 2-qubit decoherence-free gate, the system leaks out of the decoherence-free subspace, and the effects of this are studied by simulating the effect of control errors and magnetic field fluctuations. We find that the gate error remains below 1\% for errors typical of gated quantum dots in GaAs, which is below the most generous fault-tolerant threshold.

Chapter~\ref{chap:3qubit} similarly deals with a proposal to make gates, this time 3-qubit Toffoli and Fredkin gates, using linear arrays of qubits. Such gates are not only useful for quantum circuits, but are also a useful primitive for classical reversible computation. We focus on using only a single pulse, with interactions that are as uniform as possible. We find one method to perform a Toffoli gate, and two methods to perform a Fredkin gate. Experimental schemes for realising two of these gates in ion traps and donors in silicon are also proposed, along with an assessment of how viable these schemes could be, for both quantum and classical computation, using current technology. The reachable fidelities of some of these gates with current technology are lower than the quantum error correction threshold, suggesting that they could be a useful technological innovation for either quantum or classical computation, perhaps providing a stepping stone en route to full quantum computation.

Chapter~\ref{chap:AGQC} combines the traditional measurement-based quantum computation (MBQC) protocol with adiabatic evolution, in an extension of the adiabatic cluster-state model. We show that any MBQC pattern which satisfies a set of graphical rules called \emph{gflow} can be converted into a holonomic quantum computation where the measurements are replaced by adiabatic evolutions, such that the computation is the same. We investigate how the trade-offs in MBQC manifest in this adiabatically driven model, and explore which orderings of operations are possible. 

In Chapters~\ref{chap:Wigner} \&~\ref{chap:NNN}, the ability of spin chains to transfer information is studied. The viability of using a 1D (or quasi-1D) Wigner crystal as a quantum communication channel is investigated using the semi-classical instanton approximation, finding that high average fidelities are possible for the particular couplings that we would expect to be present. This is followed by a study of the effects of next-neighbour interactions on state transfer through an ideal Heisenberg spin chain, and preliminary results for using edge locking and ground state correlations to transfer information.


\section{Quantum Computation and notation}\label{sec:CircuitModel}

In this section we give a broad introduction to quantum computing, giving a brief overview of the nomenclature and notation used in the ensuing chapters of this thesis. The fundamental building block of quantum computation is the quantum bit, or \emph{qubit}, in which quantum information can be stored. A qubit can be any two-level quantum system, such as two energy levels in an atom, two spin states of an electron, or two polarisations of a photon, and in this thesis all of the results will use qubits (although computation is of course possible using higher dimensional systems).

Qubits are represented by a two-dimensional normalised complex vector (more formally, it is a vector in the Hilbert space $\mathcal{H} = \mathbb{C}^2$, where $\mathbb{C}$ is the set of complex numbers). The two states of the qubit are usually denoted $\ket{0}$ and $\ket{1}$ in analogy to the bits in a classical computer. Unlike classical bits, the postulates of quantum mechanics permit arbitrary superpositions of these qubit states to exist, written as $\ket{\phi} = \alpha \ket{0} + \beta\ket{1}$, where $|\alpha|^2 + |\beta|^2 = 1$. Several qubits can be combined together using the tensor product $\otimes$, obtaining a $2^n$ dimensional complex vector for $n$ qubits (and a corresponding Hilbert space $\mathcal{H} = \mathbb{C}^2 \otimes \mathbb{C}^2 \otimes ... \otimes \mathbb{C}^2 = (\mathbb{C}^2)^{\otimes n}$) written as e.g. $\ket{\phi}\otimes \ket{\psi}$ or $\ket{\phi \psi}$. A particularly important basis of many qubits is the computational basis, defined as all possible tensor products of $\ket{0},\ket{1}$ over $N$ qubits, i.e. $\{ \ket{000...0},\ket{100...0},\ket{010...0}...,\ket{111...1} \}$.

A more general quantum state is represented by a Hermitian, non-negative, trace one complex density matrix, allowing for the possibility that the preparation of a quantum state may be uncertain, and so is a classical ensemble of quantum states. In general these states have the form $\rho = \sum_n p_n \ket{\phi_n}\bra{\phi_n}$, where $p_n$ is the probability that the system is in state $\ket{\phi_n}$. When the density matrix can be written $\rho = \ket{\phi}\bra{\phi}$ (or equivalently $\text{tr} (\rho^2) = 1$), it is said to be a pure state, and otherwise it is said to be a mixed state ($\text{tr} (\rho^2) < 1$).

A quantum computation then typically proceeds in the following steps
\begin{itemize}
\item[1)] Qubits are prepared in an initial pure state, which is the input to the computation.  
\item[2)] A sequence of operations are performed on these qubits.
\item[3)] The qubits are measured at the end of the computation, giving the outcome of the algorithm with suitably high probability.
\end{itemize}

This is the essence of the most commonly used model of quantum computation, the \emph{circuit model}, first formulated by Deutsch~\cite{Deutsch1989}. Such computations are typically represented as circuit diagrams, where qubits are represented as horizontal lines, and operations on these qubits are shapes on these lines, with time going forward from left to right. For example, a circuit with two qubits prepared in states $\ket{\phi}$, $\ket{0}$ which performs a two qubit $\textsc{cnot}$ gate between them is
\[
\Qcircuit @C=1em @R = 2.5em @!R  {
\lstick{\ket{\phi}} & \qw  &    \ctrl{1}      &    \qw      & \qw     \\
\lstick{\ket{0}}     &  \qw  &   \targ{U}   &    \qw     & \qw     \\
}
\]

Operations are performed by turning on interactions between qubits for times prescribed by the algorithm. In the standard formalism of quantum mechanics, the interactions in the system are given by the (generally time-dependent) Hamiltonian $H(t)$, and the evolution of the qubits due to this Hamiltonian is given by the unitary matrix $U(t)$
\begin{align}
U(t) = \mathcal{T} \exp \left( -\frac{i}{\hbar} \int_0^t H(t') dt' \right)
\end{align}
where $\mathcal{T}$ is the time-ordering operator. Single qubit rotations can be achieved by turning on a Hamiltonian consisting of one of the 3 Pauli matrices:
\begin{align}
X = \left(\begin{array}{cc}
0&1\\
1&0
\end{array}\right)\text{,  }
Y = \left(\begin{array}{cc}
0&-i\\
i&0
\end{array}\right)\text{,  }
Z = \left(\begin{array}{cc}
1&0\\
0&-1
\end{array}\right).
\end{align}

To perform rotations about the $x$, $y$ or $z$ axis respectively, a Hamiltonian $h_x X,h_y Y$ or $h_zZ$ must be turned on, resulting in the unitary rotations
\begin{align}
U_x(\theta_x)  &= \exp \left(i\theta_x X/2 \right)     = \cos\left(\frac{ \theta_x}{2} \right) \idop + i\sin\left(\frac{ \theta_x}{2} \right)X\\
U_y(\theta_y)  &= \exp  \left(i\theta_y  Y /2  \right) =\cos\left(\frac{ \theta_y}{2}\right)\idop + i\sin\left(\frac{ \theta_y}{2}\right)Y\\
U_z(\theta_z)  &= \exp \left( i \theta_z  Z/2\right)    =\cos\left(\frac{ \theta_z}{2} \right) \idop + i\sin\left(\frac{ \theta_z}{2} \right)Z,
\end{align}
where $\theta_\alpha = -h_\alpha t/\hbar$, with $h_\alpha$ as the strength of the interaction, and the equality follows since $X^2 = Y^2 = Z^2 = \idop$. Using any two of these rotations is enough to perform arbitrary rotations of a single qubit, as a general rotation can be written as e.g.\ $U_x(t_1)U_z(t_2)U_x(t_3)$~\cite{NielsenChuang}. We will use the shorthand $X_n$, $Y_n$, $Z_n$ to denote single qubit operations acting on qubit $n$ only and identities elsewhere, so for example $Z_1 = Z\otimes \idop$. We also define the Pauli group on $n$ qubits, denoted $\mathcal{G}_n$, as $\mathcal{G}_n = \langle X_1,Y_1,Z_1,X_2,Y_2,Z_2,...,X_n,Y_n,Z_n\rangle$, where the notation $\langle \cdot \rangle$ indicates that these operators generate the full group. We will often use the Heisenberg exchange interaction between qubits, which we denote $E_{nm} := X_n X_m  + Y_n Y_m + Z_n Z_m$.

Other important single qubit operations that will be used extensively in this thesis are the Hadamard ($\text{H}$) and raising and lowering operators $\sigma^\pm$ defined as
\begin{align}
\text{H} = \frac{1}{\sqrt{2}}\left(\begin{array}{cc}
1&1\\
1&-1
\end{array}\right), \; \sigma^\pm = \frac{1}{2} ( X \pm iY).
\end{align}

Clearly to perform general quantum computations, we will also need interactions between qubits. To perform universal quantum computation, it is only necessary to be able to perform any entangling two-qubit gate plus single qubit rotations~\cite{Barenco95,DiVincenzo1995,Bremner2002}. In fact not even general single qubit rotations are necessary, but only certain single qubit operations are required. The two most commonly entangling two-qubit gates are the controlled-\textsc{not} (\textsc{cnot}) and controlled-$Z$ (\textsc{cz}) gates. Here the two qubits are split into a control and target qubit, and the gate implements a Pauli $X$ (\textsc{cnot}) or $Z$ (\textsc{cz}) on the target qubit when the control qubit is in the $\ket{1}$ state. In matrix form they are
\begin{align}
\textsc{cnot} = \left( \begin{array}{cccc}
1&0&0&0\\
0&1&0&0\\
0&0&0&1\\
0&0&1&0
\end{array}\right)\text{,  }
\textsc{cz}= \left( \begin{array}{cccc}
1&0&0&0\\
0&1&0&0\\
0&0&1&0\\
0&0&0&-1
\end{array}\right).
\end{align}

Two other important gates, which also act as universal gates when combined with single qubit rotations, and which are universal for classical reversible computation are the Toffoli (controlled-controlled-\textsc{not}) gate and Fredkin (controlled-\textsc{swap}) gate~\cite{Fredkin1982}:
\begin{align}
T = \left( \begin{array}{cccccccc}
1&0&0&0&0&0&0&0\\
0&1&0&0&0&0&0&0\\
0&0&1&0&0&0&0&0\\
0&0&0&1&0&0&0&0\\
0&0&0&0&1&0&0&0\\
0&0&0&0&0&1&0&0\\
0&0&0&0&0&0&0&1\\
0&0&0&0&0&0&1&0\\
\end{array}\right)\text{, }
F = \left( \begin{array}{cccccccc}
1&0&0&0&0&0&0&0\\
0&1&0&0&0&0&0&0\\
0&0&1&0&0&0&0&0\\
0&0&0&1&0&0&0&0\\
0&0&0&0&1&0&0&0\\
0&0&0&0&0&0&1&0\\
0&0&0&0&0&1&0&0\\
0&0&0&0&0&0&0&1\\
\end{array}\right).
\end{align}
An important subgroup of all possible quantum operations is the \emph{Clifford group}. An important property of this group is that conjugation of any member of the Pauli group by a Clifford gives another member of the Pauli group, or more formally
\begin{align}
MPM^\dagger \in \mathcal{G}_n \quad \forall M \in \mathcal{C}_n, P \in \mathcal{G}_n,
\end{align}
where $\mathcal{C}_n$ is the Clifford group for $n$ qubits. Because of this property, any circuit made up entirely of Clifford gates can be simulated efficiently by a classical computer, a result known as the Gottesman-Knill theorem~\cite{NielsenChuang}. The Clifford group can be generated by $\langle \textsc{cz}, H, \sqrt{Z} \rangle$, and contains the Pauli group.

\section{Noise and decoherence}

In the previous section, we saw the basic framework for quantum computation in the absence of noise. In reality, nature is never as idealised as our models, and inevitably the control we have over the quantum states is imprecise and is never completely isolated from the outside world, so there are always ways for unwanted errors to occur. Thus a crucial part of computation is devising a scheme by which we can keep the effects of errors below a tolerable level. Before we discuss how to deal with errors, we first review the tools which are typically used to model the errors that occur.

We start with a simple example which demonstrates how quantum information stored in a qubit is affected by interactions with its surroundings. Consider a single qubit prepared in a pure state $\ket{\phi(0)} = (\alpha\ket{0} + \beta \ket{1})$ (which we call the `system') coupled to another qubit (the `environment') prepared in a $\ket{+}_E$ state. Consider coupling these two qubits via a \textsc{cz} interaction with strength $J$ for time $t$. Following the evolution the system + environment ends up in a state $\ket{\phi(t)}$
\begin{align}
\ket{\phi(t)} &= e^{-i J\; \textsc{cz} \; t / \hbar} (\alpha\ket{0} + \beta \ket{1}) \otimes \ket{+}_E := e^{-i \theta \; \textsc{cz} } (\alpha\ket{0} + \beta \ket{1}) \otimes \ket{+}_E \nonumber\\
& =  (\cos \theta \idop -i\sin \theta \textsc{cz})(\alpha\ket{0} + \beta \ket{1}) \otimes \ket{+}_E \nonumber\\
&=  \cos \theta (\alpha\ket{0}\ket{+}_E + \beta \ket{1}\ket{+}_E)  -i \sin \theta (\alpha\ket{0}\ket{+}_E + \beta \ket{1} \ket{-}_E),
\end{align}
where we have defined $\theta = J t / \hbar$. 

We can see that evolving the system + environment for times such that $\theta \neq n\pi$, there is entanglement between the system and the environment, so that to recover the information originally stored in the system we need a combined measurement of both system and environment. However, since the environment is by definition inaccessible, we cannot do this and instead have to reflect our ignorance of the environment by tracing it out altogether (effectively averaging out the effects of the environment). After tracing out the environment, we find that the system is in the state
\begin{align}\label{eqn:DecEx}
\rho_S(t) = \text{tr}_E (\ket{\phi(t)} \bra{\phi(t)}) = | \alpha|^2 \ket{0} \bra{0}  + | \beta |^2 \ket{1} \bra{1}  + \alpha^* \beta \cos^2\theta \ket{1}\bra{0} + \alpha \beta^* \cos^2\theta \ket{0}\bra{1}.
\end{align}

The result has off-diagonal elements that vary with $\theta$, and go to zero when $\theta = \pi/2$, and the purity of the system varies as $\text{tr} (\rho_S(t)^2) = |\alpha|^4 + |\beta|^4 + 2 |\alpha|^2 |\beta|^2 \cos^4 \theta$. 
Thus, depending on the strength of the system-environment coupling, the relative phase between the $\ket{0}$ and $\ket{1}$ states is affected, and the state becomes more mixed. This example is slightly contrived in that the information about the initial state of the system $\ket{\phi}$ is still accessible at times where $\theta = n\pi$, but in a realistic situation the environment will be large and there will be randomness present in the couplings $J$, so that this type of process leads to an irreversible degradation of the information.

This is one way in which connection to an external environment can degrade the quantum information; another way is where the diagonal terms are affected. This is covered in more detail in the next subsection where we look at a useful formalism that captures the behaviour of the system-environment interaction.

\subsection{The operator sum representation}
\label{sec:OSR}

In general we will have a system coupled to a large environment that we have no access to. For any system coupled to an environment, the Hamiltonian $H$ governing the evolution of the qubits plus environment can generally be written
\begin{align}
H = H_{sys} \otimes \idop_E + \idop_{sys}\otimes H_{E} + H_I,
\end{align}
where $H_{sys}$, $H_E$ and $H_I$ are the system, environment and system-environment interaction Hamiltonians, respectively. If we assume that that our starting state has no entanglement between system and environment then the initial state is $\rho(0) = \rho_S \otimes \rho_E$ (this is a reasonable assumption if we assume that it is possible to prepare the system in a pure state, which is a standard requirement for quantum computation anyway). Following the evolution due to $H$, the state at time $t$ is $\rho(t) = e^{-iHt/\hbar} (\rho_S \otimes \rho_E)e^{iHt/\hbar}$. For an observer who only has access to the system, the state at time $t$ is
\begin{align}\label{eqn:rhoS}
\rho_S(t) &= \mbox{tr}_E \left( e^{-iHt/\hbar} (\rho_S(0) \otimes \rho_E(0))e^{iHt/\hbar} \right),
\end{align} 
where $\mbox{tr}_E$ indicates tracing out the environment degrees of freedom, appropriate since we assume no access to the information stored in the environment, and so we must average over the possible states of the environment. The transformation in (\ref{eqn:rhoS}) is cumbersome, but through some manipulation we can arrive at a much more useful form. 

Assume that the environment is initially in a pure state $\rho_E(0) = \ket{\nu_0} \bra{\nu_0}$ - this is without loss of generality since if it was not in a pure state we could always purify it by adding another system~\cite{NielsenChuang}. We now rewrite equation (\ref{eqn:rhoS}) in terms of $\ket{\nu_0}$ and a complete orthonormal basis of the environment $\{ \ket{\mu} \}$
\begin{align}
\rho_S(t) &=  \sum_{\mu} \bra{\mu} e^{-iH t/\hbar} \rho_S(0)  \otimes \ket{\nu_0}\bra{\nu_0} e^{iH t/\hbar} \ket{\mu} =  \sum_{\mu} \bra{\mu} e^{-iH t/\hbar}\ket{\nu_0} \rho_S(0) \bra{\nu_0}e^{iH t/\hbar} \ket{\mu}\nonumber \\
& :=  \sum_{\mu} E_{\mu}(t) \rho_S(0) E_{\mu }^{\dagger}(t),
\end{align} 
where $E_{\mu }(t) = \bra{\mu} e^{-iH t/\hbar}\ket{\nu_0}$. 
This is the \emph{operator sum representation} (OSR) of the evolution of $\rho_S$, and is useful in that it reduces the overall system-bath dynamics to the operators $E_{\mu}$ (usually called Kraus operators) acting only on the system. This transformation is usually written $\rho_S(t) = \mathcal{E} (\rho_S(0) )$, and is a completely positive trace-preserving map, which means that $\sum_\mu E_{\mu}(t)^\dagger E_{\mu }(t) = \idop$ and $\mathcal{E}$ must preserve the non-negativity of the density matrix, even when applied to a subsystem~\cite{NielsenChuang}.

Using this formalism, we can now introduce many different types of noisy channels without worrying about the specific dynamics of the environment. We can think of the qubit interaction with the environment as being like a noisy quantum communication protocol, in which $\rho_S(0) $ is input in one end and $\rho_S(t)$ is output at the other. 

One important type of noise is \emph{amplitude damping}, which has Kraus operators~\cite{NielsenChuang}
\begin{align}
&E_0 =\sqrt{p} \left[ \begin{array}{cc}
1 & 0 \\
0 & \sqrt{1 - \gamma}
\end{array} \right]\text{, }
E_1 = \sqrt{p}\left[ \begin{array}{cc}
0 & \sqrt{\gamma} \\
0 & 0
\end{array} \right]\text{, }
E_2 = \sqrt{1-p}\left[ \begin{array}{cc}
\sqrt{1-\gamma} &  \\
0 & 1
\end{array} \right] \nonumber\\
&E_3 = \sqrt{1-p}\left[ \begin{array}{cc}
0 & 0 \\
\sqrt{\gamma} & 0
\end{array} \right],
\end{align}
and which has the effect of lowering the population of the $\ket{1}$ state with probability $\gamma$, corresponding to a process such as interaction with a thermal environment (in which case $p$ corresponds to the Boltzmann occupation factor). $\gamma$ typically depends on time, and allows a characteristic time $T_1$ for the decay to be found. If $\gamma$ is an exponential or Gaussian decay then $T_1$ is the time at which $\gamma = 1/e$. In general, the way of experimentally defining $T_1$ will depend on the particular form of the decay. 

Another important channel is the phase damping channel, which has Kraus operators
\begin{align}
E_0 =\left[ \begin{array}{cc}
1 & 0 \\
0 & \sqrt{1 - \gamma}
\end{array} \right]\text{, }
E_1 = \left[ \begin{array}{cc}
0 & 0 \\
0 & \sqrt{\gamma}
\end{array} \right]\text{.}
\end{align}
This is in fact the operator-sum form of the example seen in the example in (\ref{eqn:DecEx}), and describes a process where the phase between $\ket{0}$ and $\ket{1}$ is affected (this can happen by e.g.\ random shifting of the energy levels due to the environment). Like the amplitude damping example above, $\gamma$ determines a characteristic dephasing time $T_2$, which is usually taken as the $1/e$ time for Gaussian or exponential decay. $T_2 \leq 2T_1$, and typically $T_1 \gg T_2$, so $T_2$ is the limiting factor for realising quantum computers~\cite{Ladd2010}.

In order to compare different processes later on in this thesis, a slightly modified form of the operator sum representation is needed. First the Kraus operators can be expressed in terms of a complete basis $\{ A_n \}_{n=1}^{2^N}$ which form an orthogonal basis under the Hilbert-Schmidt inner product, i.e.\ $\text{tr}({A}_m^{\dagger} {A}_n) = \delta_{mn}$. For example, we could choose a basis formed from outer products of a set of orthogonal vectors$\{ A_n \} = \{ \ket{l}\bra{m} \}$, or the Pauli group $\mathcal{G}_N$. Expressing the Kraus operators as $E_{\mu} = \sum_n a_{n\mu} A_n$, where $a_{n \mu} = \text{tr}( A_n^\dagger E_{\mu})$, a map $\mathcal{E}$ can be written
\begin{align}
\mathcal{E} (\rho) &=\sum_\mu {E}_\mu \rho {E}_\mu^{\dagger} = \sum_\mu \sum_{mn} (a_{m\mu} a_{n \mu}^{*}) {A}_m \rho {A}_n^{\dagger} := \sum_{mn} \chi_{mn} {A}_m \rho {A}_n^{\dagger},
\end{align}
where in the last line we have defined the \emph{process matrix} $\chi_{mn} := \sum_{\mu} a_{m\mu} a_{n \mu}^{*}$.

\subsection{Quantum error correction}
\label{sec:QEC}

Having introduced some of the tools of quantum error correction, we will look at some of the methods to deal with quantum errors. There are two approaches to this problem: the `hardware' approach, in which we try to minimise the amount of errors that occur in the first place by building the computer in a way that is insensitive to the particular noise, or improving the quality of the components. There is also the `software' approach, where we encode our information in such a way that any errors can be recognised and corrected. 

For an illuminating example of the differences between the software and hardware approaches, consider an `$n$-bit-flip' channel through which we can send an $N$-bit message, where the channel flips $n$ bits of the message with probability $p$. One software approach would be to use these $N$ bits to encode a single bit, using a repetition code. Provided $n < N/2$, at the other end of the channel we can just decode the noisy information by taking a majority vote. However, if the channel flips \emph{all} of the bits with probability $p$, then this repetition code is no longer useful. Instead of using a repetition code, we can encode the logical bit into the parity of the $N$ qubits, so that if the channel flips all of the bits the parity is conserved (assuming that $N$ is even, but if $N$ is odd the same can be achieved by ignoring the $N^{th}$ bit). This is the hardware approach, where the encoding of the information itself is naturally immune to errors.

Clearly the sensible tactic in making quantum computers is to develop both approaches simultaneously, since environments will rarely be as predictable as the example give above. In this section we will introduce aspects of both approaches that are relevant to this thesis. We begin with the software approach; one of the simplest examples of this approach is copying the information, and hoping that all copies of the information will not suffer the same errors (a technique familiar to anyone who backs up their computer). Whilst this is a good starting point for classical error correction, in quantum computation we are stymied by the no-cloning theorem~\cite{Dieks1982,Wootters1982}, which prohibits copying of arbitrary quantum states. Fortunately, quantum error correcting codes are possible; Shor developed a quantum analogue of the classical repetition code~\cite{Shor1995} whilst Steane developed a quantum analogue of the classical Hamming code~\cite{Steane1996}. Following these developments, a five-qubit code was discovered by Bennett et.\ al.~\cite{Bennett1996}, and Calderbank, Shor and Steane discovered a scheme whereby any classical code can be made into a quantum code~\cite{Calderbank1996,Steane1996a}. A more general framework for error correction was developed by Gottesman~\cite{Gottesman1997} using stabilisers, which are discussed in more detail in Sec.~\ref{sec:MBQCStab}. 

The general feature of error-correcting codes is that there are a number of \emph{physical} qubits, which are the basic physical systems that we have at our disposal (e.g. ions, electrons), and which can be combined together to give \emph{encoded} or \emph{logical} qubits, which are 2-level systems existing in a subspace or subsystem of the physical qubits. As an example, the 3-qubit bit flip code has 3 physical qubits for every encoded qubit: $\ket{0}_L  = \ket{000}\text{, } \ket{1}_L = \ket{111}$, and can withstand bit-flip errors on 1 physical qubit.

 Using these error correcting codes, we can derive approximate bounds on how accurate each logic gate must be such that the accumulation of errors during a computation is not too large. To calculate this, we can consider a fault-tolerant circuit to construct a fault tolerant encoded \textsc{cnot} gate on two encoded qubits. The gate itself is performed using many imperfect \textsc{cnot} gates acting between physical qubits inside the encoded qubit. For example, a \textsc{cnot} gate for the 3-qubit bit flip code between two encoded qubits $Q_1$, $Q_2$ is:
 \[
\Qcircuit @C=1em @R = 1em @!R  {
                  &             &       &&&&    &                   & \ctrl{4}    &    \qw      &    \qw         & \qw     \\
\lstick{Q_1}&\ctrl{4}  &\qw &&&&   &\lstick{Q_1} &\qw          &  \ctrl{4}   &    \qw         & \qw     \\
                  &             &       &&&&   &                   & \qw         &   \qw      &    \ctrl{4}     & \qw      \\
                  &             &       &&:=&& &                   &                &               &                    &           \\
                  &             &       &&&&   &                   & \targ{X}   &  \qw       &    \qw          & \qw     \\
\lstick{Q_2}&\targ{X} &\qw &&&&    & \lstick{Q_2}& \qw         &  \targ{X}  &    \qw         & \qw     \\
                  &             &       &&&&    &                  & \qw         &  \qw        &    \targ{X}   & \qw    
\gategroup{2}{2}{6}{2}{.7em}{--}
}
\]

 Following the gate, we perform a syndrome measurement to detect any errors, followed by a recovery step to correct these errors:
 \[
\Qcircuit @C=1em @R = 2em @!R  {
\lstick{Q_1} & \ctrl{1}  &    \qw      &    \gate{\text{Syndrome}}      & \qw  &  \gate{\text{Recovery}}  & \qw & \qw  & \qw  \\
\lstick{Q_2}  & \targ{X}      &  \qw   &    \gate{\text{Syndrome}}      & \qw  &  \gate{\text{Recovery}}   & \qw & \qw  & \qw \gategroup{1}{2}{2}{2}{.7em}{--}
}
\]

The probability of such a code failing can be calculated given the probability of each individual gate on the physical qubits failing, and finding the number of ways in which two or more errors could occur during the gate, syndrome or recovery steps (see~\cite{NielsenChuang} for an example of this). The result is that the total gate error is $c p^2$, where $c$ is a factor accounting for the number of possible ways two errors on an encoded qubit can occur.

Given that we will need to combine many gates to construct a quantum computer, the question is whether or not a particular gate on encoded qubits is scalable, so that a large number of these gates does not produce an uncorrectable error. In fact it is possible to show~\cite{Preskill1998,Kitaev1997,Aharonov1997,Knill1998,Knill2005} that scalable quantum computation is possible provided the physical gates have an error below a certain value; this is the \emph{threshold theorem}. The particular value of the threshold will depend on the encoding used, the scheme used to combine gates together, and the error model, and such a scheme is called a \emph{fault-tolerant} circuit. The currently known thresholds range from $p \lesssim 10^{-6}$ to up to around $0.01$~\cite{Preskill1998,Kitaev1997,Gottesman1997,Aharonov1997,Knill1998,Steane2003,Knill2005}.

\subsection{Decoherence-free subspace}
\label{sec:DFSintro}

Above we have seen a brief description of the `software' approach to correcting quantum errors. We will now discuss an example of the `hardware' approach. We will assume that we have some noisy physical qubits, and we will encode a single encoded bit of information over several of these physical qubits such that the errors have little or no effect on the information. Such an encoding is called a \emph{decoherence-free subspace} (DF subspace) or more generally a \emph{decoherence-free subsystem} (DF subsystem)~\cite{ChuangYamamoto1995,DuanGuo1997,Lidar1998,ZanardiRasetti1997}. 
Consider a system-environment evolution that has OSR operators $E_{\mu} = \bra{\mu} e^{-iH_I t/\hbar}\ket{\nu_0}$, where $H_I$ is the system-environment interaction. If there exists a subspace $\mathcal{S}$ such that for all of the OSR operators
\begin{align}
E_{\mu} \ket{j} = e^{i\phi_{\mu}}\ket{j} \; \; \forall \; \; \ket{j} \in \mathcal{S},
\end{align}
then any state encoded in this subspace at time $t=0$ will not be affected by the system-environment interaction, since the only effect is a global phase which has no noticeable effect. This is a \emph{decoherence-free subspace}, where the full Hilbert space is partitioned into logical states and non-logical states, and every logical state is represented by one orthonormal state. An example of a decoherence free subspace would be the bit flip channel described above which flips all bits together with probability $p$; if $N=2$ then one choice of decoherence-free subspace would be spanned by $\{\frac{1}{2}(\ket{01} + \ket{10}),\frac{1}{2}(\ket{00} + \ket{11} )\}$. 

Since we also want to do useful computations on this subspace, we also require that the Hamiltonian used to control the encoded qubit preserves the subspace, which will be true provided the eigenstates of this Hamiltonian are either completely inside or completely outside the decoherence free subspace~\cite{BaconThesis}.

This is not the most general way of encoding information; the most general way to encode information is in a \emph{subsystem}, in which there are also gauge degrees of freedom which can change without affecting the encoded information (so one logical state can be represented by many states of the physical qubits). 

The most trivial choice of subsystem encoding for 2 qubits is simply encoding the information in one physical qubit and having the second qubit in an arbitrary state. So $\ket{0_L} := \ket{0}\ket{\phi}$ and $\ket{1_L} :=\ket{1}\ket{\phi}$ are logical operators in the subsystem of qubit 1, and the gauge degree of freedom is the arbitrary choice of the state of qubit 2. Another example is for the `all-bit-flip' channel described above; if $N=2$, the logical $\ket{0}$ states is any combination of the states $\{ \ket{00}, \ket{11} \}$ which have even parity whilst the $\ket{1}$ states are any combination of the states $\{ \ket{10}, \ket{01} \}$ with odd parity. In this case the logical space is the parity of the qubits, whilst the gauge freedom is in the actual values of the physical qubits. For a slightly more abstract example, consider the Bell states on 2 qubits: $\ket{\Phi^{\pm}} = (1/\sqrt{2})(\ket{01} \pm \ket{10} )$, $\ket{\Psi^{\pm}} = (1/\sqrt{2})(\ket{00} \pm \ket{11} )$. We can rewrite these in a tensor product structure as $\ket{\chi^{\lambda}} = \ket{\chi} \ket{\lambda}$, where $\chi = \Phi,\Psi$, $\lambda = +,-$, and then we can think of storing information in the $\chi$ degree of freedom, so that $\lambda$ is the gauge degree of freedom. A subsystem $\mathcal{H}_L$, in which a quantum state $\ket{\phi}$ can be encoded in the state $\rho_{\phi}$ is a \emph{decoherence-free subsystem} if, for all of the OSR operators $E_{\mu}$
\begin{align}
\mbox{tr}_{G} \left[ E_{\mu} \rho_{\phi} (E_{\mu})^\dagger \right] = \ket{\phi} \bra{\phi},
\end{align}
where $\mbox{tr}_{G}$ means tracing out all of the gauge degrees of freedom. 

In Chapter~\ref{chap:DFS}, we will focus on a particular kind of system-environment interaction where each physical qubit interacts identically with the environment (e.g.\ by being close enough together with respect to the environment). Such a decoherence model is called \emph{collective decoherence}. We consider collective decoherence acting on $N$ spin-$\frac{1}{2}$ particles, such that the system-environment interaction Hamiltonian $H_I$ can be written as
\begin{align}
H_I = \sum_{\alpha = x,y,z} a_{\alpha} \mathbf{S}_{\alpha}\otimes B_{\alpha},
\end{align}
where $\mathbf{S}_{\alpha} := \sum_{n=1}^N \frac{1}{2}\alpha_{n}$, and ${\alpha}_{n} \in \{ X_n,Y_n,Z_n\}$ is a Pauli operator acting on the $n^{th}$ physical qubit, and $B_{\alpha}$ are operators acting on the environment. Strictly speaking, the decoherence-free subspace/subsystem should be defined using the operator-sum representation, but the algebra of the operator-sum representation for $H_I$ is generated by $\{ \mathbf{S}_{\alpha} \}_{\alpha = x,y,z}$ plus $\idop$~\cite{BaconThesis}, so these definitions cover the operator sum form as well. We have already seen an example of collective decoherence, in the `n-bit flip' channel discussed in Sec.~\ref{sec:QEC}. For the case where the channel probabilistically flips all of the qubits, this is equivalent to collective decoherence where $a_y = a_z = 0, a_x \neq 0$. This form of decoherence, where the collective spin operator only acts along one direction, is called \emph{weak} collective decoherence. The more general case, where $a_x,a_y,a_z \neq 0$ is called \emph{strong} collective decoherence~\cite{BaconThesis}. We will refer to a DF subspace (subsystem) that protects against weak or strong collective decoherence as a weak or strong DF subspace (subsystem) respectively. Clearly a strong DF subspace or subsystem can also be used as a weak DF subspace or subsystem.

To explore how to construct weak and strong DF subspaces and subsystems we first briefly review some relevant spin physics, restricted to spin-$\frac{1}{2}$ particles for this thesis, but the arguments can be extended to other systems. A system with spin is characterised by two quantum numbers: the total spin $s$ and the projection of this spin along one of the $x,y$ or $z$ axes, denoted $m_{x},m_y,m_z$ respectively. Since the spin operators $X$, $Y$ and $Z$ don't commute, only one projection can be specified at a time, and we take this to be $m_z$ without loss of generality. $m_z \hbar$ is the eigenvalue of the operator $\frac{1}{2} Z$ and can take any of the $(2s+1)$ values in the range $-s,-s+1,...,s-1,s$, whilst $s(s+1)\hbar^2$ is the eigenvalue of the operator $\frac{1}{4}(X^2+Y^2+Z^2)$. When combining a number of particles with total spin $s_1,s_2,s_3,...,s_N$ together the resultant total spin $S$ can take any of the values of the combinations of the spins, $S = | s_1 \pm s_2... \pm s_N|$. Just as for single spins, for a collection of spins there are spin operators $\mathbf{S}_{x},\mathbf{S}_y,\mathbf{S}_z$ (defined above) and $\mathbf{S}^2 = \mathbf{S}_x^2 + \mathbf{S}_y^2 + \mathbf{S}_z^2$, so that states can be labelled as $\ket{S,m_S}$ such that $\mathbf{S}_z \ket{S,m_S} = \hbar m_S$ and $\mathbf{S}^2 \ket{S,m_S} = \hbar^2 S(S+1) \ket{S,m_S}$.

\begin{figure}[h]
\begin{center}
	\includegraphics[width=0.45\textwidth]{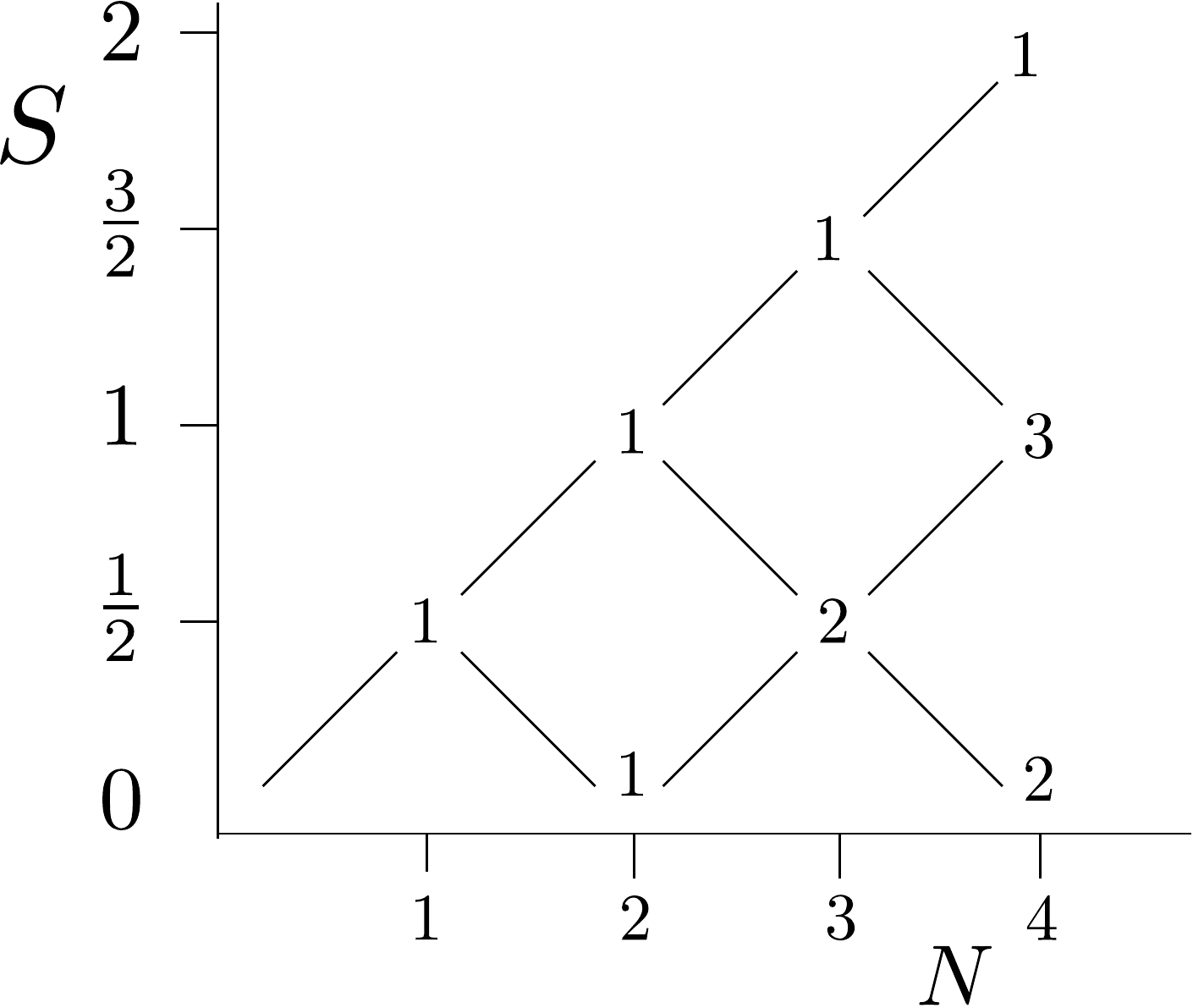}
	\caption{\label{fig:SpinAddition} An illustration of the degeneracies in total spin value $S$ when combining $N$ spins. }
\end{center}
\end{figure}

There can also be more than one way to reach the same value of $S$, and so there can be degeneracy in the system with respect to the $S$ value. A useful way of depicting the degeneracies from combining spin in this way was given in~\cite{BaconThesis}, whereby addition of spin is represented as addition of vectors (see Fig.~\ref{fig:SpinAddition}), and the degeneracies of each spin number $S$ is given by Pascal's triangle. Note that for each value of $S$, there are $(2S+1)$ states with different values of the quantum number $m_S$.

To construct a weak DF subspace, where there is only one operator such as $\mathbf{S}_z$ in the system-environment interaction, all that is required are states with the same value of $m_S$. So for example we can construct a 2-qubit weak DF subspace using the $\ket{S=1,m_S=0} = \frac{1}{\sqrt{2}}(\ket{10} + \ket{01})$ and $\ket{S=0,m_S=0} = \frac{1}{\sqrt{2}}(\ket{10} - \ket{01})$ states, and for three qubits we could use the following states used in~\cite{DiVincenzo2000}:
\begin{align}
&\ket{S = 1/2, m_S = 1/2 , S_{1,2} = 0  } = \ket{\psi^-}_{12}\ket{0}_3 \nonumber\\
&\ket{S =1/2,m_S =1/2,  S_{1,2} = 1} = \frac{1}{\sqrt{3}}( \sqrt{2} \ket{T_+}_{12}\ket{1}_3 - \ket{T_0}_{12}\ket{0}_3)
\end{align}
Here we have used the singlet states on qubits $a$ and $b$, defined as $\ket{\psi^-}_{ab} := (\ket{01}_{ab} - \ket{10}_{ab})/\sqrt{2}$ and the triplet states $\ket{T_+}_{ab} = \ket{00}_{ab}$, $\ket{T_0}_{ab} = (\ket{01}_{ab} + \ket{10}_{ab})/\sqrt{2}$. Note that, since the states are degenerate with respect to the $S$ and $m_S$ value, we distinguish between them using a third quantum number $S_{1,2}$ defined as the total spin quantum number of spins 1 and 2 only.

Finding strong DF subspaces and subsystems is more difficult. With all three of $\mathbf{S}_x$, $\mathbf{S}_y$ and $\mathbf{S}_z$ present in the system-environment interaction, since $[ \mathbf{S}_\alpha, \mathbf{S}_\beta] = i \epsilon_{\alpha \beta \gamma} \mathbf{S}_\gamma$ and $[ \mathbf{S}^2, \mathbf{S}_\alpha] = 0$ for $\alpha,\beta, \gamma \in \{ x,y,z\}$, the noise acts to mix states with the same values of $S$ but different values of $m_S$. Thus the 2- and 3-qubit encodings shown above are not suitable, as leakage into other states needs to be taken into account. Note that also degenerate states with the same value of $S$ are not mixed, since the total spin number $S_{q_1,q_2,...,q_n}$ on some subset of size $n <N$ of the qubits will differ between different degenerate states, and following the same commutation arguments the eigenvalue of $\mathbf{S}_{q_1,q_2,...,q_n}$ will not be changed by collective decoherence.

To construct a strong DF subspace requires two states with the same value of $m_S$, with no other states that can be evolved to under the collective Pauli operators. This will only be the case when $S=0,m_S=0$, and one can see from Fig.~\ref{fig:SpinAddition} that $N=4$ is the lowest $N$ for which there are two $S=0,m_S=0$ states, so this is the smallest number of spins over which a qubit can be encoded in a strong DF subspace, with logical states of the following form:
\begin{align}\label{eqn:4States}
\ket{\bar{0}^{(4)}} & :=  \ket{ S = 0, m_S=0, S_{1,2} = S_{3,4} = 0} = \ket{\psi^-}_{12} \ket{\psi^-}_{34}\nonumber\\
\ket{\bar{1}^{(4)}} & :=  \ket{ S = 0, m_S=0, S_{1,2} = S_{3,4} = 1} = \frac{1}{\sqrt{3}} \left[ \ket{T_+}_{12} \ket{T_-}_{34} - \ket{T_0}_{12} \ket{T_0}_{34}+ \ket{T_-}_{12} \ket{T_+}_{34}\right].
\end{align}
This 4-qubit encoding has one additional desirable property; it also functions as a \emph{supercoherent qubit}~\cite{Bacon2001}. Supercoherence would allow resistance to errors acting on individual physical qubits, with a mechanism as follows: When an error along any direction is applied to the physical qubits in the $S=0$ states in eqn.\ (\ref{eqn:4States}), it is accompanied by a change in the $S$ value by 1~\cite{Bacon2001}. In order to use this to create a supercoherent qubit, we could switch on the Hamiltonian ${H}_{SC}$, defined as
\begin{equation}\label{eqn:Hsc}
{H}_{SC} = J_{SC}\sum_{m,n} {E}_{mn},
\end{equation}
where ${E}_{mn} :=  X_m X_n +Y_m Y_n + Z_m Z_n$,  and the sum is over all pairs of the 4 physical qubits. With this Hamiltonian switched on, the $S=0$ states are degenerate and lowest in energy, with an energy gap between the $S=0$ states and any other states. Thus any decoherence process acting on individual physical qubits in the $S = 0$ state involves an increase in energy of the encoded qubit, and will lead to a transfer of energy from the environment to the system, which we can inhibit by cooling the environment. Thus supercoherent qubits would be very useful as quantum memories, and it was argued in~\cite{Bacon2001} that computation with supercoherent qubits could be performed provided the interaction strength between qubits was small enough compared to $J_{SC}$ (leading to a trade-off between the speed of operations and the robustness against errors). In this chapter, we will not aim to make our interactions supercoherent as well (i.e.\ we envisage a protocol in which we use the supercoherent mechanism as a means to reliably store information, but turn off the supercoherent Hamiltonian ${H}_{SC}$ when we interact encoded qubits). 

\begin{figure}[h]
\begin{center}
	\includegraphics[width=0.8\textwidth]{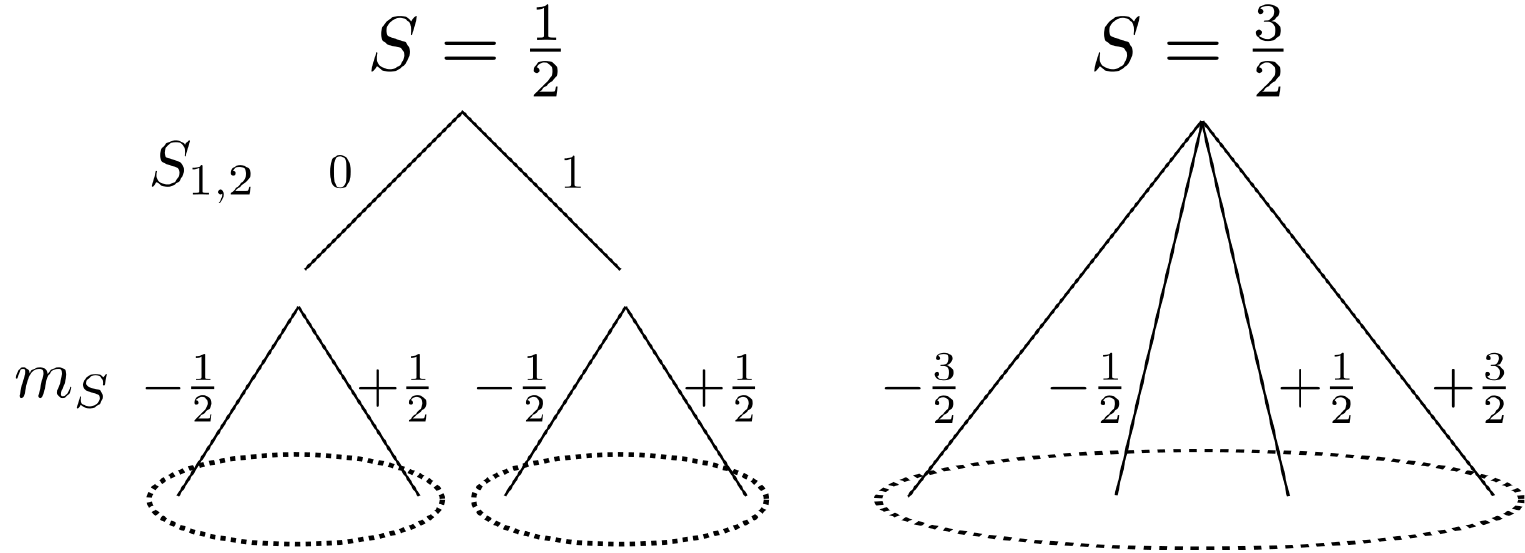}
	\caption{\label{fig:DFsubsystem} An illustration of the action of strong collective decoherence on the degenerate $S = \frac{1}{2}$ states and the non-degenerate $S = \frac{3}{2}$ states, for 3 qubits. The split lines indicate the different quantum numbers characterising different states, and the dashed ellipsoids indicate which states are mixed by the action of strong collective decoherence.}
\end{center}
\end{figure}

In~\cite{BaconThesis} it is shown that strong DF subsystems can be constructed whenever there is a degeneracy in the $S$ value. From Fig.~\ref{fig:SpinAddition} we can see that the smallest number of qubits with this property is $N=3$, for which there are two $S = \frac{1}{2}$ states. Thus we can encode a qubit into the strong DF subsystem on the following states:
\begin{align}\label{eqn:3states}
&\ket{\bar{0}^{(3)}_{+1} }=\ket{S = 1/2, m_S = 1/2, S_{1,2} = 0 } = \ket{\psi^-}_{12}\ket{0}_3\nonumber\\
&\ket{\bar{0}^{(3)}_{-1}}=\ket{S =1/2,m_S =-1/2, S_{1,2}=0} = \ket{\psi^-}_{12}\ket{1}_3\nonumber\\
&\ket{\bar{1}^{(3)}_{+1}}=\ket{S =1/2,m_S =1/2, S_{1,2}=1} = \frac{1}{\sqrt{3}}( \sqrt{2} \ket{T_+}_{12}\ket{1}_3 - \ket{T_0}_{12}\ket{0}_3)\nonumber\\
&\ket{\bar{1}^{(3)}_{-1}}=\ket{S =1/2,m_S =-1/2, S_{1,2}=1} = \frac{1}{\sqrt{3}}( \ket{T_0}_{12}\ket{1}_3 - \sqrt{2} \ket{T_{-}}_{12}\ket{0}_3).
\end{align}
where $\ket{T_-}_{ab} = \ket{11}_{ab} $. The logical zero state $(\ket{\bar{0}^{(3)}})$ in this 3-qubit subsystem is defined to be an arbitrary combination of the first two states, i.e.\ $\ket{\bar{0}^{(3)}} := \zeta |\bar{0}^{(3)}_{+1} \rangle   + \gamma | \bar{0}^{(3)}_{-1} \rangle $, whilst the logical one state $(\ket{\bar{1}^{(3)}})$ is a superposition of the last two states with the same coefficients, $\ket{\bar{1}^{(3)}} := \zeta |\bar{1}^{(3)}_{+1} \rangle + \gamma |\bar{1}^{(3)}_{-1}\rangle$. The action of collective decoherence on this encoding will act identically on the two states, and so will change the values of $\zeta$ and $\gamma$ identically for both logical states, but will not mix the different logical states since they have different values of $S_{1,2}$. The arbitrary choice of $\zeta$ and $\gamma$ is the gauge degree of freedom, and any transformation which only changes the values of $\zeta$ and $\gamma$ is called a gauge transformation (in this case a gauge transformation is an operation which couples to the $m_S$ degree of freedom). We refer to the separate subspaces with different $m_S$ values as {\it gauge subspaces}. An illustration of this subsystem encoding is shown in Fig.~\ref{fig:DFsubsystem}, comparing the the action of strong collective noise on all of the possible states with 3 physical qubits.

It has been shown~\cite{Lidar1998,ZanardiRasetti1997,Bacon2000} that universal quantum computation can be performed inside a DF subspace or subsystem. To do this, we must be able to perform certain single qubit rotations as well as gates between two encoded qubits (such as a controlled-Z gate)~\cite{Barenco95}. For collective decoherence, explicit gate sequences for two-qubit gates have been found for the 3-qubit DF subsystem and 4-qubit DF subspace~\cite{Bacon2000,DiVincenzo2000,BaconThesis,Hsieh2003,Fong2011}. Searching for these gates will be the subject of Chapter~\ref{chap:DFS}. For a more in depth discussion of DF subspaces and subsystems, the reader is referred to~\cite{Lidar1998,Lidar2003,BaconThesis,DuanGuo1998,PalmaSuominen1996,Kempe2001}.

\section{Distance measures}
\label{sec:DistMeas}
Later on in this thesis we will be attempting to create two-qubit and three-qubit gates, and so it will be useful to have some measures to evaluate how close a quantum state is to another quantum state, and how close a given unitary is to an ideal gate. Performing these two different tasks can be done using some of the same machinery we have introduced above for error correction.

One of the standard measures of distance between two quantum states $\rho_1$ and $\rho_2$ is the fidelity $F(\rho_1,\rho_2) $, which measures how close these quantum states are:
\begin{align}
F(\rho_1,\rho_2) =  \left( \text{tr} \sqrt{ \sqrt{\rho_1} \rho_2 \sqrt{\rho_1} } \right)^2,
\end{align}
which is 1 iff $\rho_1 = \rho_2$. For pure states $F(\ket{\phi_1}\bra{\phi_1},\ket{\phi_2}\bra{\phi_2}) = |\sprod{\phi_1}{\phi_2}|^2$. Note that we have squared the trace, which is different from the form seen in e.g.~\cite{NielsenChuang}. Both forms with and without the square are used in the literature, but when we later on discuss how to construct useful metrics for gate distance, this squaring will allow the fidelity to have the interpretation of a probability~\cite{Gilchrist2005}, which is more intuitive than the square root of probability. In cases where the literature uses the non-squared version, we will also use it.

The second important measure is the trace distance, which measures how far apart two quantum states $\rho_1$ and $\rho_2$ are
\begin{align}
D(\rho_1,\rho_2) =  \frac{1}{2} \| \rho_1 - \rho_2 \|_{tr},
\end{align}
where $\| A \|_{tr} := \text{tr}(\sqrt{A A^\dagger})$ is the trace norm.

To use these metrics to measure the distance between two gates, we can use the Jamiolkowski isomorphism~\cite{JamiolKowski1972}. If $\mathcal{E}$ is a quantum operation acting on a $d$-dimensional system, and $\ket{\Phi} = \frac{1}{\sqrt{d}} \sum_{j=1}^{d }\ket{j}\ket{j}$ is an entangled state of two $d$-dimensional systems ($d=2$ for qubits), then we can construct a state $\rho_{E}$ where $\rho_{\mathcal{E}} =[ \idop \otimes \mathcal{E} ](\ket{\Phi}\bra{\Phi})$, i.e.\ where the first system is left untouched but the second system transforms by $\mathcal{E}$. If we express $\mathcal{E}$ in terms of a basis $A_n = \ket{j_n}\bra{j'_n}$ such that $(\ket{j_n},\ket{j_n'})$ is a unique pairing of the possible orthogonal computational basis states of $N$ qubits, then $\mathcal{E}(\rho) = \sum_{mn} \chi_{mn} A_m \rho A_n^\dagger$ (see Sec.~\ref{sec:OSR}), and $\rho_{\mathcal{E}}$ becomes
\begin{align}
\rho_{\mathcal{E}} &= \frac{1}{d} \sum_{jkmn}\chi_{mn} (\idop \otimes A_m) \ket{jj}\bra{kk} (\idop \otimes A_n^\dagger) \nonumber\\
&=\frac{1}{d} \sum_{jkmn}\chi_{mn} \delta_{j j'_m} \ket{jj_m}\bra{kk_n} \delta_{k k_n'}= \frac{1}{d} \sum_{mn}\chi_{mn}  \ket{j'_m j_m}\bra{k'_n k_n}.
\end{align}

Since $(j_n ,j_n')$ and $(k_n,k_n')$ are unique pairs, by construction, this means that each element of $\chi$ is mapped to a unique element in $\rho_{\mathcal{E}}$. If in addition $A_n = \ket{x_1 x_2 ... x_N}\bra{x_{N+1}x_{N+2}...x_{2N} }$ such that the binary string $x_1 x_2 ... x_Nx_{N+1}x_{N+2}...x_{2N}$ is $n$ in binary notation, then $(\rho_{\mathcal{E}})_{mn} = \frac{1}{d}\chi_{mn}$.

Using this isomorphism, we can then compare the process matrix $\chi_{id}$ of the ideal process with the actual process $\chi$ using the fidelity and trace distance~\cite{Gilchrist2005}. Thus we define the process fidelity $F_{pro}$ between two process matrices $\chi_1$, $\chi_2$ as
\begin{align}
F_{pro} = \frac{1}{d^2} \left( \text{tr} \sqrt{ \sqrt{\chi_1} \chi_2 \sqrt{\chi_1} } \right)^2,
\end{align}
whilst the process distance is defined as
\begin{align}
D_{pro} = \frac{1}{d}\| \chi_1 - \chi_2 \|_{tr}.
\end{align}
$D_{pro}$ and $1- F_{pro}$ are particularly useful as they act as upper bounds on $\bar{p}_e$, the average probability that the evolution does not produce the desired output~\cite{Gilchrist2005}.

Often it is only important to know when two processes are equivalent up to some local single-qubit operations (since these are typically easy to perform). Such gates which are equivalent up to local single-qubit operations are called \emph{locally equivalent}. $D_{pro}$ and $F_{pro}$ are not appropriate for detecting gates that are locally equivalent, so in certain situations the number of search parameters can be significantly reduced if we can use a distance measure which ignores local operations. One method to do this due to Makhlin~\cite{Makhlin2002}. For a two qubit operation $M$, we first transform ${M}$ into the Bell basis, ${M}\rightarrow {M}_B = {Q}^{\dagger}{M} {Q}$ where
\begin{align}
{Q} = \frac{1}{\sqrt{2}} \left( \begin{array}{c c c c} 1 &0& 0& i \\ 0 &i &1 &0 \\ 0 &i &-1& 0\\ 1& 0& 0& -i \end{array} \right).
\end{align}
Makhlin showed that for two gates $L$ and $M$ expressed in the Bell basis as $L_B$ and $M_B$, the gates are equivalent up to local operations iff the spectra of $L_B^T L_B$ and $M_B^T M_B$ are the same. This follows since local operations correspond to orthogonal operations in the Bell basis, and so for any matrix $M_B$ in the Bell basis the spectrum of $M_B^T M_B$ after a local transformation $M_B \to O_1 M_B O_2$ is conserved. The spectrum can also be characterised by two quantities $m_1$ and $m_2$ defined as:
\begin{align}
m_1({M}) &= [\textrm{tr}\; (M_B^{T}M_B)]^2/16\det {M}^{\dagger},\nonumber \\
m_2({M}) &=[(\textrm{tr}\; (M_B^{T}M_B))^2 -\textrm{tr}((M_B^{T}M_B)^2) ]/4\det {M}^{\dagger}.
\end{align}
so that two gates $M$ and $L$ are locally equivalent iff $\{m_1(L),m_2(L)\}=\{m_1(M),m_2(M)\}$. To use these invariants, it is useful to combine them to get a single number which is 0 for identical gates and larger for very different gates. To do this we define
\begin{equation}
f_m(L,M) = \sum_{i=1}^2 \left|m_i(L) - m_i(M) \right|.
\end{equation}
This distance measure compares the spectra of the two different gates $M$ and $L$ up to some local operations, so can be thought of in a similar manner to the trace distance. The same measure of distance between gates has been used in e.g.~\cite{DiVincenzo2000}.

\section{Measurement-based quantum computation}\label{sec:MBQC}

So far we have discussed quantum computation in the `standard' circuit model picture, where operations are performed sequentially on some systems containing the information, and the result is read out at the end. This is analogous to the standard way in which classical computation is performed, and is perhaps the most intuitive way to envision a quantum computer. Unsurprisingly, quantum computation allows computations to be performed in ways that have no analogy in classical computation. For instance, instead of measurements only occurring after the logical operations have been performed, we can use the measurements to perform the computation. This is the paradigm of \emph{measurement-based quantum computation} (MBQC)~\cite{Raussendorf2001}, in which measurements on a highly entangled state are used to perform computations, rather than unitary gates.

As an illuminating example, consider the simplest MBQC setup: a system of two qubits, A and B. A contains some unknown quantum information $\ket{\chi} = \alpha \ket{0} + \beta \ket{1}$, whilst qubit B is prepared in the state $\ket{+}$. To ensure the measurement on A has an effect on B, the qubits are first entangled using a \textsc{cz} gate, resulting in the state $\ket{\psi_{AB}}= \alpha \ket{0}\ket{+} + \beta \ket{1} \ket{-}$. This state $\ket{\psi_{AB}}$ can be rewritten as
\begin{eqnarray}
\label{eqn:2qubitEx}
\ket{\psi_{AB}}= &\frac{1}{\sqrt{2}} \ket{+_{\phi}}( \alpha\ket{+} + e^{-i\phi} \beta  \ket{-}) + \frac{1}{\sqrt{2}} \ket{-_{\phi}}( \alpha\ket{+} -e^{-i\phi} \beta  \ket{-}) \nonumber\\
&= \frac{e^{-i\phi/2} }{\sqrt{2}} \ket{+_{\phi}} \text{H} U_z(\phi) \ket{\chi} +\frac{e^{-i\phi/2} }{\sqrt{2}} \ket{-_{\phi}} X \text{H} U_z(\phi) \ket{\chi},
\end{eqnarray}
where $ \ket{\pm_{\phi}} := \frac{1}{\sqrt{2}} ( \ket{0} + e^{i\phi} \ket{1})$ and $\text{H}$ is a Hadamard gate. Now consider performing a measurement on qubit A, such that the measurement has eigenstates $\ket{\pm_{\phi}}=1/ \sqrt{2} ( \ket{0} \pm e^{i\phi} \ket{1})$. If the outcome is $\ket{+_{\phi}}$, the state of qubit B is $e^{-i\phi/2}\text{H} U_z(\phi) \ket{\chi}$, and if the outcome is $\ket{-_{\phi}}$ the state of qubit B is $e^{-i\phi/2} X \text{H} U_z(\phi) \ket{\chi}$. This can be written in a more compact form by assigning a variable $m$ to the outcome of the measurement on $A$, such that $m=0$ if the outcome is $\ket{+_{\phi}}$ and $m=1$ if the outcome is $\ket{-_{\phi}}$. Then the overall result is that the information is moved to site $B$ and transformed by $e^{-i\phi/2} X^{m}\text{H} U_z(\phi) $. By applying a Pauli $X^m$ correction on qubit $B$, both outcomes are the same, and so a deterministic $\text{H} U_z(\phi)$ gate can be obtained (up to a global phase) despite the randomness of the measurement. Note that by preparing A in state $\ket{\chi}$ instead of $\ket{+}$, information can be encoded in the system, so we call A the input, and since B is the system where the information will be at the end of the computation, we call this the output. 

Thus we have seen how to perform the operation $\text{H} U_z(\phi)$ on the encoded information in this 1D chain whilst teleporting the information from one qubit to the other. This simple example shows the main features of the MBQC protocol:
\begin{itemize}
\item[1.] Non-input qubits are prepared in $\ket{+}$ states.
\item[2.] Entangling \textsc{cz} gates are performed between certain pairs of qubits.
\item[3.] Non-output qubits are measured. The measurement basis determines which operations are performed.
\item[4.] To make the outcome deterministic, corrections that depend on previous measurement outcomes are needed.
\end{itemize}

\begin{figure}
\begin{center}
\subfloat[]{\includegraphics[width=0.2\textwidth]{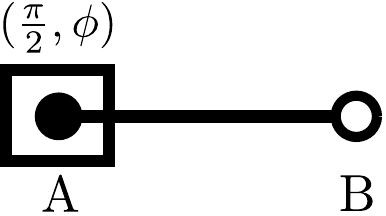}}
\qquad
\subfloat[]{\includegraphics[width=0.5\textwidth]{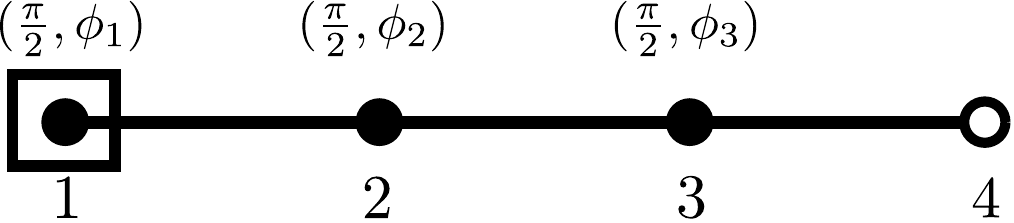}}
\caption{\label{fig:SingleRotation} a) A graphical depiction of the state in eqn.\ (\ref{eqn:2qubitEx}). b) A graphical depiction of a state that can be used to perform single qubit rotations.}
\end{center}
\end{figure}

To make the following discussion easier to understand, we introduce a graphical representation of MBQC which is commonly used in the literature. We represent qubits as vertices $V$ in a graph $G$. The input qubits $I$ are represented as vertices contained within squares, whilst output qubits $O$ are represented as hollow circles. Non-inputs are prepared in the $\ket{+}$ state, and the edges of this graph $E$ represent which pairs of qubits have been acted on by the \textsc{cz} gate (we use the notation $v \sim w$ to indicate that $v$ and $w$ are connected by an edge). The state resulting from these operations is called a {\it graph state}. We label each non-output vertex with a pair of angles $(\theta,\phi)$ corresponding to the type of measurement performed on that qubit (where applicable); the angles give the measurement axis in the Bloch sphere (so e.g.\ $(\pi/2,0)$ denotes measurement in the $X$ basis, $(\pi/2,\pi/2)$ in the $Y$-basis, etc.). An illustration of the simple 2-qubit example expressed as a graph is shown in Fig.~\ref{fig:SingleRotation}. After the measurement, the entanglement between qubits A and B is `used up', so that there is no longer an edge between them.

The 2-qubit example above can be extended to perform arbitrary single qubit rotations. Starting with a linear chain of four qubits, entangled to nearest-neighbours only, the first three qubits are measured, resulting in the transformation
\begin{align}
&X^{m_3} \text{H} U_z(\phi_3)X^{m_2}\text{H} U_z(\phi_2)X^{m_1}\text{H} U_z(\phi_1) = X^{m_3}H U_z(\phi_3)X^{m_2}U_x(\phi_2)Z^{m_1}U_z(\phi_1) \nonumber\\
&=X^{m_1+m_3}Z^{m_2}H U_z((-1)^{m_2}\phi_3)U_x((-1)^{m_1}\phi_2)U_z(\phi_1),
\end{align}
where we have used the identities $\text{H}\text{H} = \idop$, $\text{H}X = Z\text{H}$, $U_z(\phi)X = XU_z(-\phi)$, $U_x(\phi)Z = ZU_x(-\phi)$. In this last line, we recognise the form of a general rotation in the Bloch sphere, with some added terms to account for the randomness of the measurement.

 The factors such as $(-1)^{m_1}$ mean that the measurements must be performed \emph{adaptively}, and also it imposes an order in which the measurements can be done ($1 \to 2 \to 3$). This is a key property of MBQC, which prohibits simply measuring all the qubits at the same time, and we call the particular order of measurements on a graph state a \emph{measurement pattern}. The terms such as $X^{m_1}$ amount to corrections on the output state, and can be treated as extra classical processing that must be done in order to get a deterministic output.

\begin{figure}[h]
\begin{center}
	\subfloat[]{\includegraphics[width=0.25\textwidth]{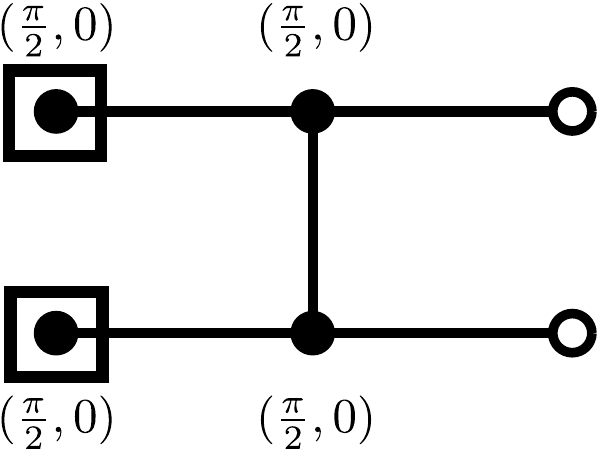}} \hspace{5mm}
	\subfloat[]{\includegraphics[width=0.5\textwidth]{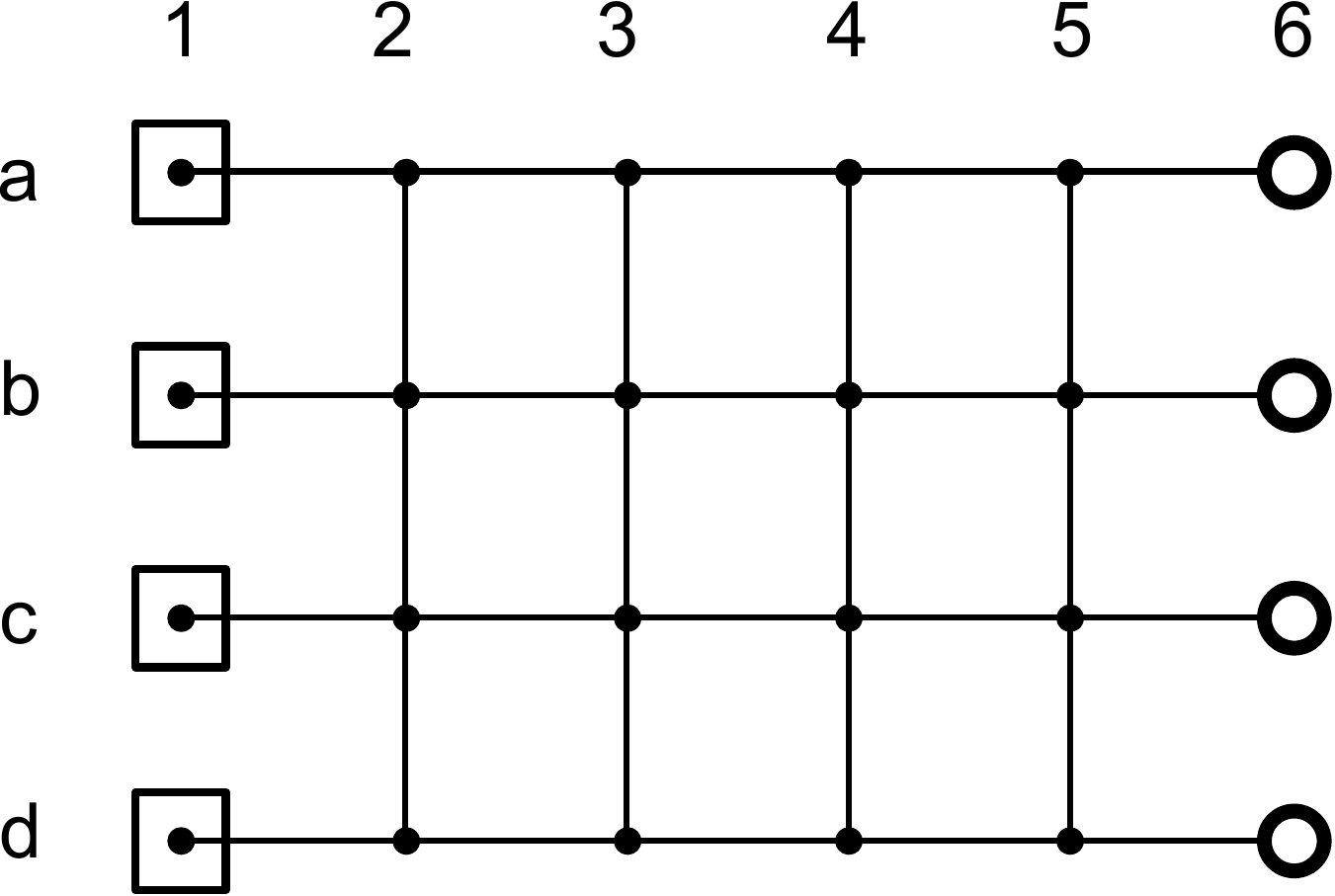}}
	\caption{\label{fig:CNOT1} a) A graph state to achieve a \textsc{cnot} gate. b) A cluster state.}
\end{center}
\end{figure}

A $\textsc{cnot}$ gate can be performed starting with qubits arranged as shown in Fig.~\ref{fig:CNOT1}. After measuring the non-output qubits from left to right, a $\textsc{cnot}$ gate is performed, plus some Hadamard operations (see~\cite{Browne2011} for details). We therefore have building blocks with which universal quantum computation is possible, and so by piecing these blocks together a graph-state can be built that is a universal resource for quantum computation. Clearly any graph state which can be made by entangling this universal resource with extra qubits is also universal; for instance starting with a graph which is a 2-dimensional grid of qubits, it is possible to strip away some of the qubits (by measuring them and performing corrections after these measurements) to realise a universal resource. This 2-dimensional rectangular graph-state is known as a \emph{cluster state}, and is the most commonly used universal resource for MBQC, due to its simplicity. There are many other regular graphs that are known to be universal, such as hexagonal and triangular graphs~\cite{VanDenNest2006}, however a complete classification of all graph states which are universal does not exist, and is part of the motivation for this work. We will see in the next section that there are tools that allow us to efficiently determine whether certain graphs can used for universal deterministic MBQC or not. 

\subsection{\emph{Flow} and \emph{gflow}}\label{sec:f&g}
\label{sec:FlowGflow}

In the previous section we have demonstrated that measurements can be performed adaptively on certain simple graph states, such that the outcome of the computation will always be the same regardless of the measurement outcomes (i.e.\ the outcome is deterministic). There are many more possible graphs, however, and it is not easy to  tell whether a given graph will allow such a procedure or not. Fortunately, there are simple graphical tools called \emph{flow}, or more generally, \emph{generalised-flow} (\emph{gflow})~\cite{Danos2006,Browne2007}, which allow an efficient characterisation of whether or not there is an adaptive measurement pattern on a graph state that produces a deterministic outcome (although they do not reveal whether or not the graph allows universal quantum computation). \emph{Flow} and \emph{gflow} provide sufficient (although not necessary) conditions for the existence of such a measurement pattern, and have proved incredibly useful in studying parallelism~\cite{Broadbent2009,Browne2011}, the translation between MBQC and the circuit model~\cite{Broadbent2009,Silva13} and causal order in MBQC~\cite{Raussendorf2011}.

For the purposes of this thesis, we will only use \emph{gflow} since it incorporates \emph{flow}. \emph{Gflow} has two components: a time ordering over the vertices (using the notation $v > w$ to represent that vertex $v$ is measured after vertex $w$, and $v=w$ to indicate that $v$ and $w$ can be measured at the same time), and a \emph{gflow function} $g$ for each vertex. $g(v)$ is a list of the vertices that are affected by the measurement outcome of vertex $v$ (and is therefore a list of the qubits that can be used to correct for this measurement outcome). We say that a set of vertices $U$ is oddly (evenly) connected to a vertex $v$ if there is an odd (even) number of edges connecting $U$ and $v$. The definition of \emph{gflow} is then~\cite{Browne2007}
\begin{defin}[Gflow]
Given an open graph state $G$ with inputs $I$, outputs $O$, edges $E$ and vertices $V$, we say it has \emph{gflow} if there exists a \emph{gflow function} $g$ and a time ordering $<$ over $V$ such that, for all $v \in V$ which are not outputs:
\begin{itemize}
\item[(G1)] All qubits $w$ in $g(v)$ are in the future of $v$, i.e. $v < w$ for all $w \in g(v)$.
\item[(G2)] if $w \leq v$, and $v \ne w$, then $w$ is evenly connected to all qubits in $g(v)$.
\item[(G3)]
\begin{itemize}
            \item For measurements in the $(X,Y)$ plane: $v\notin g(v)$, and $g(v)$ is oddly connected to $v$.
            \item For measurements in the $(X,Z)$ plane: $v\in g(v)$, and $g(v)$ is oddly connected to $v$.
            \item For measurements in the $(Y,Z)$ plane: $v\in g(v)$, and $g(v)$ is evenly connected to $v$.
\end{itemize}
\end{itemize}
\end{defin}
In this thesis, we will only use measurements in the $(X,Y)$ plane, as results for measurements in other planes should follow in a similar fashion. Note that finding out whether or not a graph has \emph{gflow} can be done in polynomial time~\cite{Mhalla}. 

\begin{figure}[h]
\begin{center}
	\includegraphics[width=0.45\textwidth]{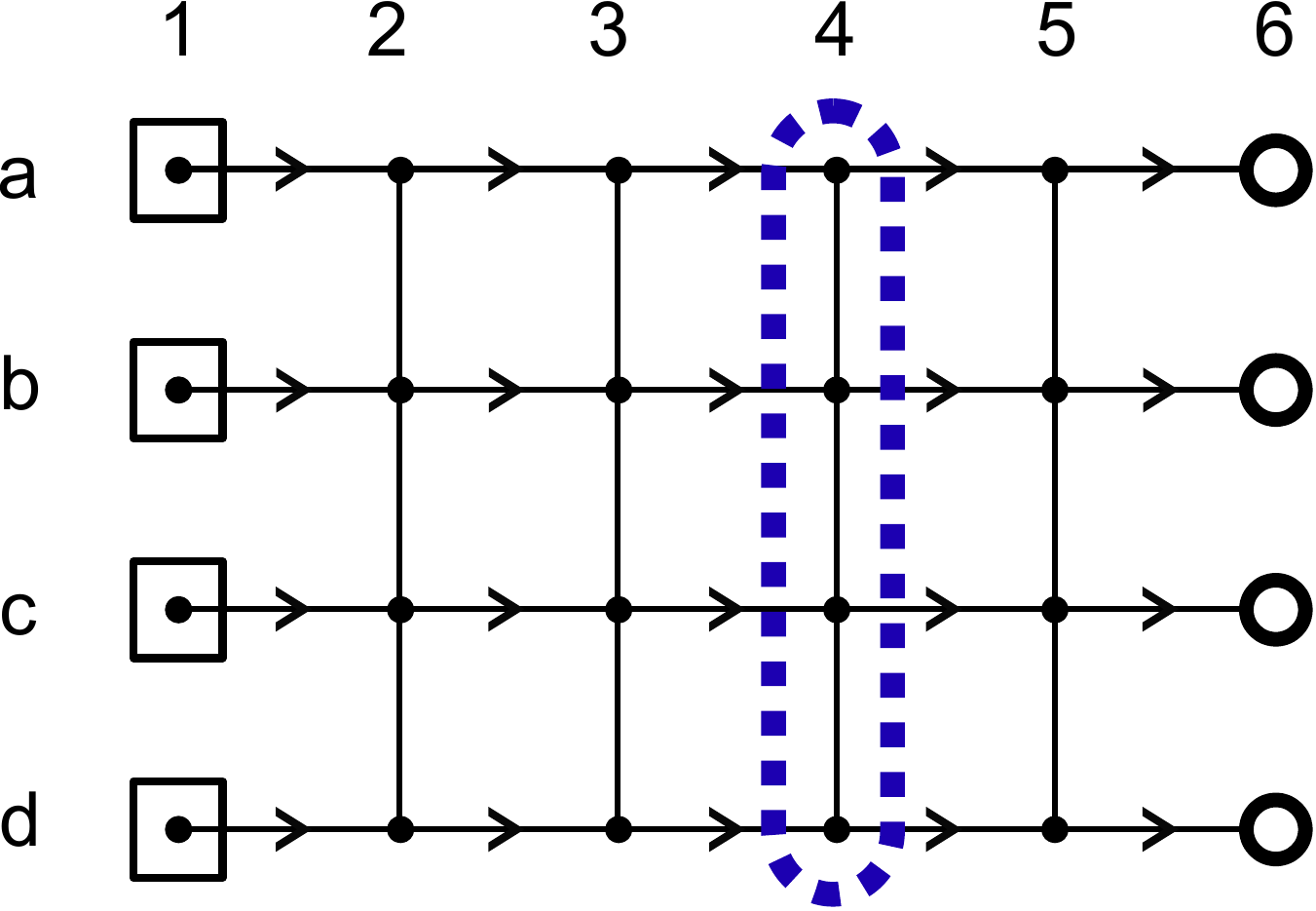}
	\caption{\label{fig:Example} An illustration of the definitions in Sec.~\ref{sec:f&g}, applied to a cluster state with depth 5. Inputs are represented by vertices with squares, and outputs are vertices with hollow circles. Arrows indicate \emph{gflow} lines and the dotted circle indicates a single layer of qubits.}
\end{center}
\end{figure}

To represent \emph{gflow}, arrows can be superimposed on a graphical representation of a graph state, such that an arrow connects $v$ to $w$ if $w \in g(v)$. We call these arrows \emph{gflow lines} (see for example Fig.~\ref{fig:gflow}). 
We have seen above that \emph{gflow} requires a time ordering in the measurement pattern, which leads to the definition of {\em layers}, which are groups of vertices which can be measured at the same time:
\begin{defin}[Layers]

 A layer of a computation is defined as any non-output qubits in a measurement pattern which can be measured at the same time.
\end{defin}
We denote the layers as $L_k$, and we use $L(v)$ to denote the layer that vertex $v$ is in. For example, for the \emph{gflow} defined on the graph in Fig.~\ref{fig:gflow}, the layers are $L_1 = \{ 1\}$, $L_2 = \{2\}$, $L_3 = \{3\}$. In Fig.~\ref{fig:Example}, the layers are given by $L_k = \{ a_k,b_k,c_k,d_k,e_k\}$, for $k < 6$, and $L(a_k) = L(b_k) = L_k$ etc.

Using the definition of layers, we can also define the depth for MBQC;
\begin{defin}[Depth]

The depth of an MBQC with \emph{gflow} is the number of rounds of measurements in the measurement pattern, or equivalently the number of layers in a measurement pattern.
\end{defin}

 In general this depth will be different depending on which \emph{gflow} we are using (there can be more than one - indeed we will see below an example where many \emph{gflow}s can be realised). Since we can think of \emph{gflow} as a directed graph superimposed on an undirected graph, an equivalent and perhaps more intuitive definition is that the depth is the longest possible path along the \emph{gflow} lines. For example, in Fig.~\ref{fig:Example} the depth is 5.
 
\subsection{Graph states in the stabiliser formalism}
\label{sec:MBQCStab}
Graph states can also be described using \emph{stabilisers}~\cite{Gottesman1997}. Stabilisers form a subgroup of the Pauli group $\mathcal{G}$ (excluding $- \idop$) which act as identity on a particular state or subspace. We say a set of stabilisers $\mathcal{K}_{\mathcal{S}}$ stabilise the subspace $\mathcal{S}$ if
\begin{eqnarray} \label{eqn:GSStabEQN}
K \ket{\phi} = \ket{\phi} \text{	} \forall K \in \mathcal{K}_{\mathcal{S}}, \text{  	} \forall \ket{\phi} \in \mathcal{S}.
\end{eqnarray}
From this definition we can see that stabilisers must also commute with each other, so they form an Abelian subgroup of the Pauli group. The smallest set of stabilisers that generate the group of stabilisers for a particular subspace are called the \emph{generators} of that group. Each stabiliser generator constrains the space by a half; with $n-k$ generators for the stabiliser the dimension of the codespace is therefore $2^n / 2^{n-k} = 2^k$, i.e.\ the codespace contains $k$ qubits~\cite{PreskillNotes}. Note that the choice of generators of a stabiliser group are not unique.

The stabiliser generators for a graph state are
\begin{eqnarray} \label{eqn:GSStab}
K_{v}  &= X_{v} \prod_{w \sim v} Z_w,\text{  } \forall v \notin I,
\end{eqnarray}
where $X_v$, $Y_v$, $Z_v$ are the Pauli matrices acting on site $v$. In this case, since we do not have stabilisers with indices $v \in I$ where $I$ are the input vertices, the number of encoded qubits equals the number of input qubits, as required. This type of graph, where the inputs are unconstrained is referred to as an \emph{open} graph-state. 

The evolution of a computation can be followed in terms of these stabiliser operators, rather than by following the evolution of the state itself. This is the Heisenberg picture of quantum computation, and provides an alternative perspective on quantum computation allowing some results, such as the Gottesman-Knill theorem, to be readily apparent. 
To follow how MBQC progresses in the stabiliser picture we first define some logical operators that determine what state the system is prepared in. In the case of open graph states, logical $X$ operators can be taken from the stabilisers acting on the inputs that were left out in (\ref{eqn:GSStab}). For example, a 1D cluster state, has stabiliser generators 
\begin{align}
K_v  &=Z_{v-1} X_{v} Z_{v+1} \quad v = 2,...,N-1; \; \;K_N =Z_{N-1} X_{N}.
\end{align}
The logical $X$ operator can then be made from the missing stabiliser generator $X_L := K_1 = X_1 Z_2$. $Z_L$ can be chosen as $Z_1$ so that it obeys the commutation relations with $X_L$ and the other stabiliser generators ($Y_L$ is usually left out since it can be inferred from the product $X_LZ_L$). Preparing the input in a logical $\ket{\pm}_L$ state is then the same as including $\pm X_L$ as one of the stabilisers. 

Now suppose we perform a measurement $M$ on this cluster state in the $X-Y$ basis; this measurement will not commute with all of the logical operators, and so the form of the logical operators will not be preserved after this measurement. However, since stabilisers act as identity on the codespace, the logical operators can be multiplied by any stabilisers which anticommute with the measurement, to put them in a form which is conserved during the measurement. So for example, consider measuring the first qubit in the 1D cluster state in the $X$-basis, so that $\{Z_L,M\}=\{Y_L,M\}=0$. The stabiliser $K_2$ also anticommutes with $M$ so multiplying $Z_L$ and $Y_L$ by $K_2$  gives new logical operators $\tilde{Z}_L = X_2 Z_3$ and $\tilde{Y}_L = X_1 Y_2 Z_3$ which do commute with $M$, and so are conserved during the measurement (i.e.\ if the system is prepared in a particular eigenstate of $\tilde{Z}_L$, it will remain in this eigenstate after the measurement). Overall the transformation of the stabilisers is
\begin{align}
X_L \to X_1 Z_2\text{,  }Y_L \to X_1 Y_2 Z_3\text{,  }Z_L \to X_2 Z_3.
\end{align}
Qubit 1 will either be in a $\ket{+}$ state or a $\ket{-}$ depending on the outcome of the measurement. We can therefore replace $X_1$ by $\pm 1$, and then following a $X_2^m$ correction, the logical operators are
\begin{align}
X_L \to Z_2\text{,  }Y_L \to  Y_2 Z_3\text{,  }Z_L \to X_2 Z_3,
\end{align}
 so we see that the result of the measurement is to move the information down the chain, and apply a Hadamard gate to it, as seen in the beginning of this section. More general measurements in the $\{ \frac{\pi}{2},\phi  \}$ basis which neither commute or anticommute with the logical operators can be dealt with in a similar way, by multiplying any non-commuting terms by stabilisers, so e.g.\ $\exp(i\theta Z_2) \to \exp( i\theta Z_2 K_{g(2)})$, where $g$ is the \emph{gflow} function defined earlier. We will see examples of this in Chapter~\ref{chap:AGQC}. 
 
An alternative and equivalent picture of MBQC is to start with a cluster state in which the qubits are prepared in $\ket{\pm_\theta}$ states, and all measurements are performed in the $\{ \ket{+} \bra{+} , \ket{-}\bra{-} \}$ basis. Thus the measurement angles are encoded in the graph state instead, and such a graph state is called a \emph{twisted} graph state. Using the notation $ X_{v}^{\theta_v }:= e^{-i\theta_v Z_v /2} X_{v}e^{i\theta_v Z_v /2}$, the stabiliser generators for the twisted graph state are 
\begin{eqnarray} \label{eqn:GSStabTwist}
K_{v}^{\theta_v}  &= X_{v}^{\theta_v} \prod_{w \sim v} Z_w,\text{  } \forall v \notin I,
\end{eqnarray}
where $\theta_v$ is the measurement angle on qubit $v$.

\subsubsection{Examples}
\label{sec:Example}

\begin{figure}[h]
\begin{center}
	\includegraphics[width=0.25\textwidth]{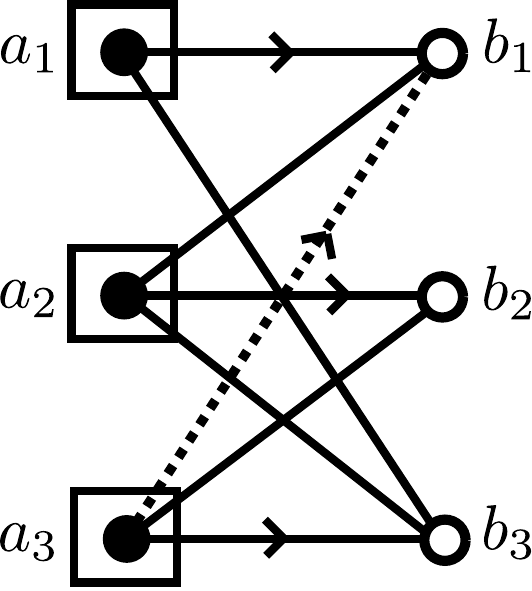}
	\caption{\label{fig:gflow} The \emph{gflow} is given by $g(a_1) = \{b_1\}, g(a_2) = \{ b_2\}, g(a_3) =\{b_1, b_3\}$ and the time ordering is $\{a_1\} < \{a_2\} < \{a_3\}$. The gflow is represented graphically as gflow lines (arrows), and the dotted line represents a gflow line not in the original graph, since there is no edge connecting vertices $a_3$ and $b_1$ but $g(a_3) = \{b_1,b_3\}$. }
\end{center}
\end{figure}

We now look at some examples of graphs, to illustrate the definitions above and since they will be useful in Chapter~\ref{chap:AGQC}. The first is the graph in Fig.~\ref{fig:gflow} from ref.~\cite{Browne2007}. The \emph{gflow} that satisfies all the \emph{gflow} rules is $g(a_1) = \{b_1 \}, g(a_2) = \{ b_2 \}, g(a_3) = \{ b_3,b_1\}$ with an ordering of $a_1 < a_2 < a_3$.

\begin{figure}[h]
   	\begin{center}
	\includegraphics[width=0.25\textwidth]{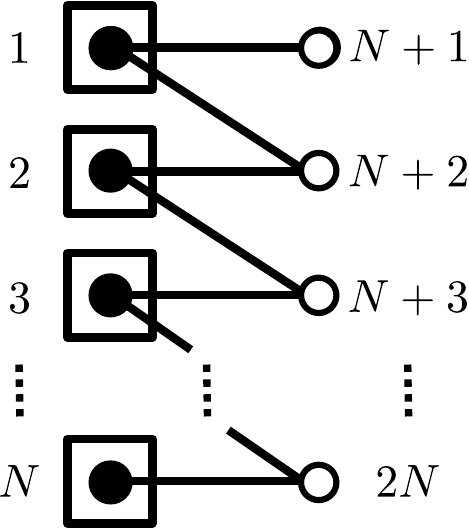}
	\caption{\label{fig:ZigZag} A graph for which many different \emph{gflows} are applicable.}
	\end{center}
\end{figure}

The second is the graph in Fig.~\ref{fig:ZigZag}, also studied in~\cite{Browne2007}. Many different possible \emph{gflows} can be defined on this graph. For example, a family of \emph{gflows} can be defined as
\begin{eqnarray}
g^r(v) = \left\{
\begin{array}{c c}
\{N+v,...,N+v+r -1 \}, & \mbox{if} \hspace{2mm}  v+r-1 \le N  \\
\{N+v,...,2N \},  & \mbox{if}  \hspace{2mm}  v+r-1 >N,
\end{array} \right.
\end{eqnarray}
where $1 \le r \le N$. If we were to perform measurements on this graph in MBQC, we can use any of these \emph{gflows}, provided we perform the right corrections on the outputs. For $g^r$, $r$ measurements can be performed simultaneously, interspersed by classical processing. Corrections on qubits will be of the form $X^{s_1 + s_2 + ... s_m}$ where the $s_m$ are binary variables accounting for the measurement outcomes. Thus the classical processing involves evaluating the binary sum $(s_1 + s_2 + ... s_m)$ of $r$ measurement outcomes, and so takes time $O(\log r)$~\cite{Furst84}. Since we can perform $r$ measurements simultaneously, the measurement depth $d_r$ is given by
\begin{eqnarray}
d_r = \left\lceil \frac{N}{r} \right\rceil.
\end{eqnarray}
The two extreme cases are where $r=1$ or $r=N$. The former is just where we can perform each measurement one-by-one ($d_1 =N$), with no addition of binary variables in between. The latter is where we can perform all measurements simultaneously ($d_N = 1$), but we must perform corrections on the outputs which take time $O(\log N)$. $g^N$ is called the \emph{maximally delayed flow} associated with this graph~\cite{Mhalla}. Note also that the zig-zag graph is not a universal resource for quantum computation, but combining several layers of the zig-zag graph we can make a universal resource.

\begin{figure}[h]
   	\begin{center}
	\includegraphics[width=0.25\textwidth]{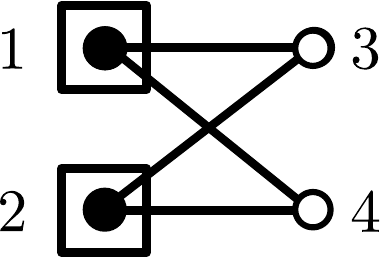}
	\caption{\label{fig:NoGflow} A graph which violates the rules of \emph{gflow}.}
	\end{center}
\end{figure}

Finally, to aid the discussions later on, an example of a graph without \emph{gflow} is shown in Fig.~\ref{fig:NoGflow}. It can easily be verified that there are no assignments of $g(v)$ and ordering that satisfy the rules of \emph{gflow}.

\section{Quantum computation through adiabatic evolution}
\label{sec:AQC}

In the previous section we saw how computation can proceed through non-deterministic adaptive measurements on an entangled resource state. We will now a method of computation which is the opposite of this; quantum computation through \emph{adiabatic} evolution, in which operations are performed slowly and deterministically.

\subsection{The adiabatic theorem}
\label{sec:AdTheor}

The word `adiabatic' has a long history in thermodynamics, where it is taken to mean any process that occurs without heat passing into or out of a system (the word \emph{adiabatic} in Greek means `not passable'). For a physical process such as compression of an ideal gas, we might achieve adiabaticity by performing the compression fast enough that the system has not had time to equilibrate with its surroundings, or by insulating it from the outside world.

The way the word adiabatic is used in quantum mechanics typically is more analogous to (isothermal) reversible adiabatic processes. For a process to be truly reversible in classical thermodynamics, infinitesimal changes must be made such that the system is always equilibrated. In other words, the process of change must be slower than the process of equilibration. Of course, to be perfectly reversible the change must take an infinite amount of time, but the dynamics can be reversible up to some error if the dynamics of equilibration is fast enough. 

As an example, imagine pulling a tablecloth from a table covered in glasses. If the cloth is pulled incredibly slowly (assuming zero friction between cloth and table and infinite friction between cloth and glasses), such that the glasses only wobble a negligible amount and so are constantly in equilibrium, then the glasses slide along with the table cloth; if the cloth is not pulled slowly enough, then the glasses wobble, with energy being released irreversibly in the form of heat as they rock from side to side on the table. Here what is `slow enough' is dictated by the properties of the glasses, and how fast the wobbling motion is damped.

Quantum adiabatic processes are much the same: the word `adiabatic' has connotations with a process that is performed \emph{slowly} with respect to the energy scale of the system, such that the probability of finding a system in the $n^{th}$ highest energy level is small during the process. Thus it is much like an isothermal reversible adiabatic process in which work can be performed to change the energy levels of the system, but the occupancy of the levels does not change (the entropy is fixed).

 So what determines what is `slow enough' for a quantum system? The first treatment of adiabatic processes in quantum mechanics dates back almost 100 years ago, to Ehrenfest's work on adiabatic invariants~\cite{Ehrenfest1916} followed by Born and Fock~\cite{Born1928}. Since then, it has  found applications in quantum computation~\cite{Farhi2001}, molecular dynamics~\cite{Born1927} and the quantum Hall effect~\cite{Niu1984,Avron1987}. 
 
We first consider the simplest form of the adiabatic theorem, valid for systems where the energy levels are non-degenerate. Consider a Hamiltonian $H(t)$ with non-degenerate eigenvectors $\ket{E_n(t)}$ and energies $E_n(t)$. If we prepare the system in state $\ket{\psi(0)} = \ket{E_m(0)}$ and evolve it with the time-dependent Hamiltonian for some time $\tau$, then the system will end in the state $\ket{E_m(\tau)}$ with probability $1-p_{err}$, where~\cite{Messiah}
 \begin{align}\label{eqn:AdiabC1}
p_{err} =  O \left(  \max_{m \neq n} \left| \frac{\max_{0 \leq s \leq 1} \bra{E_m(s)} \frac{dH(s)}{ds} \ket{ E_n (s)}}{  \min_{0 \leq s \leq 1}|E_m(s) - E_n(s)|^2}  \right|^2, \right)
\end{align}
where $s = t /\tau$. In 1950, Kato derived a more general adiabatic condition that dropped the assumption of non-degenerate eigenvectors, and without assuming anything about the spectrum other than how it behaved near the region of interest~\cite{Kato1950}. This laid the foundations for the proof by by Albert Messiah~\cite{Messiah}, and provided the framework upon which later generalisations were built. A particularly useful result from the point of adiabatic quantum computation was the result in~\cite{Reichardt2004}, building on the work in~\cite{Avron1987}, which gives an expression for the error of an adiabatic transformation for the case when the energy levels can be degenerate (again using the notation $s = t /\tau$):
\begin{theorem}[Adiabatic theorem~\cite{Reichardt2004}]
Let $H(s)$ be a finite-dimensional, $(\kappa+1)$-times differentiable Hermitian matrix, with $\kappa \ge 1$, and with eigenvalues $E_j(s)$. Let $U(s)$ be the unitary evolution due to $H(s)$. Let $P(s)$ be the projection onto the (possibly degenerate)  subspace of $H(s)$, with eigenvalue $E_0(s)$. Then after the evolution $U$, the amount by which a state initialised in the ground subspace of $H(0)$ has leaked into the excited states at time $s$ is given by
\begin{align}
p_{err} =\Vert (\idop - P(s)) U P(0) \Vert \le c(\kappa) \left( \left. \frac{1}{\tau} \frac{h(s)}{\Delta(s)^2}\right|_{b.c.}  + \max_{0 \le s \le 1} \frac{1}{\tau^\kappa} \frac{h(s)^{\kappa+1} }{\Delta(s)^{2\kappa+1}} \right),
\end{align}
where $\Delta(s) = \min_{n \ne 0} |E_n(s) - E_0(s) |$, $h(s)$ satisfies $\Vert \left(\frac{d}{ds}\right)^l H(s) \Vert \le h(s)$ for all $l \le \kappa$ and $|_{b.c.}$ indicates that these quantities are evaluated at the boundaries $s=0,1$.
\end{theorem}
The norm used here is the spectral norm, which for a square Hermitian matrix $X$ is just the largest absolute eigenvalue of $X$. $p_{err} = \Vert (\idop - P(s)) U P(0) \Vert^2$ gives the probability of having leaked out of the subspace of $E_0$ at time $\tau$, and we can see a strong resemblance to the adiabatic theorem in (\ref{eqn:AdiabC1}). In this work, we consider the simplest case, where $H(s)$ is a linear interpolation of the form $(1-s)H_0 + sH_p$, so that $H(s)$ is infinitely differentiable. It is tempting then to set $\kappa \to \infty$, however this isn't possible without raising the adiabatic time to $\infty$, since $c(\kappa) = O(2^{\kappa}\kappa!)$~\cite{Reichardt2004}, which goes to $\infty$ as $\kappa \to \infty$. So we consider $\kappa$ as begin some fixed, large finite number. We also consider $\varepsilon$ to be fixed. With a linear interpolating Hamiltonian, $\Vert \dot{H}(s) \Vert  = \Vert H_p - H_0 \Vert$ is independent of $s$, and the error in the adiabatic evolution becomes
\begin{align}
p_{err} \le c(\kappa) \left( \left. \frac{1}{\tau} \frac{\Vert \dot{H}(s) \Vert}{\Delta(s)^2}\right|_{b.c.}  + \frac{\Vert \dot{H}(s) \Vert^{\kappa+1}}{\tau^k\Delta(s)^{2\kappa+1}_{min}}  \right),
\end{align}
where $\Delta(s)_{min} := \min_{0 \le s \le 1}\left( \Delta(s) \right)$. By choosing a runtime that satisfies 
\begin{align}\label{eqn:RunTime}
\tau \ge \left( \frac{ c(\kappa) \Vert \dot{H}(s) \Vert^{1+1/\kappa}}{\varepsilon\Delta(s)_{min}^{2+1/\kappa}} \right),
\end{align}
the probability of error becomes
\begin{align}
p_{err} \le \left. \frac{\varepsilon \Delta(s)_{min}^{2+1/\kappa} }{\Vert \dot{H}(s) \Vert^{1/\kappa} \Delta(s)^2 } \right|_{b.c.}  + \frac{\varepsilon^\kappa }{c(\kappa)^{\kappa-1}}.
\end{align}
In all of the situations considered in this work, $\Vert \dot{H}(s) \Vert \geq \gamma$, and $\Delta(s) \leq \gamma$, where $\gamma$ is the energy scale set by the interactions in the particular system we implement the computation in. This means that $\Delta(s)_{min} / \Vert \dot{H}(s) \Vert \leq 1 $, and since $\Delta(s)_{min} \leq \Delta(s)$,
\begin{align}
p_{err} \le \varepsilon + \frac{\varepsilon^\kappa }{c(\kappa)^{\kappa-1}}.
\end{align}
So with a time scale of the form in (\ref{eqn:RunTime}), the error is of the order $\varepsilon$ (assuming that $c(\kappa) >1$, which is true for large enough $\kappa$). Following the terminology used in~\cite{Aharonov2008}, we say that the final state is $\varepsilon$ close to a state in the ground subspace of $H_p$, and ignoring the terms which are independent of $N$ we write the adiabatic runtime as
\begin{align}
\tau = \Omega \left( \frac{  \Vert \dot{H}(s) \Vert^{1+\delta}}{\Delta(s)_{min}^{2+\delta}} \right),
\end{align}
where $\delta := 1/\kappa$, and we use the `big Omega' notation $f(n) = \Omega(g(n) )$ to indicate that there is some constant $c$ and value $n_0$ such that $f(n) \geq cg(n)$ for $n > n_0$. Note also that in some cases the energy gap may close but properties of the Hamiltonian may prevent mixing between the degenerate states, so the adiabatic runtime will only diverge if the gap closes and transitions between the degenerate states are possible. We will see an example of this behaviour in chapter~\ref{chap:AGQC}.

The form of the adiabatic theorem given above is in fact not the most general, but it will be appropriate for all of the situations we consider in this thesis. A more general proof has been done by Jansen, Ruskai and Seiler in~\cite{Jansen2007}, generalising to situations where the projector $P(s)$ can project into several different energy levels, all of which can be degenerate. As a historical note, it is worth noting the inconsistency of the adiabatic theorem as pointed out by Marzlin and Sanders~\cite{Marzlin2004}, in which they consider adiabatic evolution including a time-varying field. Following this there was significant efforts to derive necessary and sufficient conditions~\cite{Tong2010,Boixo2010,Sarandy2004,Ortigoso2012}, with sufficient criteria subsequently defined in~\cite{Chueng2011}. 
 
\subsection{Adiabatic quantum computation}

Using the adiabatic theorem, it is possible to perform computation in a very different way to the circuit model or measurement-based models we have seen before: \emph{Adiabatic quantum computation}. The standard adiabatic quantum computation (AQC) protocol~\cite{Farhi} starts with an initial Hamiltonian $H_0$, which is easy to prepare, such as a uniform magnetic field $\sum_n X_n$, where $X_n$ is a Pauli X operator acting on the $n^{th}$ qubit. The system is prepared in the ground state of this Hamiltonian, $\ket{E_0(0)}$, and then the Hamiltonian is slowly changed to a new `problem Hamiltonian' $H_p$ which is typically non-uniform and encodes the problem to be solved. Provided the evolution time obeys the adiabatic theorem, the system will be in the ground state of $H_p$ with high probability.

A natural example is solving a satisfiability problem~\cite{Farhi}, where we have a boolean formula made up of logical clauses (OR, AND,$\neg$) which we would like to be satisfied, e.g. $(x_1 \; \mbox{OR} \;  x_2 )$ is satisfied by $(x_1,x_2) = (1,0),(0,1),(1,1)$, and $(x_1 \; \mbox{OR} \;  x_2 ) \mbox{AND} (x_1 \; \mbox{OR} \; \neg x_2 )$ is satisfied by the assignment $(x_1,x_2) = (1,0),(1,1)$ etc. Such a problem can be framed as finding the ground state of a Hamiltonian, by replacing the logical clauses with energy penalties, that penalise when the clause is not satisfied. If we replace the boolean states $0,1$ with qubit states $\ket{0},\ket{1}$, then satisfying $(x_1 \; \mbox{OR} \;  x_2 )$ is equivalent to finding the lowest energy configuration of $1 - \frac{1}{4}(1 + Z_1)(1+Z_2)$, and satisfying $(x_1 \; \mbox{OR} \;  x_2 ) \mbox{AND} (x_1 \; \mbox{OR} \; \neg x_2 )$ is equivalent to finding the lowest energy configuration of $[1 - \frac{1}{4}(1 + Z_1)(1+Z_2)] +[1 - \frac{1}{4}(1 + Z_1)(1- Z_2)]$. Using this method, any satisfiability problem can be framed as finding the ground state of an Ising Hamiltonian, although this doesn't mean that an adiabatic quantum computer could solve all satisfiability problems as the energy gap may shrink faster than polynomially (for example, certain satisfiability problems are NP complete, and it is not generally believed that a quantum computer could solve NP-hard problems efficiently).

AQC has been shown to be \emph{polynomially equivalent} to the standard circuit model of quantum computation, in that it is possible to implement any quantum circuit using an adiabatic protocol, such that the inverse of the energy gap is polynomially equivalent to the number of gates in the circuit~\cite{Aharonov2008}. AQC has also gained much recent attention due to the appearance of D-wave, billing itself as the world's first \emph{quantum annealer} (quantum annealing has a similar philosophy to AQC, except the system is not assumed to be closed). However, whether or not this machine truly is a quantum annealer, or will provide any quantum speed up, is a subject of much debate (see e.g.~\cite{Boixo2014,Shin2014}).

The time to perform the computation in AQC is sensitive to the particular method of interpolating between the simple initial Hamiltonian $H_0$ and the problem Hamiltonian $H_p$. Often we will assume a linear interpolation of the form $H(t)  = (1-t/T)H_0 + (t/T)H_p$, but clearly this cannot be optimal; we would expect the best algorithm to depend on the particular form of the gap as a function of $t$. Such a protocol, where the evolution speed can change in response to the time-dependent energy gap, is called a \emph{local} evolution (in contrast to the \emph{global} evolution we have considered so far). These differences lead to significant changes in computation speed; for instance, in~\cite{Roland2002} it was found that an adiabatic implementation of Grover's algorithm only achieved the $O(\sqrt{N})$ scaling if local adiabatic evolution was used. However, in this thesis the distinction between local and global evolution will not be important, so a linear interpolation will be used.

\subsection{Holonomic quantum computation}\label{sec:Holon}

Our discussion of the adiabatic theorem so far has focused on the probability of ending in a particular state. However, much interesting behaviour can also occur due to the phase the system picks up during an adiabatic evolution. Consider a system with Hamiltonian $H(t)$, that is prepared in the ground state $\ket{E_0(0)}$ at $t=0$ and evolved adiabatically. At some time $t$ later, the state is generally $\ket{\psi(t)} = e^{i\phi(t)} \ket{E_0(t)}$, assuming perfect adiabaticity. Inserting this into the Schr\"{o}dinger equation $i\hbar | \dot{\psi}(t)\rangle = H \ket{\psi(t)}$ gives
\begin{align}
(E_0(t) + \hbar \dot{\phi}(t) ) e^{i\phi(t)} \ket{E_0(t)} - i \hbar e^{i\phi(t)} | \dot{E}_0(t) \rangle = 0,
\end{align}
which can be rearranged, after left-multiplying by $\bra{E_0(t)}$ to give
\begin{align}\label{eqn:Berry1}
\phi(t) = i \int_0^t \sprod{E_0(t')}{\dot{E}_0(t')} dt' - \frac{1}{\hbar} \int_0^t E_0(t') dt'.
\end{align}
The term on the right is called the dynamical phase, whilst the term on the left is the \emph{Berry} or \emph{geometrical} phase~\cite{Berry1984}. This type of phase contribution was first pointed out by Pancharatnam in the context of classical optics~\cite{Pancharatnam} (and for this reason the Berry phase is also sometimes called the \emph{Pancharatnam phase}), and we can understand the form of $\gamma_n$ in equation (\ref{eqn:Berry1}) as being the combination of many small contributions due to the slight difference in phase between states at adjacent times. The Berry phase produces observable physical effects, and is used in describing many phenomena in condensed matter systems, such as topological insulators~\cite{Thouless1982,Kane2005}.

Berry phases can also be extended to degenerate Hamiltonians. If instead the Hamiltonian has a $d$-fold degenerate ground space of energy $E_0$ with states labelled as $\ket{E_0^{\alpha}}$, $1 \leq \alpha \leq d$, then following an adiabatic transformation, rather than a phase being applied the ground space is transformed by the $d \times d$ matrix $U_{\alpha \beta}$~\cite{Wilczek1984}, where
\begin{align}
U_{\alpha \beta}(t) = \mathcal{T} \exp \left( i\gamma^{\alpha \beta}(t) \right)  = \mathcal{T} \exp \left( - \int_0^t \bra{E_0^{\alpha}(t')}\frac{d}{dt'} |E_0^{\beta}(t') \rangle dt' \right).
\end{align}
$\gamma_{\alpha \beta}$ is the \emph{Wilczek-Zee connection}, and $U_{\alpha \beta}$ is called a \emph{holonomy}. Thus a state prepared in a state $\ket{E_n^\alpha(0)}$ in the degenerate subspace will be transformed into a state $ e^{ \frac{-i}{\hbar} \int_0^t  E_0(t) dt' }U_{\alpha \beta}(t) \ket{E_n^\alpha(t)}$, in which the holonomy creates transitions between different states in the degenerate subspace (or in other words, it becomes a non-Abelian generalisation of the Berry phase, since transformations no longer necessarily commute with one another). Provided the system is adequately controllable, it is possible to harness this so that information encoded in the degenerate energy levels of $H(0)$ can be arbitrarily transformed using these holonomies, and so universal quantum computation can be performed. This is known as \emph{holonomic quantum computation} (HQC)~\cite{Zanardi1999,Pachos1999}. The holonomy can be generated either using a closed loop in parameter space such that $H(0) = H(\tau)$, or can more generally have $H(0) \neq H(\tau)$, in which case it is usually called \emph{open-loop} HQC~\cite{Kult2006}. Also note that, although HQC is commonly discussed in terms of adiabatic evolutions, this is not a necessary requirement, and schemes for achieving holonomies via non-adiabatic processes have been proposed~\cite{Aharanov1987,Sjokvist2012,XiangBin2001,Burgarth2013}.

HQC has desirable properties which lead a natural `hardware' resistance to certain types of errors. To see this, consider a Hamiltonian $H(\mathbf{R})$ that depends on a set of parameters $\mathbf{R}[t] = (r_1(t),r_2(t),...,r_n(t))$. Then the Berry phase around a path $\Gamma$ becomes
\begin{align}
\gamma_n &=  i \oint_\Gamma  \bra{E_n(\mathbf{R})} \nabla_{\mathbf{R}} \ket{E_n(\mathbf{R})} \cdot d\mathbf{R} := \int \int_{S} \mathbf{F}_n \cdot d \mathbf{\sigma},
\end{align}
where we have used Stokes' theorem, and defined $\mathbf{F}_n := \nabla_{\mathbf{R}} \times \mathbf{A}_n $, $\mathbf{A}_n = \bra{E_n(\mathbf{R})} \nabla_{\mathbf{R}} \ket{E_n(\mathbf{R})}$, and $S$ is the surface contained by the path $\Gamma$. In this form, the phase has the interpretation as the integral of a curvature over an area, which highlights one of the most useful and interesting thing about these acquired phases; they only depend on the area contained within the evolution path. Thus quantum computations using such evolutions could be naturally resistant to control errors that change the path taken in parameter space but which don't change the area. This property has been tested theoretically, confirming that HQC operations are less sensitive to dephasing noise~\cite{Carollo2003, DeChiara2003}, and error processes where the gate time is either much longer or much shorter than the noise correlation time~\cite{Solanis2004}. Further protection against noise can in principle be achieved with HQC in decoherence-free subspaces and subsystems as shown in~\cite{Wu2005,Oreshkov2009prl}. HQC does not only provide `hardware' improvements; a scheme for fault-tolerant HQC on stabiliser codes has also been developed~\cite{Oreshkov2009prl2,Oreshkov2009} (unlike AQC, where developing fault tolerant computation is an ongoing problem~\cite{Young2013}), making HQC a promising method to perform quantum computation.

\section{Spin chains for communication}
\label{sec:IntroSpinChain}
As well as being able to initialise, store, and process quantum information in a quantum computer, it is also important to be able to transport quantum information between different parts of the computer, or perform gates between distant qubits, since geometrical constraints will prevent all qubits from being directly connected. Methods to do this with minimal loss of coherence are thus incredibly important. Many proposals to perform this task involve transferring information from one physical system to another (e.g.\ from a trapped ion to a photon) or shuttling a physical system from one part of the computer to another. But in certain situations, it may be beneficial to avoid transfer between mediums or moving qubits around, as these can be error-prone processes. Instead we could create communication channels such that the information is stored and transferred in the same medium, by creating chains of qubits along which quantum information can travel. Then to transfer information from one qubit to another, all that is required is to turn on interactions with the chain, and then turn them off again when the information has arrived at the other end. This is the concept behind using chains of qubits, or `spin chains' as a quantum wire to transfer information in quantum computers~\cite{Bose03}. A spin chain is a 1-dimensional chain of spins where the spin degree of freedom of nearby physical systems are coupled. In this thesis we will just consider spin chains formed of spin-$\frac{1}{2}$ particles, so really they can be thought of as a 1 dimensional array of qubits, but we use the terminology spin chain since it is commonly used in the literature. The interactions are typically confined to nearest or next-nearest neighbours, and typically have the form
\begin{eqnarray}
H=  \sum_{m<n} J_{mn} (X_m X_{n}  +Y_m Y_{n}+ \Delta_{mn} Z_n Z_n) + \sum_m B_m Z_m,
\end{eqnarray}
where $X, Y,Z$ are the Pauli matrices, $\Delta_{mn}$ is the anisotropy, $B_m$ is the magnetic field strength, and $J_{mn}$ is the coupling parameter (negative for ferromagnetic interactions which favour alignment of spins, and positive for anti-ferromagnetic interactions which favour anti-alignment). Some particularly important and well-used spin chains are where $\Delta_{mn}=1$ (isotropic or Heisenberg spin chain), $\Delta_{mn} = 0$ (XY spin chain) and $\Delta \to \infty, J<0$ (Ising spin chain). Spin chains appear in nature~\cite{Hammar1999}, and can also be realised in physical systems such as Josephson arrays~\cite{Romito2005} or atoms in optical lattices~\cite{Bloch2011}. 
Whilst it would be trivial to take a chain of qubits and perform perfect state transfer by turning on a \textsc{swap} interaction between successive pairs of qubits, we are more interested in the case where there is minimal control over the chain, and where the interactions are time-independent and ideally uniform. This is firstly since if we did have a large degree of control over the chain we may as well use it as part of the quantum computer, and secondly since minimal control means less contact with the outside world and thus less decoherence. 

The first spin chain communication protocol in~\cite{Bose03} used $\Delta_{mn} = 1$, $J_{mn} = J<0$ and nearest-neighbour interactions. The qubits in the chain are initialised in the $\ket{\mathbf{0}} = \ket{000...0}$ state, and at time $t=0$ qubit $s$ is encoded with a quantum state $\ket{\psi} = \cos \theta/2 \ket{0} + e^{i\phi} \sin \theta/2 \ket{1}$. At some time $\tau$ later, the state of the chain is $\ket{\Psi(\tau)}$, and we calculate the fidelity of a receiving qubit $r$ with the original information, where the fidelity is given by
\begin{align}
F_\psi = \bra{\psi} \text{tr}_{\hat{r}} ( \ket{\Psi(\tau)} \bra{\Psi(\tau)} ) \ket{\psi},
\end{align}
where $\text{tr}_{\hat{r}}$ means tracing out all qubits except qubit $r$, and for now we do not square the fidelity as in Section~\ref{sec:DistMeas}, to fit with the conventions used in the spin chain literature. 
Since the fidelity is sensitive to the particular state $\ket{\psi}$ input at $s$, and we require a general metric of how good this channel is, $F_\psi$ must be averaged over all possible input states, giving the overall averaged fidelity $F_{av}$
\begin{align}
F_{av}= \frac{1}{4 \pi} \int F_\psi \; \sin \theta d \phi d\theta.
\end{align}
At some time $\tau_{opt}$, $F_{av}$ will be maximised (with a cut-off after a time $\tau_{max}$, chosen as $4000/J$ in~\cite{Bose03}). When $s=1$ and $r=N$, this simple protocol can achieve near perfect state transfer for $N=4$ spins, and achieves $F_{av} > 2/3$ for up to 80 spins (which is an important benchmark since it is the maximum possible average fidelity achievable by purely classical means~\cite{Horodecki1999}). 

After this initial proposal, there followed many more schemes, all using constantly coupled qubits but with different Hamiltonians.
Particularly important results are the perfect state transfer channel~\cite{Christandl04}, which achieves perfect transfer by using a mirror symmetric spin chain with engineered couplings $J_{n,n+1} = \sqrt{N(N-n)}$. Other notable protocols include coupling two qubits weakly to the chain such that the chain acts as an effective qubit~\cite{Plenio2005}, optimised couplings at the ends of an $XY$-chain to make the excitations fall in a region of linear dispersion~\cite{Banchi2011}, and a heralded dual rail protocol~\cite{Burgarth05}. For a more in depth discussion of spin chains for quantum communication, see~\cite{Bose2008}.

In order to measure how well the channel communicates information, we will often use the average fidelity defined above, which can be put in a particularly simple form that is easier to calculate~\cite{Horodecki1999}. For a channel that transforms the input information by $\mathcal{E}$, with systems of dimension $d$, the average fidelity is
\begin{align}\label{eqn:FavrgHorod}
F_{av}[\mathcal{E}] = \frac{d F_e[\mathcal{E}] +1 }{d +1},
\end{align}
where $F_e[\mathcal{E}] = \bra{\psi} [ \idop \otimes \mathcal{E} ](\ket{\psi}\bra{\psi} ) \ket{ \psi}$, and $\ket{\psi}$ is a maximally entangled state such as $\ket{\psi} = \frac{1}{\sqrt{2}} (\ket{01} - \ket{10} )$ (this is often called the \emph{singlet fraction}). For a classical channel, through which only product states can be sent, the largest overlap between a singlet and a product state is $1/\sqrt{d}$, so a classical channel has $F_{av}[\mathcal{E}] \leq \frac{2}{3}$.

Apart from communicating quantum information, using a spin chain also serves as a useful method of sharing entanglement between two different qubits (and in fact a channel that is good for sending quantum information is also a good entanglement sharer~\cite{Bayat10,Horodecki1999}). The protocol follows in a similar fashion as above, except qubit $s$ is maximally entangled with another ancilla qubit $a$. After evolving in time, $a$ will be entangled with qubit $r$. If the protocol is not perfect, it is possible to repeat it many times and use \emph{entanglement distillation} to convert multiple copies of non-maximally entangled states into one maximally entangled state~\cite{Bennett1996}. Once such an entangled state is shared between two parties, it can be used to teleport quantum information between them (provided there is additional classical communication, which is usually considered as a plentiful resource). For a pure state $\ket{\psi_{AB}}$ over two qubits $A$ and $B$, the measure of bipartite entanglement is the Von Neumann entropy $S(\ket{\psi_{AB}})$~\cite{Popescu1997}, defined as
\begin{align}
S(\ket{\psi_{AB}})  = - \text{tr} (\rho_A \log_2 \rho_A) = - \text{tr} ( \rho_B \log_2 \rho_B),
\end{align}
where $\rho_A:=\text{tr}_{B} (\ket{\psi_{AB}}\bra{\psi_{AB}})$ and similarly for $\rho_B$. 

In the case when the two qubits are in a mixed state $\rho_{AB}$, a suitable measure of entanglement is the \emph{entanglement of formation}~\cite{Bennett1996,Wootters98}. First, note that there can be many decompositions of $\rho_{AB}$ into an ensemble of pure states, $\rho_{AB} = \sum_n p_n^{\xi} \ket{\psi_n}^{\xi}\bra{\psi_n}^{\xi}$, where the $\xi$ superscript denotes a particular decomposition. The entanglement of formation is then defined as
\begin{align}
E_f(\rho) = \min_{\xi} \sum\nolimits_n p_n^{\xi} S(\ket{\psi_n}^\xi).
\end{align}
In other words, it is the minimum of the expected entanglement over all the possible pure state ensembles that could make up $\rho_{AB}$. 
A particularly useful form of $E_f(\rho)$ was derived in~\cite{Wootters98}. If we define $\tilde{\rho} = (Y_1 \otimes Y_2) \rho^* (Y_1 \otimes Y_2)$, then $E_f(\rho) = \Lambda(C)$, where 
\begin{align}
\Lambda(C) = H\left(\frac{1 + \sqrt{1-C^2}}{2} \right), \; H(x) = -x \log_2 x - (1-x) \log_2 (1-x)
\end{align}
and $C = \max \{ 0, \lambda_1 - \lambda_2 - \lambda_3 -\lambda_4 \}$, where $\{ \lambda_n \}$ are the eigenvalues of $\sqrt{ \sqrt{\rho} \tilde{\rho} \sqrt{\rho} }$. $C$ is usually called the \emph{concurrence}.

This use of spin chains is not only restricted to transferring information or entanglement in a quantum computer. One dimensional systems can exhibit a rich variety in behaviour, and testing how quantum information is transferred through certain systems could reveal information about this. The quality of transfer could also be used to probe the `quantumness' of the individual qubits in the chain: if any members of the chain behave classically, then we would expect the information transmitting ability of the chain to be impaired.

\section{Some experimental candidates for a quantum computer}

Having seen some of the theoretical ideas behind a quantum computer, we now discuss the more practical side of creating a quantum computer, introducing some important experimental proposals and reviewing some of the experimental progress. The standard framework by which experimental proposals are judged is the \emph{DiVincenzo criteria}~\cite{DiVincenzo2000}, which is essentially a checklist of desirables for an experimental scheme. The checklist is as follows:

\begin{itemize}
\item The qubits should be scalable and well-characterised (meaning that the couplings to other qubits and external influences should be known).
\item It must be possible to initialise qubits in a pure state such as $\ket{00...0}$ (for computation and error correction).
\item The qubits must have decoherence times significantly longer than gate operation times.
\item It must be possible to perform a universal set of gates on the qubits.
\item The state of the qubits must be measurable. 
\item It must be possible to transmit qubits between specific locations.
\end{itemize}

In this section we will focus on two particular proposals which could fulfil these criteria; ion traps and donors in silicon. There are of course many more experimental systems which are also promising candidates, such as NMR~\cite{Cory1997,Gershenfeld1997}, linear optics~\cite{Knill2001b}, quantum dots~\cite{Loss1998}, and superconducting qubits~\cite{Devoret2013}, but we will focus on ion traps and donors because of their relevance to Chapter~\ref{chap:3qubit}. For a more comprehensive overview of the current candidates for a quantum computer, see e.g.~\cite{NielsenChuang,Buluta2011,Ladd2010,RoadMap}

\subsection{Ion traps}
\label{sec:IntroIonTraps}

The first scheme for using trapped ions to process quantum information was by Cirac and Zoller~\cite{Cirac1995}. The essential idea is to trap ions in an electromagnetic field, and encode information in internal states of these ions. Typically we consider ions trapped in a linear Paul trap~\cite{Paul1953}, with a combination of an axial DC field and a radial AC field creating a fluctuating potential that approximates a harmonic confining potential $\frac{1}{2}M \nu_x^2 x^2 + \frac{1}{2}M \nu_y^2 y^2 + \frac{1}{2}M \nu^2 z^2$ for each ion, such that $\nu_x,\nu_y \gg \nu$ ($\nu$ is often called the secular frequency of the trap). Note that it is not possible to simply create a static confining potential due to Earnshaw's theorem~\cite{Earnshaw1842}. 

\begin{figure}[h]
\begin{center}
	\includegraphics[width=0.65\textwidth]{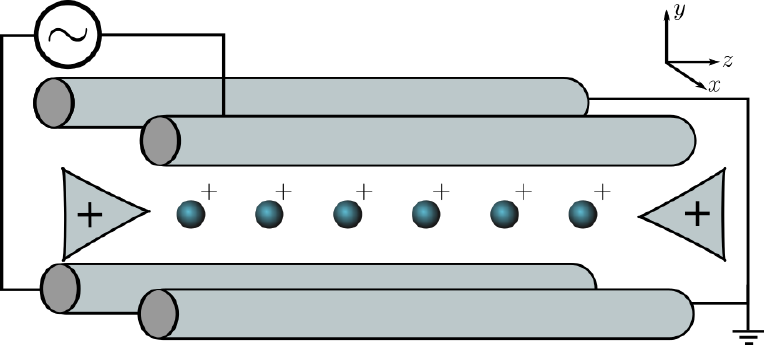}
	\caption{\label{fig:IonTrap} A schematic diagram of a linear quadrupole Paul trap. Two diagonally opposed electrodes produce a radio frequency oscillating electric field which approximates a harmonic trapping potential. The ions are confined in the $z$-axis by two endcap electrodes. }
\end{center}
\end{figure}

Initialisation of the trapped ion qubits could be achieved using the other excited levels of the ion; if there are excited states which can be reached from the qubit states by application of a suitable laser field, and which both decay into the qubit $\ket{0}$ state, then applying laser fields which are resonant with all transitions from the $\ket{1}$ state result in the qubit being prepared in the $\ket{0}$ state. 

Measurement of the qubit state can be performed by applying laser radiation that is resonant with a transition from the $\ket{1}$ state to an excited state that decays back into the $\ket{1}$ state, so that fluorescence is observed only if the qubit is in the $\ket{1}$ state. Single qubit rotations can be performed by pairs of lasers focused to a spot significantly smaller than the separation of the ions; single qubit $X$ rotations can be performed using a resonant laser pulse to drive Rabi oscillations, whilst single qubit $Z$ rotations can be performed using off-resonant laser pulses resulting in an AC Stark shift. $T_2$ times of ions can be up to $\sim 10 \text{ms}$, whilst $T_1$ times can be on the order of minutes~\cite{Buluta2011}.

To perform gates between qubits, the proposal in~\cite{Cirac1995} uses the motional degree of freedom of the ion chain. When the system is sufficiently isolated and cooled so that $k_BT \ll \hbar \nu$, the ion motion  in the $z$-direction is restricted to the ground state, which is a coherent phonon corresponding to coordinated oscillation of all of the ions. When the motion is confined in this way, we call this phonon mode the `bus qubit'. We label states of the ion/bus system as $\ket{m}_I\ket{n}_B$, where $m=0,1$ is the state of the ion and $n=0,1$ corresponds to the phonon mode being unoccupied/occupied respectively. When ions absorb a photon, there is a resultant change in momentum of the ion, which translates to an excitation of the phonon mode. If the energy splitting between all of the ion qubit states is different along the chain (which can be achieved by e.g.\ applying a magnetic field gradient) then using a global field we can couple individual ions to the bus qubit. 

This can then be used to perform a gate between two ions in the chain (say ions $1$ and $2$). To begin, the internal states of ion $1$ is swapped with the bus qubit by selecting a laser frequency that causes oscillations between the $\ket{0}_1\ket{1}_B$ and $\ket{1}_1\ket{0}_B$ states. Then it is possible to select a laser frequency that is resonant with a transition from the $\ket{1}_2\ket{1}_B$ state to another (non-computational) state, resulting in a $-1$ phase applied to state $\ket{1}_2\ket{1}_B$ but no evolution of the other states. The bus qubit can be swapped back into ion $1$, thereby achieving a \textsc{cz} gate between the two qubits, mediated by the bus qubit.

An alternative method to create interactions between ions, which we will use later on in Chapter~\ref{chap:3qubit}, is magnetic gradient induced coupling~\cite{Mintert2001}. The essential idea is to use a magnetic field gradient to create an effective Ising interaction between two ions, rather than using a laser field to swap ions in and out of the bus qubit. In this magnetic field gradient, the energy of the ion depends on the internal state of the ion, and so the equilibrium position of the ion will also depend on the internal state of the ion. Thus the two qubit states yield different equilibrium positions of the ions, $\bar{z}_n$ and $\bar{z}_n + d_n$. The difference in equilibrium position $d_n$ leads to conditional dynamics; roughly speaking, when the $n^{th}$ ion is in a $\ket{1}_n$ state, it becomes closer to the $(n+1)^{th}$ ion and thus has an effect on the state of this ion. Thus the internal states of the ions affect the interactions between the ions.

The full analysis is given in~\cite{Mintert2001,Wunderlich2001}, in which it is shown that the magnetic field gradient produces couplings of the form $\frac{1}{2}\hbar J_{zz}^{ij} Z_i Z_j$ with coupling strength given by
\begin{align}\label{eqn:IonJzz}
J_{zz}^{ij} = \sum_{n=1}^N \nu_n \varepsilon_{ni} \varepsilon_{nj},
\end{align}
where $\varepsilon_{ni}$ is the effective Lambe-Dicke parameter and $\nu_n$ is the angular frequency of the $n^{th}$ vibrational mode of the ion chain. The effective Lambe-Dicke parameter here depends on the shift in equilibrium position of the $j^{th}$ ion $d_j$ relative to the spatial extent of the $n^{th}$ harmonic oscillator wavefunction, $\Delta z_n = \sqrt{\hbar/2m\nu_n}$, and is given by
\begin{align}
\varepsilon_{nj} =S_{nj} \frac{\Delta z_n \partial_z \omega_j}{\nu_n},
\end{align}
where $\omega_j := \omega_0(z_j)$. The matrix $S$ is a matrix whose columns are the normal modes of the chain, which take into account the extent to which each ion is involved in the $n^{th}$ vibrational mode. 

Aside from the above two methods of coupling qubits, the two most prominent alternatives are M{\o}lmer-S{\o}rensen gates~\cite{Sorensen1999} using detuned laser fields, and using a state dependent shifting of the ion wavepacket~\cite{Cirac2000}. Both of these schemes have the additional advantage of not depending on being in the motional ground state, so are less susceptible to decoherence of the bus qubit.

The linear trap architecture is suitable for performing computations on up to $\sim 100$ qubits~\cite{Zhu2006}. Thus scaling up to thousands of qubits or more requires a different architecture. The most prominent proposals to date include shuttling ions between linear traps where logic gates can be performed~\cite{Kielpinski2002}, using a dedicated movable ion to mediate interactions between different ions~\cite{Cirac2000}, or linking distant traps using photons~\cite{Cirac1997},\cite{Pellizzari1997}. To realise these schemes, planar ion traps are a promising candidate~\cite{Chiaverini2005,Seidelin2006}, in which ions are trapped above the surface of a 2D micro-fabricated electrode array.

The first successful implementation of a \textsc{cz} gate on ion traps was by Monroe et.\ al.~\cite{Monroe1995}. Since then there have been significant advances such as realisation of high fidelity two-qubit gates~\cite{Benhelm2008}, Toffoli gates~\cite{Monz2009}, demonstration of a 7-qubit topological error correcting code~\cite{Nigg2014}, demonstrations of ion shuttling~\cite{Blakestad2011,Moehring2011}, and demonstrations of micro-fabricated planar traps~\cite{Moehring2011,Amini2010,Merrill2011,Sterling2014}, making ion traps a strong contender for a quantum computer, or at least a candidate for an early prototype. For a more in depth review of ion traps, see e.g.~\cite{Steane1996c,Buluta2011,Ladd2010,Monroe2013,Haffner2008}.

\subsection{Donors in silicon}
\label{sec:DonIntro}

The use of dopants in silicon to perform quantum information processing was first proposed by Kane~\cite{Kane1998}. The central idea in this scheme is to replace a silicon atom in crystalline silicon by a group V atom such as phosphorus or bismuth. Group V elements have a spare electron, which is donated to the conduction band at room temperature (and hence is called a donor), a technique used in current electronics to negatively dope silicon to create n-type semiconductors. At low temperatures ($\lesssim 15 K$) the electron stays close to the donor nucleus, allowing information to be stored and processed by manipulating this electron. Such a proposal benefits from the vast research efforts into silicon-based technology for classical computation, as well as the low spin-orbit coupling in silicon and low abundance of isotopes with non-zero nuclear spin. Denoting $\{ S_k^x,S_k^y,S_k^z\} = \{ \frac{1}{2} X_k,\frac{1}{2} Y_k,\frac{1}{2} Z_k\}$ and $\{I_k^x,I_k^y,I_k^z\}$ as the spin operators of the $k^{th}$ electron or $k^{th}$ nucleus respectively, the effective spin Hamiltonian for a system of two donors in the presence of a magnetic field in the $z$-direction is~\cite{Kane1998} 
\begin{align}
H_{don} = g_e \mu_B B( S_1^z +S_1^z) - g_N \mu_N B( I^z_{1} +I^z_{2} )+  A (\underline{I}_1 \cdot \underline{S}_1+  \underline{I}_2 \cdot \underline{S}_2)+  J \underline{S}_{1}\cdot \underline{S}_{2},
\end{align}
where $\underline{S}_k = S_k^x+S_k^y+S_k^z$, $\underline{I}_k = I_k^x+I_k^y+I_k^z$. The parameter $A$ is the strength of the electron-nucleus hyperfine coupling (which depends on the group V atom used), and $J$ is the strength of the electron-electron exchange coupling. Typical values for $A$ are 117 MHz in phosphorus (nuclear spin $\frac{	1}{2}$) and 1.475 GHz in bismuth (nuclear spin $\frac{9}{2}$) and J depends on the separation, but is typically on the order of GHz. Magnetic fields used in experiment tend to be up to $<1$T (up to $g_e \mu_B B < 20$GHz). We consider donors with spacings $\lesssim 20$nm, so that the electron-electron exchange is significantly larger than the electron dipole-dipole interactions and interactions between donor nuclear spins~\cite{Zwanenburg2013}.

\begin{figure}[h]
\begin{center}
	\includegraphics[width=0.7\textwidth]{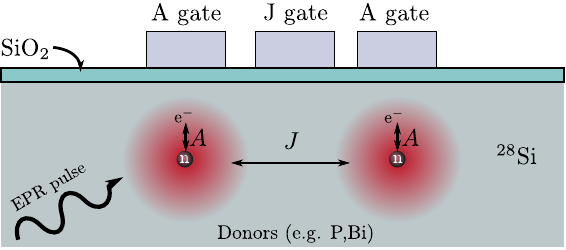}
	\caption{\label{fig:Donors} A schematic diagram of the Kane proposal for a quantum computer made from donors in silicon. }
\end{center}
\end{figure}

In the Kane proposal, the hyperfine coupling between each electron and their respective nucleus is controlled by modifying the electron wavefunction overlap with the nucleus via a single electrode placed above each donor (the `A gate', see Fig.~\ref{fig:Donors}) as recently demonstrated in~\cite{Wolfowicz2014}. By manipulating the hyperfine coupling in this way, the energy splitting of the electron spin changes, which performs a $Z$ rotation. $X$-rotations could then be achieved by applying an AC field which matches the resonant frequency of the electron spin. Individual addressability of spins could be achieved by varying the resonances of the electron spins, by e.g.\ applying a magnetic field. Measuring the electron spin could be done using spin-dependent tunnelling~\cite{Morello2009}, which has recently been demonstrated~\cite{Morello2010}, and could also be used to initialise the spins. The interactions between donor electrons are exchange interactions which depend on the wavefunction overlap between the electrons, and could be controlled using an electrode between the donors (also called a `J' gate, see Fig.~\ref{fig:Donors}) which controls this overlap and thus allows the two-qubit interactions to be switched on and off. Alternative schemes for performing two-qubit gates also exist, such as using a control atom that can be optically excited to create an interaction between two donor electrons~\cite{Stoneham2003}, or using different nuclear spin states to create interactions that are activated by an AC field~\cite{Kalra2014}.

Measurements of decoherence times for an ensemble of P atoms in $^{28}\text{Si}$ have shown nuclear $T_2$ times on the time scale of minutes, using dynamical decoupling~\cite{Saeedi2013,Steger2012} and electron $T_2$ times of up to $10\text{s}$ using dynamical decoupling~\cite{Tyryshkin2012}. For individual donors, the most recent study has shown nuclear $T_2$ times of around 30s and electron $T_2$ times of up to 0.5s using dynamical decoupling in isotopically pure silicon~\cite{Muhonen2014}. Since the nuclear spin of the donor has a much larger coherence time than the electron, the optimal strategy would most likely be to use the nuclei for long term storage of information, and use the hyperfine interaction to transfer information to the electrons so that it can be read out or manipulated quickly (as demonstrated in~\cite{Morton2008}).

For the work in this thesis, we will focus on bismuth donors in silicon, because of its large nuclear spin and hyperfine coupling relative to the other group V elements, which will will prove useful in Chapter~\ref{chap:3qubit} for creating quantum gates. Another advantage of bismuth is the presence of optimal working points \cite{Mohammady2010,Mohammady2012,Balian2012,Morley2013,Balian2014}, at which $T_2$ times can be enhanced up a time scale of seconds~\cite{Wolfowicz2013}. However in high magnetic fields, where optimal working points cannot be reached, the best achieved $T_2$ times are $\sim 0.5 \text{ms}$ in natural silicon~\cite{Morley2010,George2010} or around 700ms for isotopically pure silicon~\cite{Wolfowicz2012} For a more in depth review of donors in silicon, see~\cite{Mohammady2013,Morley2014,Zwanenburg2013,Simmons2013}.

%
\chapter{Creating gates for decoherence-free qubits}
\label{chap:DFS}

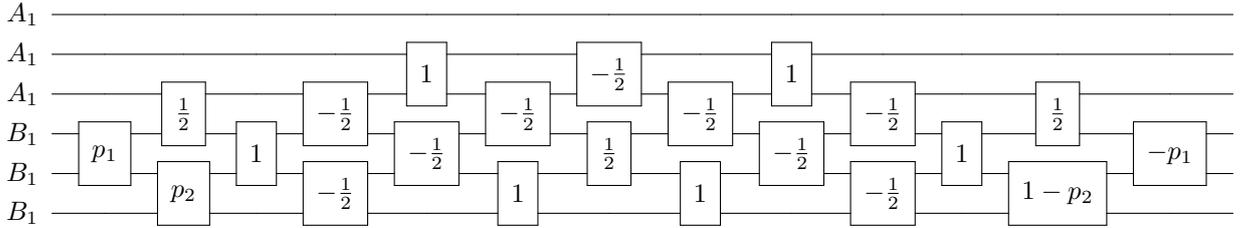
\begin{figure}[h]
\[
\Qcircuit @C=1em @R = 0.6em @!R  {
\lstick{A_1}&  \qw    &    \qw   &     \qw    & \qw  & \qw  	  & \qw     & \qw 		& \qw & \qw & \qw   & \qw  & \qw & \qw & \qw\\
\lstick{A_1}&  \qw   &    \qw     &     \qw    & \qw   & \multigate{1}{1} & \qw   																							 &  \multigate{1}{-\frac{1}{2}}  & \qw                                                                             & \multigate{1}{1}  & \qw   & \qw  & \qw& \qw & \qw\\
\lstick{A_1}&  \qw  &    \multigate{1}{\frac{1}{2}}   &     \qw     &  \multigate{1}{-\frac{1}{2}}     & \ghost{1}    & \multigate{1}{-\frac{1}{2}} & \ghost{-\frac{1}{2}}  	                                                   & \multigate{1}{-\frac{1}{2}} & \ghost{1}   & \multigate{1}{-\frac{1}{2}}  & \qw   &\multigate{1}{\frac{1}{2}}& \qw & \qw\\
\lstick{B_1}&  \multigate{1}{p_1}  &    \ghost{\frac{1}{2}}     &     \multigate{1}{1}     & \ghost{-\frac{1}{2}}   & \multigate{1}{-\frac{1}{2}}   & \ghost{-\frac{1}{2}}           & \multigate{1}{\frac{1}{2}}    & \ghost{-\frac{1}{2}}        & \multigate{1}{-\frac{1}{2}}  &   \ghost{-\frac{1}{2}} &\multigate{1}{1}  &  \ghost{\frac{1}{2}} & \multigate{1}{-p_1} & \qw\\
\lstick{B_1}&  \ghost{p_1}  	 &    \multigate{1}{p_2}   &     \ghost{1}  &  \multigate{1}{-\frac{1}{2}}     & \ghost{-\frac{1}{2}} & \multigate{1}{1}    & \ghost{\frac{1}{2}}             										& \multigate{1}{1}			 &  \ghost{-\frac{1}{2}} &   \multigate{1}{-\frac{1}{2}} & \ghost{1}  & \multigate{1}{1-p_2}& \ghost{-p_1} & \qw\\
\lstick{B_1}&  \qw   	 &    \ghost{p_2}   &     \qw  & \ghost{-\frac{1}{2}}   & \qw  & \ghost{1}    & \qw  & \ghost{1} & \qw  &  \ghost{-\frac{1}{2}}  &\qw  & \ghost{1-p_2}& \qw & \qw\\
%
}
\]
\caption{\label{fig:ExchGate} The sequence of exchange gates used in~\cite{Fong2011} to realise a 2-qubit gate between 3-qubit DF subsystem qubits. The number in each box represents the time for which the exchange gate is applied.}
\end{figure}

In Section~\ref{sec:DFSintro} we introduced the concept of decoherence-free subspaces and subsystems as ways of encoding qubits to reduce the detrimental effects of the environment on the information. The possible ways of encoding qubits to protect against strong and weak collective noise were also discussed, revealing that the 4-qubit strong DF subspace and a 3-qubit strong subsystem are the smallest encodings. Whilst these protected qubits could be useful as a quantum memory, it is desirable to perform universal quantum computation using them. To achieve this, it must be possible to perform single qubit rotations and interactions between pairs of qubits. Performing single qubit rotations in the 4-qubit DF subspace and 3-qubit DF subsystem can be realised using exchange interactions between pairs of physical qubits in the encoded qubit (see e.g.~\cite{BaconThesis,Hsieh2003}), however creating two-qubit gates is harder; the methods found so far to perform two-qubit gates involve either interactions that could be difficult to create in an experiment~\cite{BaconThesis,Bacon2000}, or involve large numbers of gates to be switched on and off sequentially (e.g.\ 22 gates in 13 time steps for the 3-qubit DF subsystem in~\cite{Fong2011}), or use perturbative/complicated control sequences to create these gates~\cite{Jiang2009}. As an example of why we might want to simplify the existing two-qubit gates in the 4-qubit DF subspace, consider one of the interactions used in~\cite{BaconThesis,Bacon2000} to create a gate:
\begin{align}
H = 3 {E}_{12} + \frac{2}{3}({E}_{24} + {E}_{23} + {E}_{34}).
\end{align}
Given the asymmetry in these interactions, this could be a challenging gate to realise (note that although this only acts within one encoded qubit, this Hamiltonian is turned on whilst we are out of the logical subspace, and so is not simply a local unitary operation). The main other alternative to realising this gate is shown in Fig.~\ref{fig:ExchGate}; here many 2-body exchange interactions are used. Whilst this gate requires very simple interactions, it also requires many steps, so is perhaps susceptible to control errors, and could perhaps be simplified by using less constrained interactions.

\begin{figure}[h]
\begin{center}
\includegraphics[scale=0.35]{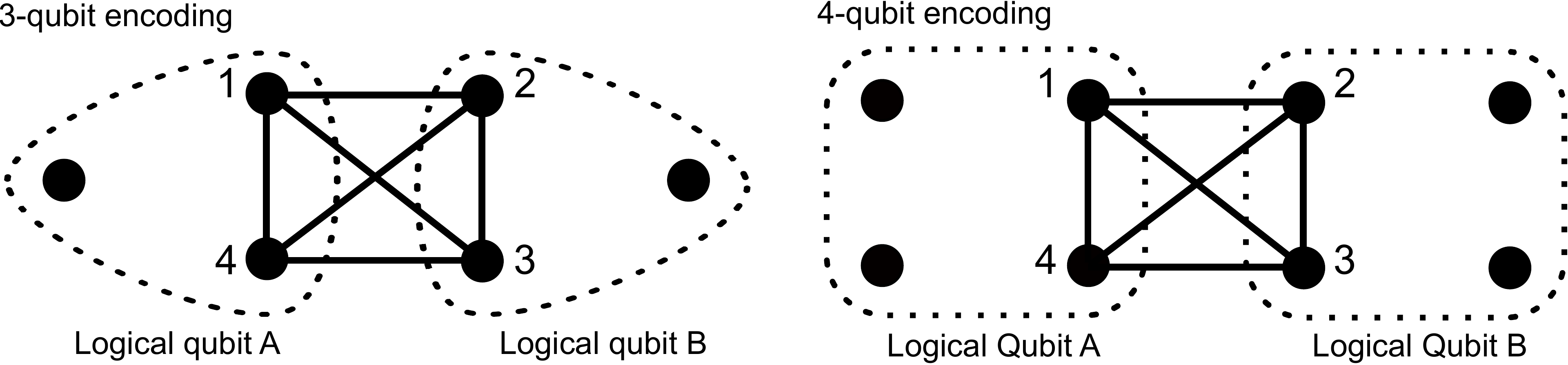}
\caption{An illustration of the two geometries of qubits we consider when constructing logical qubits. Filled circles represent physical qubits, and solid lines illustrate the kind of interactions we consider when looking for a gate in this chapter. The top diagram shows the layout for the 4-qubit encoding and the bottom diagram shows the layout for the 3-qubit encoding. }
\label{fig:Plaquette}
\end{center}
\end{figure}

In this chapter we outline alternative ways to perform 2-qubit gates in the 3-qubit DF subsystem and 4-qubit DF subspace, using less operations that require minimal control, so that realisation of decoherence-free qubits might be more attainable in an experiment (motivated by the recent experimental advances in realising quadruple quantum dots in a square configuration~\cite{Thalineau2012}).
With practicality in mind, we also only consider physical qubits arranged in a regular formation; three spins in an equilateral triangle for the 3-qubit DF subsystem, and four spins in a square for the 4-qubit DF subspace. The interactions we consider for both encodings are interactions between the middle 4 spins (see Fig.~\ref{fig:Plaquette} for an illustration of this). This work is a more detailed account of the work published in~\cite{AntonioPRA2013}.

\subsection{Ring Exchange Interactions} \label{sec:Rexchange}
When constructing a two-qubit DF gate, we will include `ring exchange' interactions. These interactions, which are used to explain  excitations in La$_2$CuO$_4$~\cite{Katanin02,Coldea01} and become important in electrons forming a Wigner crystal~\cite{Bernu2001,Voelker2001}, appear as corrections in the exchange Hamiltonian due to higher order hopping processes between different physical qubits in the extended Hubbard Hamiltonian~\cite{Takahashi1977}. They have also been investigated in the context of quantum computing~\cite{Scarola2005,Mizel2004,Mizel2004B}, and it is clear from these works that ring exchange processes should not be ignored. We will not derive the full ring exchange Hamiltonian here, but will present the main results. Consider electrons confined to quadratic potentials; when a single band tight-binding model is appropriate we can describe the physics of the system using an extended Hubbard model with fermions
\begin{align}
H = -\sum_{ \substack{ i < j,\\ \sigma,\sigma' \in \{\uparrow, \downarrow\} }} t_{ij} c^\dagger_{i,\sigma}c_{j,\sigma'}^{} + U \sum_{i} n_{i,\uparrow}n_{i,\downarrow}^{} + V \sum_{i; \sigma, \sigma' \in \{ \uparrow , \downarrow \} }n^\dagger_{i,\sigma} n_{j,\sigma'},
\end{align}
where $c^\dagger_{i,\sigma}$ ($c_{i,\sigma}$) is the creation (annihilation) operator for an electron with spin $\sigma$ at site $i$, $t_{ij}$ is the tunnelling coefficient between sites $i$ and $j$, $U$ is the on-site Coulomb interaction, and $V$ is the inter-site repulsion. In the regime where the tunnelling is weak compared to the on-site repulsion, and where the system is half-filled (i.e.\ one electron per site),  the Hamiltonian for four spin-1/2 particles located at sites 1,2,3 and 4 is~\cite{Takahashi1977}:
\begin{align}\label{Rexchange1}
{H}&= \sum\nolimits_{i < j}  J_{ij} {E}_{ij} +C_{1234} [ {E}_{12} {E}_{34} + {E}_{14} {E}_{23}  -  {E}_{13} {E}_{24}] \nonumber \\
  &+ C_{1324} [ {E}_{13} {E}_{24} + {E}_{14} {E}_{23}  -  {E}_{12} {E}_{34}]+C_{1342} [ {E}_{13} {E}_{24} + {E}_{12} {E}_{34}  -  {E}_{14} {E}_{23}]  +O \left(\frac{t^5}{U^4} \right),
\end{align}
where $t = \max_{i,j}(t_{ij})$, $E_{ij}$ is the exchange interaction defined in Sec.~\ref{sec:CircuitModel}, and 
\begin{align}
J_{ij} = \frac{t_{ij}^2}{U} - \frac{4 t_{ij}^4}{U^3}  + \sum_k \frac{t_{ik}^2 t_{kj}^2}{U^3} + \sum_{ \substack{ k\neq l,i,j \\ l \neq i,j \\ i \neq j}} \frac{t_{ik}t_{kl}t_{lj}t_{ji}}{U^3};\text{        }C_{ijkl} =\frac{t_{ik}t_{kl}t_{lj}t_{ji}}{U^3}.
\end{align}
From the form of $C_{ijkl}$ we can see that ring exchange terms will appear when there are exchange terms present which form a loop (i.e.\ when a physical qubit is indirectly coupled to itself). In~\cite{Scarola2005}, it was also found that the presence of magnetic fields changes the coefficients $C_{ijkl}$ and introduces three-body terms, with couplings that depend on the magnetic flux passing through 3 or 4-site loops. Here we assume that these magnetic fields are low enough so that the magnetic flux has negligible impact and three body terms can be ignored, and we leave the effects of non-zero magnetic fields to later work.

With four qubits arranged on a square, it is possible to make all of the exchange couplings uniform, and since this will simplify things considerably, whenever we need to include ring-exchange terms in this chapter we will assume that this is the case. Taking uniform interactions in eqn.\ (\ref{Rexchange1}) results in a `symmetric' version of the Hamiltonian, ${H}_S$, which takes the same form as derived in~\cite{Mizel2004}:
\begin{align}\label{Rexchange2}
{H}_S &= J_{\Box}{H}_{\Box} + J_{\times} {H}_{\times} + J_{\circlearrowleft} {H}_{\circlearrowleft} = J ({H}_{\Box} + {H}_{\times}) + J_{\circlearrowleft} {H}_{\circlearrowleft}  =J \left( {H}_{\Box} + {H}_{\times} + \alpha {H}_{\circlearrowleft} \right) \nonumber\\
 &= J \left(  \sum_{n=1}^4 {E}_{n,n+1} + \sum_{n=1}^2 {E}_{n,n+2} + \alpha [{E}_{12} {E}_{34} + {E}_{14} {E}_{23}  + {E}_{13} {E}_{24}] \right),
\end{align}
where ${H}_{\Box}$, ${H}_{\times}$, ${H}_{\circlearrowleft}$ represent the nearest-neighbour, next-nearest-neighbour and ring-exchange Hamiltonians, respectively, and $J_{\Box}$, $J_{\times}$, $J_{\circlearrowleft}$ are the corresponding interaction strengths for these Hamiltonians. Since we have equal coupling between all sites, $J_{\Box} = J_{\times} = J$ in this equation, and in the final line we have defined $\alpha := J_{\circlearrowleft} /J_{\Box} \equiv  J_{\circlearrowleft} /J$ as the ratio of ring exchange terms to the nearest/next-nearest neighbour terms. Restricting ourselves to symmetric Hamiltonians of this form simplifies things considerably, since ${H}_{\circlearrowleft}$ commutes with ${H}_{\Box}$ and ${H}_{\times}$, and as argued in~\cite{Mizel2004}, this form of Hamiltonian contains enough degrees of freedom to fix all of the eigenvalues, and so we do not need to take into account any higher order terms. Since single qubit rotations of the encoded qubits given in (\ref{eqn:3states}) and (\ref{eqn:4States}) can be performed by exchange interactions between physical qubits, a two-qubit interaction on these encoded qubits is equivalent to a 4-body interaction of the form $E_{ij}E_{kl}$, and since the ring exchange interactions contain terms similar in nature to these, it seems plausible that we can use these to simplify the 2-qubit gates acting on the encoded qubits. Investigating whether or not the presence of ring-exchange interactions can lead to simpler gates on the encoded qubits forms part of the motivation for this work.

\section{Two-qubit gates}\label{sec:Two}

We now look at creating a controlled-${Z}$ gate ($\textsc{cz}$) between two encoded qubits, which, along with certain single qubit gates (e.g.\ Hadamard and $U_z(\pi/4)$ gates), enables us to perform universal quantum computation \cite{Barenco95}. An important point to note is that, in this case if we can find a two-qubit gate which works for the 3-qubit DF subsystem, this may also be a valid gate for the 4-qubit DF subspace. To see this, first observe that we can rewrite the states in (\ref{eqn:4States}) in this form:
\begin{align}\label{eqn:App1}
\ket{\bar{0}^{(4)}} = \frac{1}{\sqrt{2}} \left[ |\bar{0}^{(3)}_{+1}\rangle\ket{1} - |\bar{0}^{(3)}_{-1}\rangle\ket{0} \right], \;\ket{\bar{1}^{(4)}}  = \frac{1}{\sqrt{2}} \left[ |\bar{1}^{(3)}_{+1}\rangle\ket{1} -|\bar{1}^{(3)}_{-1}\rangle \ket{0} \right].
\end{align}
A valid pulse sequence for the 3-qubit DF subsystem will perform a gate which is (locally equivalent to) a $\textsc{cz}$ gate on the two 3-qubit states up to a gauge transformation. In general, this gauge transformation may mean that this pulse sequence does not work for the 4-qubit DF subspace (it may not amount to a simple local rotation in the 4-qubit case). However, in certain situations, a valid pulse sequence for the 3-qubit DF subsystem will work for the 4-qubit DF subspace as well. For example, since exchange interactions commute with $\mathbf{S}^2$ and $\mathbf{S}_z$ over all the qubits, any interactions made up of exchange coupling only (such as the gate found in~\cite{Fong2011}, shown in Fig.~\ref{fig:ExchGate}) will not couple different gauge states. So pulse sequences for the 3-qubit DF subsystem which are made up of exchange interactions enable us to perform a $\textsc{cz}$ gate on the 4-qubit DF subspace as well.

The converse is also not necessarily true; a valid pulse sequence on the 4-qubit DF subspace will not necessarily work on the 3-qubit DF subsystem, simply because in the 4-qubit case, interactions can be over 8 physical qubits rather than 6. However, if all interactions are confined to 3 physical qubits on each logical qubit (which is the case in our protocol) then a valid pulse sequence for the 4-qubit DF subspace is also a valid gate on the 3-qubit DF subsystem. This can be seen since any gate locally equivalent to a $\textsc{cz}$ gate is also locally equivalent to the interaction $\bar{Z}_A \bar{Z}_B$ acting between two logical qubits $A$ and $B$, where $\bar{Z}_k$ is the logical $Z$ operator acting on logical qubit $k$. Since the logical $Z$ operator for the 4-qubit encoding is the same as for the 3-qubit encoding (an exchange interaction between two qubits, e.g.\ $-E_{1,2}$ using the above definitions of the states), then this gate is also locally equivalent to a $\textsc{cz}$ gate for the 3-qubit encoding. This simplifies the search considerably, as only one pulse sequence needs to be found for both of these encodings.

Performing two qubit gates is, predictably, not as straightforward as performing single qubit gates, mainly because the system can explore a much larger Hilbert space as soon as interactions between the two qubits are turned on. Since $[ \mathbf{S}^2,{E}_{ij}] = 0$ (provided $i$,$j$ are both qubits that $\mathbf{S}^2$ operates on), if two encoded qubits are initialised in the logical subspace with a total spin number $S$, then when they are interacted together using exchange interactions the system will still have overall spin number $S$, so we can use this property to restrict calculations to a small region of the full Hilbert space, to speed up calculations. We also constrain ourselves to couplings which are realistically possible; we only couple sites which can realistically be placed near each other (by e.g.\ placing two encoded qubits side-by-side, see Fig.~\ref{fig:Plaquette}). The method used to search for a quantum gate, once we had chosen a certain set of interactions to use, uses the invariant quantities found by Makhlin~\cite{Makhlin2002} (see Sec.~\ref{sec:DistMeas}).

\begin{figure}[h]
\begin{center}
\includegraphics[scale=0.3]{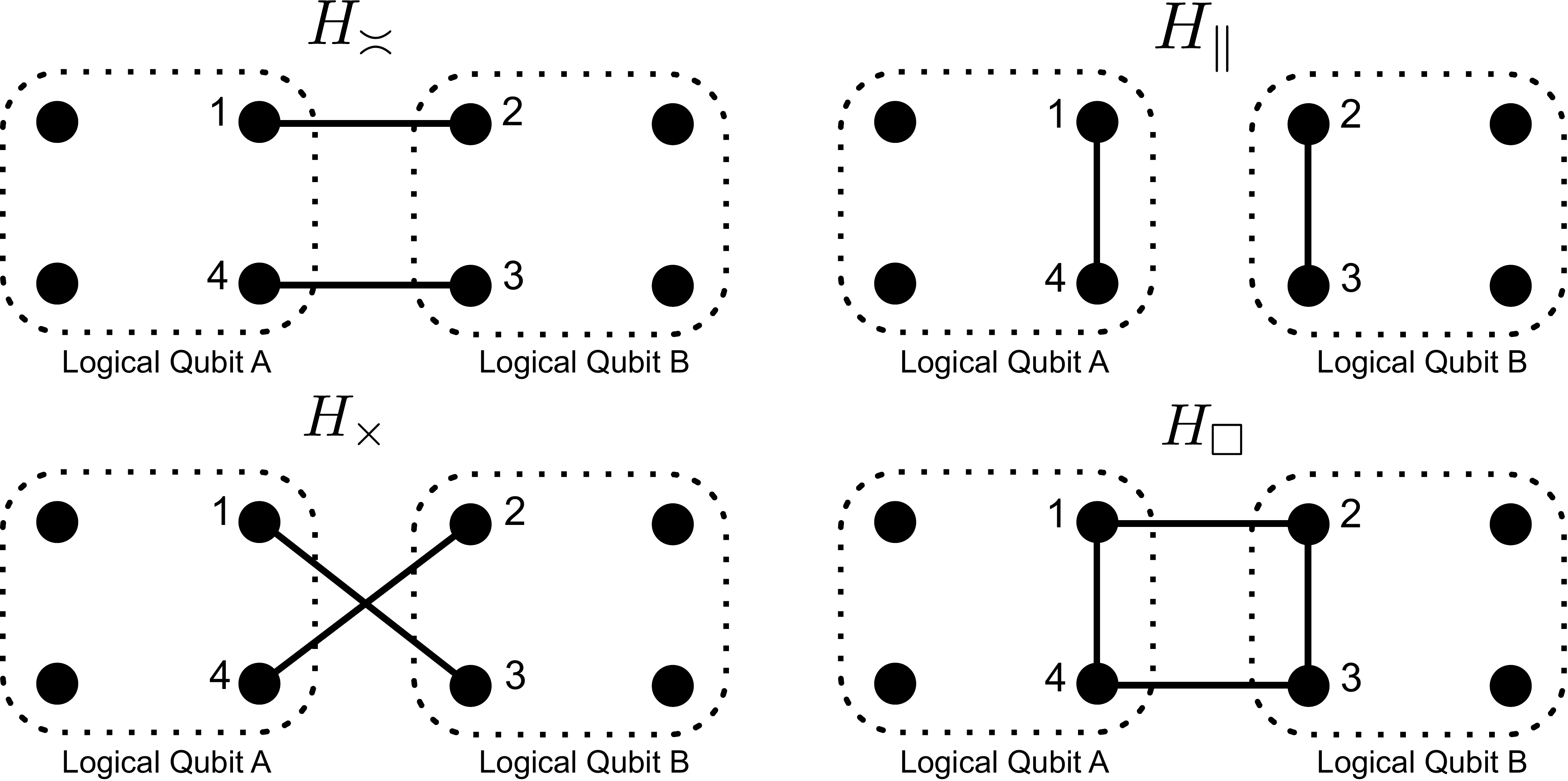}
\caption{An illustration of the interactions used to construct the gates, excluding ring exchange interactions which are difficult to represent in this form. Filled circles represent physical qubits, and solid lines represent exchange interactions.}
\label{fig:AllHams}
\end{center}
\end{figure}

The Hamiltonians that we used to construct a gate with are the following:
\begin{align} \label{HamSeq}
{H}_{\asymp} &:= {E}_{12} + {E}_{34}, \text{   } {H}_{\parallel} := {E}_{14} + {E}_{23}, \text{   }{H}_{\times} := {E}_{13} + {E}_{24}, \hspace{2mm}\nonumber\\
{H}_{\Box}& := {H}_{\asymp} + {H}_{\parallel}, \text{   } {H}_{\circlearrowleft} :={E}_{12} {E}_{34} + {E}_{14} {E}_{23}  + {E}_{13} {E}_{24}.
\end{align}
The convention for numbering the qubits is shown in Fig.~\ref{fig:AllHams}, along with an illustration of these Hamiltonians (excluding $H_{\circlearrowleft}$ which is hard to represent in pictorial form). For each Hamiltonian in this set we define a corresponding unitary operator:
\begin{align}
 {U}_{n} &= \exp \left(-i \frac{J_n  {H}_{n} \tau_n}{ \hbar} \right) \equiv \exp \left(-i  {H}_{n} \theta_n \right)\text{, e.g. }  {U}_{\asymp} = \exp(-i  {H}_{\asymp} \theta_{\asymp}), \; \mbox{etc.} \nonumber
\end{align}
where $\theta_n := J_n \tau_n / \hbar$. Note that all of the $\theta_n$ are phases, so we are free to add multiples of $2 \pi$. However, since $\theta_{n} = J_n t_{n}/\hbar$, the trade-off is that increasing $\theta_{n}$ corresponds to either increasing $J_n$ or increasing $t_{_n}$, and under the assumption that we are using the strongest couplings possible, this really means increasing the gate time by $2\pi \hbar / J$ every time we add a multiple of $2 \pi$. This is not ideal, but at least provides some more flexibility. the full gate operator is then arrived at by multiplying all of these unitaries together:
\begin{align}
{U}_{tot} = \prod_n {U}_n = \prod_n \exp(-i  {H}_{n} \theta_{n}).
\end{align}
Since the method described above requires a $4 \times 4$ matrix to compare with the ideal $\textsc{cz}$ gate, the unitary evolution $U_{tot}$ must first be projected into the logical subspaces of the 3- or 4-qubit encodings, so $U_{tot} \to U'_{tot} = P_j^{\dagger} U_{tot}P_j$, where $P_j$ is a projector into the logical subspace $\mathcal{L}_j$ of the two encoded qubits, with the subscript $j$ indicating whether it is the 3-qubit encoding ($j=3$) or the 4-qubit encoding ($j=4$). We define these projection operators as
\begin{align}
P_4 = \sum_{x,y \in \{0,1\}} |\bar{x}^{(4)} \rangle | \bar{y}^{(4)} \rangle \langle \bar{x}^{(4)}|\langle \bar{y}^{(4)}|,\text{  }P_3 = \sum_{i,j = \pm 1 } \sum_{x,y \in \{0,1\}} |\bar{x}^{(3)}_{i} \rangle | \bar{y}^{(3)}_{j} \rangle \langle \bar{x}^{(3)}_{i}| \langle \bar{y}^{(3)}_{j}|.
\end{align}
Note that when combining two 3-qubit states with $S = \frac{1}{2}$ together, there are four possibilities; three $S=1$ states and one $S=0$ states. Each of these states is 4-fold degenerate, due to the gauge choices, so $P_3$ projects into a 16-dimensional subspace overall. We can define the projectors onto each of these 4-dimensional subspaces as
\begin{align}
&P_3^{(1,1)} = \sum_{x,y \in \{0,1\} } |\bar{x}^{(3)}_{+1}\rangle |\bar{y}^{(3)}_{+1} \rangle \langle \bar{x}^{(3)}_{+1} | \langle \bar{y}^{(3)}_{+1}|,\; P_3^{(1,0)} = \frac{1}{2}\sum_{i,j= \pm 1}\sum_{x,y \in \{0,1\} } | \bar{x}^{(3)}_{i}\rangle |\bar{y}^{(3)}_{-i}\rangle \langle \bar{x}^{(3)}_{j}| \langle \bar{y}^{(3)}_{-j}| \nonumber\\
&P_3^{(1,-1)} := \sum_{x,y \in \{0,1\}} |\bar{x}^{(3)}_{-1} \rangle | \bar{y}^{(3)}_{-1} \rangle \langle \bar{x}^{(3)}_{-1} | \langle \bar{y}^{(3)}_{-1}| \nonumber\\
&P_3^{(0 ,0)} = \frac{1}{2} \sum_{i,j = \pm1}\sum_{x,y \in \{0,1\} } (-1)^{\frac{i+j+2}{2}}  | \bar{x}^{(3)}_{i} \rangle | \bar{y}^{(3)}_{-i} \rangle \langle \bar{x}^{(3)}_{j}| \langle \bar{y}^{(3)}_{-j} |.
\end{align}
where $P_3^{(S,m_S)}$ projects into the subspace with quantum numbers $S,m_S$. In order to find the Makhlin invariant we must project into a 4-dimensional subspace; for the 3 -qubit encoding, we could project onto each of the $S=1$ subspaces and the $S=0$ subspace individually, and verify that the gate works in each subspace. However, this is not necessary, since $[\sum_{i=1}^N \sigma_i^\pm,E_{nm}] = 0$ if  $n,m \in \{ 1,2,...,N\}$, which means that if we have a $\textsc{cz}$ gate which works in one of the $S=1$ subspaces, then this gate also works in any of the other two $S=1$ subspaces. So if we can find a gate which is locally equivalent to a $\textsc{cz}$ in any one of the $S=1$ subspaces and also the $S=0$ subspace, it will be locally equivalent to a $\textsc{cz}$ gate when acting on the overall 3-qubit DF subsystem up to some gauge transformation.

As shown above, any parameters which work for the 4-qubit encoding will also work for the 3-qubit encoding (since we use exchange interactions restricted to the middle four physical qubits), so rather than searching for two separate parameters for the 4- and 3-qubit encodings, we can just search for gates which work for the 4-qubit encoding, which simplifies this process.
So we take the projected unitary $ U'_{tot} = P_4^{\dagger} U_{tot}P_4$, find the corresponding Makhlin invariants $m_1({U}'_{tot}),m_2({U}'_{tot})$, and compare these to the Makhlin invariants $m_1(\textsc{cz}),m_2(\textsc{cz})$ of an ideal $\textsc{cz}$ gate using the function $f_m(\textsc{cz},U_{tot})$ defined in Sec.~\ref{sec:DistMeas}, which gives a measure of how close $U_{tot}'$ is to a $\textsc{cz}$ gate, excluding local unitary rotations. Minimising over $f_m$ gives possible ways to implement a gate (or will give evidence that it isn't possible for the considered type of interactions). The local operations required to transform our result to a $\textsc{cz}$ gate will be easy to find compared to the difficulty of minimising over $f_m$, and so we focus on finding sequences which minimise $f_m$ without searching for the local operations. We also need to consider how far out of the logical subspace the system is. One way to do this is to take a state inside the logical subspace $\mathcal{L}$, evolve it with the unitary $U_{tot}$, and then find what the probability of being in the logical subspace is. This can be written $\sum\nolimits_{\ket{m} \in \mathcal{L}} |\bra{m} U_{tot} \ket{n}|^2$ for some input state $\ket{n} \in \mathcal{L}$. Since we are interested in how the gate behaves for all possible input states, this can be turned into an average probability $\bar{p}$ of staying in the logical space by summing over the inputs $\ket{n}$ and dividing by the size of $\mathcal{L}$
\begin{align}
\bar{p} := \frac{1}{|\mathcal{L}|}\sum\nolimits_{\ket{m},\ket{n} \in \mathcal{L}} |\bra{m} U_{tot} \ket{n}|^2 = \frac{1}{|\mathcal{L}|} \Vert P_\mathcal{L} U_{tot} P_\mathcal{L} \Vert_F^2,
\end{align}
where $\| U \|_F$ is the Frobenius norm of $U$ defined as $\| A \|_F = \sqrt{\sum_{i=1}^m \sum_{i=1}^n |A_{ij}|^2}$ for an $m \times n$ matrix $A$, and $P_\mathcal{L}$ projects into $\mathcal{L}$. Then we define the leakage parameter as $L = 1-\bar{p}$, giving for the 3 and 4-qubit encodings respectively:
\begin{align}
\label{eqn:Leakage}
L_3 := 1 - \frac{1}{16}\| P_3 {U}'_{tot} P_3 \|_F^2,\text{  } L_4 := 1 - \frac{1}{4}\| P_4 {U}'_{tot} P_4 \|_F^2.
\end{align}
Note the division by 16 and 4 since we are considering systems of two encoded qubits. Using these measures, we searched for suitable parameters to make a \textsc{cz} gate using a genetic algorithm (see e.g.~\cite{GeneticBook}), followed by a Nelder-Mead simplex search~\cite{NelderMead} once the search had been narrowed down to a sufficient level. The same method has been used in~\cite{DiVincenzo2000} and~\cite{Hsieh2003}. 

Finally we note an identity which makes this search over parameters easier to make. As noted in Section~\ref{sec:Rexchange}, we consider ${H}_{\circlearrowleft}$ such that it commutes with ${H}_{\times}$ and ${H}_{\Box}$ (and ${H}_{\times}$ and ${H}_{\Box}$ also commute), which allows us to rearrange  ${U}_{\times} {U}_{\Box} {U}_{\circlearrowleft}$ as:
\begin{align}\label{eqn:DFSIden1}
 {U}_{\times} {U}_{\Box} {U}_{\circlearrowleft} &= e^{-i {H}_{\times} \theta_{\times}} e^{-i {H}_{\Box} \theta_{\Box}} e^{-i {H}_{\circlearrowleft} \theta_{\circlearrowleft}} = e^{-i ({H}_{\times} \theta_{\times} + {H}_{\Box} \theta_{\Box} + {H}_{\circlearrowleft} \theta_{\circlearrowleft})} \nonumber \\ 
&= e^{-i ({H}_{\times} (\theta_{\times} - \theta_{\Box}))} e^{(-i  {H}_{\times}\theta_{\Box}  + {H}_{\Box}\theta_{\Box} + {H}_{\circlearrowleft}\theta_{\circlearrowleft} )} = e^{-i ({H}_{\times} (\theta_{\times} - \theta_{\Box}))} e^{-i J\tau_{\Box}( {H}_{\times}  + {H}_{\Box} + (J_{\circlearrowleft}/J) {H}_{\circlearrowleft} )} \nonumber\\
&= e^{-i ({H}_{\times} (\theta_{\times} - \theta_{\Box}))} e^{-i \theta_{\Box} ( {H}_{\times}  + {H}_{\Box} + \alpha {H}_{\circlearrowleft} )} = {U}_{\times}' {U}_S,
\end{align}
where
\begin{align}
{U}_S  = e^{-i[{H}_{\Box} + {H}_{\times} + \alpha {H}_{\circlearrowleft} ]} \equiv e^{-i {H}_S \theta_S},\text{   }{U}_{\times}'  = e^{-i ({H}_{\times} (\theta_{\times} - \theta_{\Box}))}  =e^{-i {H}_{\times} \theta_{\times}'} .
\end{align}
In the above we set $\tau_{\Box} = \tau_{\circlearrowleft}$ since $H_{\Box}$ and $H_{\circlearrowleft}$ operate at the same time, and recall that $\alpha = J_{\circlearrowleft}/J $ and $J_{\Box} = J_{\times} = J$ in order to use the symmetric form of the ring exchange interaction, $H_{\circlearrowleft}$. Then using this identity, if there are within the gate sequential applications of the Hamiltonians ${H}_{\times}$, ${H}_{\Box}$ and ${H}_{\circlearrowleft}$ with parameters $\theta_{\Box}$, $\theta_{\times}$, $\theta_{\circlearrowleft}$ respectively, these can be expressed as the Hamiltonian in eqn.\ (\ref{Rexchange2}) plus an additional next-nearest-neighbour term ${U}_{\times}' $ with parameter $\theta_{\times}'$ out in front (if $\theta_{\times}'$ is negative we can just add $2 \pi$ since it is just a phase). Also note that, since we have set $t_{\Box} = t_{\circlearrowleft}$, then $\alpha = \theta_{\circlearrowleft} / \theta_{\Box}$. This identity means that for the purposes of the minimisation we can treat ${H}_{\times}$, ${H}_{\Box}$ and ${H}_{\circlearrowleft}$ as if they were separate interactions with independent parameters and then combine them together at the end using the identity in eqn.\ (\ref{eqn:DFSIden1}), and when we do combine them the value of $\alpha$ is set by $\theta_{\circlearrowleft} / \theta_{\Box}$.


\subsection{Results}

Starting with the simplest Hamiltonian which might result with ring exchange interactions, we tried many combinations of the Hamiltonians given above, performing a genetic search for each one followed by a Nelder-Mead search (typically with around 200 iterations of genetic search, then to within tolerance of $10^{-12}$ with the Nelder Mead search, and stopping if nothing was found after 100 attempts). Using the following sequence of interactions did not yield any gates with $f_m < 0.001$ 
\begin{align}
{U}_{1} = {U}_{\times} {U}_{\Box} {U}_{\circlearrowleft}, \; {U}_{2} = {U}_{\asymp}^{(1)} {U}_{\times} {U}_{\Box} {U}_{\circlearrowleft} {U}_\asymp^{(2)}.
\end{align}
The reason for choosing these sequences in particular is that we are interested in forming a gate from the ring exchange, and these are the simplest sequences that contain a ring exchange interaction (with local rotations before and after the ring exchange). By adding one more interaction, we can find a sequence which does work:
\begin{align} \label{Usequence}
{U}_{gate}= {U}_{\asymp}^{(1)} {U}_{\parallel} {U}_{\times} {U}_{\Box} {U}_{\circlearrowleft} {U}_\asymp^{(2)} = {U}_{\asymp}^{(1)} {U}_{\parallel} {U}_{\times}' {U}_S  {U}_{\asymp}^{(2)},
\end{align}
with corresponding parameters $\{ \theta_{\asymp}^{(1)}, \theta_{\parallel}, \theta_{\times}, \theta_{\Box}, \theta_{\circlearrowleft}, \theta_\asymp^{(2)} \}$, where the superscripts (1) and (2) are used to differentiate between the interactions used at the beginning and at the end of the gate. This is a gate which requires only $5$ separate pulses to perform (since identity (\ref{eqn:DFSIden1}) can be used). Note that we also attempted to form a gate using the sequence in (\ref{Usequence}) but with $\theta_\times = \theta_\Box$ in an effort to reduce the number of pulses, but no solutions were found.

Using this combination of interactions, with the parameters given in Table~\ref{thetaVals}, we were able to find a gate with a value of $f_m = O(10^{-16})$ and leakage $L = O(10^{-16})$ (i.e.\ both around the machine precision of $O(10^{-16})$, suggesting the existence of an exact solution). If we assume that for each interaction the coupling strength is limited to some maximum possible value $J_{max}$, then we can find the total gate time $T$ in units of $\hbar /J_{max}$, as an indicator of how long this gate would take compared to other gates. Note that we do not simply add the parameters in Table~\ref{thetaVals}, since we apply the identity in eqn.\ (\ref{eqn:DFSIden1}) first, so in fact the true gate time $T$ is
\begin{align}
T =\left( \sum_n \theta_n \right) - \theta_{\times} + (\theta_{\times}-\theta_{\Box})\textrm{mod}\;{2\pi}.
\end{align}
This gives a gate time of $16.7\; \hbar /J_{max}$. The gate in~\cite{Fong2011} has a total time of 9.9 in these units, and although we found other parameter sets which yielded similar gate times to this, we have picked the parameters with the most realistic ring exchange couplings (see Sec.~\ref{sec:Constraints}).

\begin{table}[htbp]
\begin{centering}
\begin{tabular}{c c c c c c | c c  c}
$\theta_{\asymp}^{(1)}$ & $\theta_{\parallel}$ & $\theta_{\times}$ & $\theta_{\Box}$& $\theta_{\circlearrowleft}$& $\theta_{\asymp}^{(2)}$& $f_m$ & $L$ &  $\alpha$ \\[5pt] \hline 
2.748894 &  4.319690 &  2.552544  &  3.730678 &  0.589049 &  0.785361  & $O(10^{-16})$ & $O(10^{-16})$ &  0.158
\end{tabular}
 \caption{\label{thetaVals}Parameters which realise a $\textsc{cz}$ gate, up to local rotations.}
\end{centering}
\end{table}

\section{Constraints on the ring exchange strength} \label{sec:Constraints}

We are constrained in our choices of parameters, as the relative sizes of the nearest-neighbour couplings, $J$ and ring-exchange coupling $J_{\circlearrowleft}$, are set by the ratio $\alpha = J_{\circlearrowleft}/J = \theta_{\circlearrowleft} /\theta_{\Box}$, since as we have seen in eqn.\ (\ref{eqn:DFSIden1}) we end up turning on the  Hamiltonian $ {H}_{\times}+{H}_{\Box}+ \alpha {H}_{\circlearrowleft}$ for some time $t_{\Box} =t_{\circlearrowleft}=\theta_{\Box} \hbar /  J$. This means we are constrained to situations where we can set $\alpha$ to $\theta_{\circlearrowleft} /\theta_{\Box}$. We would expect $\alpha \lesssim 0.17$ (see e.g.~\cite{Katanin02,Coldea01,Mizel2004}), which is why we have chosen to use the particular parameter set shown in Table~\ref{thetaVals} which has $\alpha = 0.158$.

We are not completely constrained to this value, since all of the $\theta$ values are just phases. As discussed in Sec.~\ref{sec:Two}, we are free to add factors of $2 \pi$ to any of them, with the gate time increasing by $2\pi \hbar / J$ every time we add a multiple of $2 \pi$. However, even with this freedom, we still seem to be tied down to a few precise values of $\alpha$. To get around this, we notice that we can split up the ring exchange into two parts:
\begin{align}
{U}_S &= \exp \left(-i {H}_{\times} \theta_{\Box} \right) \exp \left(-i {H}_{\Box} \theta_{\Box} \right) \exp \left(-i {H}_{\circlearrowleft} \theta_{\circlearrowleft} \right) \nonumber \\
&= \exp \left(-i {H}_{\times} (\theta_{\Box}^a + \theta_{\Box}^b) \right) \exp \left(-i {H}_{\Box} (\theta_{\Box}^a + \theta_{\Box}^b)\right) \exp \left(-i {H}_{\circlearrowleft} (\theta_{\circlearrowleft}^a + \theta_{\circlearrowleft}^b) \right) \nonumber\\
&= \exp \left(-i {H}_{\times} \theta_{\Box}^a \right)\exp \left(-i {H}_{\Box} \theta_{\Box}^a \right) \exp \left(-i {H}_{\circlearrowleft} \theta_{\circlearrowleft}^a \right) \exp \left(-i {H}_{\times} \theta_{\Box}^b \right)\exp \left(-i {H}_{\Box} \theta_{\Box}^b \right) \exp \left(-i {H}_{\circlearrowleft} \theta_{\circlearrowleft}^b \right) \nonumber \\
&= {U}_{\times}^a {U}_{\Box}^a {U}_{\circlearrowleft}^a  {U}_{\times}^b{U}_{\Box}^b {U}_{\circlearrowleft}^b:={U}_S^a {U}_S^b 
\end{align}
so we now have two ring exchange interactions, ${U}_S^a$ and ${U}_S^b$ which have the same form as $H_S$ but with different values of $\alpha$. Now since $\theta_n := J_n t_n / \hbar$, and since ${U}_S^a{U}_S^b = {U}_S$ and all of the terms commute, this means that:
\begin{align}\label{eqn:Constr}
&J^a t_a + J^b t_b = \theta_{\Box}, \; J_{\circlearrowleft}^a t_a + J_{\circlearrowleft}^b t_b = \theta_{\circlearrowleft}.
\end{align}
where $\theta_{\Box}$, $\theta_{\circlearrowleft}$ are the parameters we found in the search in Sec.~\ref{sec:Two}, and we have defined $\theta_{\Box}^a = J^a t_a$, $\theta_{\Box}^b = J^b t_b$. For simplicity, we take $J^a = J^b = J$, without loss of generality, since we are free to scale these parameters as we wish, provided we scale the corresponding $J_{\circlearrowleft}$ values correctly. Since $\theta_{\Box}$ and $\theta_{\circlearrowleft}$ are phases, we are free to add multiples of $2\pi$ to these values, so we replace $\theta_{\Box}$ with $\theta_{\Box}^{(n)}$, and rearrange equation (\ref{eqn:Constr}) to give
\begin{align}
&t_a = \frac{\theta_{\Box}^{(n)}}{J}\left[ \frac{\alpha^{(n)} - \alpha_b}{\alpha_a - \alpha_b} \right] , \;t_b = \frac{\theta_{\Box}^{(n)}}{J}\left[ \frac{\alpha_a- \alpha^{(n)} }{\alpha_a  - \alpha_b} \right],
\end{align}
where $\alpha_a := \theta_{\circlearrowleft}^a / \theta_{\Box}^a =J_{\circlearrowleft}^a / J^a $, $\alpha_b := \theta_{\circlearrowleft}^b / \theta_{\Box}^b = J_{\circlearrowleft}^b / J^b$, and $\alpha^{(n)}$ is defined above. For $t_a$ and $t_b$ to be positive, we need couplings such that $\alpha_a > \alpha^{(n)}  >\alpha_b$. So this tells us that if we are able to control the relative strengths of the ring and nearest-neighbour terms, and if we could get them such that $\alpha_a >\alpha^{(n)} >\alpha_b$ is satisfied, then regardless of what the actual values of $\alpha_a$ and $\alpha_b$ are, we can create the $\textsc{cz}$ gate (at the expense of adding more interactions and thus increasing the time of the gate).


\section{Including noise}\label{sec:Performance}
We now look at the performance of this gate under the influence of noise. The two types of noise we consider are errors in coupling strengths or timing of the gates (i.e.\ random errors in the coupling parameters when implementing each of the Hamiltonians in (\ref{Usequence})) and fluctuations in magnetic fields acting on the qubits during the gate implementation (which would normally be protected by the decoherence-free or supercoherent properties, but are not while the gates are being implemented). The reasons for picking these particular types of errors are that one of the most promising systems in which to implement these encoded qubits is in arrays of quantum dots, and there have been several advances towards achieving these, e.g.~\cite{Laird2010,Thalineau2012}. These quantum dots are susceptible to errors in exchange coupling due to charge fluctuation~\cite{Burkard1999} (which we are modelling by fluctuations in the $\{ \theta \}$ values) and fluctuations in external magnetic field due to the nuclear spin bath or stray magnetic fields (see e.g.~\cite{Taylor2007}). When looking at these magnetic fluctuations, we look at two cases; in the first case we assume that magnetic field fluctuations are roughly uniform over each encoded qubit (so that there is collective decoherence acting on it), but the magnetic fields acting on different encoded qubits are different in magnitude and direction. In the second case, we consider magnetic field fluctuations acting independently on each physical qubit (i.e.\ a situation where a supercoherent qubit would be more appropriate). For all errors, we assume that the time scale for the fluctuations is large compared to the time to perform the gate, which is typically the case~\cite{Hu2006,Merkulov2002,Taylor2007}. We also compare the susceptibility of the 3-qubit and 4-qubit encodings to errors, since there may be difference.

To measure the effects of noise, we calculate the gate fidelity using the techniques of quantum process tomography discussed in Section~\ref{sec:OSR}. The unitary transformation applied to the encoded qubits is $e^{i\Phi} {V}^{\dagger}{U} {V}$, where $V$ is a local rotation, $\Phi$ is a global phase, and $U$ is the unitary found in Sec.~\ref{sec:Two} which is locally equivalent to a $\textsc{cz}$ gate. To find the local operations $V$, we take the gate with zero noise ${U}$ corresponding to the gate sequence found in Sec.~\ref{sec:Two}, and find the matrix ${V}$ which diagonalises it, and then find $\Phi$ such that e.g. $e^{i\Phi} \bra{\bar{0}^{(4)}}{V}^{\dagger}{U} {V} \ket{ \bar{0}^{(4)}} =1$ for the 4-qubit encoding. Once we have a gate with noise added, ${U}'$, ${V}$ and $\Phi$ are still used to convert ${U}'$ into a noisy $\textsc{cz}$ gate (i.e. $e^{i\Phi} {V}^{\dagger}{U}' {V} \approx \textsc{cz}$), since we are restricted to always using the same local operations (i.e.\ we cannot change our local operations since we don't know what the noise is doing to our system). 

Following this unitary evolution, we imagine performing a measurement on each of the encoded qubits to test whether or not they are still in the logical subspace, which corresponds to applying the projectors $P = P_3 $ or $P_4$ depending on whether we are considering the 3- or 4-qubit encoding. The probability of the qubit being in the logical subspace is given by the leakage defined in (\ref{eqn:Leakage}). Overall then, a state $\rho$ is transformed by this process according to $\mathcal{E} (\rho) = \sum_i {E}_i \rho {E}_i^{\dagger}$ where the Kraus operators take the form
\begin{align}
E_1= \sqrt{1-L}\text{ } P e^{i\Phi} {V}^{\dagger}{U} {V}, \; E_2=\sqrt{ L} \text{ } (1-P) e^{i\Phi} {V}^{\dagger}{U} {V}
\end{align}
where $P = P_3 $ or $P_4$.

Rather than finding the fidelity of this process as a whole, it is more useful to decompose the analysis of this noisy gate into two parts: the probability $L$ of being outside the logical subspace, and the process fidelity of the gate when projected into the logical subspace, since we don't mind what happens to states outside the logical subspace / subsystem. Therefore we construct a process matrix $\chi$ from a process $\mathcal{E}'$ involving $E_1$ alone:
\begin{align}
\mathcal{E}' (\rho) &=  \frac{1}{(1-L)}{E}_1 \rho {E}_1^{\dagger} = \sum_{mn} \chi_{mn} {A}_m \rho {A}_n^{\dagger} 
\end{align}
where $\{ {A}_m \}$ is a set of $4 \times 4$ matrices that form an orthogonal basis, and we have divided by $(1-L)$ to account for the probabilistic nature of the measurement. We use the process fidelity $F_p(\chi,\chi_{\textsc{cz}})$ to compare this to the process matrix of an ideal \textsc{cz} gate, and then $(1-(1-L)F_{p})$ has the interpretation as the upper bound of the average failure probability $\bar{p}_e$~\cite{Gilchrist2005}, making this a natural choice for measuring the accuracy of the gate. Note that a subtlety arises when dealing with the 3-qubit DF subsystem, in that $P_3$ projects onto a 16-dimensional space, so we cannot directly compare this to the 4-dimensional $\textsc{cz}$ gate. To get around this, we can instead project onto only one of the gauge degrees of freedom, so for instance we could replace $P_3$ with $P_3^{(1,1)}$ (defined in Sec.~\ref{sec:Two}),
provided $P_3^{(1,1)} U P_3^{(1,1)} \ne 0$, otherwise we could project onto any of the other 3 gauge subspaces. Then instead of normalising $\chi$ by dividing by $(1-L)$, we can just divide by $\textrm{tr}(\chi)$ since $\textrm{tr}(E_1 \rho E_1) = \textrm{tr}(\chi)$, and the effect is the same. Note that the leakage is still calculated in the same way as in Sec.~\ref{sec:Two}, independently of the method used to project into the 4-dimensional subspace.

\subsection{Coupling Errors}

\begin{figure}[h]
\centering\includegraphics[width = 0.6\textwidth]{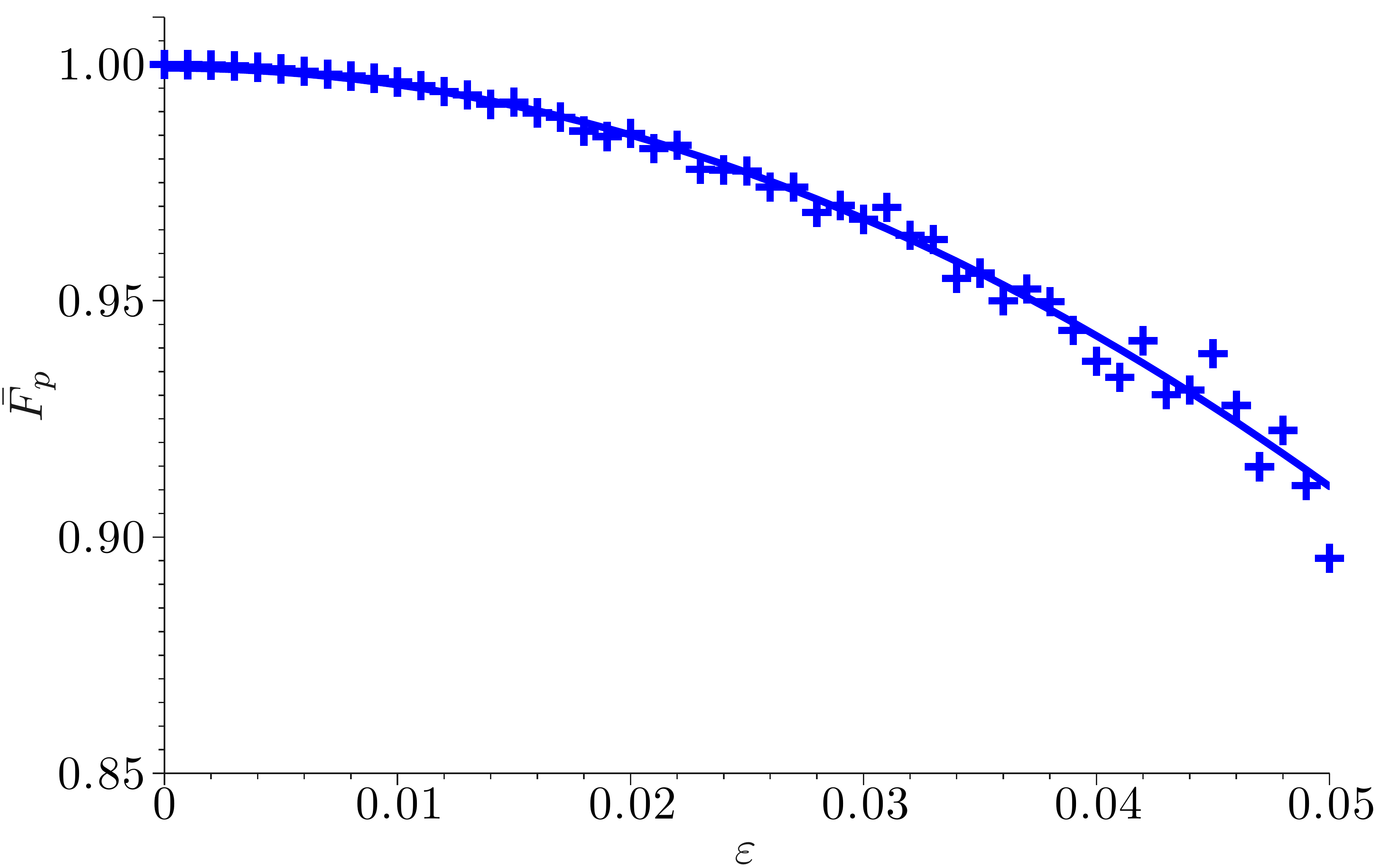}
\caption{The average process fidelity $\bar{F}_p$ when performing the gate with random fluctuations in gate times, where the fluctuations have standard deviation $\varepsilon$. Only results for the 4-qubit encoding are shown, since both encodings showed a similar behaviour. A fit to a curve of the form $y = 1 - c \varepsilon^2$ is also shown, with $c=35.4$.}
\label{TimingErrors}
\end{figure}

To simulate random errors in values of the couplings between physical qubits, charge fluctuations or timing when implementing gates, we added random noise to each of the $\theta_n$ parameters:
\begin{equation}
\theta_n \rightarrow \theta_n + \delta\theta_n,
\end{equation}
where $\delta\theta_n$ is sampled from a Gaussian distribution. Over the range $\varepsilon \in [0,0.05]$ of $\varepsilon$, we calculated the process fidelity over 250 iterations taken from a Gaussian distribution with standard deviation $\varepsilon$ and mean 0, finding the average over all of these iterations (note that all the interactions commute with $\mathbf{S}^2$, so the system is confined to a subspace of constant $S$). The results are shown in Fig.~\ref{TimingErrors}, with only one set of results shown since both the 3- and 4-qubit encoding give very similar results. The process fidelity falls off slowly and stays above $0.9$ for $\epsilon \lesssim 0.05$. A reasonable estimate of these fluctuations in gate couplings would be around $0.01$~\cite{Burkard1999}, at which point both gates have fidelity $\sim 0.99$, so we can see that these gates still have high fidelity even with this level of noise. Over the entire range of $\epsilon$, the leakage stayed below $0.003$, and the leakage at $\epsilon \sim 0.01$ is around $O(10^{-6})$. So overall we can achieve an overall average gate failure probability of $\bar{p}_e \lesssim 0.01$ even with a reasonable level of coupling errors. 

\subsection{Magnetic fluctuations}

 \begin{figure}[h!]
  \centering
 \subfloat[]{\includegraphics[width = 0.7\textwidth]{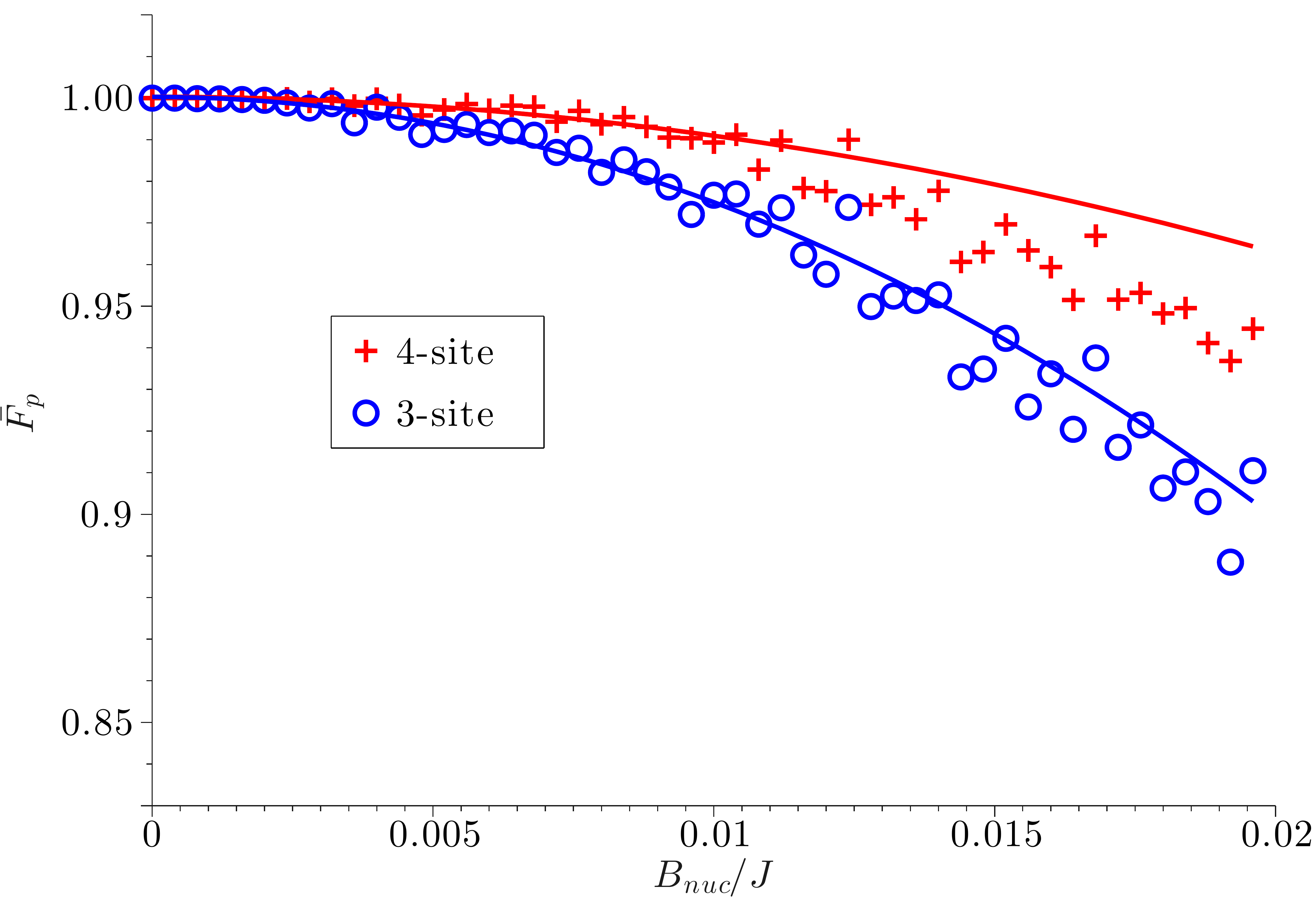}} \\
 \subfloat[]{\includegraphics[width = 0.7\textwidth]{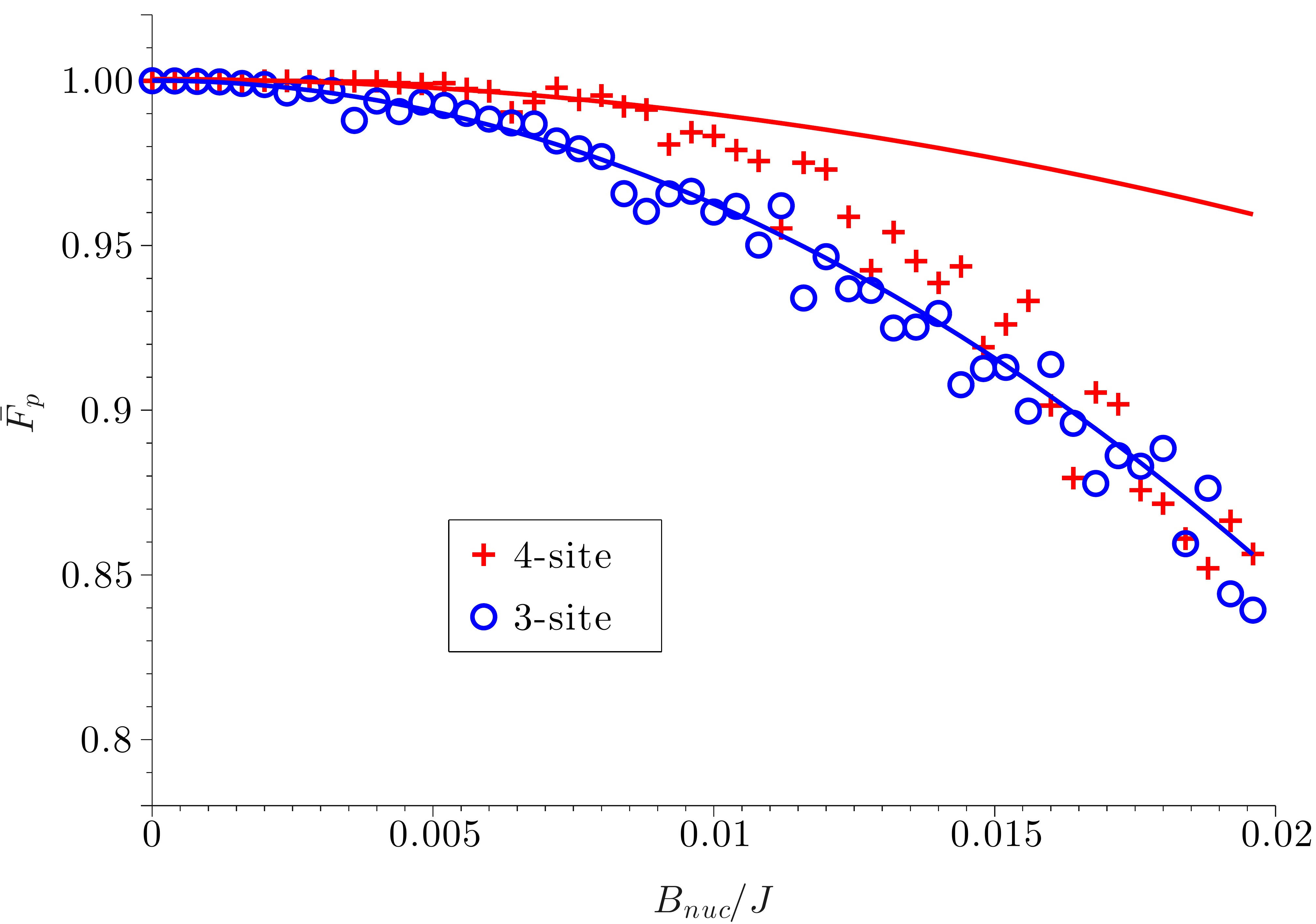}}
 \caption{\label{fig:MagEvolveLong}Average process fidelity $\bar{F}_p$ as a function of $B_{\textrm{nuc}} / J$, the strength of magnetic field fluctuations relative to the exchange coupling, with a) random uniform fields acting on each logical qubit and b) random fields on each physical qubit. The range is over $B_{\textrm{nuc}} / J= 0 \rightarrow 0.02$, and results corresponding to a 3-qubit or 4-qubit encoding are both shown. A fit of all data up to $B_{\textrm{nuc}} / J = 0.01$ to a curve of the form $y = 1 - c ( B_{\textrm{nuc}} / J)^2$ is also shown, with $c_4=95.6$, $c_3=252.9$ for the 4-qubit and 3-qubit encoding respectively. Note that we use this fit to see roughly when the fidelity starts to deviate from a quadratic decay, not as a best fit to the data.}
 \end{figure}

If we were to implement this supercoherent qubit in a quantum dot, then there could be random magnetic field fluctuations due to the nuclear spins in the substrate material, or stray magnetic fields. We studied two scenarios which could occur; in one case we looked at the effects of having random field fluctuations which are uniform over all physical qubits, but which may vary between encoded qubits (so that the collective decoherence assumption is valid for single encoded qubits, but when we interact two of these the assumption is not valid). In the other case we considered having independent random magnetic field fluctuations on each physical qubit. For both of these cases, we followed the arguments in~\cite{Merkulov2002, Taylor2007}, using the quasi-static approximation in which the magnetic field from the nuclei $\vec{B}$ stays constant over the time we perform the gate, has random direction and has magnitude $| \vec{B} |$ following a Gaussian probability distribution
\begin{align}
P(| \vec{B} |) = \frac{1}{(2 \pi  B_{\textrm{nuc}}^2)^{3/2}} \exp(- | \vec{B} |^2 / 2B_{\textrm{nuc}}^2),
\end{align}
 where $B_{\textrm{nuc}}$ is the standard deviation in fluctuations of magnetic field. We took the average over 250 iterations, with each iteration having a different magnitude and direction of magnetic field sampled from the above Gaussian distribution. The magnetic field strength was taken relative to the nearest-neighbour coupling strength $J$. In each case we found how the process fidelity varied for the 3-qubit and 4-qubit encoding. The results are shown in Fig.~\ref{fig:MagEvolveLong} a) and b), together with a fit of all data up to $B_{\textrm{nuc}} / J= 0.01$ to a curve of the form $\bar{F}_p = 1 - c \left( B_{\textrm{nuc}} / J  \right)^2$ (we use this fit to see roughly when the fidelity starts to deviate from a quadratic decay, not as a best fit to the data).

 Based on the exchange values given in~\cite{Loss1998} and the values for $B_{\textrm{nuc}}$ given in~\cite{Taylor2007}, we would expect $B_{\textrm{nuc}}/J$ to be at most $\sim 0.01$. With this kind of nuclear field present, the 4-qubit encoding achieves a process fidelity of $\bar{F}_p \sim 0.99$ and a leakage probability of $L \sim 0.002$ if there are errors across the encoded qubits, or $\bar{F}_p \sim 0.97, L \sim 0.002$ if there are errors on each individual physical qubit. The 3-qubit encoding achieves $\bar{F}_p\sim 0.97,L \sim 0.002$ if there are errors across encoded qubits, or $\bar{F}_p \sim 0.95,L \sim 0.002$ if there are errors on physical qubits. 
 
So overall, as we might expect, the qubits are much more robust to errors which are uniform on the physical qubits, but less robust to errors which vary between physical qubits. Also, the gate on the 3-qubit encoding performs slightly worse than the 4-qubit one, which we are currently unable to explain. Using a 4-qubit encoding in the presence of magnetic fluctuation across encoded qubits, with a strength we might expect in a realistic system, we can achieve process fidelities of $\bar{F}_p > 0.99$ and leakages of $L \sim 0.002$ (leading to an overall average gate failure probability of $\bar{p}_e \lesssim 0.01$). The results for the situation with errors on physical qubits are unsurprisingly worse, but we can still achieve $\bar{F}_p \sim 0.97$, $L \sim 0.002$, giving an overall average gate failure probability of $\bar{p}_e \lesssim 0.03$. These results could be improved if we reduced the effects of fluctuations in nuclear spin, such as the methods presented in~\cite{Imamoglu2003} and~\cite{Reilly2008}.


\section{Conclusions}
We have demonstrated a simple way to implement a \textsc{cz} gate in the 4-qubit decoherence-free subspace and the 3-qubit decoherence-free subsystem, using a sequence of 5 operations (excluding local operations), and including ring exchange interactions. The gate we have found minimises the Makhlin invariant function $f_m$ to within machine precision, suggesting the existence of an exact solution. The simplicity of this gate also suggests that interactions involving more than 2 physical qubits can be used to achieve simplified gates, which we might intuitively expect since a direct $\textsc{cz}$ gate on these encoded qubits would involve a four-body interaction, which is present in the ring exchange terms. Investigating this more rigorously could pose an interesting question for future research. 

We introduced errors when performing these gates, to simulate errors in coupling strength or gate times, and to simulate fluctuations in magnetic field due to some external environment, e.g.\ nuclear spins in a quantum dot. The 4-qubit gate maintained an average failure probability of $\bar{p}_e \lesssim 0.01$ even with nuclear fluctuations of around $1\%$ of $J$ over the encoded qubits, or with timing errors of up to around $1 \%$ of $J/\hbar$, where $J$ is the strength of the nearest-neighbour exchange coupling. Whilst this maintains the gate below the most generous threshold condition~\cite{Knill2005}, it is likely that it will be advantageous to keep these errors errors significantly lower in order to implement this gate in a fault tolerant manner. Techniques such as dynamical decoupling and leakage reduction, previously applied to 3-qubit encoded qubits in~\cite{West2012,Fong2011}, could perhaps be used to strengthen the performance of this gate in the presence of noise.

Such a gate could be useful in systems where ring exchange is particularly prominent, or in situations where it is particularly important to keep the number of pulses to a minimum, or where the control is limited. Further studies could be done to investigate the effects of magnetic flux on the couplings as reported in~\cite{Scarola2005}, or using more general forms of the ring exchange interaction rather than the symmetric one we have used here.

\chapter{Three qubit gates}
\label{chap:3qubit}

In this chapter, we describe different ways to create 3-qubit Toffoli and Fredkin gates, which are universal for classical \emph{reversible} computation. The origins of reversible computation stem from Landauer's erasure principle; every bit of information erased results in $kT \log 2$ heat being irreversibly lost to the environment~\cite{Landauer1961}. It follows that in order to perform computation without dissipating heat, no information must be erased, and so only a reversible computation will not dissipate heat. In 1973, Bennett~\cite{Bennett1973}, building on work by Lecerf~\cite{Lecerf1963}, showed that reversible computation can be performed just as efficiently as irreversible computation. The field was then further enhanced by Fredkin and Toffoli~\cite{Fredkin1982} (around the same time that quantum computation was being first thought of), in which they introduced the Fredkin and Toffoli gates.

The Toffoli and Fredkin gates are also useful primitives for quantum computation. The Toffoli gate is universal for quantum computation with the addition of any basis-changing single qubit gate such as a Hadamard~\cite{Shi}, and is used in some quantum error correction schemes (e.g.~\cite{Sarovar,Cory}). The Fredkin gate can be useful in fusing W states~\cite{Ozaydin2014}, implementing Deustch's algorithm~\cite{Chuan1995}, and is also used in error-correcting schemes~\cite{Barenco1997}. Although Toffoli and Fredkin gates can be synthesised by two-qubit gates and single qubit operations, being able to perform these gates in a more direct way may reduce the size or complexity of certain circuits (e.g.\ a Toffoli gate requires a minimum of five 2-qubit gates~\cite{Yu2013}).

Given the ever-increasing density of components on a chip, and the apparent absence of a Moore's law for cooling techniques, it seems likely that using a reversible architecture for classical computation will become important over the next 20 years, so that chip heating rates and power usage do not become unmanageable. Not only \emph{logical} reversibility, but \emph{physical} reversibility will be important to reduce the heat dissipated per logic gate. Quantum computation is of course inherently reversible, since transformations occur through reversible unitary transformations. However, in the absence of any operational quantum computers in the near future, it may be useful instead to use quantum gates to perform classical logic; this would replace the traditional irreversible dissipative components such as CMOS technology (which can be made physically reversible but at the expense of increasing the gate operation time~\cite{Merkle1993}) by inherently physically reversible quantum components. In addition, since the requirements for classical computation are less stringent (there are only bit flip errors, and gates only have to be coherent for the length of the gate, not the length of the computation), creating classical reversible gates with quantum components could be a good halfway point en route to full quantum computation, perhaps allowing faster gates or more miniaturised components.

With this as motivation, we consider methods to create Toffoli and Fredkin gates. We propose three different ways of constructing 3-qubit gates, using time-independent Hamiltonians (in the sense that the Hamiltonian does not change during the operation of the gate). Proposals already exist to implement Toffoli or Fredkin gates, such as qubits interacting through cavities~\cite{Chen2012,Chen2006,Xiao2007,Shao2013}, using photon interference~\cite{Deng2007}, or using systems with extra `hidden' states or higher dimensional Hilbert spaces~\cite{Fedorov2012,Lanyon2008,Zheng2013,Ivanov2011}, or using extra `mediator' qubits~\cite{Fei2012}. Toffoli gates and Fredkin gates have also been successfully experimentally implemented using `hidden' states in superconductors~\cite{Fedorov2012}, and using interactions controlled by laser pulses in ion traps~\cite{Monz2009}. In our approach, we aim to implement the gate in a single step, using interactions that are fixed over the length of the gate, in the hope that minimal control of the system will enable better isolation of the system, and thus limit decoherence. We also focus on performing these gates on systems of qubits, rather than higher-dimensional systems, and without the aid of an external systems such as cavities or ancillas. In addition, we aim to find gates that can be easily implemented experimentally, and we will give examples of how these gates could be implemented in ion traps or in doped silicon

This chapter is structured as follows; in Sec.~\ref{sec:Toff} we describe a method to implement a (locally equivalent) Toffoli gate using Ising interactions and magnetic fields. In Sec.~\ref{sec:FredHeis} we describe a method to perform a locally equivalent Fredkin gate using Ising and Heisenberg exchange interactions and magnetic fields. In Sec.~\ref{sec:XX} we describe a method to perform a locally equivalent Fredkin gate using Ising and XX interactions with magnetic fields.

\section{Toffoli gate using uniform Ising interactions}
\label{sec:Toff}

\begin{figure}[b]
\begin{center}
\includegraphics[scale = 0.5]{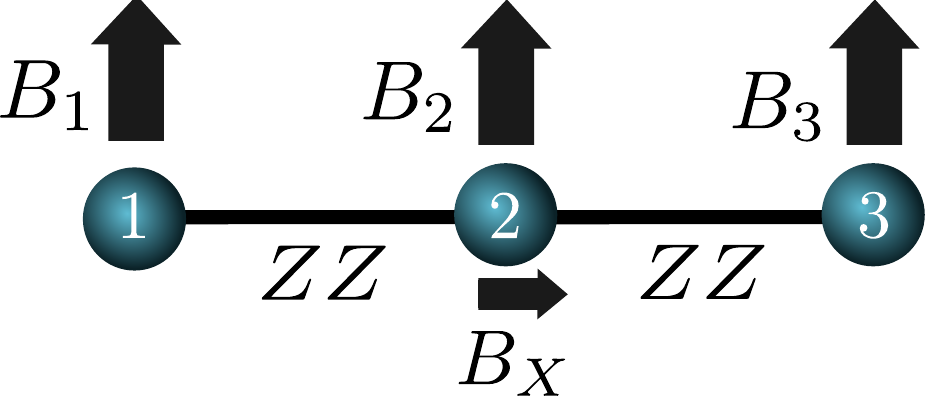}
\caption{Setup for creating a 3-qubit Toffoli gate.}
\label{fig:ZZ}
\end{center}
\end{figure}

In this section we will describe how to implement a Toffoli gate using Ising interactions, with a transverse field on the target qubit (see Fig.~\ref{fig:ZZ}). A Toffoli gate performs a controlled-controlled-\textsc{not} operation $\ket{ij}\ket{k} \to \ket{ij}\ket{k \oplus ij}$. To construct this gate using a single fixed Hamiltonian, with qubits 1 and 3 as controls and qubit 2 as target, we consider a Hamiltonian of the form
\begin{align}\label{eqn:HamToff1}
H_{\textsc{tof}} = \frac{1}{2}J_{zz} ( Z_1Z_2 + Z_2 Z_3) + \frac{1}{2}\sum_n \omega_n Z_n +  \frac{1}{2}\Omega X_2.
\end{align}
The Hamiltonian is assumed to be in units of frequency, and for the rest of this chapter we will work in these units. This is similar to the type of Hamiltonian used in~\cite{Kumar2013} to implement an $N$-qubit controlled unitary, however this protocol requires non-uniform interactions and interactions between all qubits, which we would like to avoid. This Hamiltonian has the convenient property that it commutes with both $Z_1$ and $Z_3$, so that the eigenstates are formed from pairs of states with the same $z$-value on qubits 1 and 3 but different value on qubit 2 (e.g. $\{ \ket{010} , \ket{000} \}$. 

To produce a Toffoli gate, the gate needs to flip the target qubit (qubit 2 in this case) when each of the control qubits (1 and 3) are in the $\ket{1}$ state. For this to be possible it is sufficient that $H_{\textsc{tof}}$ has eigenstates of the form $\frac{1}{\sqrt{2}}\ket{1} (\ket{0} \pm e^{i\phi}\ket{1})\ket{1}$. By multiplying $H_{\textsc{tof}}$ by states of this form, it can easily be confirmed that the states $\frac{1}{\sqrt{2}}\ket{1} (\ket{0} \pm \ket{1})\ket{1}$ are eigenstates when $\omega_2 = 2J_{zz}$ (intuitively, the effect of the Ising coupling and local $Z_2$ field cancel out, so that the only remaining operator acting on qubit 2 is $X_2$).

The eigenstates and energies of the full Hamiltonian were then found using Mathematica with $\omega_2 = 2J_{zz}$, giving
\begin{align}
\ket{\pm}_{ab} &= \frac{1}{\mathcal{N}^{\pm}_{ab}}\text{ } \ket{a}  \left[(d_{ab} \pm \sqrt{1 + d_{ab}^2})\ket{0} + \ket{1} \right] \ket{b},\\
E_{ab}^{\pm} /\hbar &= \frac{1}{2} \left[  (-1)^{a}\omega_1 +  (-1)^{b} \omega_3  \pm \Omega \sqrt{d_{ab}^2 + 1} \right],
\end{align}
where $a,b\in \{0,1\}$ and we have introduced the parameters $d_{11} = 0$, $d_{01} = d_{10} = \omega_2/\Omega$, $d_{00} = 2\omega_2 /\Omega$, and $\mathcal{N}^{\pm}_{ab}$ are normalising factors. 
The $\ket{101} \leftrightarrow\ket{111}$ swapping will occur when $\exp\left[ -iH_{\textsc{tof}}t\right] \ket{101} = e^{i\theta} \ket{111}$, which occurs at a time $t = \tau_n = (2n+1)\pi \hbar /|E_{110}^{+} - E_{110}^{-}| = (2n+1)\pi / \Omega $ (where $n$ is an integer, and assuming without loss of generality that $\Omega$ is positive).

The remaining computational states $\{ \ket{000},\ket{010},\ket{001},\ket{011},\ket{100},\ket{110} \}$ will also evolve during this time, so we now look at how to engineer the Hamiltonian such that this swapping happens an even number of times during the gate time $\tau_n$. The fidelities of these swapping events occurring is measured by the fidelity $f_{lmn \rightarrow xyz} := \bra{xyz} e^{-i H_{\textsc{tof}}\tau_n } \ket{lmn}$, which we obtain by first expressing each computational state in terms of the eigenstates of the system, then multiplying each eigenstate $\ket{\pm}_{ab}$ by $e^{-iE_{ab}^{\pm}  \tau_n / \hbar}$ and simplifying using Mathematica and the Quantum Mathematica add-on~\cite{QuantumMathematica}. Following this process, and recalling that the Hamiltonian commutes with $Z_1$ and $Z_3$ so that transitions only occur between states with the same values of $m_z$ on qubits 1 and 3, we obtain the following fidelities:
\begin{align}\label{eqn:Fid}
&f_{abc \rightarrow a \bar{b} c} =  \frac{-ie^{-i\phi_{ac}} (2n+1) \pi}{2}\mbox{sinc} \left( \frac{(2n+1)\pi}{2} \sqrt{d_{ac}^2 + 1} \right) \nonumber\\
&f_{abc \rightarrow abc} = -ie^{-i\phi_{ac}} \left[ \cos \left(\frac{(2n+1)\pi}{2} \sqrt{d_{ac}^2 + 1} \right)  - \frac{(2n+1)  i \pi}{2} d_{ac} \text{ }\mbox{sinc} \left( \frac{(2n+1)\pi}{2} \sqrt{d_{ac}^2 + 1} \right) \right],
\end{align}
where $a,b,c \in \{0,1\}$, $(a,c) \neq (1,1)$, $\bar{b} := b \oplus 1$, and 
\begin{align}
\phi_{ac} &= \frac{(2n+1)\pi (E^+_{ac} + E^-_{ac})}{4 \hbar \Omega} = \frac{(2n+1)\pi ((-1)^a \omega_1 + (-1)^c \omega_3)}{2\Omega}.
\end{align} 
Note that $|f_{101\leftrightarrow 111}| = 1$ by our choice of gate time.

The condition these fidelities need to satisfy to realise a Toffoli gate is $|f_{a0b}\rightarrow f_{a1b}| = 0$ for $(a,b)\ne (1,1)$. To achieve this the phase inside the sinc function in (\ref{eqn:Fid}) must be an integer multiple of $\pi$ for each fidelity. Noting that the phases for $f_{100 \rightarrow 110}$ and $f_{001 \rightarrow 011}$ are the same, this condition leads us to the following two coupled equations;
\begin{align}
\frac{1}{2} \sqrt{\frac{\omega_2^2}{\Omega^2} + 1} = \frac{m_1}{(2n+1)}, \label{eqn:Diop1}\\
\frac{1}{2} \sqrt{\frac{4\omega_2^2}{\Omega^2} + 1} = \frac{m_2}{(2n+1)}, \label{eqn:Diop2}
\end{align}
where $m_1,m_2$ are non-zero integers. Eliminating $\frac{\omega_2}{\Omega}$ from these equations gives an equation that $m_1$, $m_2$ and $n$ must satisfy for a Toffoli gate to be possible:
\begin{align}
16 m_1^2 - 4m_2^2 = 3(2n+1).
\end{align}
The left hand side of this equation is even, and the right side is odd, so there is no assignment of integer $m_1$, $m_2$ and $n$ which satisfies this. 

Whilst this is proof that there is no choice of $\omega_2,$ $\Omega$ that achieves a perfect gate, we can still find parameters which achieve  an approximate gate. Assume that $n=0$ for simplicity (i.e.\ only one swap of $\ket{101} \leftrightarrow \ket{111}$ occurs during the gate). Then rearranging eqn.\ (\ref{eqn:Diop1}) for $\frac{1}{2} \sqrt{\frac{\omega_2^2}{\Omega^2} + 1} =m_1$ with $m_1 \geq 1$ and inserting into eqn.\ (\ref{eqn:Diop2}) gives
\begin{align} \label{eq:Cond1}
&\frac{1}{2} \sqrt{\frac{4\omega_2^2}{\Omega^2} + 1} = 2m_1 - \frac{3}{16m_1} + O\left( \frac{1}{m_1^3} \right).
\end{align}
Therefore we find that $m_2 \approx 2m_1$ with this choice of parameters, so that the phases inside the sinc functions in (\ref{eqn:Fid}) are all either zero or approximately zero up to order $1/m_1$. Substituting these expressions into (\ref{eqn:Fid}), we obtain the fidelities (making $m_1$ even for simplicity, but the calculation for odd $m_1$ is similar and doesn't change the form of the error terms)
\begin{align} \label{eqn:Fid2}
&f_{000 \rightarrow000} =f_{010 \rightarrow010} = -ie^{-i\phi_{00}}  \left[ \cos \left( \pi \sqrt{16 m_1^2 - 3}\right)  - i \pi \sqrt{4 m_1^2 - 1} \text{ sinc}  \left( \pi \sqrt{16 m_1^2 - 3} \right)  \right]\nonumber\\
&f_{000 \rightarrow 010} = f_{010 \rightarrow 000} =\frac{-ie^{-i\phi_{00}}  \pi}{2} \text{sinc} \left( \pi \sqrt{16 m_1^2 - 3} \right),\nonumber\\
& f_{001 \rightarrow 001} =f_{011 \rightarrow 011} = -i e^{- i \phi_{01} },\nonumber\\
& f_{001 \rightarrow 011} = f_{011 \rightarrow 001} = 0, \nonumber\\
&f_{100 \rightarrow 100} = f_{110 \rightarrow 110}=-i e^{- i \phi_{10} },\nonumber\\
& f_{100 \rightarrow 110}  =  f_{110 \rightarrow 100}  = 0,\nonumber\\
&f_{101 \rightarrow101}  = f_{111 \rightarrow111} = 0,\nonumber\\
&f_{101 \rightarrow 111} = f_{111 \rightarrow 101} = -ie^{-i\phi_{11}}.
\end{align}

To evaluate how close this approximation brings us to an exact Toffoli gate, we use the process trace distance, which for a 3-qubit system is defined as (see Sec.~\ref{sec:DistMeas} and~\cite{Gilchrist2005}):
\begin{align}
\mathcal{D}_{pro} = \frac{1}{16} \| \chi(U) - \chi(T) \|_{tr},
\end{align}
where $T$ is an ideal Toffoli gate, $U$ is the gate we can achieve with the above setup, and $\| X \|_{tr} = \mbox{tr} ( \sqrt{X^\dagger X} )$ is the trace norm. For simplicity, we ignore any phases and define $U_{|f|}$ such that $ \bra{y}U_{|f|}\ket{x} = |f_{x \rightarrow y }|$ so this will only measure how closes we are to a Toffoli gate apart from some local operations, or how close we are to a classical gate. By inspecting the phases in (\ref{eqn:Fid2}), we can see that the local rotations needed to get rid of the phases on the leading order terms are $i\exp [ i \pi \omega_1 Z_1 / \Omega] \exp [ i \pi \omega_3 Z_2 / \Omega ]$.
Using the terms in (\ref{eqn:Fid2}), and defining $f_0 := f_{000 \rightarrow010}$, $U_{|f|}$ is then
\newlength{\mycolwd} 
\settowidth{\mycolwd}{10000em} 

\begin{align*}
U_{|f|} = \left(\begin{array}{C{\mycolwd} C{\mycolwd} C{\mycolwd} C{\mycolwd} C{\mycolwd} C{\mycolwd} C{\mycolwd} C{\mycolwd} }
\sqrt{1 -  |f_{0}|^2 } & |f_{0}| & 0 & 0 & 0 & 0& 0 & 0\\
|f_{0}| & \sqrt{1 -  |f_{0}|^2 }  & 0 & 0 & 0 & 0& 0 & 0\\
0 & 0 & 1& 0 & 0 & 0& 0 & 0\\
0 & 0 & 0 & 1 & 0 & 0& 0 & 0\\
0 & 0 & 0 & 0 & 1&0& 0 & 0\\
0 & 0 & 0 & 0 & 0 & 1 & 0 & 0\\
0 & 0 & 0 & 0 & 0 & 0& 0 & 1\\
0 & 0 & 0 & 0 & 0 & 0& 1 & 0
\end{array} \right).
\end{align*}
Calculating $\chi(U_{|f|})$ can be done straightforwardly by flattening $U_{|f|}$ into a $64 \times 1$ vector $V$ containing the elements of $U_{|f|}$, and defining $\chi_{mn} =  V_m^* V_n$. This was evaluated using Mathematica, giving $\mathcal{D}_{pro} =\frac{1}{4} \sqrt{6 + |f_0|^2 - 6 \sqrt{1 - |f_0|^2}}$. Expanding $|f_0|$ in a Taylor series for $m_1>1$, and using $\text{sinc}^2(2 m_1 \pi + \delta ) = \frac{\delta^2}{2 m_1^2 \pi^2} + O \left( \frac{\delta^3}{m_1^3} \right)$:
\begin{align}
&|f_0|^2 = |f_{000 \rightarrow010}|^2 = \pi^2 \mbox{sinc}^2  \left( \frac{\pi}{2} \sqrt{16m_1^2 - 3 } \right)= \pi^2 \mbox{sinc}^2  \left( 2 m_1 \pi\sqrt{1 - \frac{3}{16m_1^2} } \right) \nonumber\\
&=  \pi^2  \mbox{sinc}^2  \left( 2m_1 \pi - \frac{3 \pi}{16m_1}  +  O \left( \frac{1}{m_1^3} \right)  \right) =  \frac{9\pi^2 }{1024 m_1^4} + O \left( \frac{1}{m_1^6} \right).
\end{align}
Substituting this into the expression for $\mathcal{D}_{pro}$, we find that $\mathcal{D}_{pro} = \frac{3 \pi}{16 m_1^2} + O\left( \frac{1}{m_1^4} \right)$. $\mathcal{D}_{pro}$ gives an upper bound on the average probability $\bar{p}_e$ that the gate fails~\cite{Gilchrist2005}, so
\begin{align}
\label{eqn:p}
\bar{p}_e \lesssim \frac{3 \pi}{16m_1^2} \approx  \frac{3 \pi}{4} \left(\frac{\Omega}{\omega_2} \right)^2  = \frac{3 \pi}{16} \left(\frac{\Omega}{J_{zz}} \right)^2  .
\end{align}
In summary, a gate locally equivalent to a Toffoli gate can be achieved with average failure error of $\bar{p}_e \lesssim  \frac{3 \pi}{16} \left(\frac{\Omega}{J_{zz}} \right)^2$, provided that $J_{zz} = \omega_2/2$ and $\frac{\omega_2}{\Omega} =\sqrt{4 m_1^2 - 1}$, with $m_1 \geq 1$.

\subsection{Sequences of gates}

\begin{figure}[h]
\begin{center}
\includegraphics[scale = 0.45]{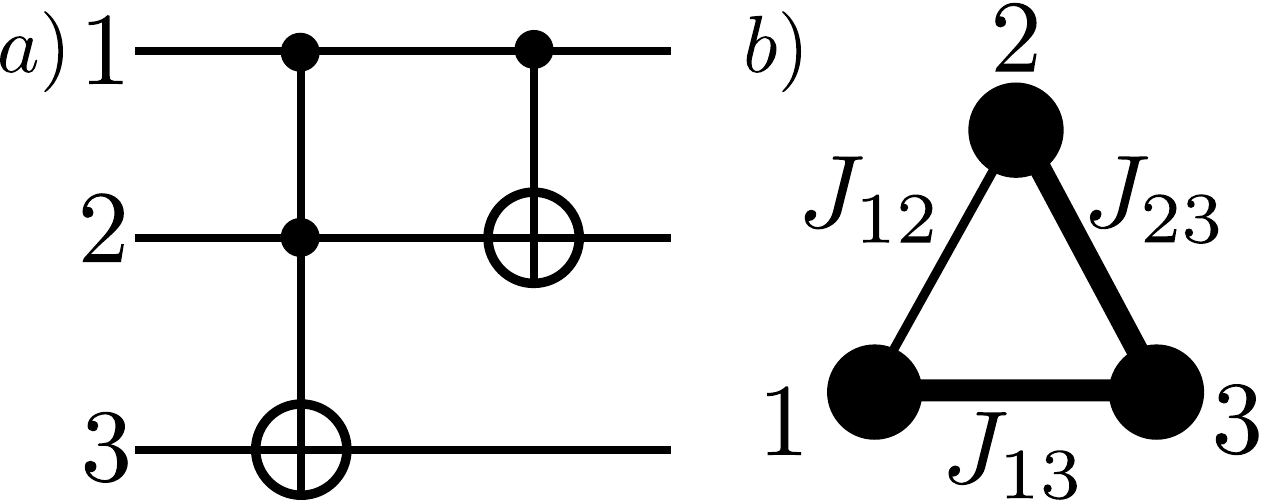}
\caption{a) A half-adder circuit. b) Setup for creating a half adder, using two pulses and with $J_{13} = J_{23} \neq J_{12}$.}
\label{fig:HAcirc}
\end{center}
\end{figure}

We can take advantage of the fact that the Toffoli gate outlined above is activated by an external field to compose it together with other gates. One of the simplest circuits that could be performed is a Toffoli gate followed by a \textsc{cnot} gate, which performs a half-adder circuit (Fig.~\ref{fig:HAcirc}). To perform this, starting with an arrangement of qubits as shown in Fig.~\ref{fig:HAcirc}, we can first perform the Toffoli gate in Section~\ref{sec:Toff} with qubits 1 and 2 as controls. Then to apply the controlled-\textsc{not} gate, we can apply fields such that the field is resonant with qubit 2 only if qubit 1 is in a $\ket{1}$ state, but is independent of the state of qubit 3. This can be done by applying a Hamiltonian of the form
\begin{align}
H = \frac{1}{2}\sum_{n=1}^3 \omega_n Z_n+ \frac{1}{2}\sum_{mn} J_{mn} Z_m Z_n +  \frac{1}{2}\Omega_1 X_2 \cos ( \omega_+ t ) + \frac{1}{2}\Omega_2 X_2 \cos (\omega_+ t),
\end{align}
where $\omega_{\pm} =  \frac{1}{2}( \omega_3 -J_{13} \pm J_{23}) $, and $\omega_n$ is the resonant frequency of qubit $n$. The first of these fields flips qubit 2 when qubits 1 and 3 are in a $\ket{10}_{12}$ state, and the second field flips when 1 and 3 are in the  $\ket{11}_{12}$ state, overall resulting in a \textsc{cnot} gate.

\section{Fredkin gate with Ising and Heisenberg interactions}
\label{sec:FredHeis}

\begin{figure}[b]
\begin{center}
\includegraphics[scale = 0.5]{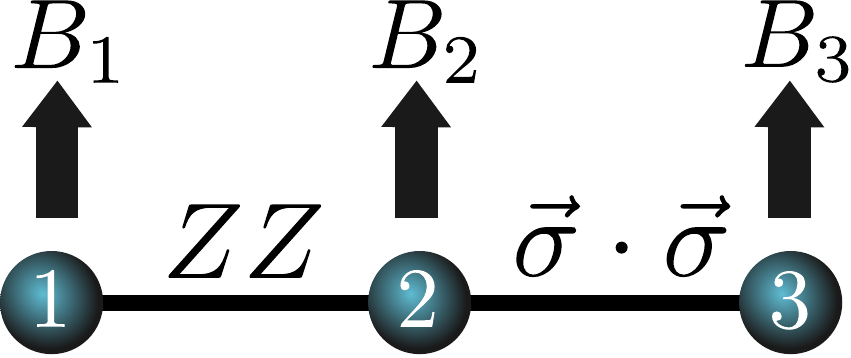}
\caption{Setup for creating a 3-qubit Fredkin gate, using Ising and Heisenberg coupling.}
\label{fig:FredHeis}
\end{center}
\end{figure}

We now consider creating a quantum Fredkin gate (controlled-\textsc{swap}), using Ising and anisotropic Heisenberg interactions. A Fredkin gate performs the operation $\ket{k}\ket{lm} \to \ket{k}\textsc{swap}^k\ket{lm}$. To realise this gate we consider a Hamiltonian of the form
\begin{align}\label{eqn:HamFred}
H_{\textsc{fred}} = \frac{1}{2} J_{12} Z_1 Z_2 + \frac{1}{2} J_{23} (X_2 X_3 + Y_2 Y_3 + Z_2 Z_3)  +\frac{1}{2}\sum_{j=1}^3 \omega_j Z_j,
\end{align}
where qubit 1 is the control qubit, and qubits 2 and 3 are to be swapped (see Fig.~\ref{fig:FredHeis}). The intuition is that the swapping induced by the Heisenberg interaction between qubits 2 and 3 will only occur when qubit 1 is in a state that makes the energy splitting of qubits 2 and 3 match.
As in the previous section, we need to choose parameters such that states of the form $\frac{1}{\sqrt{2}} (\ket{110} \pm \ket{101} )$ are eigenstates of the Hamiltonian, which requires $J_{12} = \omega_2 - \omega_3$. With this settings $\ket{100},\ket{111},\ket{011},\ket{000} $ are all eigenstates, and the only eigenstates of the Hamiltonian which are not computational basis states are
\begin{eqnarray}
&\ket{\psi}_{110}^{\pm} = &\frac{1}{\sqrt{2}} (\ket{110} \pm \ket{101} )\nonumber\\
& \ket{\psi}_{010}^{\pm} = &\frac{1}{\mathcal{N}^{\pm}_{010}} \left[ (J_{12} \pm \sqrt{J_{12}^2 + J_{23}^2}) \ket{001} + J_{23}\ket{010} \right],
\end{eqnarray} 
where $\mathcal{N}^{\pm}_{010}$ are normalising factors. The eigenenergies of these states are
\begin{align}
&E_{110}^{\pm} /\hbar= - \frac{1}{2} \left[ \omega_1 - J_{23} \pm 2 J_{23} \right],\; E_{010}^{\pm}  /\hbar = \frac{1}{2} \left[ \omega_1 -  J_{23} \pm 2 \sqrt{J_{12}^2 + J_{23}^2} \right].
\end{align}
The swap $\ket{110} \leftrightarrow \ket{101}$ is complete at a time $\tau_n = (2n+1) \pi \hbar/|E_{110}^+ - E_{110}^-| = (2n+1)\pi /2J_{23}$ (with $n \in \{0,1,2,...\}$, and assuming $J_{23}> 0$, without loss of generality). By expanding $\ket{001},\ket{010}$ in terms of the eigenstates $\ket{\psi}_{010}^{\pm}$, multiplying these by $e^{-iE_{010}^{\pm} \tau_n /\hbar}$ and simplifying we find that the fidelity for swapping $\ket{010} \leftrightarrow \ket{001}$ at time $\tau_n$ is
\begin{align}
f_{010 \leftrightarrow 001} =\frac{i \pi e^{-i \phi }}{ J_{23}} \mbox{sinc} \left[ \frac{ (2n+1)\pi \sqrt{J_{12}^2 + J_{23}^2} }{2J_{23}} \right],
\end{align}
where $\phi = (J_{23}-\omega_1)(2n+1) /2 J_{23}$. For this fidelity to be zero, we need $J_{23}^2 \left(\frac{m^2}{(2n+1)^2}-1 \right) =J_{12}^2 = (\omega_2 - \omega_3)^2$, where $m$ is an integer and $\frac{m}{(2n+1)} > 1$.

\section{Fredkin gate using XX interactions combined with Ising interactions}
\label{sec:XX}

Now we consider performing a Fredkin gate, based on the swapping protocol in~\cite{Benjamin2003}, where a swap operation is achieved between the ends of an $XX$ spin chain. Somewhat intuitively, a Fredkin gate can be achieved simply by adding a qubit connected the middle of this chain, with couplings that effectively fix the middle of the chain and block the swap. We start with the simplest case, three qubits coupled by an XX interaction with an Ising coupling of the middle qubit to an external qubit A
\begin{align}
H_{\textsc{xx}} = \frac{1}{2}J_{xx} (X_1X_2 +Y_1Y_2 + X_2X_3 + Y_2 Y_3) +\frac{1}{2} J_{zz} Z_A Z_2 + \frac{1}{2}\omega_A Z_A + \frac{1}{2}\omega_2 Z_2.
\end{align}

\begin{figure}[h]
\begin{center}
\includegraphics[scale = 0.5]{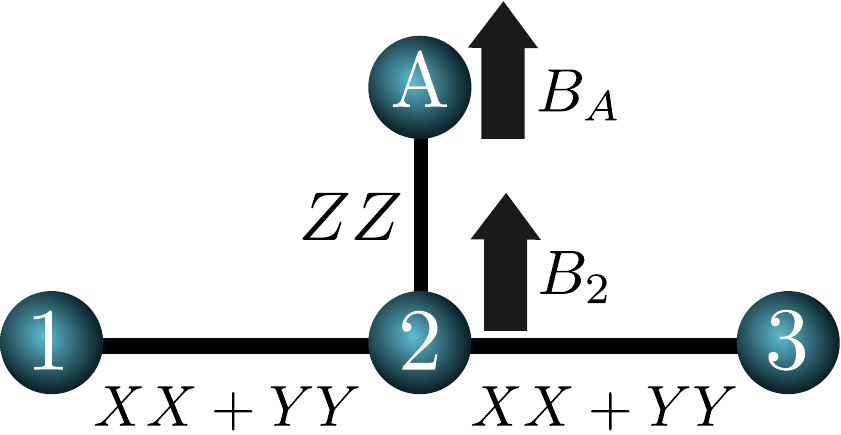}
\caption{Setup for creating a Fredkin gate, using Ising and $XX+YY$ interactions.}
\label{fig:FredXX}
\end{center}
\end{figure}

See Fig.~\ref{fig:FredXX} for an illustration of this. Here we define qubits 1 and 3 as the qubits to be swapped, whilst qubit A is the control, and the qubits are written in the order $A,1,2,3$ so e.g. $\ket{0000} = \ket{0}_A \ket{0}_1 \ket{0}_2 \ket{0}_3$. We will also only consider protocols in which qubit 2 is initialised in the $\ket{0}$ state. For simplicity, we will express all quantities relative to $J_{xx}$. Then defining $\alpha = (\omega_2 + J_{zz})/J_{xx}$, $\beta = (\omega_2 - J_{zz})/J_{xx}$, $F(x) = \sqrt{8 + x^2}$, the eigenstates of the Hamiltonian are
\begin{align}
\begin{array}{c c}
\ket{\phi}_{110} = \frac{1}{\sqrt{2}} (\ket{1100} - \ket{1001}), \; &\ket{\phi}_{010} = \frac{1}{\sqrt{2}} (\ket{0100} - \ket{0001})  \\
\ket{\pm}_{110} = \frac{2\ket{1001} - (\beta \pm F(\beta))\ket{1010} + 2\ket{1100}}{\sqrt{8 + (F(\beta) + \beta)^2}} , \; &\ket{\pm}_{010} = \frac{2\ket{0001} - (\alpha \pm F(\alpha))\ket{0010} + 2\ket{0100}}{\sqrt{8 + (F(\alpha) + \alpha)^2}} \\
\ket{\pm}_{111}  = \frac{2\ket{1011} + (\beta\pm F(\beta) )\ket{1101} + 2\ket{1110}}{\sqrt{8 + (F(\beta) + \beta)^2}}, \; &\ket{\pm}_{011}  = \frac{2\ket{0011} + (\alpha\pm F(\alpha) )\ket{0101} + 2\ket{0110}}{\sqrt{8 + (F(\alpha) + \alpha)^2}} \\
\ket{\phi}_{100} = \ket{1000}, \; &\ket{\phi}_{000} = \ket{0000}.
\end{array}
\end{align}
The energies of these states are, respectively:
\begin{align}
&E_{110} = -\frac{\omega_A}{2J_{xx}} + \frac{1}{2}\beta, \; E_{110}^{\pm} = -\frac{\omega_A}{2J_{xx}} \mp \frac{1}{2}F(\beta), \; E_{010} = \frac{\omega_A}{2J_{xx}} + \frac{1}{2}\alpha, \; E_{010}^{\pm} = \frac{\omega_A}{2J_{xx}} \mp \frac{1}{2}F(\alpha) \nonumber\\
&E_{111}^{\pm} = -\frac{\omega_A}{2J_{xx}}  \pm \frac{1}{2}F(\beta), \; E_{011}^{\pm} = \frac{\omega_A}{2J_{xx}}  \pm \frac{1}{2}F(\alpha) , \; E_{100} = -\frac{\omega_A}{2J_{xx}} + \frac{1}{2}\beta, \; E_{000} = \frac{\omega_A}{2J_{xx}} + \frac{1}{2}\alpha.
\end{align}
We are now interested in finding the how these states evolve under the Hamiltonian. 
Similarly to Sec.~\ref{sec:Toff}, we define the fidelities as $f_{x \leftrightarrow y} := \bra{y} e^{-iH_{\textsc{xx}} t} \ket{x} $ where $x,y$ are 3-digit binary strings containing the states of qubits A,1, and 3 (qubit 2 is initialised in the $\ket{0}$ state so is left out). By expressing each of the possible input states $\ket{0001},\ket{0100},\ket{1001},\ket{1100}$ in terms of the eigenstates of $H_{\textsc{xx}} $, and evolving these in time, we find that the relevant fidelities are (excluding phases, and excluding transitions involving $\ket{0000}$ and $\ket{1000}$ as they are eigenstates)
\begin{align}\label{eq:Fid}
&|f_{010 \leftrightarrow 001}|^2 =  G^+(\alpha,t) , \; |f_{010\rightarrow 010}|^2= |f_{001\rightarrow 001}|^2 =G^- (\alpha,t) \nonumber\\
&|f_{110\leftrightarrow 101}|^2 =   G^+(\beta,t) , \; |f_{110\rightarrow 110}|^2= |f_{101\rightarrow 101}|^2= G^- (\beta,t) \nonumber\\
&|f_{011 \rightarrow 011} |^2=  \frac{1}{F(\alpha)^2} [\alpha^2 + 8\cos^2 F(\alpha)t ], \; |f_{111\rightarrow 111} |^2 =  \frac{1}{F(\beta)^2} [\beta^2 + 8\cos^2 F(\beta)t ] \nonumber\\
&f_{|100 \rightarrow 100}|^2 = |f_{000\rightarrow 000} |^2= 1,
\end{align}
where
\begin{align}
&G^{\pm} (\omega,t) := \frac{1}{2} \left(  1 \mp \cos \omega t \cos F(\omega) t  - \frac{\sin F(\omega)t}{F^2(\omega)} \left[ 4 \sin F(\omega)t  \pm \omega F(\omega) \sin \omega t  \right] \right).
\end{align}

To construct a gate locally equivalent to a Fredkin gate, we require $|f_{1xy \leftrightarrow 1yx}|^2 = 1$, $|f_{0xy \leftrightarrow 0yx}|^2 = 0$,  $|f_{0xy \leftrightarrow 0xy}|^2 = 1$ for all $x,y \in \{0,1\}$. These will be satisfied when $G^+(\alpha,t) = G^-(\beta,t) = 0$, $G^-(\alpha,t) = G^+(\beta,t) = 0$, $\cos^2 F(\beta)t=\cos^2 F(\alpha)t =1$. We look for solutions to these conditions in 3 regimes:
\begin{itemize}
\item[(1)] $\omega_2, J_{zz} \sim J_{xx}$  ($\alpha,\beta \sim 1$).
\item[(2)]  $|\omega_2| \approx |J_{zz}| >> J_{xx}$ ($|\alpha| >> 1, \beta = 0$ or $|\beta| >> 1, \alpha = 0$).
\item[(3)] $|\omega_2| >>|J_{zz}|$ or $|J_{zz}| >> |\omega_2|$ ($|\alpha| >> 1, |\beta| >> 1$). 
\end{itemize}
Note that this is not intended to cover all of the possible conditions, but covers a few regimes where results can be found.

\subsection{Regime 1: $\omega_2, J_{zz} \sim J_{xx}$}

We first consider the regime where $\omega_2, J_{zz} \sim J_{xx}$, so that $\alpha,\beta$ are small, and we look for parameters which yield exact solutions. To make $|f_{011 \rightarrow 011} |^2 = |f_{111\rightarrow 111} |^2 = 1$, we must set $F(\alpha)t = n_{\alpha} \pi$, $F(\beta)t = n_{\beta} \pi$. Substituting these into $G^{\pm}(\alpha,t)$ and $G^{\pm}(\beta,t)$ gives:
\begin{align}
&G^{\pm} (\alpha,t) = \frac{1}{2} \left[ 1 \mp (-1)^{n_\alpha}\cos \alpha t \right] , \;G^{\pm}(\beta,t) = \frac{1}{2} \left[ 1 \mp (-1)^{n_\beta}\cos \beta t \right].
\end{align}
To satisfy the remaining conditions, we require $G^+(\alpha,t) =G^-(\beta,t) = 0$, $G^-(\alpha,t) = G^+(\beta,t) = 1$, which occurs when
\begin{align}\label{eqn:Parity}
&\alpha t = m_\alpha \pi, \; \beta t = m_\beta \pi , \;  P(m_\alpha) = P(n_\alpha), \; P(m_\beta) = 1 \oplus P(n_\beta),
\end{align} 
where $P$ is a parity function, which equals 1 if a number is even and 0 if odd. These conditions are not satisfiable, since
\begin{align}
&F(\alpha)^2 t^2 - F(\beta)^2 t^2  = \alpha^2 t^2 - \beta^2 t^2\;  \longrightarrow \;  n_\alpha^2 - m_\alpha^2 = n_\beta^2 - m_\beta^2,
\end{align}
which cannot be satisfied by integers that also satisfy the conditions in (\ref{eqn:Parity}). So there are no exact solutions of $\alpha$ and $\beta$ which create a Fredkin gate. In this regime, there may still be an approximate way to make a Fredkin gate, however to date the author has been unsuccessful in finding one.

\subsection{Regime 2: $|\omega_2| \approx |J_{zz}| >> J_{xx}$}

We now investigate the regime where $|\omega_2| \approx |J_{zz}| >> J_{xx}$, so that either $|\alpha| >> 1, \beta \sim 0$ or $|\beta| >> 1, \alpha \sim 0$. To simplify the analysis, we will consider only $|\beta| >> 1, \alpha \sim 0$, since we expect both scenarios to have similar behaviour. To satisfy the condition $|f_{011 \rightarrow 011} |^2 = 1$, we pick $\alpha = 0$ (i.e. $\omega_2 = - J_{zz}$) so that $F(\alpha) = \sqrt{8}$. Under these conditions, $G^\pm(\alpha,t)$ becomes
\begin{align}
G^{\pm}(\alpha=0,t)= \frac{1}{2} [ \cos 2t\sqrt{2}  \mp 1].
\end{align}
The conditions $G^+(\alpha = 0,t)=0$ and $G^-(\alpha=0,t) = 1$ are satisfied at times $t = \tau_n = (2n+1) \pi/ 2\sqrt{2}$, so we consider evolution at integer multiples of $\tau_n$. With $|\beta| \gg 1$, $F(\beta) \approx |\beta| + O(\frac{1}{\beta})$, and
\begin{align}
G^{\pm}(\beta,t) &= \frac{1}{2}(1\mp \cos(F(\beta) - \beta)t) +  O\left( \frac{1}{\beta^2} \right),
\end{align}
where $\cos( F(\beta) - \beta)t) $ has not been expanded since $t$ may also be large. Substituting these expression into eqn.\ (\ref{eq:Fid}) gives
\begin{align}
&|f_{110\leftrightarrow 101}  |^2 = \sin ^2\frac{(F(\beta) - \beta)\tau_n}{2}  - O\left( \frac{1}{\beta^2} \right), \;|f_{110\rightarrow 110}|^2 = \cos ^2\frac{(F(\beta) - \beta)\tau_n}{2}  + O\left( \frac{1}{\beta^2} \right)\nonumber\\
&|f_{111 \rightarrow 111}|^2 = 1 - O \left(\frac{1}{\beta^2} \right), \; |f_{101\rightarrow 101}|^2 = |f_{110\rightarrow 110}|^2.
\end{align}
This is satisfied for values of $\beta = \frac{2 \tau_n}{(2m+1) \pi} - \frac{(2m+1) \pi}{\tau_n}$, for some integers $m,n$ such that either $\tau_n \gg m \pi$ or $m \pi \gg \tau_n$. Thus up to an error of $O(1/\beta^2) = O(  J_{xx}^2 / (2J_{zz})^2 )$ we can achieve approximate solutions.

\subsection{Regime 3: $|\omega_2| >>|J_{zz}|$ or $|J_{zz}| >> |\omega_2|$ }

We now consider the regime where  $|\omega_2| >>|J_{zz}|$ or $|J_{zz}| >> |\omega_2|$, such that $|\alpha| >> 1, |\beta| >> 1$ and $F(\alpha) \approx |\alpha| + O(\frac{1}{\alpha})$, $F(\beta) \approx |\beta| + O(\frac{1}{\beta})$. Following a similar analysis to the previous subsection, the fidelities under these conditions are
\begin{align}
&|f_{010 \leftrightarrow 001}|^2 = \sin^2\frac{(F(\alpha) - \alpha)t}{2}  + O\left( \frac{1}{\alpha^2} \right), \; |f_{110\leftrightarrow 101}|^2=  \sin^2 \frac{(F(\beta) - \beta)t}{2}   -  O\left( \frac{1}{\beta^2} \right)  \nonumber\\
&|f_{010\rightarrow 010}|^2 = \cos^2\frac{(F(\alpha) - \alpha)t}{2}  - O\left( \frac{1}{\alpha^2} \right), \; |f_{110\rightarrow 110}|^2 = \cos ^2\frac{(F(\beta) - \beta)t}{2}  + O\left( \frac{1}{\beta^2} \right)\nonumber\\
&|f_{011 \rightarrow 011}|^2 = 1  -  O\left( \frac{1}{\alpha^2} \right), \; |f_{111\rightarrow 111}|^2  = 1  -  O\left( \frac{1}{\beta^2} \right)\nonumber\\
&|f_{001\rightarrow 001} |^2 = |f_{010\rightarrow 010}|^2, \; |f_{100 \rightarrow 100}|^2 = |f_{000\rightarrow 000}|^2 = 1.
\end{align}
These can be satisfied by $\alpha = \frac{2t}{m\pi} - \frac{4m\pi}{t}$, $\beta = \frac{4t}{(2n+1)\pi} - \frac{(2n+1)\pi}{2t}$, where $m$ and $n$ are integers such that $|\alpha|,|\beta| \gg 8$.

\section{Experimental implementations}

Having seen three theoretical proposals for 3-qubit gates, we now focus on how experimentally attainable these would be using current technology. We focus on using linear ion traps and bismuth donors in silicon, although in there are many systems that these gates could be applied to.

\subsection{Toffoli gate using ion traps}

Here we propose a possible way to achieve the Toffoli gate in Sec.~\ref{sec:Toff}, using trapped ions. We consider three $^{171}\mbox{Yb}^+$ ions in a linear Paul trap with secular frequency $\nu = 2\pi \times 100  \mbox{kHz}$. The qubits are encoded in the ground state hyperfine levels which are separated by approximately $12.6 \text{GHz}$~\cite{Monroe2013}. To create Ising interactions between the qubits, we consider using the scheme presented in~\cite{Mintert2001} and discussed in Sec.~\ref{sec:IntroIonTraps}. For three ions in a linear trap with secular frequency $\nu$, the vibrational eigenstates of $H_D$ are $\frac{1}{\sqrt{3}}(1,1,1)$, $\frac{1}{\sqrt{3}}(-1,0,1)$, $\frac{1}{\sqrt{6}}(1,2,1)$ with eigenvalues $\nu_1 = \nu$, $\nu_2 =\nu\sqrt{3}$ and $\nu_3 =\nu\sqrt{29/5}$ respectively~\cite{James1998}. Inserting these motional eigenstates into equation (\ref{eqn:IonJzz}), and with an axial magnetic field gradient of $250 \text{Tm}^{-1}$ one can realise exchange couplings $J_{12} =J_{23} = 2\pi \times 9.98 \mbox{kHz} $, $J_{13} =2\pi \times 7.07 \mbox{kHz}$. The extra $J_{13}$ coupling results in introducing extra phases to the gate, which makes it more complicated to make this into an exact quantum Toffoli gate but has no effect if the gate is used for classical computation.

The $X_2$ field can be achieved by applying a resonant microwave pulse, giving rise to a Hamiltonian of the form
\begin{align}\label{eqn:IonTrap1}
H &= \frac{1}{2}\sum_{j=1}^3 \omega_j Z_j +  \frac{1}{2}J_{zz}( Z_1 Z_2 + Z_2 Z_3) + \frac{1}{2}\Omega \cos( \omega_x t )X_2.
\end{align}
We can transform into the interacting frame with respect to $H_{0} = \frac{1}{2}\sum_{j=1}^3 \omega_j Z_j  - J_{zz} Z_2$. States in this rotating picture evolve according to the Hamiltonian 
\begin{align}
&H_I = e^{-i H_{0}t} (H-H_{0}) e^{i H_{0}t}  \nonumber\\
&= J_{zz} Z_2 + \frac{1}{2}J_{zz}( Z_1 Z_2 + Z_2 Z_3)  +  \frac{1}{2}\Omega e^{-i(\omega_2  - 2J_{zz})Z_2 t } \cos(\omega_x t )X_2 .
\end{align}
Setting $\omega_x = \omega_2 - 2J_{zz}$ and applying identity~\ref{eqn:Iden1} from Appendix~\ref{app:IntPic} to this, we find that the evolution due to eqn.\ (\ref{eqn:IonTrap1}) is approximately the same as the evolution due to this Hamiltonian:
\begin{align}
\label{eqn:HIntIon}
H_{I} =  J_{zz}  Z_2 +\frac{1}{2} J_{zz} ( Z_1 Z_2 + Z_2 Z_3) + \frac{1}{4}  \Omega X_2  +O\left( \frac{\Omega}{2(\omega_2  -2J_{zz})} \right).
\end{align}

 With $\Omega = 2 \pi  \times 3.5 \mbox{kHz} $, the systematic gate error is around $0.02$ and the gate time is $\sim0.2\text{ms}$, giving an error due to decoherence of around $(1- e^{-t_{gate}/T_2})  = 0.02$ ($T_2 \approx 10 \mbox{ms}$~\cite{Piltz2013} using dynamical decoupling). The error in the approximation in (\ref{eqn:HIntIon}) is negligible since $\omega_2 \sim 12.6 \text{GHz}$. Overall we therefore expect gate with error $\sim 0.04$ to be achievable with current technology, which does not fall below the thresholds required for fault-tolerant quantum computation.

\subsection{Toffoli gate using donors in Silicon}

To create the Toffoli gate in Sec.~\ref{sec:Toff} we can consider three donors arranged in a line. To achieve Ising couplings between two qubits, we propose placing the donors close to each other but preparing the nuclear spins in different states~\cite{Morley2010}, resulting in different hyperfine couplings at each site which convert the natural Heisenberg coupling to an approximate Ising coupling (see e.g.~\cite{Benjamin2003}) and allow individual addressability of the electrons. This is a similar method to that proposed in~\cite{Kalra2014}, where the nuclear states of P atoms are set in different states to realise a two-qubit entangling gate. Such an initialisation of the nuclear spins could be performed starting from a nuclear spin polarised sample (polarisations of up to 90\% were reported in~\cite{Sekiguchi2010}). The process of flipping the nuclear spin from $\frac{9}{2}$ to $-\frac{9}{2}$ would take 9 steps of $\sim$10$\mu s$ each, so around $90\mu s$. However, the nuclear spins are stable on a timescale of hours~\cite{Feher1959,Castner1962}, so this process could be done once for many operations of the gate, and so does not significantly affect the gate time. 

We must also apply a magnetic field much larger than the hyperfine splitting to ensure that the nuclear spins do not evolve. As mentioned in Sec.~\ref{sec:DonIntro}, optimal working points~\cite{Mohammady2010,Mohammady2012,Balian2012,Morley2013} cancellation resonances~\cite{Mohammady2010} and clock transitions~\cite{Wolfowicz2013} are not useful here because these schemes only work at lower magnetic fields. The nuclear spins of the donors are initialised in different states $I_1^{z},I_2^{z},I_3^{z}$, and an AC field is applied on qubit 2 in the $x$-direction (which could be applied globally since the hyperfine splittings are different). The effective spin Hamiltonian is then (see Sec.~\ref{sec:DonIntro})
\begin{align}
H &= \omega_L \sum_{n=1}^3 S_n^z + \sum_{n=1}^3 A I_n^z S_n^z + \Omega \cos ( \omega_x t )S_2^x+ \sum_{n=1}^2 J \bm{S}_n \cdot \bm{S}_{n+1},
\end{align}
where the $\sum_{n=1}^3 A I_n^{z} Z_n$ terms account for the different hyperfine splittings and the $\omega_L$ terms account for the large magnetic field in the $z$-direction. In a similar manner to the previous section, we switch to the interaction picture of this Hamiltonian with respect to $H_0 :=\omega_L\sum_{n=1}^3 S_n^z - JS_2^z + \sum_{n=1}^3 A I_n^z S_n^z $:
\begin{align}
H_I &= e^{-i H_0t} (H - H_0) e^{iH_0 t}\nonumber\\
&=  \frac{1}{2} e^{-i\tilde{\omega}_2Z_2t}\Omega \cos ( \omega_x t )X_2 + \frac{1}{4} J \sum_{n=1}^2 \left[  e^{-i(Z_n\tilde{\omega}_n + Z_{n+1}\tilde{\omega}_{n+1})t} ( X_n X_{n+1} +  Y_n Y_{n+1} )  + Z_n Z_{n+1}  \right]
\end{align}
where $\tilde{\omega}_1 :=  \omega_L + A I_1$, $\tilde{\omega}_2 := \omega_L + A I_2 - 2J$ and $\tilde{\omega}_3 :=\omega_L+ A I_3 $. By setting $\omega_x = \tilde{\omega}_2$ and using the identities from Appendix~\ref{app:IntPic}, the evolution in this interaction picture is approximately equivalent to the evolution by the following Hamiltonian:
\begin{align}\label{eqn:Rot1}
&H_I = \frac{1}{2}J Z_2 + \frac{1}{4}J(Z_1 Z_2 + Z_2 Z_3) +\frac{1}{4} \Omega X_2 + O\left(\sum_{n=2}^3 \frac{J}{A I_n^z - A I_{n-1}^z \pm 2J} + \frac{\Omega}{2(\omega_L -2J + AI_2^z)} \right),\end{align}
which is in the same form as the Hamiltonian in (\ref{eqn:HamToff1}). Note that the additional factor of $-2JZ_2$ added into the rotating frame means that a different local $z$-rotation must be applied to achieve an exact Toffoli gate. 

To achieve the minimum error we set $I_1^z = \frac{9}{2}$, $I_2^z=-\frac{9}{2}$, and $I_3^z =\frac{9}{2}$. The large nuclear spin and hyperfine coupling for bismuth (A=1.475 GHz) justifies the use of bismuth over other donors. $\Omega$ must be small enough that the errors in (\ref{eqn:Rot1}) and the systematic error in (\ref{eqn:p}) are small, and large enough to allow a gate time within the decoherence time of the bismuth donors ($\sim 0.5 \text{ms}$ in natural silicon at the high magnetic fields where the nuclear spin is a good quantum number~\cite{Morley2010,George2010}, or 700ms in isotopically pure silicon~\cite{Wolfowicz2012}). 

\begin{figure}[h]
\begin{center}
\includegraphics[scale = 0.45]{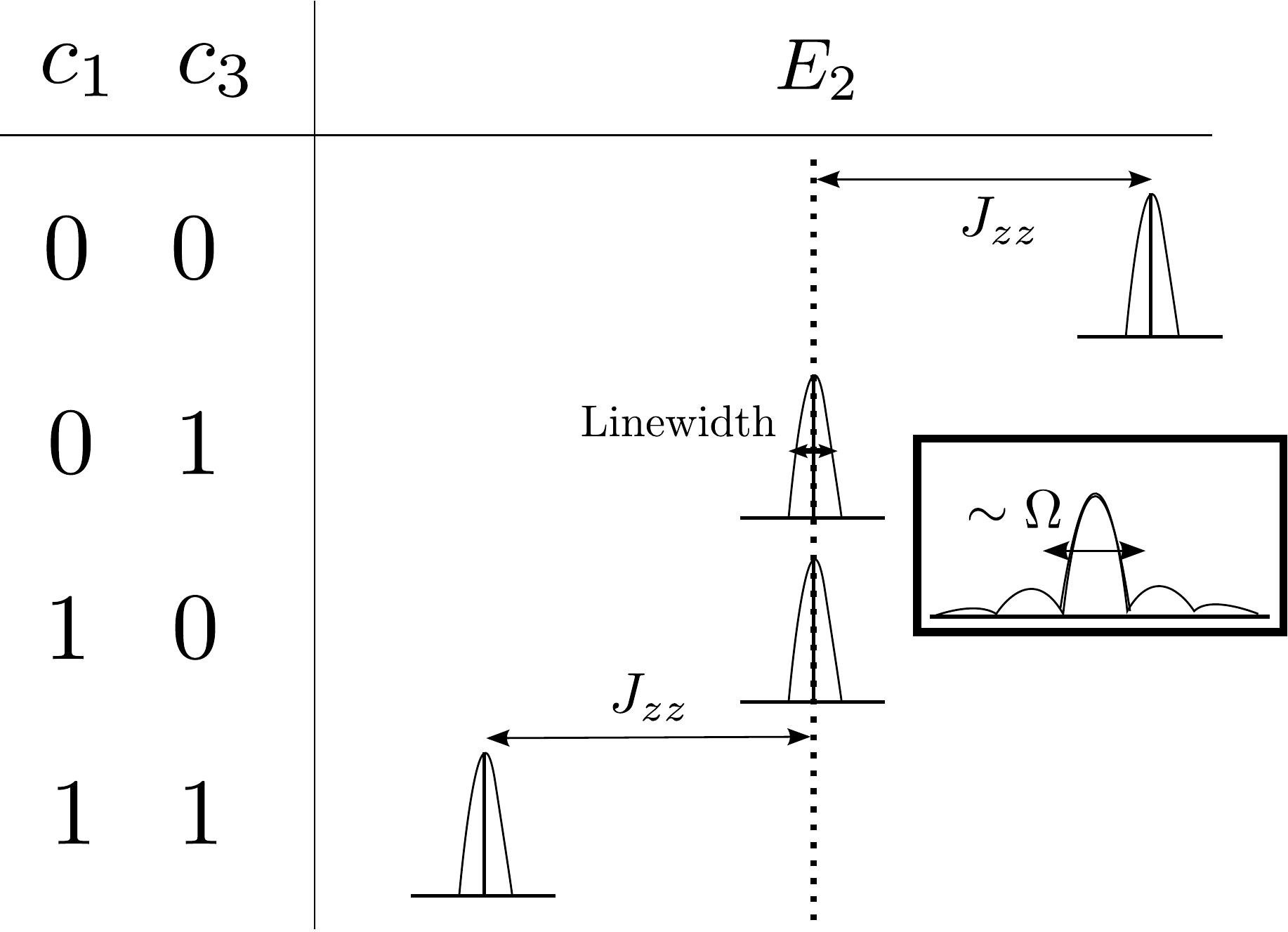}
\caption{Visual guide to the energy levels of the Toffoli gate. $c_1$ and $c_3$ indicate the states of qubits 1 and 3, and $E_2$ indicates the corresponding shift in the resonance of qubit 2. The inset illustrates the frequency profile of the square $X$ pulse, with bandwidth $\sim\Omega$. }
\label{fig:ToffEnergy}
\end{center}
\end{figure}

$J$ must be small enough to give a small error in (\ref{eqn:Rot1}), but large enough to reduce the systematic error in (\ref{eqn:p}). The bandwidth of the $X_2$ pulse must be large enough compared to the linewidth of the dopant energy levels (around 0.2MHz in natural silicon~\cite{Tyryshkin2003} and around 20kHz in isotopically pure silicon~\cite{Wolfowicz2012}) so that it excites the whole level. It also must be small enough that it can accurately resolve only one resonance and so only flip qubit 2 under the right conditions (this second condition is already implicitly taken into account in the systematic gate error in (\ref{eqn:p})). In this scheme we assume a square pulse, giving a sinc function of width $1 /t_{gate}$ in the frequency domain. See Fig.~\ref{fig:ToffEnergy} for an illustration of the energy levels of the Toffoli gate and the linewidth and bandwidth.

With settings of $\Omega = 1$MHz $J = 30$MHz, the gate time is $2\mu s$. The systematic gate error term in (\ref{eqn:p}) for these settings is $\sim 10^{-3}$, the errors due to decoherence are roughly $1-e^{-t_{gate}/T_2} \sim 10^{-6}$ and the errors in the Hamiltonian in (\ref{eqn:Rot1}) become $\sim 10^{-3}$, so we expect a gate with probability of failure $\lesssim 10^{-3}$ to be achievable. The bandwidth is around 4 MHz which is much larger than the linewidth of pure silicon. At the expense of greater gate error, the gate time can be as low as 50ns whilst still keeping all of the errors below 1\% (e.g. with $\Omega = 40MHz$, $J = 300MHz$). Since fault tolerant thresholds can be as high as 1\%~\cite{Knill2005}, and fault tolerant thresholds for classical computation are larger, we see that fault-tolerant gates could be achieved using today's technology for both classical and quantum computation. For an exact quantum Toffoli gate, local rotations would need to be performed, however we would not expect these operations to change the result significantly since they can in principle be applied in a short time.

\subsection{Fredkin gate using bismuth donors in silicon}

To create the Fredkin gate in Sec.~\ref{sec:FredHeis} we consider starting with a similar Hamiltonian to the Toffoli gate
\begin{align}
H &= \omega_L \sum_{n=1}^3 S_n^z + \sum_{n=1}^3 A I_n^z S_n^z +\sum_{n=1}^2 J \bm{S}_n  \cdot \bm{S}_{n+1},
\end{align}
Transforming into the interaction picture with respect to  $H_0= \omega_L \sum_{n=1}^3 S_n^z + A I_1^z S^z_1 + A I_2^z(S_2^z + S_3^z)$ gives
\begin{align}
e^{-iH_0t}(H-H_0)e^{-iH_0t} &=\frac{1}{4} J_{12}e^{-it ( \tilde{\omega}_1 Z_1 + \tilde{\omega}_2 Z_2 )} (X_1 X_2 + Y_1 Y_2)+\frac{1}{4} J_{23}e^{-it ( \tilde{\omega}_2 Z_2 + \tilde{\omega}_3 Z_3)}( X_2 X_3 + Y_2Y_3) \nonumber\\
&+ \frac{1}{4} J_{12}Z_1 Z_2 + \frac{1}{4} J_{23} Z_2 Z_3 + \frac{1}{2}A(I_3 - I_2)Z_3
\end{align}
where $\tilde{\omega}_1 := \omega_L + I_1^z A$, $\tilde{\omega}_2 =\tilde{\omega}_3 := \omega_L + I_2^z A$. Using identity (\ref{eqn:Iden2}), this Hamiltonian behaves the same as
\begin{align}\label{eqn:Rot2}
H_I &= \frac{1}{4} J_{12} Z_1 Z_2 +\frac{1}{4} J_{23}(X_2 X_3 + Y_2 Y_3+ Z_2 Z_3 ) + \frac{1}{2}(A I_3^z - A I_2^z)\sigma_3^z +O\left( \frac{J_{12}}{|A I_1^z - A I_2^z|}\right),
\end{align}
which is similar to the Hamiltonian in (\ref{eqn:HamFred}). To satisfy the conditions of the gate found in Sec.~\ref{sec:FredHeis}, $J_{12} = 2(A I_3^z - A I_2^z)$ and $J_{23}^2 (\frac{m^2}{(2n+1)^2}-1) =4(A I_3^z - A I_2^z)^2$.  For the minimal gate time ($n=1$), this would give $J_{23} = \frac{1}{\sqrt{3}} J_{12}$, or alternatively $J_{23}$ could be tuned to different fractions of $J_{12}$ by altering $m$ and $n$. For example, we can achieve $J_{23} = J_{12}(1+10^{-6} )$ by setting $n = 675$ and $m = 2340$. The resulting gate time with $J_{12} = 2(A I_3^z - A I_2^z)$ would be $675\times 0.5\text{ns} \sim 0.5 \mu\text{s} = 10^{-3}T_2$, so such tuning still would not lead to large decoherence errors. 

\begin{figure}[t]
\begin{center}
\includegraphics[scale = 0.5]{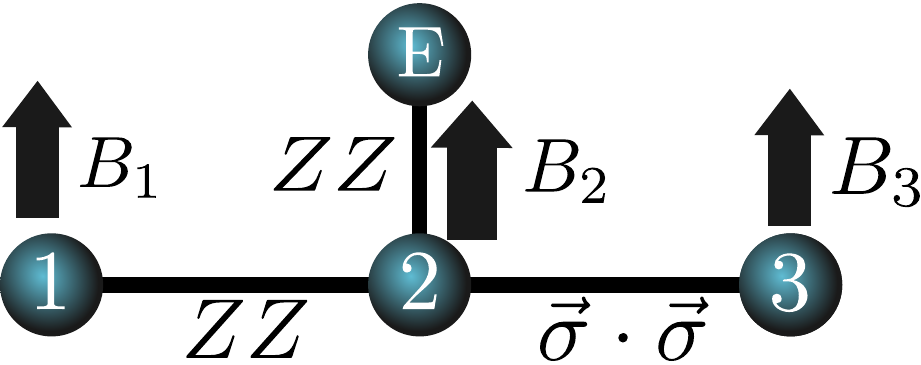}
\caption{Adding a fourth qubit to the original setup, in order to relax the constraints on $J_{23}$ and provide a control to turn the gate on and off.}
\label{fig:FredHeisT}
\end{center}
\end{figure}

With $J_{12}=2(A I_3^z - A I_2^z)$, the error in (\ref{eqn:Rot2}) is $O(  2(I_2^z - I_3^z) / (I_1^z -  I_2^z))$, which is minimal when $ I_1^z = -\frac{9}{2}, I_2^z = \frac{9}{2}, I_3^z = \frac{7}{2}$ ($I_2^z \neq I_3^z$ since this would mean $J_{12} = 0$). To decrease this error, a fourth qubit could be included, as shown in Fig.~\ref{fig:FredHeisT}. Adding this qubit with Ising coupling $J_{E2}$ effectively adds another local magnetic field to qubit 2 and so changes the resonance condition to $J_{12} +J_{E2}= \omega_3 - \omega_2$ (assuming that qubit E is prepared in the $\ket{0}$ state). Note however that this coupling also has an error associated to it, and so the optimal situation is with $J_{E2} = J_{12} =(A I_3^z - A I_2^z) = 1.475$GHz. This results in an error of around 11\%. Adding yet another control qubit $E'$ with coupling $J_{E'2}$ and choosing couplings such that $J_{12} = J_{E2} = J_{E'2} = \frac{2}{3}(A I_3^z - A I_2^z)$ results in errors of around 7\%  (adding any more becomes unrealistic as the control qubits may begin to interact significantly). The gate time in this situation would be $\sim1.5\text{ns}$ which is still significantly smaller than $T_2$.

Additionally, setting $J_{12} = 2(A I_3^z - A I_2^z)$ is not straightforward, since the donor position is not continuously tunable. There are oscillations in the exchange interaction that depend on the separation between donors and the orientation of the donors relative to the crystal~\cite{Koiller2001}, although these oscillations can be minimised if donors are aligned along a particular axis, or if strain is applied~\cite{Koiller2002}. Atomically-precise positioning of the donors is difficult but possible~\cite{OBrien2001,Schofield2003,Fuechsle2012}, so a possible approach is to position the donors at separations of around 15-20nm such that $J_{12} \approx 2(A I_3^z - A I_2^z)$ and then use a magnetic field gradient to change the resonances of ions 2 and 3 to match $J_{12}$ (this magnetic field would also decrease the error in (\ref{eqn:Rot2})). Since it would be possible to bring a magnetic tip close to the sample, large magnetic field gradients of up to $10^7 \text{ Tm}^{-1}$ should be possible~\cite{Mamin2007}, which allows a tuning of up to $\pm 0.5\text{ GHz}$, although this could introduce extra heating of the qubits. Alternatively, we could adopt the method in~\cite{Kane1998}, and use electric gates to tune the inter-donor couplings and increase hyperfine interactions, however this might introduce extra noise due to charge fluctuations.

\subsection{Experimental implementation of T-shaped Fredkin gate}

Achieving an experimental implementation of this gate using ion traps or donors is more challenging. One option using ions would be to use different pulses which rapidly oscillate qubit $A$ and then rapidly oscillate qubits $1$ and $3$. Such a rapid oscillation has the effect of decoupling a qubit from its neighbouring spins, allowing us to perform the $Z_AZ_2$ interaction followed by a $X_1X_2 + Y_1Y_2+ X_2X_3 +Y_2Y_3$ (also obtained via stroboscopic means, by rapidly changing the orientation of the field), which approximates to the desired interaction. However, such a technique is cumbersome and deviates from the philosophy in this work of using just a single time-independent pulse. 

\section{Conclusions}

We have outlined three different protocols to realise 3-qubit gates, with uniform fixed interaction without any extra states or ancillas, such that these gates are implemented in one step (i.e.\ using only a single pulse). A Toffoli gate can be realised using a linear array of qubits interacting via uniform Ising interactions and with a transverse magnetic field applied to the middle qubit. A Fredkin gate can be achieved using a system of three qubits with Ising and Heisenberg interactions. Another Fredkin gate can also be achieved using four qubits in a T shape, with Ising and XY interactions. Note that all of the gates realised in this work are only equivalent to exact 3-qubit quantum gates up to local $Z$-rotations. It would be interesting and useful to find more general results on which types of uniform interactions on 3-qubits could achieve a Toffoli gate. Other approaches could also be taken such as taking logarithms of Toffoli and Fredkin gates.

We have investigated whether or not current technology can be used to perform these gates. We find that using ion traps we would not be able to create a gate that exceeds the quantum computation threshold, but by using bismuth donors in silicon fault-tolerant Toffoli gates should be possible. For the linear Fredkin gate, we find that bismuth donors in silicon would yield a gate with $\sim 10 \%$ error, which could be reduced by the addition of extra qubits to around 4\%. We have not yet discovered an appropriate experimental scheme for the T-shaped Fredkin gate. Whilst some of the experimental realisations here are limited in success probability, and gate times and gate size are similar to existing classical technology, there is no reason why technological improvements may change this in the future, and other experimental realisations could give better results.

Further work could include using more sophisticated pulses that are more selective than a square wave pulse, such as a Gaussian profile~\cite{Vandersypen2005,Vasilev2004}, and including the effects of realistic noise on the gates. It would also be interesting to see if the idea behind the T-shaped Fredkin gate in Sec.~\ref{sec:XX} could be applied to spin chains that perform perfect state transfer.

\chapter{Adiabatic graph-state quantum computation}
\label{chap:AGQC}

In Chapter~\ref{chap:Intro} we introduced the three most widely used models of computation; the circuit model, measurement-based quantum computation (MBQC) and adiabatic quantum computation (AQC). These models are all ``equivalent'' to each other in the following sense; for a given computation, the number of gates required in the circuit model is \emph{polynomially equivalent} to the number of measurements required in MBQC~\cite{Raussendorf2001}, whilst the number of gates required in the circuit model is \emph{polynomial} in the inverse energy gap of the equivalent AQC computation~\cite{Aharonov2007}. 

These results are incredibly important, but from a more practical point of view we are not only interested in polynomial equivalence, but also in the differences that each model has which may make them more attractive when building a quantum computer, and which may give insight into what makes quantum computation powerful. A striking example of these differences is for a circuit consisting of Clifford gates on $N$ qubits: In the circuit model the computation takes at least $\log N$ gates, while in MBQC the measurements can be done in a single constant time step, with a classical processing time that scales with $\log N$~\cite{Browne2011}. So, while it is not possible to use one computational model to (exponentially) outperform another, in MBQC it is possible to move computational time from one ``form'' to another. This presents a potential advantage of MBQC over the circuit model when classical processing is a cheap resource.

Bacon and Flammia proposed the direct translation of MBQC on a cluster state into an adiabatically driven evolution~\cite{Bacon2010}, which they call \emph{adiabatic cluster-state quantum computation} (ACSQC). In this model, the discontinuous measurements of MBQC are replaced with continuous adiabatic transformations. Whilst the evolution proceeds adiabatically, this is not AQC in the typical sense, as the Hamiltonian has a degenerate ground space. It is in fact a type of \emph{open-loop holonomic quantum computation} (see Section~\ref{sec:Holon}), in which the time-varying Hamiltonian performs geometric transformations on information encoded within the degenerate subspace~\cite{Zanardi1999,Pachos1999,Kult2006}. There are also several other works combining ideas from MBQC and AQC. In ancilla-controlled adiabatic evolution~\cite{Wiebe2012,Kieferova2014,Hen2014}, computation is carried out by a combination of adiabatic passage and measurement. In adiabatic topological quantum computation~\cite{Cesare2014}, defects in a topological code are adiabatically deformed to perform logical operations. There are also examples of adiabatically driven computations on non-stabiliser states such as symmetry-protected states of matter~\cite{Williamson2014} and generalised cluster states~\cite{Brell2014}. 

In this chapter we discuss the extension of ACSQC to more general graph states which have \emph{gflow}, a property that is fundamentally linked to the adaptivity of MBQC and determines if a given measurement pattern can produce a deterministic computation~\cite{Danos2006,Browne2007}. This new \emph{adiabatic graph-state quantum computation} (AGQC) allows us to investigate how the properties of MBQC translate into a continuous adiabatically driven evolution. Such a connection between two quite different models could yield interesting insight into what underlies the power of quantum computation, and allows ideas, insight and techniques from one model to be shared with the other. In particular, we mentioned above that in MBQC it is sometimes possible to decrease quantum computation time whilst increasing classical processing time, so that there is a trade-off between quantum and classical time. We are interested in how this will manifest itself in an adiabatically driven model.  

Since there is also an essential requirement in MBQC that measurements are adaptive, as measurement outcomes are non-deterministic, another natural question to ask is what happens if the adiabatic `measurements' in AGQC are performed in a different order to that prescribed in MBQC. At first glance it appears as if we are really post-selecting the correct measurement outcome by using an adiabatic process, so it is not obvious how the causal structure of the measurements should necessarily remain.

We begin by introducing the adiabatic cluster-state quantum computation model, before generalising it to adiabatic graph-state quantum computation in Section~\ref{sec:GeneralAGQC}. We then investigate what trade-off exists between the adiabatic time, the classical computing time for MBQC and the degree of the initial Hamiltonian in Section~\ref{sec:AGQCexample} and finally discuss the role of the ordering of measurements in adiabatic graph-state quantum computation in Section~\ref{sec:Order}. This chapter is based on the work that was published in~\cite{Antonio2014}.

\section{Adiabatic cluster-state quantum computation}\label{sec:ACSQC}

In Section~\ref{sec:MBQC} it was shown how it is possible to achieve universal MBQC and how stabilisers can be used to describe the measurement process. Using these tools, we will now review \emph{adiabatic cluster-state quantum computation} (ACSQC) which converts MBQC on a cluster state into an adiabatically driven holonomic quantum computation~\cite{Bacon2010}. First consider a 1D cluster state of $N$ qubits. The twisted stabiliser generators for this state are
\begin{eqnarray}\label{Stab1D}
K_v^{\theta_{v}}  &=Z_{v-1} X^{\theta_{v}}_{v} Z_{v+1} \quad v = 2,...,N-1; \; \;K_N &=Z_{N-1} X^{\theta_{N}}_{N}.
\end{eqnarray}
To fit with the notation used in~\cite{Bacon2010}, we will use a slightly different definition of these stabilisers generators. In this notation, stabiliser generators are defined as $T_v  := K^{\theta_{v+1}}_{v+1}$, i.e.\ the same as the $\{ K^{\theta_{v}}_v \}$ stabilisers except the indices are shifted by 1 site.
The first important thing to note is that, since the stabilisers by definition commute and act as identity on the cluster state, the cluster state is the ground state of the following Hamiltonian:
\begin{eqnarray}
H_0 \equiv - \gamma\sum_{v= 1}^{N-1} T_v,
\end{eqnarray}
where $\gamma$ parametrises the strength of the interactions (in this work we assume that the highest coupling strength possible is chosen, so that $\gamma$ is already set at its highest value and therefore is a constant that does not scale with the system size). The eigenspace is spanned by the two states which can be used to encode a qubit:
\begin{eqnarray}
\ket{0}_L = \prod_{v \sim w} \textsc{cz}_{(v,w)} \ket{0}_1 \bigotimes_{n > 1} \ket{+_{\theta_n}}_n, \; \; \ket{1}_L = \prod_{v \sim w} \textsc{cz}_{(v,w)} \ket{1}_1 \bigotimes_{n > 1} \ket{+_{\theta_n}}_n
\end{eqnarray}
The logical $X$, $Y$, $Z$ operators of this qubit ($X_L$, $Y_L$ and $Z_L$, respectively) are the same as in MBQC;
\begin{eqnarray}\label{eqn:LogOp}
X_L \equiv X_1 Z_2, \; Y_L \equiv Y_1 Z_2 , \; Z_L \equiv Z_1.
\end{eqnarray}
The computation begins by preparing the ground state of $H_0$, which could be done by starting in the ground state of a uniform magnetic field and adiabatically evolving to $H_0$ (encoding an arbitrary input state could then be done by adding a field $\alpha_x X_L + \alpha_y Y_L + \alpha_z Z_L$ to $H_0$). Then instead of measuring qubits as in MBQC, the first stabiliser generator in the Hamiltonian $T_1$ is adiabatically replaced with $X_1$:
\begin{align} \label{Ht}
H(s) &=-  (1-s)\gamma T_1 - s \gamma X _1 - \gamma \sum_{n =2}^{N-1} T_n
\end{align}
where $s = t/\tau$, $\tau$ is the total run time and $0 \le t \le \tau$. To see what happens after this adiabatic substitution, the logical operators can be multiplied by stabilisers (since they act as identity, as shown in Section~\ref{sec:MBQC} on MBQC) until they commute with $X_1$ (i.e.\ the logical operators are put in a form in which they are conserved during the adiabatic transformation). $X_L$ already commutes with $X_1$, but $Z_L$ doesn't so it must be multiplied by $T_1$:
\begin{align}
Z_L \rightarrow Z_L T_1 = X_2^{\theta_2} Z_3.
\end{align}
Similarly $Y_L$ also doesn't commute with $X_1$, so must be multiplied by $T_1$:
\begin{align}
&Y_L \rightarrow Y_L T_1 =(Y_1 Z_1)  (Z_2 X_2^{\theta_2}) Z_3 \nonumber\\
&=iX_1 (-i)e^{i\pi Z_2/2} e^{-i \theta_2 Z_2} X_2 Z_3= X_1 e^{-i \theta_2 Z_2} e^{i\pi Z_2/2} X_2 Z_3 = X_1 e^{-i \theta_2 Z_2} Y_2 Z_3.
\end{align}
Then, after setting $X_1 = \idop$ since the system is in the $+1$ eigenstate of $X_1$, the logical operators are transformed into
\begin{eqnarray}\label{eqn:LogOp2}
X_L \rightarrow Z_2, \;Y_L\rightarrow e^{-i \theta_2 Z_2} Y_2 Z_3,\; Z_L \rightarrow X_2^{\theta_2} Z_3.
\end{eqnarray}
If new logical operators $\{ \tilde{X}_L, \tilde{Y}_L, \tilde{Z}_L \}$ are defined in the same way as in eqn.\ (\ref{eqn:LogOp}):
\begin{eqnarray}
\tilde{X}_L \equiv X_2 Z_3, \; \tilde{Y}_L \equiv Y_2 Z_3, \;\tilde{Z}_L \equiv Z_2.
\end{eqnarray}
Then expressing the operators in (\ref{eqn:LogOp2}) in this basis gives:
\begin{align}\label{eqn:LogOp3}
X_L &\rightarrow Z_2 = \bar{U}_{2} \bar{\text{H}}_L \tilde{X}_L  \bar{\text{H}}_L\bar{U}_{2} \nonumber\\
Y_L &\rightarrow e^{-i \theta_2 Z_2} Y_2 Z_3 \equiv \bar{U}_{2} \bar{\text{H}}_L \tilde{Y}_L  \bar{\text{H}}_L\bar{U}_{2} \nonumber\\
Z_L &\rightarrow X_2^{\theta_2} Z_3.\equiv \bar{U}_{2} \bar{\text{H}}_L \tilde{Z}_L  \bar{\text{H}}_L\bar{U}_{2},
\end{align}
where $\bar{\text{H}}_L$ is a Hadamard operation acting in the logical subspace, and $\bar{U}_{v} :=\exp[ -i\theta_{v} Z_L / 2]$ is a logical $Z$ rotation. So the effect of the evolution on the logical information is to move it one step down the chain, and to apply $\bar{U}_{2} \bar{\text{H}}$ to the information (exactly the same as in the MBQC example in Sec.~\ref{sec:MBQC}). Then at the next step $T_2$ is replaced with $X_2$, and so on from left to right down the chain (see Fig.~\ref{tab:Normal} for an illustration of this). This results in the information being encoded in the $N^{th}$ qubit, with the information transformed by the operation $\bar{U}_{tot} = \bar{H} \prod\nolimits_{v = N-2}^2(\bar{U}_{v} \bar{\text{H}})$, and since $\bar{U}_{v} = \exp[ -i\theta_{v} Z_L / 2]$ this will depend on the sequence of angles $\{\theta_v \}$ used (following the convention in~\cite{Bacon2010} we set $\theta_1 = 0$). Note that, due to the Hadamard operations, this results in alternating $X$ and $Z$ rotations, which allow universal single-qubit rotations.
When replacing one stabiliser by a Pauli $X$ operator, the speed of this adiabatic transformation is limited by the ratio of the minimum energy gap and $\Vert \dot{H}(s) \Vert$ (see Sec.~\ref{sec:AdTheor}). For the above case, the minimal energy gap of $H(s)$ is $\sqrt{2}$, and $\| \dot{H} \| = 1$, which means the time taken is
\begin{eqnarray}\label{eqn:OneStep}
\tau \geq \tau_0 :=\frac{1}{\varepsilon 2^{1 + \delta/2} \gamma}
\end{eqnarray}
for some fixed $0<\delta\leq1$ and $\varepsilon$. For the remainder of this chapter we will compare adiabatic evolution time to this time $\tau_0$, so that this is our definition of one unit of time for the adiabatic computation, and we say that the adiabatic computation takes one `step' if the time satisfies the same condition as above.

\begin{table}[t]
\begin{eqnarray}
\begin{array}{l | cccccc}
\mbox{Start}& T_1 & T_2 & T_3 & \dots & T_{N-2} & T_{N-1}\\
& \downarrow &  &  &  & & \\
\mbox{Step 1} & X_1 & T_2 & T_3 & \dots & T_{N-2} & T_{N-1}\\
& & \downarrow &  &  &  & \\
\mbox{Step 2} & X_1 & X_2 & T_3 & \dots & T_{N-2} & T_{N-1}\\
& & & \downarrow  &  &  & \\
\mbox{Step 3} & X_1 & X_2 & X_3 & \dots & T_{N-2} & T_{N-1}\\
\hspace{10mm} \vdots&& & \vdots \\
& & & & & \downarrow  &  \\
\mbox{Step }  (N- 2) & X_1 & X_2 & X_3 & \dots & X_{N-2} & T_{N-1}\\
& & & & & & \downarrow \\
\mbox{Step }  (N- 1) & X_1 & X_2 & X_3 & \dots & X_{N-2} & X_{N-1}
\end{array}
\end{eqnarray}
\caption{An illustration of the method to perform single qubit operations in adiabatic cluster-state quantum computation; arrows indicate an adiabatic transition from one operator to the other.}
\label{tab:Normal}
\end{table}

\subsection{Two-qubit gates}\label{sec:CNOT}

\begin{figure}[h]
\centering
\includegraphics[scale = 0.7]{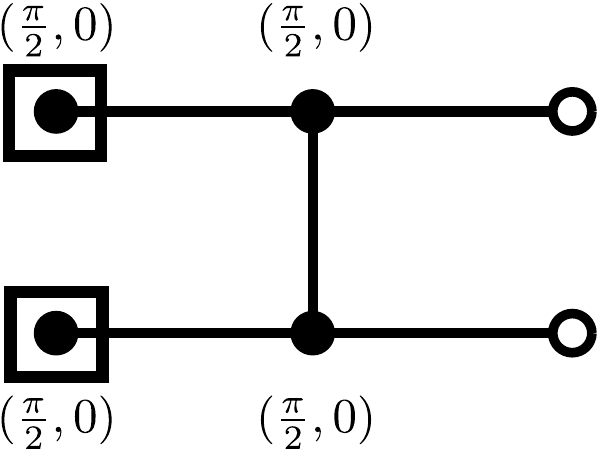}
\caption{An illustration of the graph used to perform a \textsc{cnot} gate in adiabatic cluster-state quantum computation.}
\label{fig:CNOT}
\end{figure}

We have just seen how to perform arbitrary single qubit rotations in the adiabatic cluster-state model. In order to perform universal quantum computations, an entangling gate such as a \textsc{cnot} gate must also be possible. Consider the graph discussed in Sec.~\ref{sec:MBQC}, with two rows of qubits, labelled $a$ and $b$, with 3 columns numbered from left to right (see Fig.~\ref{fig:CNOT}). The stabilisers are $T_{a1} = K_{a2}$, $T_{a2} = K_{a3}$ and similarly for $T_{b1}$, $T_{b2}$. Then the protocol starts with an initial Hamiltonian in which all angles $\{ \theta_n \}$ are 0, and which the inputs are on qubits $a_1$ and $b_1$, and the outputs are on qubits $a_3$ and $b_3$:
\begin{align}
H_{\textsc{cnot}} &= -T_{a1} -T_{a2} - T_{b1} - T_{b2} \nonumber\\
&=-Z_{a1}X_{a2}Z_{a3} Z_{b2} - Z_{a2} Z_{b1} X_{b2} Z_{b3} - Z_{a2} X_{a3} - Z_{b2 }X_{b3}.
\end{align}
Following a similar procedure as in the 1D case above, we see that the effect of adiabatically replacing all of these stabilisers with $X$ operators results in a \textsc{cnot} gate acting on the encoded information. Each replacement of a single stabiliser by a local Pauli operator still has the same adiabatic time as in the previous section, i.e.\ the adiabatic evolution time takes the form $\tau_0$ (provided these substitutions are done in the correct order). Note that also a \textsc{cz} gate can easily be achieved using an adiabatic scheme based on the gates in~\cite{Browne2007}, in which only 3 qubits are required to perform a gate on two qubits.

\section{Adiabatic graph-state quantum computation}

In the previous section we saw that universal quantum computation is possible using adiabatic substitutions, instead of measurements, on the cluster state. Here this approach is generalised to other graph states using tools from MBQC. This is an extension of adiabatic cluster-state quantum computation, which we call \emph{adiabatic graph-state quantum computation} (AGQC).

 \subsection{General translation of an MBQC pattern to a holonomic computation}
\label{sec:GeneralAGQC}
Here we will look at converting any general measurement pattern on a graph state into a holonomic computation. Consider an open graph state with vertices $V$, \emph{gflow} $(g,<)$, input vertices $I$ and output vertices $O$. For simplicity we take the number of inputs to be equal to the number of outputs $|I|=|O|$, but all statements and proofs can be easily extended to the cases $|O|>|I|$ ($|O|<|I|$ is not allowed as this would mean information is lost). The AGQC protocol starts in the ground state of the initial Hamiltonian
\begin{eqnarray}
H_0 =-\gamma \sum_{v \not\in O} T_v,
\end{eqnarray}
where the $T_v$ operators are products of the twisted stabiliser generators,
\begin{align}\label{eqn:GflowStab}
T_v := \prod_{w \in g(v)} K_w^{\theta_w},
\end{align}
$\forall v \notin O$. The ground state of $H_0$ corresponds to the open graph state. Reaching the ground state of $H_0$ could be achieved by starting in the ground state of a uniform magnetic field and adiabatically changing the Hamiltonian to $H_0$. The final Hamiltonian will be $H_f=-\gamma\sum_v X_v$, and the transition will be done in steps, as in the cluster-state case (for which $g(v)$ contains only one element and so $T_v = K_{g(v)}$, as seen in Section~\ref{sec:ACSQC}). The initial Hamiltonian can be computed efficiently from the graph and \emph{gflow}, since each $T_v$ operator can be found through multiplication of the $K_w^{\theta_w}$ generators which can be performed in a time that scales as $O(\log (\mbox{max}_v |g(v)|))$ using the methods in~\cite{Aaronson2004}.

Notice that the $T_v$ operators can cover many sites; this is an important property that we must keep track of, so we define the number of sites an operator covers as \emph{degree} of the operator:
\begin{defin}[Degree of an operator]
The degree of an operator is the number of non-trivial (i.e. non-identity) terms when the operator is written as a tensor product of 2 dimensional Pauli matrices.
\end{defin}
So, for example, the operator $Z_1 \otimes X_2 \otimes \idop_3 \otimes \idop_4$ has degree 2, and the degree of the 1D cluster state stabilisers are $\leq 3$. For a Hamiltonian which is a sum of many operators, we say the Hamiltonian $H$ has degree $k$ if all of the operators in $H$ have degree $\leq k$.
It will also be useful to define the size of a \emph{gflow}, in terms of its stabilisers
\begin{defin}[$|g(v )|$]
The size $|g( v )|$ of a \emph{gflow} is the number of qubits for which the product of the stabilisers over $g( v )$, $\prod_{v\in g(v )}K_v$, is non-trivial.
\end{defin}
E.g. for the cluster-state, there is only one vertex in $g(v)$, and since each stabiliser generator acts on up to 5 qubits, the size of the \emph{gflow} is 5.

To perform the computation, the operators $\{T_v \}$ can be replaced one-by-one in an order that doesn't violate the \emph{gflow}, or those in the same layer can be replaced all at once. One may choose instead to ignore the ordering or \emph{gflow} and replace all the stabilisers in one adiabatic step, as shown in~\cite{Bacon}. The downside of this approach is that the energy gap will most likely shrink polynomially in the the number of qubits, as suggested by the numerical results in~\cite{Bacon}, which means the gap protection is lost, and additionally analytical or numerical results become very difficult to find except for a few specific cases. Thus we consider stepwise computations with fixed energy gap, which still tell us many interesting things about adiabatically driven versions of MBQC. We would also expect that replacing all of the stabilisers will only result in non-zero energy gap when the graph has \emph{gflow}; this is supported by the re-ordering results in Sec.~\ref{sec:Order}, and can be verified for the example in Fig.~\ref{fig:NoGflow}.

Replacing the stabilisers one-by-one will take $N$ steps of time $\tau_0$. When replacing layer by layer, the adiabatic transition for the $k$th step is governed by the interpolation Hamiltonian
\begin{equation}
H_{L_k}(s) = (1-s) H_{k-1} + s H_k,
\end{equation}
where
\begin{equation}
H_k= \gamma \left(-\sum_{v\leq L_k}X_v - \sum_{v>L_k}T_v \right).
\end{equation}
What is the time needed to perform the $k^{th}$ step? To determine this we will need to find the form of the energy gap and the form of $\Vert \dot{H} \Vert$ to put into the adiabatic formula given in eqn.\ (\ref{eqn:RunTime}). First we split the Hamiltonian into two parts
\begin{align}
\label{eqn:HamSplit}
H_{L_k}(s) =  - \gamma \left( \sum_{v <  L_k}X_v + \sum_{v>L_k}T_v \right) - \gamma \left(\sum_{v\in L_k} s X_v + (1-s)T_v \right).
\end{align}
The terms in the sum on the left commute with each other, which can be seen from the definitions of the $\{ T_v \}$. Each of the summands on the right also commutes with each other and all of the summands on the left. To see this, note that from the conditions of \emph{gflow}, a product of stabiliser generators $\prod_{w \in g(v)} K_w$ has an even number of connections to vertices in the same layer or in previous layers (which means an even number of $Z$ operators, which cancel to give identity) and an odd number of connections to vertex $v$ (which means there is one $Z_v$ term in $\prod_{w \in g(v)} K_w$. This does not change when we use twisted stabiliser generators). Therefore $\{T_v,X_v\} = 0$, and $[T_w,X_v] = 0$ for all $w \geq v$. 

A full analysis of the eigenvalues of this Hamiltonian is given in Appendix~\ref{sec:LinAlg}; the result is that the eigenvalues of $H_{L_k}$ are $-|L_k|\gamma\eta,-(|L_k|-2)\gamma\eta,...,(|L_k|-2)\gamma\eta,|L_k|\gamma\eta$ plus factors of $\pm \gamma$, where $\eta := \sqrt{(1-s)^2 + s^2} $. The factors of $\pm \gamma$ do not contribute to the minimum energy gap since $\gamma \eta\le \gamma$. The energy gap between the ground state and first excited state is therefore $2 \gamma  \eta$, which is minimal at $s=\frac{1}{2}$, at which point the gap is $\sqrt{2}\gamma$.

The time-derivative of the Hamiltonian is
\begin{eqnarray}
\dot{H}(s)=\gamma\sum_{u \in L_k} ( T_u - X_u).
\end{eqnarray}
Using the same methods as in Appendix~\ref{sec:LinAlg}, the eigenvalues of this derivative are $\{ -|L_k|\gamma,-(|L_k|-2)\gamma,...,(|L_k|-2)\gamma,|L_k|\gamma \}$, so the spectral norm of $\dot{H}(s)$ is $|L_k|\gamma$. Inserting the minimal energy gap and $\Vert \dot{H}(s)  \Vert$ into equation (\ref{eqn:RunTime}) for an adiabatic evolution, we then find that after adiabatically replacing the $k^{th}$ layer using a linear interpolating function, the final state will be in the ground space with some fixed error $\varepsilon$ if the adiabatic time $\tau$ scales as
\begin{eqnarray}
\tau =   \Omega \left( \frac{  \Vert \dot{H}(s) \Vert^{1+\delta}}{\Delta(s)_{min}^{2+\delta}} \right) = \Omega \left( |L_k|^{1 + \delta}  \right)
\end{eqnarray}
for some $0 < \delta  \leq  1$.

Now consider what happens to the information when we replace all $T_v$ operators defined above with an $X_v$ operator, in the order given by \emph{gflow}. Following the method in~\cite{Bacon2010}, the effects on the information can be seen by multiplying logical operators with the $T_v$ operators in such a way that the logical operators commute with all the adiabatic `measurements'. This means that it is possible to update the logical operators $\alpha_L$, to $\tilde{\alpha}_L$ such that $[\tilde{\alpha}_L,X_v] = 0$ for all $v$, and where $\alpha = X,Y,Z$. This then tells us what state is output by the computation. For any graph state with \emph{gflow}, it is always possible to do this, since all we have to do is multiply any $Z_v$ or $Y_v$ operators by the $T_v$ defined above; the \emph{gflow} conditions guarantee that after this replacement, $Z_v \to \idop_v$ and $Y_v \to X_v$, and the new logical operator will commute with any $X_w$ such that $v\ge w$. For graphs without \emph{gflow}, such a process is not generally possible, since when multiplying a $Z_v$ or $Y_v$ operator by a $T_v$ stabiliser which violates \emph{gflow}, the result may contain an operator $Z_w$ such that $w < v$. This property can be seen in the example in Fig.~\ref{fig:NoGflow}; to make the logical operator $Z_1$ commute with $X_1$, we multiply $Z_1$ by $K_3$ giving $Z_1 \to \tilde{Z}_1 = X_3 Z_2$. Now to make this commute with $X_2$, we can only multiply by $K_4$, leaving $\tilde{Z}_1 = Z_1 X_3 X_4$ which no longer commutes with $X_1$.

To see that a holonomic quantum computation on a graph state with \emph{gflow} performs the same computation as the equivalent measurement pattern in MBQC, consider starting with a twisted graph state (i.e.\ the stabilisers are twisted by angles $\{ \theta_v \}$, and we make all measurements in the $X$ basis). If we start with the logical operators of the MBQC resource state, and update these logical operators $\alpha_L$, to $\tilde{\alpha}_L$ such that $[\tilde{\alpha}_L,X_v] = 0$ for all $v$, then if we start in the $+1$ eigenstate of a logical operator $\tilde{\alpha}_L$, after the adiabatic substitution (or measurement) we will still be in the $+1$ eigenstate of $\tilde{\alpha}_L$. This is just the same as simulating the MBQC computation in the Heisenberg picture. 

These results are summarised in the following theorem:
\begin{theorem}\label{theor:AGQCTheor}
Any measurement based pattern on a graph $G$ of $N$ qubits which has \emph{gflow} g and depth $d$ can be efficiently converted into an equivalent holonomic computation for which
\begin{itemize}
\item The holonomic computation can be done in $d$ steps, where the minimum energy gap for each step is fixed at $\sqrt{2} \gamma$, and $\Vert \dot{H} \Vert = |L_j|$ for the $j^{th}$ step. Thus the time to perform the $j^{th}$ step is $\Omega( |L_j|^{1 +
\delta})$, where $0 < \delta \leq 1$.
\item The resulting computation is the same. 
\item The maximum degree of the initial Hamiltonian $k_{\max}$ is equal to the maximum \emph{gflow} size : $k_{\max} = |g(\nu)|_{max}$,
\item The initial Hamiltonian can be computed efficiently.
\end{itemize}
 \end{theorem}
As an example, consider the graph from~\cite{Browne2007} shown in fig.~\ref{fig:gflow}, which has \emph{gflow} $g(a_1) = \{b_1 \}, g(a_2) = \{ b_2 \}, g(a_3) = \{ b_3,b_1\}$. Following the prescription given above, we start with stabilisers $T_{a_1} = K_{b_1}, T_{a_2} = K_{b_2}, T_{a_3} =K_{b_3}K_{b_1}$ which have degree at most 3. The adiabatic substitutions are then performed in the order $a_1 < a_2 < a_3$. Using these stabilisers, Theorem~\ref{theor:AGQCTheor} tells us how to carry out the AGQC in adiabatic time proportional to the depth, which is 3 in this case. Also for adiabatic cluster-state quantum computation on a rectangular graph qubits with $r$ rows and $d$ columns, $g(v) = v + r$, $T_v = K_{v+r}$, $|g(v) | \leq 5$ and there are $d$ layers, with each layer containing $r$ qubits, so the time for each layer scales as $\Omega(|r|^{1 +
\delta})$.

Note that, for all known universal graphs there is a \emph{gflow} for which $|g(v)|$ is bounded, so typically the degree will also be bounded. However this is not necessarily the case for all families of graphs; in some cases it can scale with the number of inputs (an example of this is shown in Section~\ref{sec:Example}). This raises an important question: Is this increase in degree regarded as a free resource, or is there a cost associated with it? The evidence to date would suggest that the latter is true, since typically in nature we only see 2-body interactions, with higher degree interactions resulting as a low energy approximation. We take this approach in our model, and assume that degree is bounded and such high degree Hamiltonians can only result as an approximation to a 2-body Hamiltonian. 

A few fundamental results are known about high-degree Hamiltonians. For instance, it is known that that there are no 2-body Hamiltonians for which the ground state is exactly the same as that of a $k$-body Hamiltonian, where $k >2$~\cite{Siu2005,Haselgrove2004,VanDenNest2008}. If we try to simulate this $k$-body interaction with a 2-body interaction, then it was shown in \cite{Siu2005,Haselgrove2004} that the energy gap will scale as $E_{gap} < O(1-F^2)$, where $F= \sprod{\psi}{E_0}$ is the fidelity between the actual ground state $\ket{\psi}$ and the $k$-body ground state $\ket{E_0}$ (i.e.\ the accuracy of the approximation). However, as far as the author is aware, there are no known results which say that there is a \emph{fundamental} reason why many body interactions should be difficult to create. What is known is that, for the currently known methods of simulating large-degree Hamiltonians such as perturbation gadgets~\cite{Kempe2004,Oliveira2008,Bartlett2006,Jordan2008}, the Hamiltonian has an energy gap which shrinks with the degree. For example, we can create k-local Hamiltonians $H_k$ using a perturbative Hamiltonian acting on $rk$ ancilla qubits and n computational qubits (r is the number of terms in Hamiltonian with degree $k$, which we consider as being fixed)~\cite{Jordan2008}. The result is that the effective Hamiltonian $H_{\textrm{eff}}$, apart from some overall energy shift, is
\begin{eqnarray}
H_{\textrm{eff}} = \frac{-k(-\lambda)^k}{(k-1)!} H_{k} \otimes P_{+}  + O(\lambda^{k+1}),
\end{eqnarray}
where $P_+$ is a projector on the space of $r$ ancilla qubits, projecting each one into the $\ket{+}$ state, and the perturbation converges provided that $\lambda < \frac{k-1}{4k}$. This energy gap decreases exponentially with $k$, therefore making the reasonable assumption that interactions in nature are limited to interactions between 2 (or a finite number of) bodies, then if the degree is allowed to scale with $N$, this imposes a prohibitive cost in that the minimum energy gap of the system shrinks exponentially, and so therefore the adiabatic time grows exponentially. Thus as much as possible it is best to avoid large degree Hamiltonians in adiabatically driven schemes.

Note that we have implicitly assumed that the energy scale is fixed; if it were not, then since the adiabatic runtime scales as $\Omega( | L_k|^{1+\delta} / \gamma)$ for each layer, then increasing the interaction strengths will compensate for the increase in adiabatic runtime. However, scaling the energy freely is likely to be a difficult task in any realistic system, so we consider the case where the interactions are already set at the strongest they can be.

\subsection{Trade-offs in AGQC}
\label{sec:AGQCexample}

Above we have seen that it is possible to transform any measurement pattern on a graph state with \emph{gflow} into an adiabatic computation. Here we investigate how the trade-off between quantum and classical processing in MBQC manifests itself in AGQC. To begin with, consider the graph shown in Fig.~\ref{fig:ZigZag}. Recall from Section~\ref{sec:FlowGflow} that it is possible to define many different \emph{gflows} $g^r(v)$ on this graph
\begin{eqnarray}
g^r(v) = \left\{
\begin{array}{c c}
\{N+v,...,N+v+r -1 \}, & \mbox{if} \hspace{2mm}  v+r-1 \le N  \\
\{N+v,...,2N \},  & \mbox{if}  \hspace{2mm}  v+r-1 >N
\end{array} \right.
\end{eqnarray}
where $1 \le r \le N$. The measurement depth $d_r$ is given by $d_r = \left\lceil \frac{N}{r} \right\rceil$, and the size of the \emph{gflow} is $|g^r(v)| \leq r+2$. Following the the application of Theorem~\ref{theor:AGQCTheor} to this graph, the computation can be performed in $d_r$ adiabatic steps, where the time to perform the $j^{th}$ layer scales as $\Omega( |L_j|^{1+ \delta} )$, and the maximum degree is given by the influencing volume. Thus $g^r$ is converted into an adiabatic computation which has $d^r$ steps, and each step takes $\Omega(r^{1+\delta})$ time, and the Hamiltonian degree $k = r+2$. In the most extreme case, $g^{N}$ is converted into an adiabatic computation which takes 1 step, but this step takes $\Omega(N^{1 + \delta})$ time to complete, and the Hamiltonian degree is proportional to $N$. 

Thus by playing with the \emph{gflow} in this way, there is a trade-off in this zig-zag graph between the number of adiabatic steps, the size of $\| \dot{H} \|$ and the degree of the Hamiltonian, in direct analogy to the quantum/classical trade-off seen in MBQC. This shows an interesting contrast between MBQC and AGQC: since classical computation is a cheap, low-error resource it is desirable to shift as much computation into classical processors as possible, so the optimal flow to choose for MBQC would be $g^N$ (also called the maximally-delayed gflow) since it maximises classical computation. However, in AGQC the total computation time is always approximately linearly in $N$, so there is no advantage in using large delayed \emph{gflow}, and given the expense of creating high degree operators discussed above, it even seems prohibitive to use the maximally delayed \emph{gflow}, and so the optimal situation seems to be using $g^1$ (the minimally delayed \emph{gflow}).

In MBQC, it is only possible to perform measurements in one step for examples like the zig-zag graph where the measurements are in the Pauli basis, since for more general measurements the random outcomes of the measurements cannot be corrected at the end. Given this fact, it is natural to ask whether or not the trade-off seen above for the zig-zag graph applies to all graphs, and all computations. It is interesting partly for the sake of rigorousness, but also since if it was possible to find a way to perform any computation in one constant time step at the expense of degree, this would have profound implications, giving a fundamental bound on the energy gap for simulating large degree operators (under the reasonable assumption that all computation cannot be done in constant time). Even if this isn't the result, exploring the particular way in which AGQC stops us being able to perform computations in one step is illuminating in itself.

In the following we will investigate this question whilst keeping the energy gap constant, and by manipulating the terms in the Hamiltonian. As mentioned in the previous subsection, this has the advantage of keeping the energy gap protection and allowing analytical results to be found. The numerical results in~\cite{Bacon} allude to another complementary trade-off, perhaps more intuitive, between the energy gap and the number of steps in the computation. Since it is harder to find exact results, or even numerical results, for systems other than 1-dimensional chains, we concentrate on the method below which allows analytical results.

To perform the computation in one step with constant energy gap, following the logic leading to Theorem~\ref{theor:AGQCTheor} we must find a transformation of the stabilisers $T_v \to \tilde{T}_v$ such that all of the summands in the following Hamiltonian commute:
\begin{align}\label{eqn:AGQCallatonce}
H(s) = -\gamma\sum_{v \in V \setminus O} (1-s)\tilde{T}_v + s X_v
\end{align}
These summands will commute when the $\tilde{T}_v$ stabilisers satisfy the commutation relations
\begin{eqnarray}
[\tilde{T}_v, X_w] = 0\; \mbox{for all} \; v \ne w \; \mbox{in} \; V \setminus O
\end{eqnarray}
If such a transformation is possible, then all of the stabilisers $\tilde{T}_v$ can be replaced at the same time with fixed energy gap $2\gamma \eta$, provided the adiabatic runtime scales as $\Omega(N^{1+\delta})$ (see Appendix~\ref{sec:LinAlg}). For a graph with \emph{gflow}, such a procedure is always possible by multiplying the $\tilde{T}_v$ stabilisers together, since any $Z_v$ or $Y_v$ operator can be transformed to identity or $X_v$ by multiplication by $T_v$, and this is guaranteed to commute with all vertices $w < v$. To update stabiliser $T_v$, we start in layer $L_{n+1}$, where $L_n = L(v)$, and update as follows; (1) If $T_v$ contains a $Z_w$ or $Y_w$ term in layer $L_{n+1}$, multiply by $T_w$. (2) Proceed to the next layer, as determined by the time order. Iterate until the outputs are reached. (3) The final updated stabilisers are denoted $\tilde{T}_v$.

However, creating a Hamiltonian in such a way is polynomially equivalent to simulating the computation (see e.g.~\cite{Markham13}), a fact which has already been used in the discussion of Theorem~\ref{theor:AGQCTheor}. For instance, starting with the logical operator $\bar{X}_1 = X_1 Z_2$ and multiplying this by $\tilde{T}_2$ gives a new logical operator $\bar{X}' = X_1 Z_2 \tilde{T}_2$ which commutes with all $X$ operators at non-outputs. Since it commutes with all $X$-measurements, the eigenstate of $\bar{X}'$ is preserved throughout the adiabatic evolution, so if the system starts in the $+1$ eigenstate of $\bar{X}'$ we end in the $+1$ eigenstate of $\bar{X}'$. Then to solve the computation all that is required is to ignore all the $X$ operators on non-outputs, and find the $+1$ eigenstate of the operators in $\bar{X}$ that act on the outputs.

Or to put it in a different way, we can look at how efficient the update procedure is compared to the efficiency of the computation before the procedure. For Clifford operations the procedure takes polynomial time: if we have $N$ qubits in total, we will have to perform one updating sweep per qubit, each sweep involves a search over at most $N$ stabilisers to see whether they commute or anti-commute, and the cost of testing if a stabiliser commutes or not will be $O(N)$ since each stabiliser will contain at most $N$ terms. So the procedure takes $O(N^2)$ steps. For general angles the procedure takes an exponential amount of time, since at every step a $Z_v$ operator is replaced with a $Z_vT_v = e^{-i \theta_{v+1} Z_{v+1}} X_{v+1} Z_{v+2}$ term, so every update converts a $Z$ term into 3 new terms. If we start off with a stabiliser containing $n$ $Z$ operators, then after $r$ sweeps we will have $O(3^r n)$ operators to search through.

In addition, during this update procedure, we multiply every $Z_w$ operator by $T_v = \prod_{w\in g(v)} K_{w}$. So following this procedure, each $\tilde{T}_v$ operator will have Pauli $X$'s in positions $g(w)$ for every $Z_w$ that has been have corrected for. The only parts that will contribute to the degree of $\tilde{T}_v$ are situated at vertices which can be arrived at by following a path on which a non-gflow line is preceded and followed by a gflow line. Following the discussion in Sec.~\ref{sec:GeneralAGQC} we expect the simulation of these Hamiltonians to be very prohibitive. 
 
So we have seen that the trade-off seen for the zig-zag graph can be extended to more general computations, such that the number of adiabatic steps can be decreased at the expense of increasing $\| \dot{H} \|$ and the degree of the Hamiltonian. In addition, the process of finding the initial Hamiltonian becomes inefficient for non-Clifford operations, since the description of the Hamiltonian is polynomially equivalent to the solution of the computation. Using this method, it is not possible to reduce the overall adiabatic time, only the number of steps. Perhaps surprisingly, there is also no apparent advantage in having free access to unbounded-degree Hamiltonians. Indeed, given the prohibitive cost of the known ways to achieve high-degree operators, our result suggests that if it is desirable to keep the energy gap fixed it is best to perform the computation in as many steps as possible (or in the language of MBQC, it is best to use the minimally-delayed \emph{gflow}). This result highlights the subtlety in using the adiabatic theorem; although the minimum energy gap is the same for graphs of different flow, the size of $\Vert  \dot{H} \Vert$ changes and leads to a dependence of the adiabatic time on the number of elements in a \emph{gflow}.

\section{Reordering the computation} \label{sec:Order}

In MBQC, doing measurements in the wrong order can result in random outcomes to the computation. In some cases however, such as when all qubits are measured in the Pauli basis, the correct state can still be recovered by using local Pauli corrections. Given that the non-deterministic measurements are absent in AGQC, it is natural to wonder whether or not the same ordering rules apply, and if they do, what factors collude to make it so. One approach to studying this is using the same method as in the previous sections, by manipulating terms in the Hamiltonian and keeping the energy gap fixed. However, as we have seen previously, whilst this does allow the ordering to be changed, it is generally at the expense of high degree Hamiltonians. In this section we will go back to the original formulation of adiabatic cluster-state quantum computation, and look at what order the adiabatic substitutions can be performed in, with the constraint that we will only use the stabilisers used in~\cite{Bacon2010}. As a physical motivation we may imagine that we have some experimental apparatus which is limited to only applying operators below a certain size, but we can turn them on in any combination. Firstly we will consider the most obvious case, where we replace one stabiliser by one Pauli operator, and then we consider a less constrained method, and discuss how these two approaches lead to different behaviour.

\subsection{Re-ordering the adiabatic computation with fixed number of terms in Hamiltonian}\label{sec:Order2}

We start with an $N$-qubit 1D chain with $N-1$ stabilisers (i.e.\ one qubit is encoded in the chain), and we consider replacing one stabiliser by one Pauli operator in a different order to the order in MBQC. We would like to find orders of replacements which preserve information in the 2-dimensional logical subspace, and we might expect from MBQC that changing the order will in some way disrupt the computation. To begin with, consider a 4-qubit chain with all angles set to zero (i.e.\ with un-twisted stabilisers). We start in the ground state of the initial Hamiltonian $H_0$:
\begin{eqnarray}
H_0 = -T_1 - T_2 - T_3.
\end{eqnarray}
First notice that replacing $T_2 \to X_2$ does not affect the energy gap of this Hamiltonian, since $[X_2 , T_1] = 0$ in this case. Thus we are free to go out-of-order in this way, provided $\theta_2 = n \pi$. However, if instead we start by replacing $T_3$ with $X_3$ out-of-order, the time-dependent Hamiltonian becomes
\begin{eqnarray}
H(s) = -T_1 - T_2 - (1-s)T_3 - sX_3,\; \mbox{ with } 0\le s \le 1.
\end{eqnarray}
At $s=1$, this Hamiltonian has a ground state degeneracy of 4, i.e.\ the degeneracy doubles during the evolution. However, there is a constraint in that the operator $T_1T_3$ commutes with $H(s)$ for all $s$ and so the eigenstate of $T_1T_3$ is conserved throughout the evolution. This means that since we start in the $+1$ eigenspace of $T_1T_3$, we will also end in the $+1$ eigenspace of $T_1T_3$. So although the gap closes, the transitions between these degenerate eigenstates are forbidden and so the subspace is preserved.

Now consider replacing $T_1 \to X_1$. Throughout this evolution, $T_1T_3$ no longer commutes with $H(s)$, and so unless there is a stabiliser that $T_1T_3$ can be multiplied with to make it commute with $H(s)$, the system is no longer constrained to the $+1$ eigenspace of $T_1T_3$ . We can easily check that there are no stabilisers which we could multiply $T_1T_3$ with in order to make it commute, since this requires an operator with a $Z_1$ term, which only $T_1$ contains, but we cannot multiply by $T_1$ since this leaves $T_1 T_1 T_3 = T_3$ which doesn't commute with $H(s)$. Since at the very start of the $T_1 \to X_1$ evolution, the energy gap is zero and the subspace is no longer preserved, this means the system will now leak out of the subspace unless the evolution time $\tau \to \infty$. Thus the computation fails at the last step for this ordering. Note that this happens whether or not we replace $T_2 \to X_2$; the key part was the fact that there were no stabilisers available to multiply $T_1 T_3$ with to make it commute with $H(s)$.

Extending this to larger chains, we can see that whenever $T_n$ is replaced with $X_n$ for $n > 2$, the system will still be constrained to the $+1$ eigenspace of $T_{n-2}T_n$. If this `hidden' stabiliser anticommutes with the next Pauli replacement $X_m$, then provided we can multiply by $T_{m-2}$ the subspace is still preserved. But since at some point it will be necessary to replace $T_{1}$ or $T_2$, and there are no other stabilisers to multiply with to make sure that the `hidden' stabiliser still commutes with $H(s)$, the computation fails as there will be zero energy gap and the subspace is no longer preserved.

Similar behaviour can be seen when performing $Y$ measurements, again using untwisted stabilisers. Consider a Hamiltonian on 3 qubits, $H = T_1 + T_2$. Now after replacing $T_2 \to Y_2$, the system is still confined to a 2-dimensional subspace since $[T_1T_2,H(s)]=0$ and so the subspace where $T_1T_2$ has eigenvalue $+1$ is preserved. However, after replacing $T_{1} \to Y_1$, there are no stabilisers which can be chosen to make $T_1T_2$ commute with $H(s)$, and so the computation fails at the last step. This argument can similarly be extended to larger chains, and just like in the above case we will find that once the boundary is reached the computation time must diverge to avoid losing information.

This argument can easily be extended to more general measurement angles. Consider starting with a Hamiltonian made up of untwisted stabilisers as above, and replacing the $n^{th}$ untwisted stabiliser with $X_n^{\theta_n}$. The interpolating Hamiltonian has the form
\begin{align}\label{eqn:HamReorder}
H(s) = -\sum_{j < n -2} T_j  - \left( T_{n-2} + T_{n-1} + (1-s) T_n + sX_n^{\theta_n} \right) - \sum_{k > n} T_k.
\end{align}
The terms inside the two sums commute with the terms in brackets, so we can ignore these and focus on the terms in brackets (see Appendix~\ref{sec:LinAlg}). The eigenvalues at $s =0$ and $s=1$ can be calculated in Mathematica, with the result that the degeneracy at $s=1$ is double the degeneracy at $s=0$ for $n > 2$. Just as above, this degeneracy does not initially mean leakage out of the logical subspace, as there are terms that are conserved (in this case, $T_{n-2}T_n$ and $T_{n-1}T_n$ are conserved quantities). However, at the point where the boundary stabilisers need to be replaced, there will be leakage out of the logical subspace. 

This argument works provided $n > 2$, but for $n=2$ it is in fact possible to perform a computation without leakage out of the logical subspace, for certain angles. For $n=2$, the eigenenergies are doubly degenerate for all $s$ and take the form:
\begin{eqnarray}
E_{1} (\theta_{2},s) &=\pm \sqrt{ 2(1 - s + s^2) \pm \Gamma(\theta_{2},s)},
\end{eqnarray}
where $\Gamma (\theta,s) \equiv \sqrt{ 2s^2 \cos{2\theta_{2}} + (4 - 8s + 6s^2)}$. The energy gap $\Delta_1$ is given by
\begin{eqnarray}
\Delta_1 (\theta_{2},s) &= \sqrt{ 2(1 - s + s^2) + \Gamma(\theta_{2},s)} -  \sqrt{ 2(1 - s + s^2) - \Gamma(\theta_{2},s)}.
\end{eqnarray}
For a given $\theta_{2}$, this reaches a minimum at $s = (1- \frac{1}{2}\cos \theta_{2})$. For all $\theta_{2}$ then, the minimum energy gap $\Delta_1^{min}$ is given by:
\begin{eqnarray}
\Delta_1^{min}(\theta_{2})= \Delta_1 \left(\theta_{2} ,1- \frac{1}{2}  \cos \theta_{2}\right).
\end{eqnarray}
Over all possible $\theta_{2}$ values, $\Delta_1^{min}(\theta_{2})$ is largest for $\theta_{2} = l\pi, s = 1/2$, and goes to zero when $\theta_{2}= (2l+1)\pi/2, s=1$, where $l$ is an integer.

So we see that, for angles $\theta_2 \neq \frac{(2n+1)\pi}{2}$, it is possible to perform computations out of order, in a very limited sense. 

While the information remains in a protected subspace, it is also necessary to check if the information is transformed in the expected way. To see this, we start with the logical operators
\begin{eqnarray}
X_L \equiv X_1 Z_2, \; Z_L \equiv Z_1.
\end{eqnarray}
($Y_L$ isn't included since $Y_L = iX_LZ_L$). Using the method in~\cite{Bacon2010}, these can be multiplied by stabilisers to make them commute with the `measurements'. The transformations that would be applied to the logical operators above would be
\begin{eqnarray}
&X_L \to X_L T_2 = X_1 X_3^{\theta_3}Z_4 \nonumber\\
&Z_L \to Z_LT_1 =  e^{-i\theta_2 Z_2} X_2 Z_3 \to  e^{-i\theta_2 Z_2 T_2} X_2 Z_3 =  e^{-i\theta_2 X_3^{\theta_3}Z_4} X_2 Z_3
\end{eqnarray}
Note that this is exactly the same transformation we would make in the normal computation; we have just done two steps at once. Both these logical operators are in a form which commute with $X_1$ and $X_2$, so they commute with the time-dependent Hamiltonian, and so, for example, if the system starts in the $+1$ eigenstate of $X_L$, it will end up in the $+1$ eigenstate of $X_3^{\theta_3}Z_4$.

In summary, we have seen that when replacing $T_n$ operators with $X_n$ operators one-by-one on a 1D chain, the orderings that preserve the ground subspace are limited, and reordering is only possible using measurement angles which are not odd multiples of $\frac{\pi}{2}$. In particular, in a 1D chain, replacing the stabiliser at the first site or the second site appears to be the only other allowed ordering. The information stored in the chain is transformed in the same way in both cases, however the latter only works for $\theta_2 \ne (2n+1) \frac{\pi}{2}$, and the energy gap depends on $\theta_2$ so the speed of the adiabatic substitution must vary.

\subsection{Re-ordering without a fixed number of terms in Hamiltonian}

So far we have seen that the initial approach to re-ordering the adiabatic substitutions works only for a limited case. This is perhaps expected, as the out-of-order measurements considered so far are not a proper reflection of what happens in MBQC; in MBQC measuring a qubit destroys any entanglement on edges connected to that qubit. This is clearly not true above, since the final Hamiltonians contain terms like $T_1 + T_2 + X_3$. So a more natural way to perform the out-of-order measurements would be to remove all entanglement to measured qubits, or just remove all anticommuting terms entirely~\cite{Fitzsimons}. Take the chain considered above:
\begin{eqnarray}
H_0 = -T_1 - T_2 - T_3
\end{eqnarray}
Now if we replace $T_3 \to X_3$, we also remove any other operators which anticommute with $X_3$. The result is
\begin{eqnarray}\label{eqn:ReOrder1}
H_1 = - T_2 - X_3
\end{eqnarray}
Notice that the operator $T_1 T_3$ is still conserved, just like before, but this time when we want to measure $X_1$, there is no problem, since the system is always constrained to be within a 2-dimensional subspace. This would also work if we had instead just removed all entanglement to site $3$ (except $T_2$ can be left untouched since it commutes with $X_3$), i.e.
 \begin{eqnarray}\label{eqn:ReOrder2}
 H_1 = -Z_1 X_2 - T_2 - X_3
 \end{eqnarray}
 The minimum energy gap will also be the same when we introduce $X_1$ in equation (\ref{eqn:ReOrder1}) to when we replace $Z_1X_2\to X_1$ in equation (\ref{eqn:ReOrder2}).

 Similar results hold for $Y$ measurements or the \textsc{cnot} gate, so that Clifford operations can be performed in any order. Note that more than one operator is replaced at the same time, the number of anticommuting terms increases, and based on the numerical studies in~\cite{Bacon}, if the number of anticommuting terms scales with $n$ we would expect the energy gap to be polynomial in $1/n$.
 
 In summary, we have seen that Clifford operations can be done in any order, provided we either change the stabilisers so that the entanglement to `measured' sites is destroyed, or remove any stabilisers which anticommute with the measurement operator entirely. This is in contrast to the first method in Sec.~\ref{sec:Order2}, in which the stabilisers did not reflect what happens in MBQC, and in which the re-ordering is limited to performing alternating measurements from left to right. In both of the approaches considered, we cannot perform Clifford operations in one step, since the energy gap decreases inversely with the number of simultaneous measurements made.

\section{Conclusions} \label{sec:conc}

Using the tools of MBQC, such as \emph{flow} and \emph{gflow}, we have shown that the method of implementing measurements on a cluster state in an adiabatic, continuous fashion~\cite{Bacon} can be generalised to graph states with \emph{gflow}, resulting in adiabatic graph-state quantum computation (AGQC). Specifically, any measurement pattern with \emph{gflow} can be converted into an adiabatic evolution such that the number of adiabatic steps is equal to the depth in MBQC, and each step takes time proportional to the size of each layer. 

This work opens up the possibility of future results about efficiency and trade-offs in MBQC being transferred to holonomic computation. For example, there is still little understood about how to view the efficient simulatability of Clifford gates from the perspective of \emph{gflow}. The framework developed here offers a natural route to translate future possible results in this area, which may have implications for the efficiency of AGQC as well as the simulation of many-body Hamiltonians.

We have also seen that, in analogy to the trade-off between quantum and classical time in MBQC, there is a trade-off between the number of adiabatic steps taken and the size of $\Vert \dot{H} \Vert$, together with the degree of the initial Hamiltonian. The adiabatic time is proportional to $\Vert \dot{H} \Vert$, so this means a trade-off between the number of steps and the step time. One interesting point is that the trade-off is only between the number of steps and $\Vert \dot{H} \Vert$, but does not involve the energy gap, which highlights the subtleties of using the adiabatic theorem.

This suggest that, in this model of computation, freely available high degree operators are not a useful resource, which is perhaps counter-intuitive as we might expect that access to unbounded degree operators would give some kind of advantage. Conversely, under the assumption that simulating high degree operators shrinks the energy gap (which is the case for all known methods of simulating k-body operators, e.g.~\cite{Jordan2008,Bravyi2008}), this means that the optimal way of performing AGQC is in fact with as many steps as possible (the minimally-delayed \emph{gflow}), in contrast to in MBQC.

We considered the influence of adiabatic measurements in a different order from the one corresponding to the MBQC pattern. When the original stabilisers were replaced with the same local operators as before, just in a different order, the information generally leaked out of the logical subspace and the computation failed with the exception of a few, special cases. When however the continuous measurement is performed in such a way that the terms that anticommute with the measurement, e.g.\ $X_3$, are all adiabatically removed, then all Clifford operations can be performed in any order.

Finally we note that these results do not cover all possible methods of performing a measurement-based quantum computation, as there are some graph states without \emph{gflow} which still yield deterministic computations, and more general states such as those investigated in~\cite{Gross2007}. Understanding these could lead to further insights into holonomic quantum computation.

\chapter{Quantum communication through a Wigner crystal}
\label{chap:Wigner}

\section{Quantum communication through a Wigner crystal}
\label{sec:Wigner}
In the introduction, we motivated the need for quantum communication channels to share quantum information or to distribute entanglement between different parties, and we introduced the concept of using spin chains to perform this task. In this chapter, we will investigate a novel way to realise these `quantum wires', using a \emph{Wigner crystal}~\cite{Wigner1934}. Wigner crystals form in an electron gas, in the regime where the electron density is below a certain critical value such that coulomb interactions dominate over the kinetic energy of the electrons. Under these conditions, the electrons are extremely localised near to the classical lowest energy configuration, and so are said to have formed a crystal. 

The regimes in which it is possible to see a Wigner crystal can be parameterised using the Wigner-Seitz radius $r_s$, defined as the average distance between electrons in units of the Bohr radius, which for 2D systems is $r_s^2 =  (\pi a_B^2n)^{-1}$ where $n$ is the density of electrons per unit area, $a_B$ is the Bohr radius $a_B = 4 \pi \epsilon \hbar^2 / m^* e^2$, and $\epsilon,m^*$ are the permittivity and effective mass of the material being used. To arrive at the conditions for a Wigner crystal, we can think of each electron as sitting in a potential well due to the Coulomb repulsion from other electrons. Modelling this as a harmonic potential with strength $\omega$, the electron wavefunction has width roughly $\Delta r \sim \omega^{-1/2}$. Approximating $\omega^2 \sim \frac{1}{m} \frac{\partial^2 V}{\partial r^2 }$, gives $r_s = \omega^{-2/3} $ when $V$ is a Coulomb potential. Thus $\Delta r / r_s \sim r_s^{-1/4}$, which means the Wigner crystal picture (where the spatial localisation is small compared to the separation of electrons) becomes valid for large values of $r_s$. Monte Carlo simulations support this rough argument~\cite{Tanatar1989}, suggesting that 2D Wigner crystals would occur for $r_s > 37 \pm 5$. For 1D Wigner crystals (where the electrons are strongly confined in two directions), Monte Carlo simulations suggest that the the melting point is much lower, allowing Wigner crystals at $r_s := (n_{\textsc{1d}} a_B)^{-1}> 4$ \cite{Egger1999}, where $n_{\textsc{1d}}$ is the number of electrons per unit length . 

A classical Wigner crystal was first realised experimentally on the surface of liquid $^3$He~\cite{Grimes1979}. Since then, there have been many promising experimental observations. In solid state architectures, evidence for a Wigner crystal has been observed in a Si-MOSFET~\cite{Okamoto1998}, and in a 2-dimensional hole structure formed at a GaAs/AlGaAs boundary~\cite{Yoon1999}, which also confirmed the transition to a Wigner crystal at $r_s \sim 37$ predicted in~\cite{Tanatar1989}. In addition there have been many promising advances in the fabrication of quantum wires in GaAs/AlGaAs structures, in studies investigating the Tomonaga-Luttinger phase of electrons, where spin and charge excitations are independent~\cite{Auslaender2005,Steinberg2007,Jompol2009,Laroche2014}. Carbon nanotubes suspended over a substrate are also promising candidates, as they appear to allow high values of $r_s$ to be achieved, with small effects due to impurities~\cite{Tans1997,Jarillo2004}.

There are also other proposals for quantum technologies using Wigner crystals. The most prominent is a proposal for quantum computation using electrons on the surface of liquid helium~\cite{Platzman1999}. In this scheme, metal electrodes are placed below the electron gas so that image charges form which provide a trapping potential. The two logical qubit states are similar to Rydberg states in an atom, and have different average displacements from the surface which could be used to create a 2-qubit interaction via the image charges. Another existing scheme is to use a Wigner crystal made from trapped ions~\cite{Baltrusch2011,Taylor2008}, where qubit operations are performed using a laser field. The use of Wigner crystals as quantum communication channels has been mentioned as a possibility in carbon nanotubes~\cite{Deshpande2008}, but to date as far as the author is aware there have been no studies that have explored the viability of such a proposal.

\subsection{Instanton approach to calculate exchange coupling}
\label{sec:Instanton}

For a system of $N$ electrons constrained to move in 2D, we define collective spatial and spin coordinates $\mathbf{R} = (\mathbf{r}_1,\mathbf{r}_2,...,\mathbf{r}_N)=(x_1,y_1,x_2,y_2...,x_N,y_N)$, $\underline{\sigma} = (\sigma_1,\sigma_2,...,\sigma_N)$, and we use $R_j$ to denote the $j^{th}$ element of $\mathbf{R}$, such that $R_{2j-1} = x_j$, $R_{2j} = y_j$. We assume that the electrons are in the Wigner crystal regime, and that the temperature is low enough compared to the Debye temperature so there are negligible phonon excitations, so the electrons are localised near to $N$ `lattice sites' $\mathbf{\bar{R}} = (\bar{x}_1,\bar{y}_1,...,\bar{x}_{N},\bar{y}_N )$.

Under these conditions we approximate the electron wavefunction using localised orbital wavefunctions $\{ \ket{ \phi_n(\mathbf{r})} \}$, where orbital $n$ is centred around lattice site $\bar{\mathbf{r}}_n$, and such that there is still enough overlap between adjacent orbital wavefunctions to allow the electrons to exchange. We label one configuration of the orbital wavefunction as $\ket{ \Phi( \bar{\mathbf{R}} )} := \ket{  \phi_1(\mathbf{r}_1)} \ket{ \phi_2(\mathbf{r}_2)} ... \ket{ \phi_N(\mathbf{r}_N)}$, corresponding to when electron $n$ occupies the $n^{th}$ orbital wavefunction. Clearly this is not the only way of distributing the electrons; the electron positions can be permuted in $N!$ different ways, and we denote spatial permutations of $\ket{\Phi( \bar{\mathbf{R}} )}$ as $\ket{\Phi( P^{\textsc{r}}\bar{\mathbf{R}} )} = \hat{P}^{\textsc{r}}\ket{\Phi( \bar{\mathbf{R}} )}$, where $P^{\textsc{r}}$ is a spatial permutation of the electrons. Within this space of localised orbitals, the Hamiltonian effectively performs 
permutations of the electron positions, since it only contains terms acting on the spatial degree of freedom. Thus the Hamiltonian takes the form
\begin{align}
H_\textsc{r}  \propto \sum_{\hat{P}^{\textsc{r}} \neq \idop} C_P \hat{P}^{\textsc{r}}.
\end{align}
To determine the coupling constants $C_P$ for each permutation, we follow the argument of Thouless~\cite{Thouless1965}, referring to each permutation $|\Phi( \hat{P}^{\textsc{r}} \bar{\mathbf{R}} ) \rangle$ of the electrons in orbitals as a `cavity'. We imagine that it is possible to set the system up so that no tunnelling between different cavities can occur. Then suppose we arrange the system so that tunnelling between $\ket{\Phi( \bar{\mathbf{R}} )}$ and $|\Phi( \hat{P}^{\textsc{r}}\bar{\mathbf{R}} ) \rangle$ can occur. This can be treated exactly the same as a particle in a double potential well; there are symmetric and antisymmetric combinations of the states $\ket{\psi^\pm_P} = \frac{1}{\sqrt{2}}\ket{\Phi(  \bar{\mathbf{R}} )} \pm |\Phi( \hat{P}^{\textsc{r}}\bar{\mathbf{R}} )\rangle$ with energies $E^\pm_P$. It can be shown~\cite{Thouless1965,Roger1984} that $E^+_P < E^-_P$ so that defining $J_P : = \frac{1}{2} (E^-_P - E^+_P)$ the effective Hamiltonian between the cavities $\ket{\Phi( \bar{\mathbf{R}} )}$ and $|\Phi( \hat{P}^{\textsc{r}}\bar{\mathbf{R}} )\rangle$ can be written
\begin{align}
\langle \Phi( \bar{\mathbf{R}} ) | H_\textsc{r} | \Phi( \hat{P}^{\textsc{r}}\bar{\mathbf{R}} ) \rangle = - J_P
\end{align}
Similarly for all pairs of cavities, such a term in the Hamiltonian will appear, and under the assumption that the connections between these cavities does not significantly alter the cavities themselves (i.e.\ assuming that the wavefunctions overlap without being distorted too much) the coupling for each permutation is roughly given by the singlet-triplet splitting $J_P$. Thus the effective Hamiltonian constrained to the localised orbitals can be written
\begin{align}
H_\textsc{r} \simeq - \sum_{\hat{P}^{\textsc{r}} \neq \idop}  J_P \hat{P}^{\textsc{r}}.
\end{align}
The effects of this spatial Hamiltonian can also be expressed as an effective Hamiltonian acting on the spin degree of freedom, by recalling that for fermions the permutation $P$ of fermion labels results in a sign of $(-1)^{m_P}$ being applied to the state, where $m_P$ is the number of pairwise swaps that the permutation can be expressed as. The full permutation operator $\hat{P}$ can be decomposed as $\hat{P} =\hat{P}^{\textsc{r}} \hat{P}^\sigma$, where $\hat{P}^\sigma$ is an operator that permutes the spin degree of freedom, so that $\hat{P}^{\textsc{r}} = (-1)^{m_P} (\hat{P}^\sigma)^{-1}$, and by redefining $(\hat{P}^\sigma)^{-1} \to \hat{P}^\sigma$ we find that the effective spin Hamiltonian is the multi-spin exchange (MSE) Hamiltonian (derived more formally in~\cite{Thouless1965,Roger1984}):
\begin{align}\label{eqn:HThouless}
H_\textsc{mse} \simeq  - \sum_{\hat{P}^\sigma \neq \idop} (-1)^{m_P}  \hat{P}^{\sigma} J_{P}.
\end{align}
Thus the effective Hamiltonian of the Wigner crystal can equivalently be described in terms of the spin degrees of freedom evolving by the Hamiltonian $H_{\textsc{mse}}$, and this is the Hamiltonian that we will use in this chapter to transfer quantum information encoded in the spin degree of freedom. The operators $\hat{P}^\sigma$ can be related to standard Pauli spin matrices, for example the operator that permutes spins 1 and 2 can be written $\hat{P}^\sigma_{1,2} = \frac{1}{2}(\idop + E_{12})$, and more complicated permutations can be arrived at using the identities in~\cite{Klein1980} (we will see examples of these in Section~\ref{sec:WignInfTrans}). The difficulty lies in calculating the exchange coupling $J_P$, which is generally very hard to do except by using approximations. In this study we will use the path integral instanton (or many-dimensional WKB) method (for an overview of this method, see for example~\cite{Altland2010}). This method has been used extensively in the literature, first by Roger~\cite{Roger1983,Roger1984}, which was successful in qualitatively describing the behaviour of the Weiss temperature on the surface of solid $^3$He, and subsequently by many others e.g.~\cite{Ashizawa2000,Katano2000,Voelker2001,Hirashima2001,Fogler2005,Klironomos2005,Klironomos2007,Meyer2009,Candido2011}. The instanton approach assumes that quantum effects are small and that most of the evolution occurs near the classical trajectories, so that a semiclassical description is appropriate. This reproduces path-integral Monte Carlo results for values of $r_s$ far from the Wigner crystal melting point where the dynamics is close to the classical dynamics~\cite{Bernu2001}, but is not known to be accurate near the melting point where quantum effects can become more dominant. Also it is not likely to be accurate for Wigner crystals formed as a result of impurity pinning, since these may have small $r_s$ values and thus large quantum effects.

In order to apply the path integral instanton approximation, we first derive a particularly useful expression which puts $J_P$ in terms of path integrals. We start with the \emph{propagator} $G( \bar{\mathbf{R}}; P^{\textsc{r}} \bar{\mathbf{R}}; T)$ defined as
\begin{align}\label{eqn:prop1}
G( \bar{\mathbf{R}}; P^{\textsc{r}} \bar{\mathbf{R}}; T) :=  \bra{P^{\textsc{r}}\bar{\mathbf{R}}  } e^{- i H_{\textsc{r}}  T/ \hbar} \ket{ \bar{\mathbf{R}} }.
\end{align}
where $\langle \mathbf{R} \ket{\bar{\mathbf{R}}}: = \delta(\mathbf{r}_1 - \bar{\mathbf{r}}_1) \delta(\mathbf{r}_2 - \bar{\mathbf{r}}_2)... \delta(\mathbf{r}_N - \bar{\mathbf{r}}_N)$ is an eigenstate of the position operator, where electrons lie exactly on the lattice sites. 

To simplify the propagator, we express $H_\textsc{r}$ in terms of the eigenstates $\ket{\psi^\pm_P}$ given above that mix the $\ket{\Phi( \bar{\mathbf{R}} )}$ and $\ket{\Phi( P^{\textsc{r}} \bar{\mathbf{R}} )}$ states. Expressing $ e^{- i H_{\textsc{r}}  T/ \hbar}$ in this subspace, and defining $E_0$ such that $E^+_P = E_0 - J_P$, $E^-_P = E_0 + J_P$, gives
\begin{align}\label{eqn:WigEvo}
\left. e^{-iH_\textsc{r} T / \hbar} \right|_P &= \ket{\psi^+_P}\bra{\psi^+_P} e^{-iH_\textsc{r} T / \hbar} \ket{\psi^+_P}\bra{\psi^+_P} + \ket{\psi^-_P}\bra{\psi^-_P} e^{-iH_\textsc{r} T / \hbar}\ket{\psi^-_P}\bra{\psi^-_P} \nonumber \\
&=  e^{i (J_P - E_0) T/\hbar} \ket{\psi^+_P}\bra{\psi^+_P} + e^{-i  (J_P + E_0)  T/\hbar} \ket{\psi^-_P}\bra{\psi^-_P}.
\end{align}
where we use the notation $\left. U \right|_P$ to indicate operator $U$ constrained to the subspace spanned by $\left\{ \ket{\Phi( \bar{\mathbf{R}} )} , \ket{\Phi(P^{\textsc{r}} \bar{\mathbf{R}} )} \right\}$. 

We make the assumption that the amplitude of each orbital wavefunction is negligible at other lattice sites, so $\int \delta(\mathbf{r}_k - \bar{\mathbf{r}}_l) \phi_m(\mathbf{r}_k) d \mathbf{r}_k \simeq \delta_{lm} c$, where $c$ is a complex number. From this assumption it follows that $\left\langle \bar{\mathbf{R}}  \right. \left|  \psi^\pm_P \right\rangle \simeq \left\langle \bar{\mathbf{R}}  \right. \left|  \Phi( \bar{\mathbf{R}} ) \right\rangle :=  \frac{C}{\sqrt{2}}$, $\left\langle P^{\textsc{r}} \bar{\mathbf{R}}  \right. \left|  \psi^\pm_P \right\rangle \simeq \pm \left\langle P\bar{\mathbf{R}}  \right. \left|  \Phi( P^{\textsc{r}} \bar{\mathbf{R}}) \right\rangle := \pm\frac{C}{\sqrt{2}}$. Inserting the evolution operator $\left. e^{-iH_\textsc{r} T / \hbar} \right|_P$ into the propagator in (\ref{eqn:prop1}) and simplifying using these identities gives
 \begin{align}
G( \bar{\mathbf{R}}; P^{\textsc{r}} \bar{\mathbf{R}}; T)&=   \bra{P^{\textsc{r}} \bar{\mathbf{R}}  } e^{- i H_{\textsc{r}}  T/ \hbar} \ket{ \bar{\mathbf{R}} }=  \bra{P^{\textsc{r}} \bar{\mathbf{R}}  } \left. e^{-iH_\textsc{r} T / \hbar} \right|_P \ket{ \bar{\mathbf{R}} } \nonumber\\
&\simeq e^{-i E_0 T/\hbar} |C|^2 A_P \sin \left( \frac{J_PT}{\hbar} \right),
\end{align}  
where $A_P$ accounts for permutations that are self-inverse, and so are counted twice in the Hamiltonian (for our purposes only two-body exchange processes are self-inverse, so that $A_P = 2$ for 2-body exchange and $A_P = 1$ otherwise). A similar calculation yields $G( \bar{\mathbf{R}}; \bar{\mathbf{R}}; T) \simeq  e^{-i E_0 T/\hbar} |C|^2 \cos \left( \frac{J_P T}{\hbar} \right)$. Dividing these two expressions gives
\begin{align}
\frac{G( \bar{\mathbf{R}}; P^{\textsc{r}} \bar{\mathbf{R}}; T)}{G( \bar{\mathbf{R}}; \bar{\mathbf{R}}; T) }  \simeq  A_P \tan  \left( \frac{J_P T}{\hbar} \right) .
\end{align}
When the time $T$ is sufficiently small with respect to $J/\hbar$, then $\tan  \left( \frac{J_P T}{\hbar} \right) \simeq \frac{J_P T}{\hbar}$, so that 
\begin{align}\label{eqn:J}
J_P \simeq   \frac{\hbar A_P}{T}\frac{G( \bar{\mathbf{R}}; P^{\textsc{r}} \bar{\mathbf{R}}; T) }{G( \bar{\mathbf{R}};\bar{\mathbf{R}}; T)},
\end{align}

\begin{figure}[h]
\begin{center}
\includegraphics[width = 0.5\textwidth]{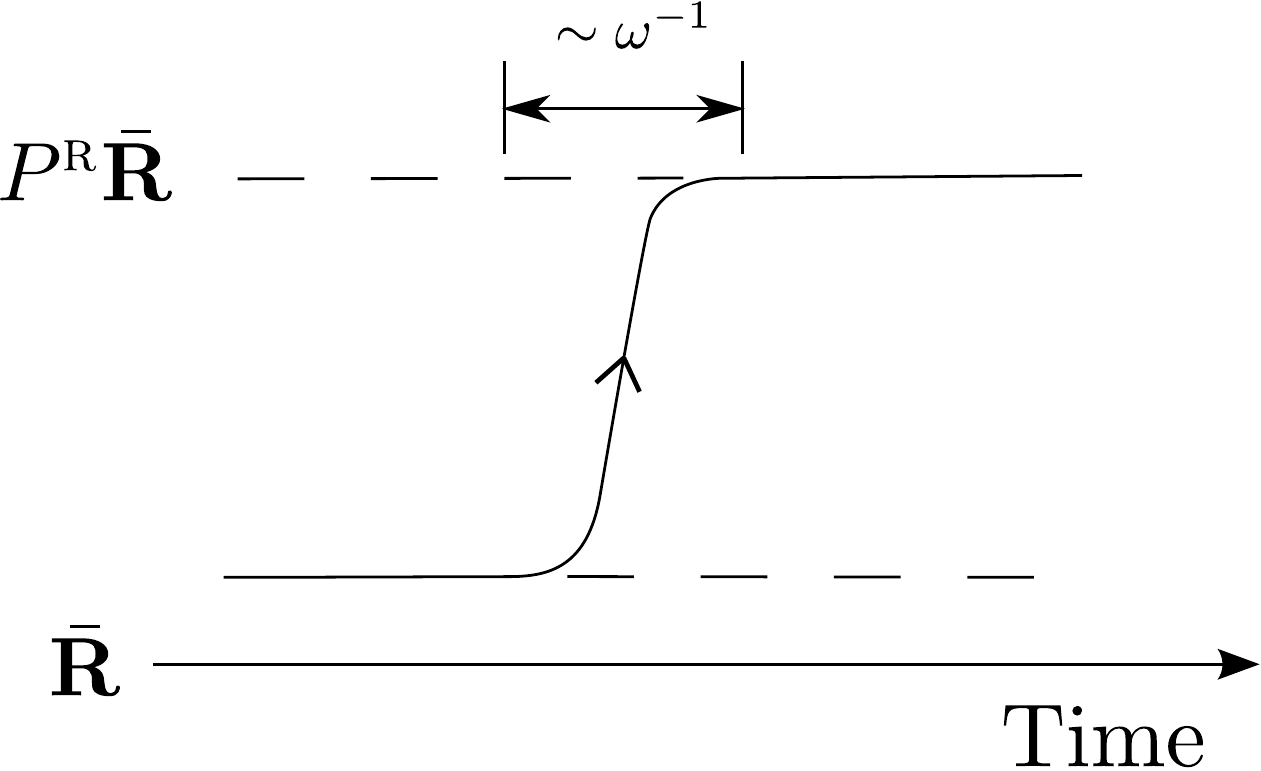}
\caption{\label{fig:Instanton} A schematic diagram of an `instanton' solution between the two configurations $\bar{\mathbf{R}}$ and $P^{\textsc{r}} \bar{\mathbf{R}}$. For simplicity the second derivative of the potential is denoted as $\omega^2$, but in general there will be many frequencies for higher dimensional problems.}
\end{center}
\end{figure}

The expression for $J_P$ in eqn.\ (\ref{eqn:J}) is in a suitable form that approximation techniques based on path integrals can be used. The path integral form of the propagator can be written, assuming a Hamiltonian of the form $H = \sum_i \frac{\mathbf{p}_i^2}{2m} + V(\mathbf{R})$ (see e.g.~\cite{Altland2010}):
\begin{align}\label{eqn:FullProp}
&G( \bar{\mathbf{R}},P^{\textsc{r}} \bar{\mathbf{R}} ; T) =\bra{P^{\textsc{r}} \bar{\mathbf{R}} } e^{-i H T} \ket{ \bar{\mathbf{R}}}  =\int_{\mathbf{R}(0) = \bar{\mathbf{R}}}^{\mathbf{R}(T) = P^{\textsc{r}} \bar{\mathbf{R}}} D\mathbf{R}\text{ exp}\left[ - \frac{i}{\hbar} \int_0^{T}  dt  \frac{m}{2} \left(\frac{d \mathbf{R}}{d t} \right) ^2+ V(\mathbf{R}) \right] \nonumber\\
&=\int_{\mathbf{R}(0) = \bar{\mathbf{R}}}^{\mathbf{R}(T) = P^{\textsc{r}} \bar{\mathbf{R}}} D\mathbf{R}\text{ exp}\left[ - \frac{i}{\hbar} S[\mathbf{R}] \right].
\end{align}
where $S[\mathbf{R}]$ is the action along the path $\mathbf{R}(t)$. To evaluate this path integral, we first perform a Wick rotation $t \to -i \tau$, after which the propagator becomes an imaginary time propagator:
\begin{align}
&G( \bar{\mathbf{R}},P^{\textsc{r}} \bar{\mathbf{R}} ; T_\tau) =\bra{P^{\textsc{r}} \bar{\mathbf{R}} } e^{- H T_\tau } \ket{ \bar{\mathbf{R}}}  =\int_{\mathbf{R}(0) = \bar{\mathbf{R}}}^{\mathbf{R}(T_\tau) = P^{\textsc{r}} \bar{\mathbf{R}}} D\mathbf{R}\text{ exp}\left[ - \frac{1}{\hbar} S[\mathbf{R}] \right].
\end{align}
We then use a semiclassical (stationary phase) approximation to approximate this imaginary-time propagator, whereby the quantum dynamics are approximated by quantum fluctuations around the classical path of least action $\mathbf{R}_{cl}(t)$ which satisfies the boundary conditions $\mathbf{R}_{cl}(0) = \bar{\mathbf{R}}$, $\mathbf{R}_{cl}(T_\tau) = P^{\textsc{r}} \bar{\mathbf{R}}$. This classical solution that connects $\bar{\mathbf{R}}$ and $P^{\textsc{r}} \bar{\mathbf{R}}$ is known as an \emph{instanton}. The name instanton reflects the fact that this solution is a kink in the temporal profile of the particles (see Fig.~\ref{fig:Instanton}), and so can be interpreted as a type of particle in time. The width of the instanton depends on the second derivative of the potential at $\bar{\mathbf{R}}$. In theory, any number of instantons connecting $\bar{\mathbf{R}}$ and $P^{\textsc{r}} \bar{\mathbf{R}}$ are solutions to the equation of motion. To get around this, we make an assumption (often called the `dilute gas' approximation), that the distance between instantons $\Delta t$ is much larger than the instanton width $\delta t$. Then if the propagator is evaluated at times $T_\tau$ that satisfy $\delta t \ll T_\tau \ll \Delta t$, we can assume that the contribution from multi-instanton solutions is negligible and so only single instanton solutions need to be taken into account. The point in time at which these instantons occur is also arbitrary, which will be an an important factor later on.

We assume that the dilute gas approximation is valid, so that the classical path of least action is a single instanton. We then consider Gaussian fluctuations about this classical instanton path; for a fluctuation $\mathbf{R} (t)=  \mathbf{R}_{cl}(t) + \mathbf{v}(t)$ such that $\mathbf{v}(0) = \mathbf{v}(T_\tau) =0$, we can Taylor expand the action to second order (see e.g.~\cite{Riley2006,Altland2010})
\begin{align}
S[\mathbf{R} ]  \simeq S[ \mathbf{R}_{cl} ] + \int_0^{T_\tau} dt \mathbf{v}^T(\tau) \cdot \left. \mathbf{A}(\tau)\right|_{\mathbf{R} = \mathbf{R}_{cl}} \cdot \mathbf{v}(\tau),
\end{align}
where
\begin{align}\label{eqn:A}
\mathbf{A}(\tau) = -m \partial_{\tau}^2 \idop + \mathbf{H}(\tau),
\end{align}
and $\mathbf{H}$ is the time-dependent Hessian matrix defined as 
\begin{align}
\mathbf{H}_{ij}(\tau) := \left. \frac{\partial^2 V}{\partial  R_i(\tau) \partial  R_j(\tau)} \right|_{\mathbf{R} = \mathbf{R}_{cl}},
\end{align}
where we use the notation $R_{2n-1 }(\tau) = x_n(\tau)$, $R_{2n}(\tau) = y_n(\tau)$.

The stationary phase approximation then approximates the full action in (\ref{eqn:FullProp}) by the second order Taylor expansion, giving
\begin{align}\label{eqn:Fdefin}
G( \bar{\mathbf{R}};P^{\textsc{r}} \bar{\mathbf{R}} ; T_\tau ) &\simeq e^{-\frac{1}{\hbar}S[\mathbf{R}_{cl}]} \int_{\mathbf{v}(0) = 0}^{\mathbf{v}({T_\tau}) = 0} D\mathbf{v}(\tau) \exp \left[ -\frac{1}{2\hbar} \int_0^{T_\tau} d\tau \mathbf{v}^T(\tau) \cdot \left. \mathbf{A}(\tau)\right|_{\mathbf{R} = \mathbf{R}_{cl}} \cdot \mathbf{v}(\tau)\right] \nonumber\\
&:= e^{-\frac{1}{\hbar}S[\mathbf{R}_{cl}]} F[\mathbf{R}_{cl} ].
\end{align}
and similarly $G( \bar{\mathbf{R}};P^{\textsc{r}} \bar{\mathbf{R}} ; T_\tau ) = F[\mathbf{R}_{cl} ]$, where $F[\bar{\mathbf{R}}]$ is defined with $\mathbf{R}(t) = \bar{\mathbf{R}}$, and we have defined $V(\mathbf{R})$ to be zero when the electrons are at positions $\bar{\mathbf{R}}$ so that the classical action for $G( \bar{\mathbf{R}} ;\bar{\mathbf{R}} ;T_\tau)$ is zero.

Putting this all into equation (\ref{eqn:J}), the expression for $J_P$ becomes
\begin{align}\label{eqn:JP}
J_P \simeq \frac{\hbar A_P}{T_\tau}\frac{  F[\mathbf{R}_{cl} ]  }{ F[\bar{\mathbf{R}} ]  } e^{-\frac{1}{\hbar}S[\mathbf{R}_{cl}]},
\end{align}
for a process which permutes the electron positions by $P^{\textsc{r}}$, and for sufficiently small $T_\tau$. It appears counter-intuitive that a factor of time should appear in the denominator, however later on we will see that this will cancel out. The factors of $F[\mathbf{R}_{cl} ]$ in eqn.\ (\ref{eqn:JP}) are still quite unwieldy and can be put in a much more convenient form by noting it is a multidimensional Gaussian integral, and so using standard functional integration results it can be written~\cite{Altland2010}
\begin{align}
F[\mathbf{R}_{cl}]  &= K \frac{1}{\sqrt{\det( \mathbf{A} )}} = K \frac{1}{\sqrt{\det( -m \partial_{\tau}^2 \idop + \mathbf{H}(\tau) )}},
\end{align}
where any constants have been absorbed into the prefactor $K$. A similar calculation for $F[ \bar{\mathbf{R}} ]$ gives 
\begin{align}
F[\bar{\mathbf{R}}  ] = K  \frac{1}{\sqrt{\det( -m \partial_{\tau}^2 \idop + \mathbf{H}^{(0)}(\tau) )}} ,
\end{align}
where $\mathbf{H}^{(0)}(\tau) := \left. \mathbf{H} \right|_{\mathbf{R}(\tau) = \bar{\mathbf{R}} }$.

We must be careful that all the eigenvalues of $-m \partial_{\tau}^2 \idop + \mathbf{H}(\tau)$ are non-zero, to avoid $F[\mathbf{R}_{cl}]$ diverging. As noted earlier, the position of the instanton in time is arbitrary, and this leads to a eigenvector with zero eigenvalue (or small eigenvalue for finite systems) which corresponds to translation of the instanton in time. Since the positions of the instantons are not important for this calculation, we can ignore this zero mode and only consider the effects of the remaining non-zero modes. This can be done by a suitable change of variables in the functional integration, introducing an extra multiplicative factor~\cite{Zinn2002}
\begin{align}
F[ \mathbf{R}_{cl}] =  \frac{ K T_\tau}{\hbar} \left( \frac{S[\mathbf{R}_{cl}]}{2 \pi \hbar m} \right)^{1/2} \frac{1}{\sqrt{\det'( -m \partial_{\tau}^2 \idop + \mathbf{H}(\tau) )}},
\end{align}
where $\det'$ indicates the determinant without the lowest eigenvalue. Combining these results all together, we find that
\begin{align}\label{eqn:JPfinal}
J_P \simeq  A_P \left( \frac{S[\mathbf{R}_{cl}]}{2 \pi \hbar m} \right)^{1/2}  \sqrt{  \frac{ \det( -m \partial_{\tau}^2 \idop + \mathbf{H}^{(0)}(\tau) )  }{ \det'( -m \partial_{\tau}^2 \idop + \mathbf{H}(\tau) )  } } e^{-\frac{1}{\hbar}S[\mathbf{R}_{cl}]}.
\end{align}
This expression is in a convenient form since it can, in principle, be calculated once the classical path of least action is known. For this study, we will find this path numerically, however before discussing these numerical methods, we first introduce the particular setup we will use to simulate the Wigner crystal, and introduce some important length scales.

\subsection{Proposed setup}

\begin{figure}[h]
\begin{center}
\includegraphics[width = 0.8\textwidth]{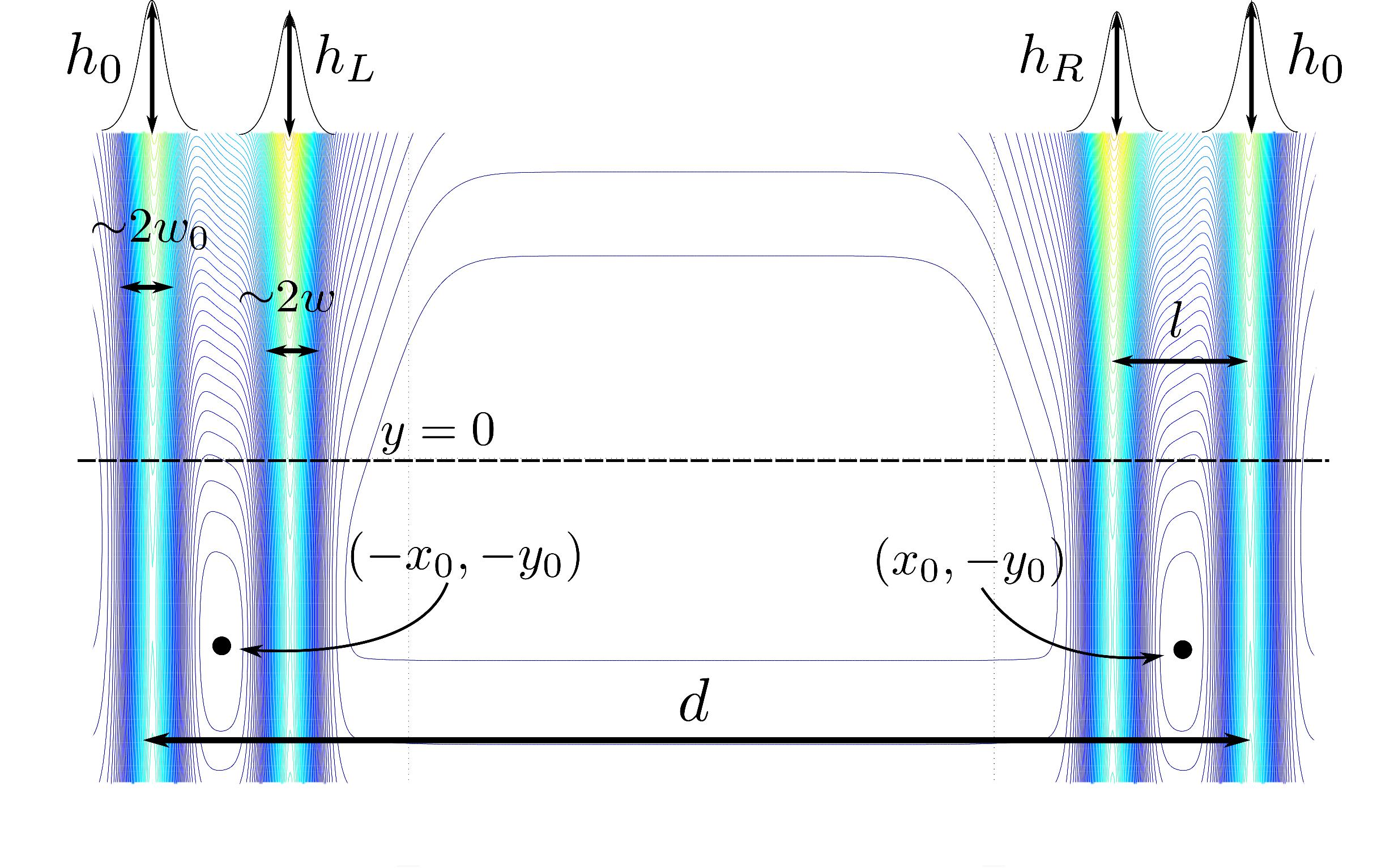}
\caption{\label{fig:Potential} An illustration of the potential given in eqn.\ (\ref{eqn:Potential})}
\end{center}
\end{figure}

For this scheme, we consider a 2-dimensional system in the Wigner crystal regime, at suitably low temperatures that the MSE model is applicable. We consider a parabolic confining potential $\frac{1}{2}m^* \Omega^2 y^2$ in the $y$ direction, with barriers placed at $y = \pm d/2$, and $N$ electrons placed between these barriers such that the density of electrons is $N/d$ (such a setup is also often called a \emph{Wigner molecule}). Since this potential is mirror symmetric around the $x$-axis, more potentials must be introduced to break this symmetry and make the ground state unique. In our scheme this is done by trapping the two end electrons such that they are moved off centre (in opposite directions or the same directions for odd and even chains respectively), and isolated from the rest of the chain using an internal barrier (see Fig.~\ref{fig:Potential}). We refer to these end electrons as quantum `dots'. Not only does this lift the degeneracy, these two end electrons isolated or coupled to the chain depending on the internal barrier height, and so can act as sender and receiver when sending quantum information through the chain.

We model the potentials as quadratic potentials modulated by Gaussians, with the full potential $V(\mathbf{R})$ given by
\begin{align}\label{eqn:Potential}
&V(\mathbf{R}) =  \sum_{i=1}^N  \left[ \sum_{j < i}\frac{e^2}{4 \pi  \epsilon |\mathbf{r}_i - \mathbf{r}_j|}  + \frac{1}{2}m^* \Lambda^2 \left( e^{-\frac{ (x_i - x_0)^2}{2 \sigma^2}}   (y_i + (-1)^{N+1} y_0)^2 +e^{-\frac{(x_i + x_0)^2}{ 2 \sigma^2}}  (y_i+y_0)^2 \right)  \right. \nonumber\\
&\left.  + \frac{1}{2}m^* \Omega^2 y_i^2 +  h_{L} e^{ - \frac{(x_i + d/2 - l)^2 }{ 2 w^2}} + h_{R} e^{ - \frac{(x_i - d/2 + l)^2 }{ 2 w^2} }+ h_{0} \left( e^{ - \frac{ (x_i - d/2)^2 }{ 2 w_{out}^2}} +  e^{ - \frac{(x_i + d/2)^2 }{ 2 w_{out}^2}} \right) \right]
\end{align}
For an illustration of this potential, see Fig.~\ref{fig:Potential}. Note that we have included different barrier heights $h_L,h_R$ for the left and right dots, to allow the coupling between these dots and the chain to be adjusted. There is also a factor of $(-1)^{N+1}$ so that the offset lifts the degeneracy appropriately for odd or even electrons. The possible experimental implementation that we envisage is a 2-dimensional electron gas in GaAs, using surface electrodes to create the potential, but our results extend to any realisation of a Wigner crystal.

The dot displacement $y_0$ is chosen so that the mirror inversion symmetry is broken sufficiently to prevent excitations into the reflected configuration. Let the two degenerate ground states be $\psi_1$ and $\psi_2$. One way to arrive at a rough condition for $y_0$ is to make sure that the probability of thermal excitation from $\psi_1$ to $\psi_2$ is sufficiently low. The probability of excitation from $\psi_1$ to $\psi_2$, in the presence of the symmetry breaking potential $\frac{1}{2}m^* \Lambda^2 (y-y_0)^2$, is roughly $\varepsilon$, where
\begin{align}
\varepsilon \sim \exp ( -\Delta E \beta) = \exp \left(- 2 m^*\Lambda^2 y_0^2 \beta \right) 
\end{align}
and $\beta$ is the inverse temperature. To obtain an estimate, we assume that the system is realised in GaAs/AlGaAs at a temperature of $T\simeq 5 \text{mK}$. The minimum size of $\hbar \Omega$ can be estimated from the band splitting in GaAs/AlGaAs, which can be as low as $0.3$meV~\cite{Daneshvar1997}. Inserting this into the expression above, we find $\varepsilon \sim \exp( -2\times 10^{19} y_0^2)$, so choosing $y_0 \sim 10$nm can give a sufficiently low error, and additionally should be within the precision available in experiments (using the characteristic length units $r_0$ defined below, 10nm$\simeq 0.1 r_0$ for GaAs).

The remaining parameters are chosen based on the electron configuration without the dot potentials present (i.e.\ when $h_L = h_R = \Lambda = 0$). For each particular situation, the minimum energy configuration $\bar{ \mathbf{R}} = (\bar{x}_1,\bar{y}_1,\bar{x}_2,\bar{y}_2,...,\bar{x}_N,\bar{y}_N)$ was found without the dot potentials, and then the parameters were chosen such that the internal barriers would sit roughly halfway between the end pairs of electrons, and such that the internal barriers have as little effect as possible on the separation between the dot electron and the chain. Thus $l = \frac{d}{2} - \frac{1}{2}(\bar{x}_1 + \bar{x}_2)$, and empirically $w = \frac{l}{8}$, $w_{out} =\frac{l}{8}$ was found to have little effect on the electron equilibrium positions. We set $\sigma = l/2$, so that the effect of the two quadratic potentials for the dots is confined to within the barriers. The size of $\Lambda$ in eqn.\ (\ref{eqn:Potential}) was chosen to be $4\Omega$, so that it is much stronger than the confining potential of the chain. This was also empirically found to help the numerical algorithm converge to one minimum, and not a local minimum.

It is also useful to introduce a characteristic length scale of the system, to ease the notation, and so that the calculations in the following sections are not dependant on the particular parameters used, but are instead more general. Following~\cite{Klironomos2007}, we define $r_0$ as the length scale at which the Coulomb repulsion between electrons is comparable to the potential energy of the parabolic confinement in the $y$-direction:
\begin{align}\label{eqn:r0}
\frac{1}{2}m^* \Omega^2 r_0^2 = \frac{e^2}{4 \pi  \epsilon r_0} \Longrightarrow r_0 = \left( \frac{ e^2}{4 \pi \epsilon  m^* \Omega^2} \right)^{1/3}
\end{align}
We also define $r_\Omega$, which is $r_0$ in units of the Bohr radius $a_B = 4 \pi \epsilon  \hbar^2 / m^* e^2$:
\begin{align}\label{eqn:rOmega}
r_{\Omega} = 2 \left( \frac{e^4 m^*}{32 \pi^2 \epsilon^2 \hbar^3 \Omega} \right)^{2/3}
\end{align}
Using these units, we also define a dimensionless density of the electrons $\nu = nr_0$ as the number of electrons in a length $r_0$. Thus in these units, $r_s = r_0/a_B \nu = r_\Omega /\nu$. 

By making a change of variables of $R_n \to R_n r_0$, and $\tau \to \tau \frac{\sqrt{2}}{\Omega}$ the action then becomes
\begin{align}
S[ \mathbf{R} ] &= \int_0^{T_\tau} d\tau \left[ \frac{m\dot{\mathbf{R}}^2}{2} + V(\mathbf{R}) \right]=\int_0^{T_\tau} d\tau \left[ \sum_j \frac{m\dot{\mathbf{r}}_j^2}{2} + \frac{1}{2}m \Omega^2 y_j^2  + \sum_{j < i} \frac{e^2}{4 \pi  \epsilon |\mathbf{r}_i - \mathbf{r}_j|} + \sum_j v(\mathbf{r}_j) \right]\nonumber\\
&\to \hbar \sqrt{r_{\Omega}} \int_0^{T_\tau} d\tau \left[ \sum_j \frac{\dot{\mathbf{r}}_j^2}{2} + y_j^2  + \sum_{j < i} \frac{1}{|\mathbf{r}_i - \mathbf{r}_j|} + \frac{1}{\hbar \sqrt{r_\Omega}}\sum_j v(\mathbf{r}_j)\right] :=\hbar \sqrt{r_{\Omega}} \text{ } \eta[\mathbf{R} ],
\end{align}
where $\eta[\mathbf{R} ]$ is the dimensionless action, $v(\mathbf{r}_j)$ contains all of the remaining terms in $V(\mathbf{R})$, and we have used the identity $\frac{1}{\sqrt{2}} m \Omega_y r_0^2  = \frac{1}{\sqrt{2}} m \Omega_y r_{\Omega}^2 a_B^2 = \hbar \sqrt{r_{\Omega}}$. Performing a similar change of variables on the prefactor $F[\mathbf{R}_{cl}]$, we obtain
\begin{align}
F[\mathbf{R}_{cl}]  =  \int_{\mathbf{v}(0) = 0}^{\mathbf{v}({T_\tau}) = 0} D\mathbf{v}(\tau) \exp \left[ -\frac{\sqrt{r_{\Omega}}}{2} \int_0^{T_\tau} d\tau \text{ }\mathbf{v}^T(\tau) \cdot \left[ -\partial_{\tau}^2\idop +  \frac{1}{\hbar\sqrt{r_{\Omega}}} \mathbf{H}(\tau) \right] \cdot \mathbf{v}(t)\right].
\end{align}
So overall in these units the exchange energy is
\begin{align}
\label{eqn:Jreduced}
J_P \simeq \frac{ e^2 }{4 \pi \epsilon a_B} r_\Omega^{-5/4} A_P \left( \frac{\eta[\mathbf{R}_{cl}]  }{2 \pi } \right)^{1/2}  \sqrt{  \frac{ \det( -\partial_{\tau}^2 \idop + \frac{1}{\hbar \sqrt{r_\Omega}}\mathbf{H}^{(0)}(\tau) )  }{ \det'( - \partial_{\tau}^2 \idop + \frac{1}{\hbar \sqrt{r_\Omega}} \mathbf{H}(\tau) )  } } e^{-\sqrt{r_\Omega} \eta[\mathbf{R}_{cl}]}.
\end{align}
This is the form of $J_P$ which will be used in the numerical simulations below. Notice that the factors of $\hbar / \sqrt{r_\Omega}$ will disappear if $\mathbf{H}^{(0)}(\tau)$ is expressed with a change of variables of $R_n \to R_n r_0$, and $\tau \to \tau \frac{\sqrt{2}}{\Omega}$. For the remainder of this chapter, we will also give coupling strengths in units of $\Omega$, unless explicitly stated otherwise.

\subsection{Calculating the exchange numerically}

We now describe the process of numerically evaluating eqn.\ (\ref{eqn:JPfinal}) for the potential given in the previous section. The first step is to find the configuration of positions $\bar{\mathbf{R}}$ that minimises $V(\mathbf{R})$, which was done numerically using the built-in MATLAB algorithm \texttt{fminunc}. The minimisation was performed first with $h_L = h_R = \Lambda=0$, to find the spacing between the electrons with a pure harmonic potential in the $y$-direction, so that appropriate values of $l,x_0$ could be inferred. Then the full minimisation was performed with these parameters, starting with the electrons evenly spaced along the $x$-axis with a slight zig-zag perturbation in the $y$-direction. 

\begin{figure}[h!]

\subfloat[]{\includegraphics[width=0.5\textwidth]{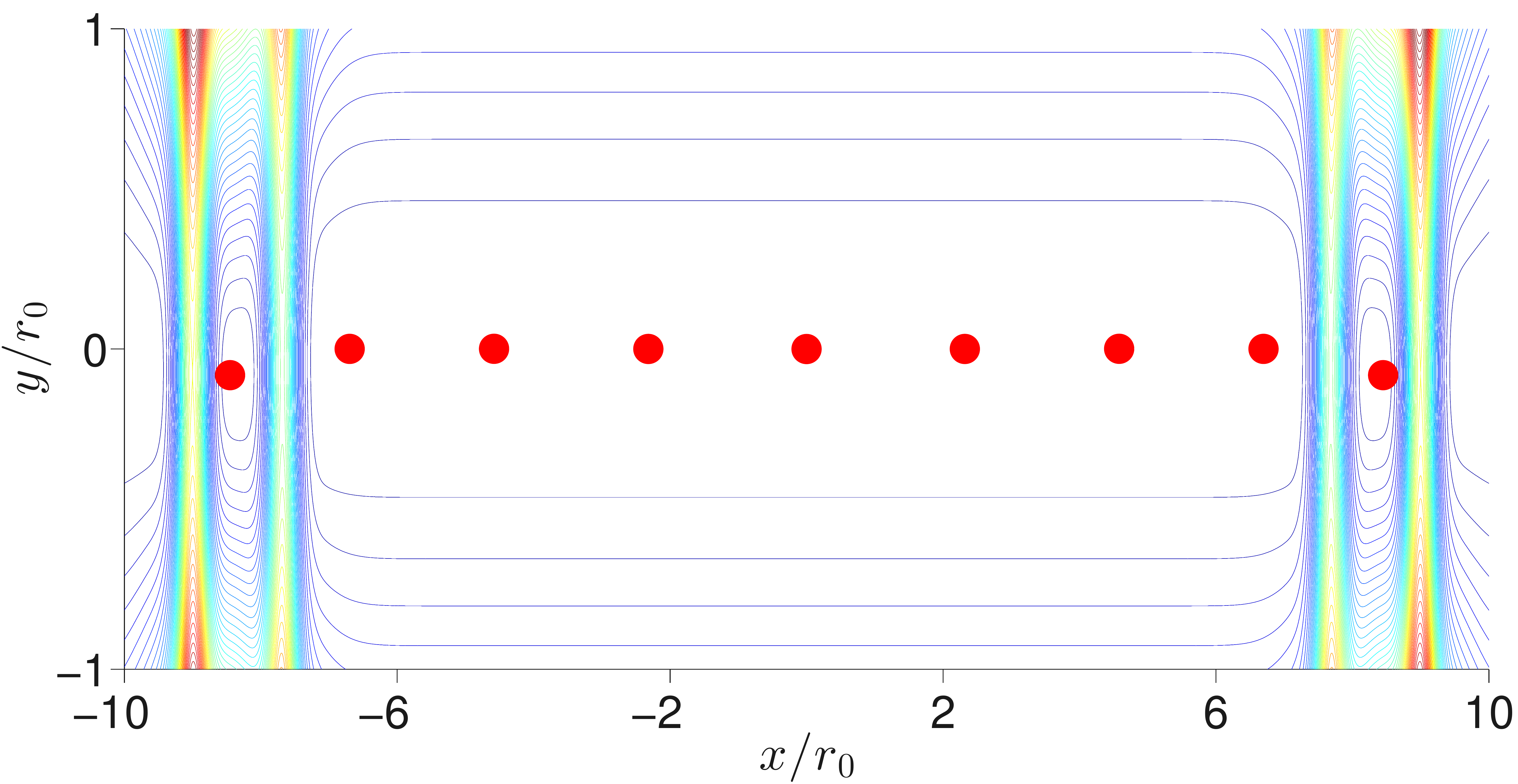}}
\subfloat[]{\includegraphics[width=0.5\textwidth]{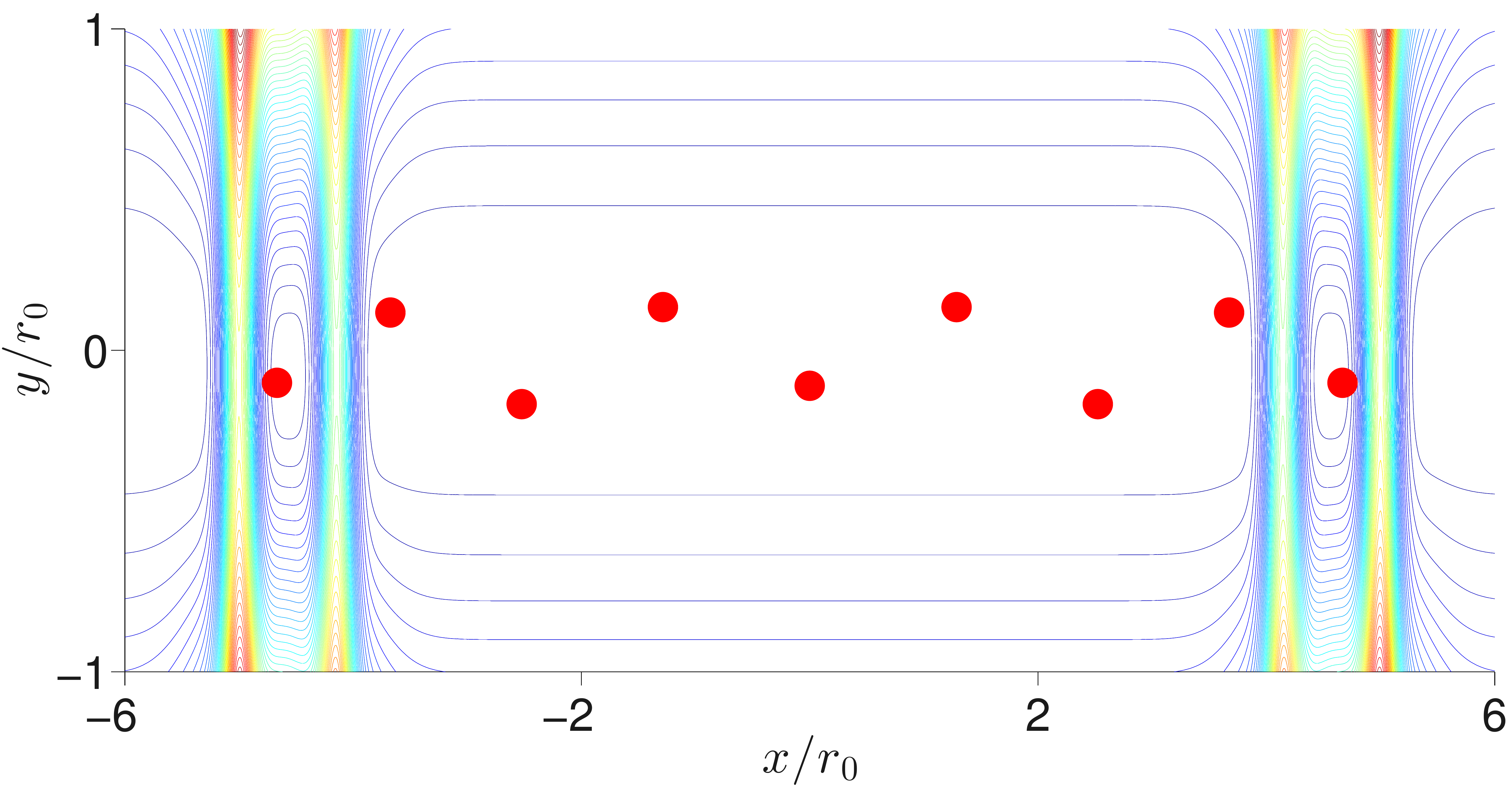}}\\
\subfloat[]{\includegraphics[width=0.5\textwidth]{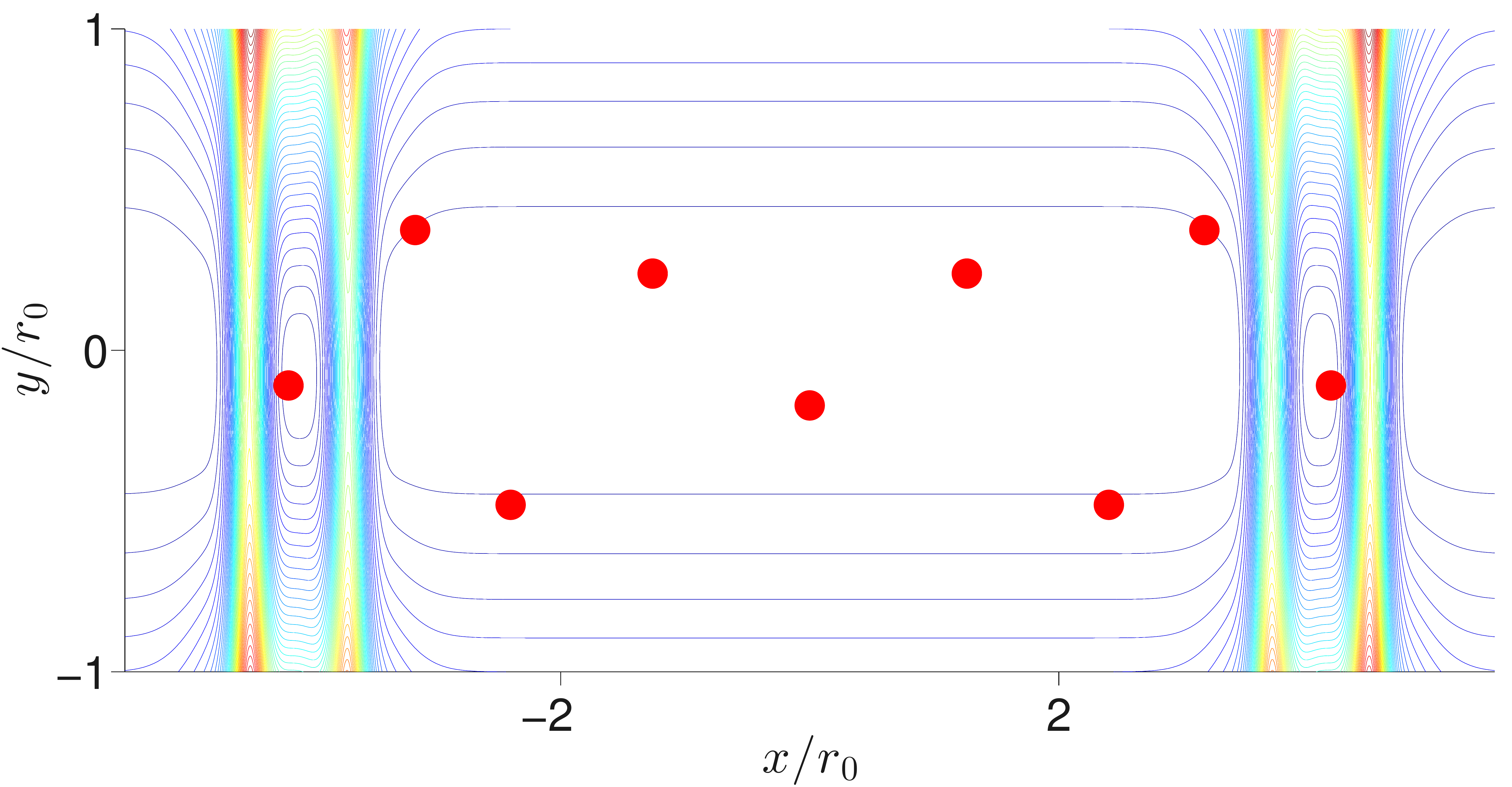}}
\subfloat[]{\includegraphics[width=0.5\textwidth]{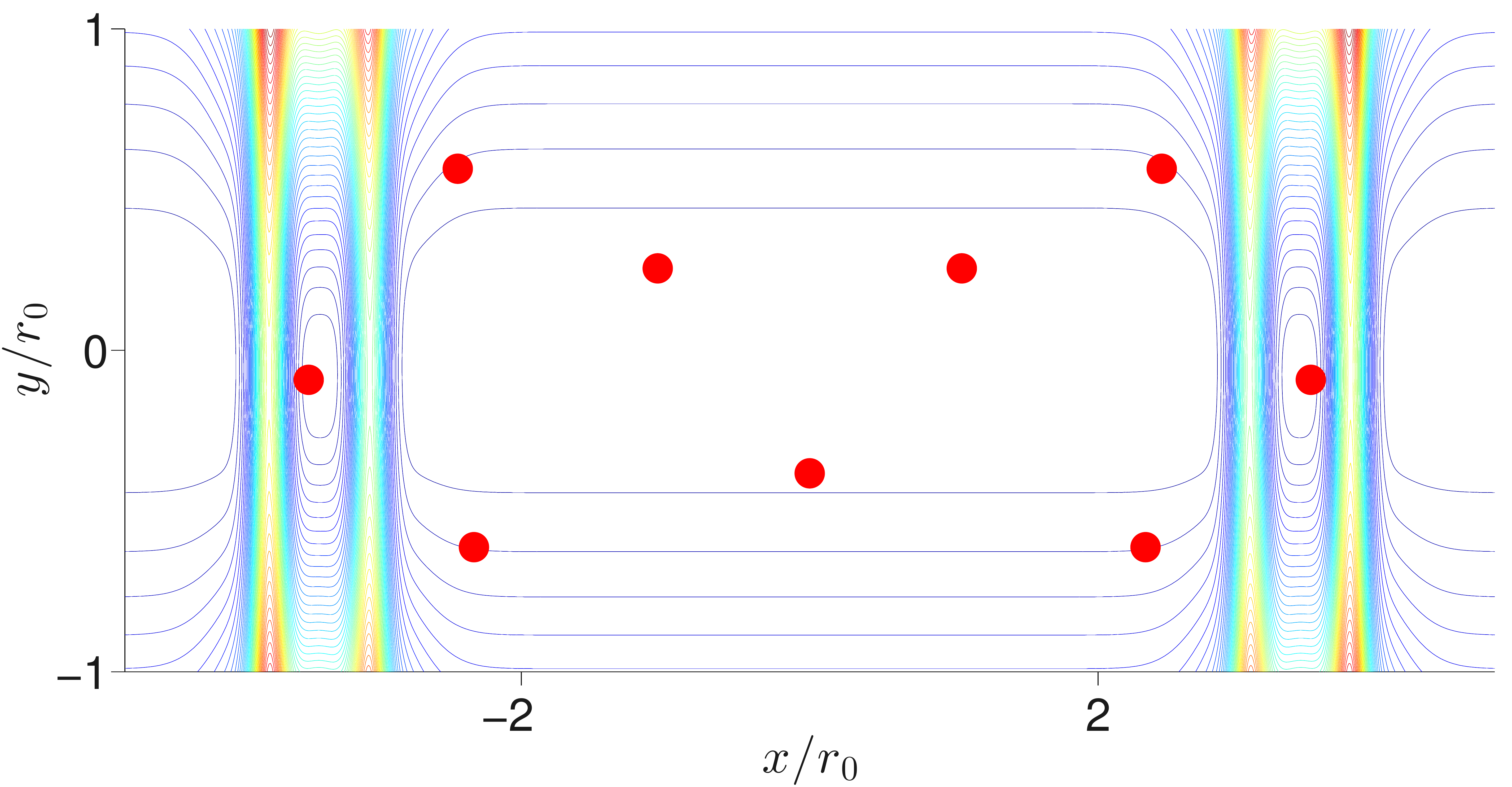}}
\caption{\label{fig:Density} Classical equilibrium positions for 9 electrons, at densities of a) $\nu =$0.5, b) $\nu =0.9$, c) $\nu = 1$ and d) $\nu = 1.2$. Distances are shown in units of $r_0$, and note the different scales on the axes as density is increased.}
\end{figure}

Results for different density are shown in Fig.~\ref{fig:Density}. Around $\nu=0.9$ there is a transition from a linear chain to a zig-zag chain (since at $r_0=1$ the strength of the confining potential becomes similar to the Coulomb repulsion), and towards higher densities we see that there are two electrons almost equidistant to each dot electron. 
\begin{figure}[h]
\begin{center}
	\includegraphics[width=0.6\textwidth]{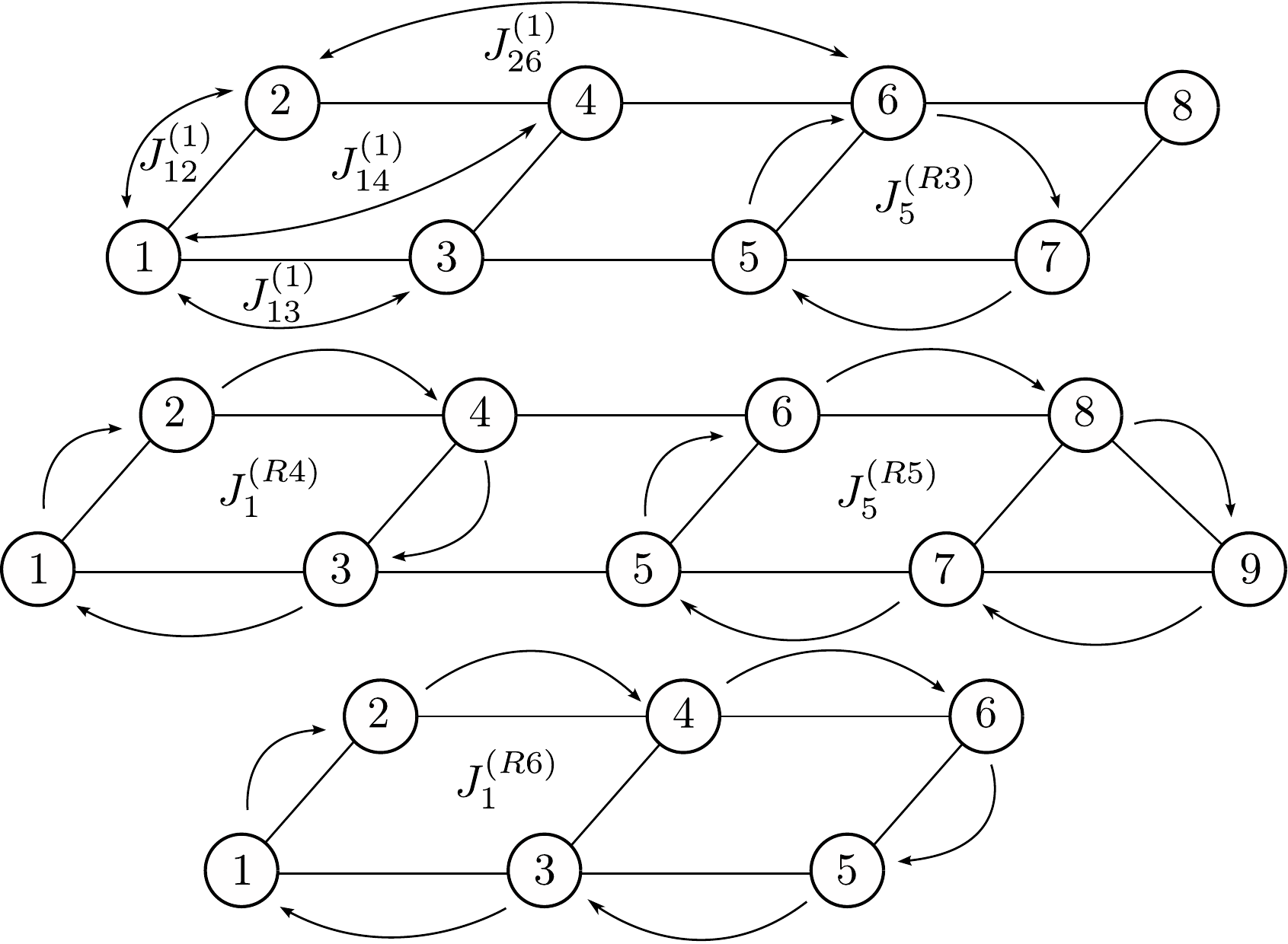}
	\caption{\label{fig:WigExch} An illustration of the types of exchange processes that we consider in this work. $J^{(1)}_{mn}$ indicates pairwise exchange between electrons $m$ and $n$. $J^{(R3}_{n}, J^{(R4)}_n$, $J^{(R5)}_n$, $J^{(R6)}_n$ indicates 3- 4- 5- and 6-body ring exchange starting clockwise from electron $n$.}
\end{center}
\end{figure}
The electrons are typically arranged near the vertices of triangles, and we considered permutation processes involving up to 6 electrons and up to 4$^{th}$-nearest neighbour pairwise exchanges, as shown in Fig.~\ref{fig:WigExch}. To find the strength of each of these permutation processes, we began by numerically finding the classical path of least action for each. To do this, we first discretised the action integral into $M$ equal time steps $\Delta \tau = T_\tau/M$, so that the action integral becomes a Riemann sum:
\begin{align}
\eta[\mathbf{R}] = \int_0^{T_\tau} d\tau \left[  \sum_j \frac{\dot{\mathbf{r}}_j^2}{2} + \frac{1}{\hbar \sqrt{r_\Omega}}V(\mathbf{R})\right] \rightarrow \sum_{m=1}^M \sum_j \frac{1}{\Delta \tau} \frac{ (\mathbf{r}_j(\tau_m) - \mathbf{r}_j(\tau_{m-1}))^2}{2} + \frac{\Delta \tau}{\hbar \sqrt{r_{\Omega}} } V(\mathbf{R}_m) 
\end{align}
where $\tau_m = m\Delta \tau$ and $\mathbf{R}_m =(r_1(\tau_m),r_2(\tau_m),...,r_{2N}(\tau_m))$. Notice that this is the same equation as the energy of $M$ particles joined together with springs of stiffness $\frac{1}{ \Delta \tau}$, in a potential $\sum_{n=1}^M \frac{\Delta \tau}{\hbar \sqrt{r_{\Omega}} } V( \mathbf{R}_n)$, and with endpoints attached to $\bar{\mathbf{R}}$ and $P^{\textsc{r}} \bar{\mathbf{R}}$. Thus minimising the action in this discretised picture is the same as minimising the energy of this chain. This can be minimised using a 2-point steepest descent algorithm~\cite{Barzilai1988} which proceeds as follows, with the time-dependent positions of all of the particles at the $n^{th}$ iteration labelled as $\mathbf{R}^{(n)} (\tau)= \{ \mathbf{R}_1^{(n)}, \mathbf{R}_2^{(n)},...,\mathbf{R}_M^{(n)} \}$:
\begin{itemize}
\item Make an initial guess $\mathbf{R}^{(0)}(\tau)$.
\item For each step, $\mathbf{R}^{(n+1)}(\tau) = \mathbf{R}^{(n)}(\tau)- \alpha_n \mathbf{G}_n$, where $\mathbf{G}_n = \nabla_{ \mathbf{R} }\eta[\mathbf{R}^{(n)}(\tau)] $, $\alpha_0$ is a small step size (we use $\alpha_0 = 0.001$) and $\alpha_n = \mathbf{G}_n \cdot (\mathbf{R}^{(n)}(\tau) - \mathbf{R}^{(n-1)}(\tau)) / \|\mathbf{G}_n\|^2$ for $n > 0$~\cite{Barzilai1988}.
\item Stop if $\| \mathbf{G}_n \| < \varepsilon_c$ (convergence to a minimum), or if $\| \mathbf{R}^{(n)}(\tau) \| < \varepsilon_p$ (convergence to a point).
\end{itemize} 
where $\| \cdot \|$ indicates the Euclidean norm. For our calculations, we used $T_\tau = 30$ and $M=70$, based on the work in~\cite{Katano2000} that shows reasonable convergence with these parameters for 27 electrons in a Wigner crystal (suggesting that for these parameters the instanton assumptions made in Sec~\ref{sec:Instanton} are valid). The accuracy of discretising time in this way has also been studied in~\cite{Richardson2011} by comparing numerical results to an analytically solvable system. Here they find that for $M=64$, $T_\tau = 30$ the error is on the order of 1\%. The tolerance used for the steepest descent algorithm was $\varepsilon_c = 1\times 10^{-6}$. 

Once the classical path $\mathbf{R}_{cl}$ has been found using this algorithm, the prefactor must be calculated by diagonalising the matrix $\mathbf{A}$ in (\ref{eqn:Jreduced}) so that the determinant can be found. This matrix is calculated along the classical path, and in discretised form becomes a $2N(M+1) \times 2N(M+1)$ matrix of the form
\begin{align}
(\mathbf{A})_{R_i(\tau_m)R_j(\tau_n)} &= -\frac{1}{\Delta \tau^2} \left( -2\delta_{mn} + \delta_{m,n-1} + \delta_{m,n+1} \right) \delta_{ij}  +  \frac{\hbar}{\sqrt{r_\Omega}}\mathbf{H}_{ij}(\tau_n)\delta_{mn}
\end{align}
This matrix takes the following form
\begin{align}
\left[ \begin{array}{c c c c c c }
\frac{\hbar}{\sqrt{r_\Omega}}\mathbf{H}_{11}(\tau_1) +\frac{2}{\Delta \tau^2} &\frac{\hbar}{\sqrt{r_\Omega}}\mathbf{H}_{12}(\tau_1)   & \dots  & -\frac{1}{\Delta \tau^2}&0 &  \\
\frac{\hbar}{\sqrt{r_\Omega}}\mathbf{H}_{21}(\tau_1)  & \frac{\hbar}{\sqrt{r_\Omega}}\mathbf{H}_{22}(\tau_1)   + \frac{2}{\Delta \tau^2} & \dots  & 0 &-\frac{1}{\Delta \tau^2} &   \\
 \vdots & \vdots & \ddots & & &   \\
 -\frac{1}{\Delta \tau^2} &  0 &  & \frac{\hbar}{\sqrt{r_\Omega}}\mathbf{H}_{11}(\tau_2)  + \frac{2}{\Delta \tau^2}& \frac{\hbar}{\sqrt{r_\Omega}}\mathbf{H}_{12}(\tau_2) &  \dots  \\
 0&  -\frac{1}{\Delta \tau^2}&  & \frac{\hbar}{\sqrt{r_\Omega}}\mathbf{H}_{21}(\tau_2)  & \frac{\hbar}{\sqrt{r_\Omega}}\mathbf{H}_{22}(\tau_2)  + \frac{2}{\Delta \tau^2} &  \dots \\
 &  & &\vdots & \vdots & \ddots
\end{array}\right]
\end{align}

\subsection{Information transfer}
\label{sec:WignInfTrans}

To send information through the chain, we start with the quantum dots isolated from the rest of the chain (i.e.\ the internal barriers are high), then the internal barriers are dropped to turn on the interactions. This could be done either suddenly (a `quench') or slowly with respect to the couplings in the chain. For this work, we will study the viability of performing a quench, and leave the study of other protocols to later work. 

\begin{figure}[h]
\begin{center}
	\includegraphics[width=0.9\textwidth]{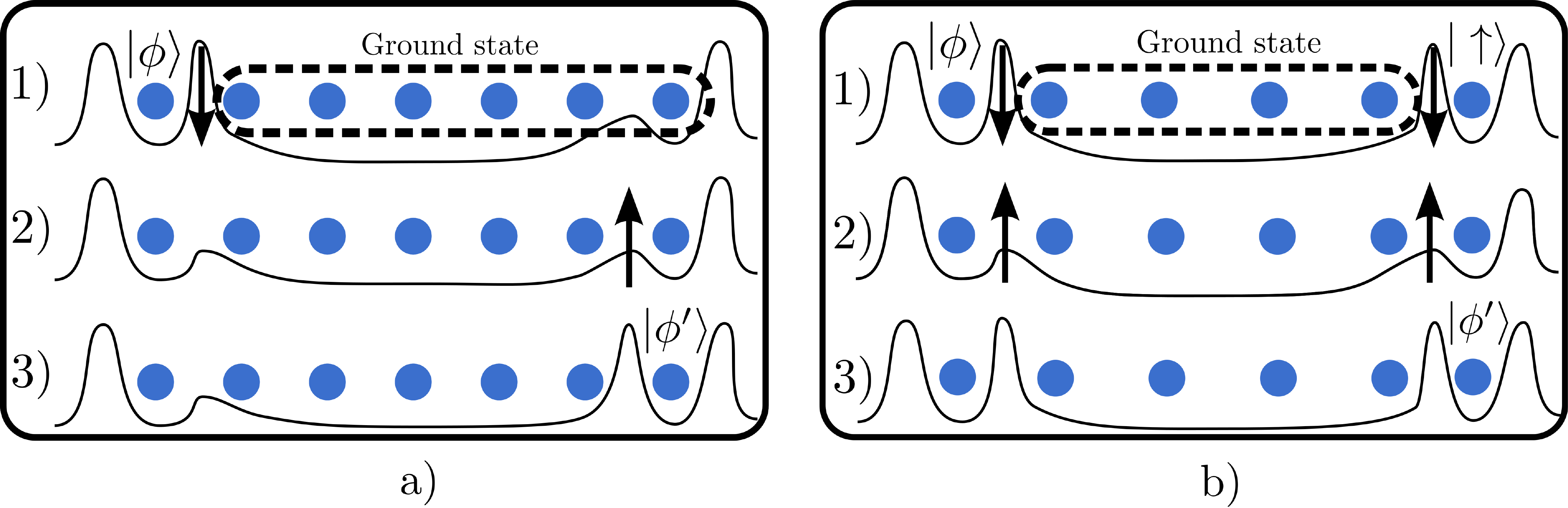}
	\caption{\label{fig:TransferProto} An illustration of the transfer protocol, a) using an odd number of electrons and b) using an even number of electrons. 1) The left barrier is up and the right barrier is down. The isolated (left) dot is prepared in a quantum state $\ket{\phi}$ and the remaining electrons are prepared in the ground state. For the even case, the opposite is also isolated and prepared in an $\ket{\uparrow}$ state. 2) The barriers are lowered, allowing information to travel through the chain. 3) The right barrier is raised at the optimal point to remove the transferred information.}
\end{center}
\end{figure}

The protocol used here depends on the number of electrons (see Fig.~\ref{fig:TransferProto}). For an odd number of electrons, the system is initialised with one barrier up and one barrier down, so that there are an even number of electrons coupled together (prepared in the ground state) and one isolated dot, which is prepared in the quantum state to be sent. The barrier is then lowered, coupling the isolated spin to the chain, and some time later the opposite barrier is raised to isolate the output electron. For an even number of electrons, the system is initialised with both barriers up, with one dot prepared in the quantum state to be transferred and the opposite dot prepared in a state such as $\ket{ \uparrow }$. The barriers are then simultaneously lowered and raised at the appropriate times. The reason for these different protocols is that intuitively we expect a chain of even electrons to have a unique ground state, since each electron can pair with one nearest neighbour, whereas for an odd chain there can be unpaired electrons which increases the degeneracy.

For a quench, the potential problem is that such a sudden change will cause the electrons to move out of equilibrium, which could cause phonon excitations leading to decoherence and fluctuations in the exchange interactions. To assess how good this approach would be, we calculated the average change in equilibrium positions of the electrons after lowering the barriers quickly, starting with a barrier height of $5\Omega$ as this was found to reduce the dot-chain coupling to around $10^{-6}$ relative to the rest of the chain. To quantify the average change in equilibrium positions after the barriers drop ($\mathbf{R}$) relative to the equilibrium positions with the barriers up ($\mathbf{R}_0$), we use the following distance measure
\begin{align}\label{eqn:bard}
\bar{d} = \frac{1}{N}\frac{ \| \mathbf{R} - \mathbf{R}_0 \|}{\Delta \mathbf{R}_0},
\end{align}
where $\|\mathbf{R} \| $ is the Euclidean norm, and $\Delta\mathbf{R}_0$ is the average nearest-neighbour separation of the electrons.

\begin{figure}[h]
\begin{center}
	\includegraphics[width=0.8\textwidth]{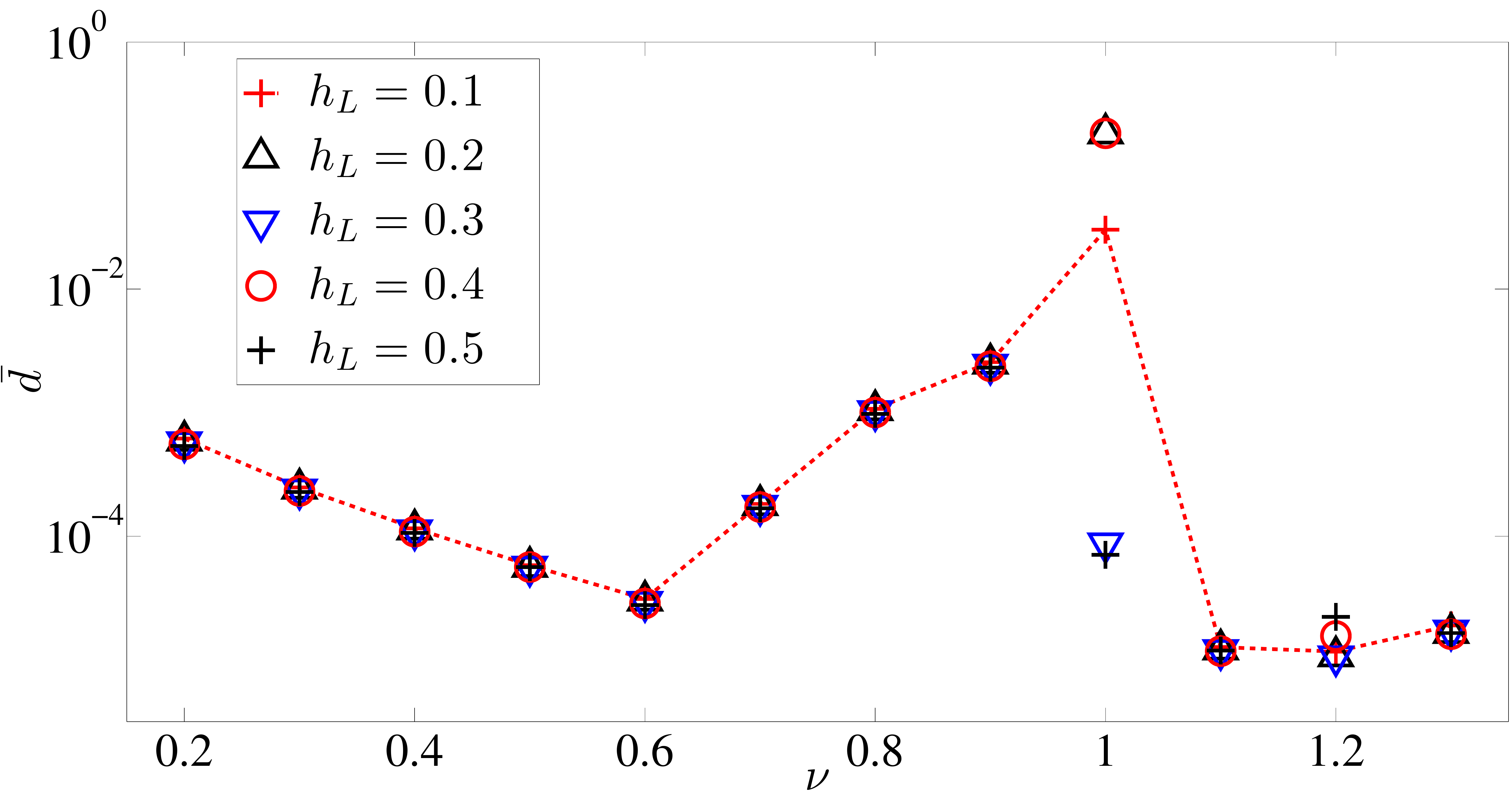}
	\caption{\label{fig:Quench9_2} $\bar{d}$ for a chain of 9 electrons, where $\bar{d}$ is defined in eqn.\ (\ref{eqn:bard}). Only the left barrier is lowered here, with $h_R=5$.}
\end{center}
\end{figure}

Results for lowering one barrier with $N=9$ are shown in Fig.~\ref{fig:Quench9_2}. In general $\bar{d}$ stays below about $10^{-2}$, and there are clear dips in $\bar{d}$ around $\nu = 0.6$ and $\nu = 1.2$, and little change over the range of $h_L$. There appears to be some instability around $\nu = 1$, which could be an artefact of the particular potentials used here or could be because this is the length scale where the coulomb repulsion becomes roughly equal to the trapping potential (from the definition of $r_0$). The results for lowering two barriers with either $N=10$ or $N=9$ were similar, but without the instability at $\nu = 1$, probably because two barriers are being lowered simultaneously, so the situation is more symmetric throughout.

The magnitude of the exchange coupling $J_P$ was then found using the instanton technique described above, with settings of $M=70$, $T_\tau = 30$, $h_R =h_L= 0.1$, $r_{\Omega} = 10$ (this value of $r_\Omega$ is chosen based on the band splitting of $\sim 0.3$meV in GaAs). Since the chains are symmetric, to speed up calculations only results for one half of the chain were taken, after confirming that the simulations were giving symmetric results (up to around 0.1 \% error, which confirms that the results are converging well). 

\begin{figure}[h]
\begin{center}
\subfloat[]{\includegraphics[width=0.49\textwidth]{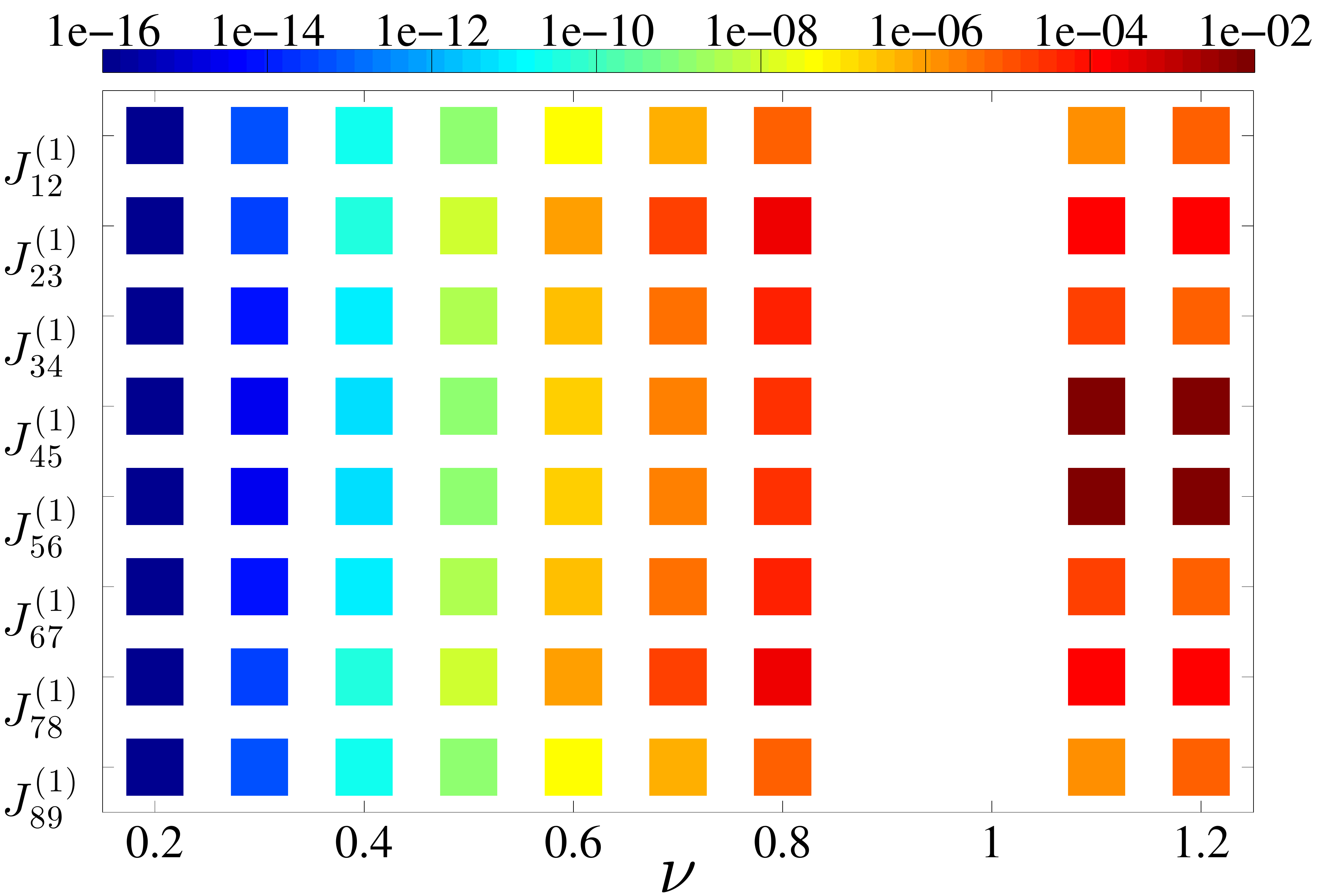}}
\subfloat[]{\includegraphics[width=0.49\textwidth]{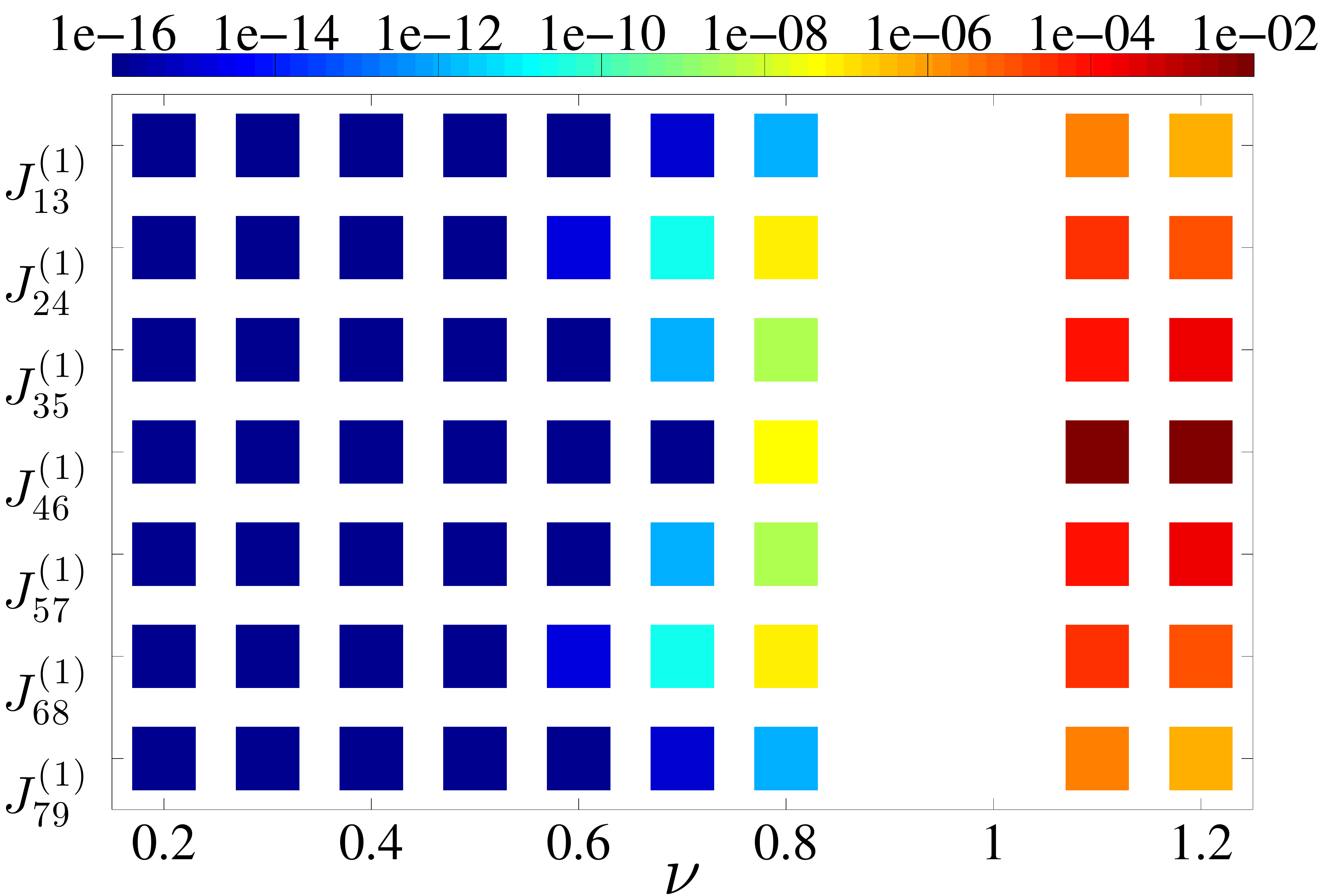}}\\
\subfloat[]{\includegraphics[width=0.49\textwidth]{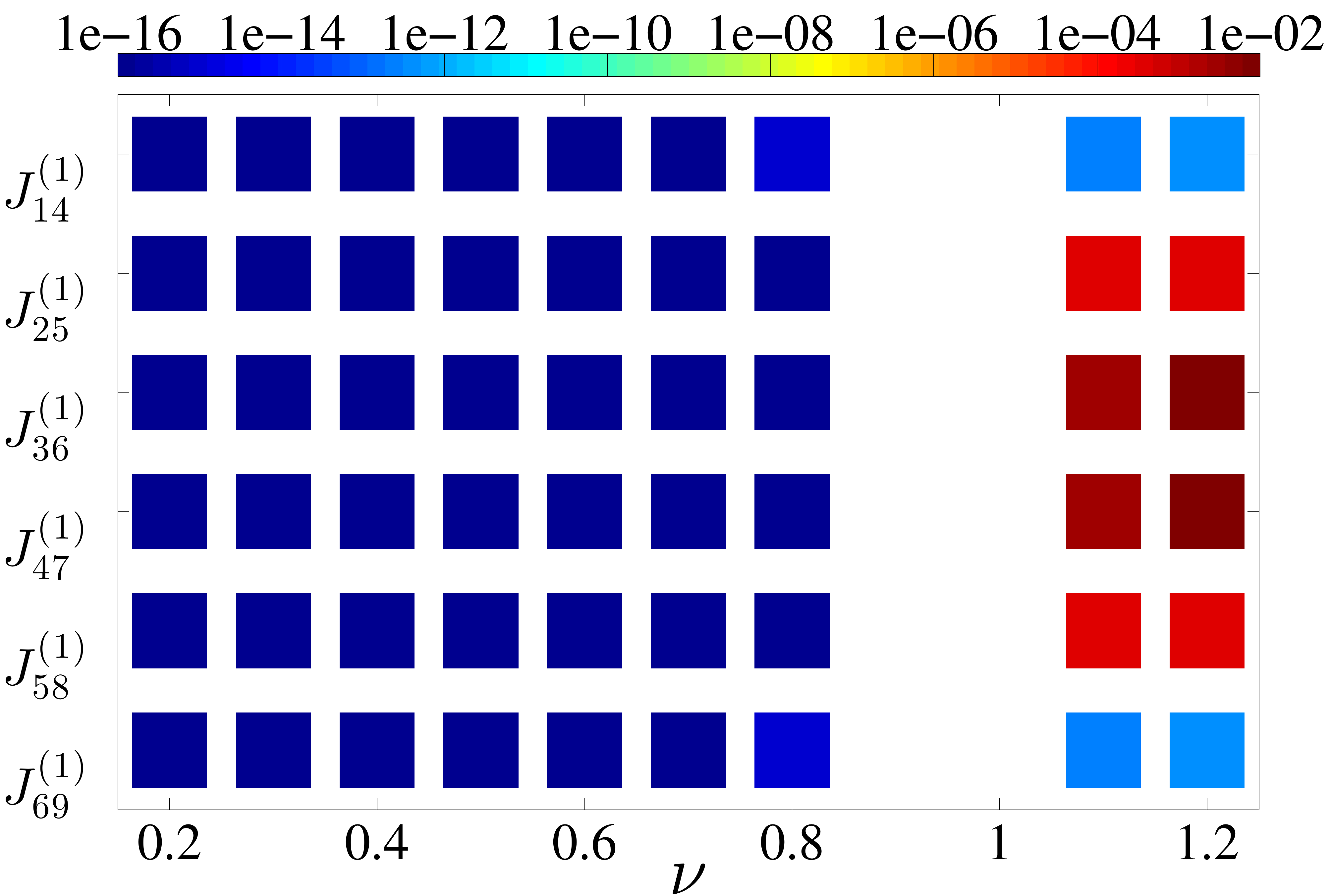}}
\subfloat[]{\includegraphics[width=0.49\textwidth]{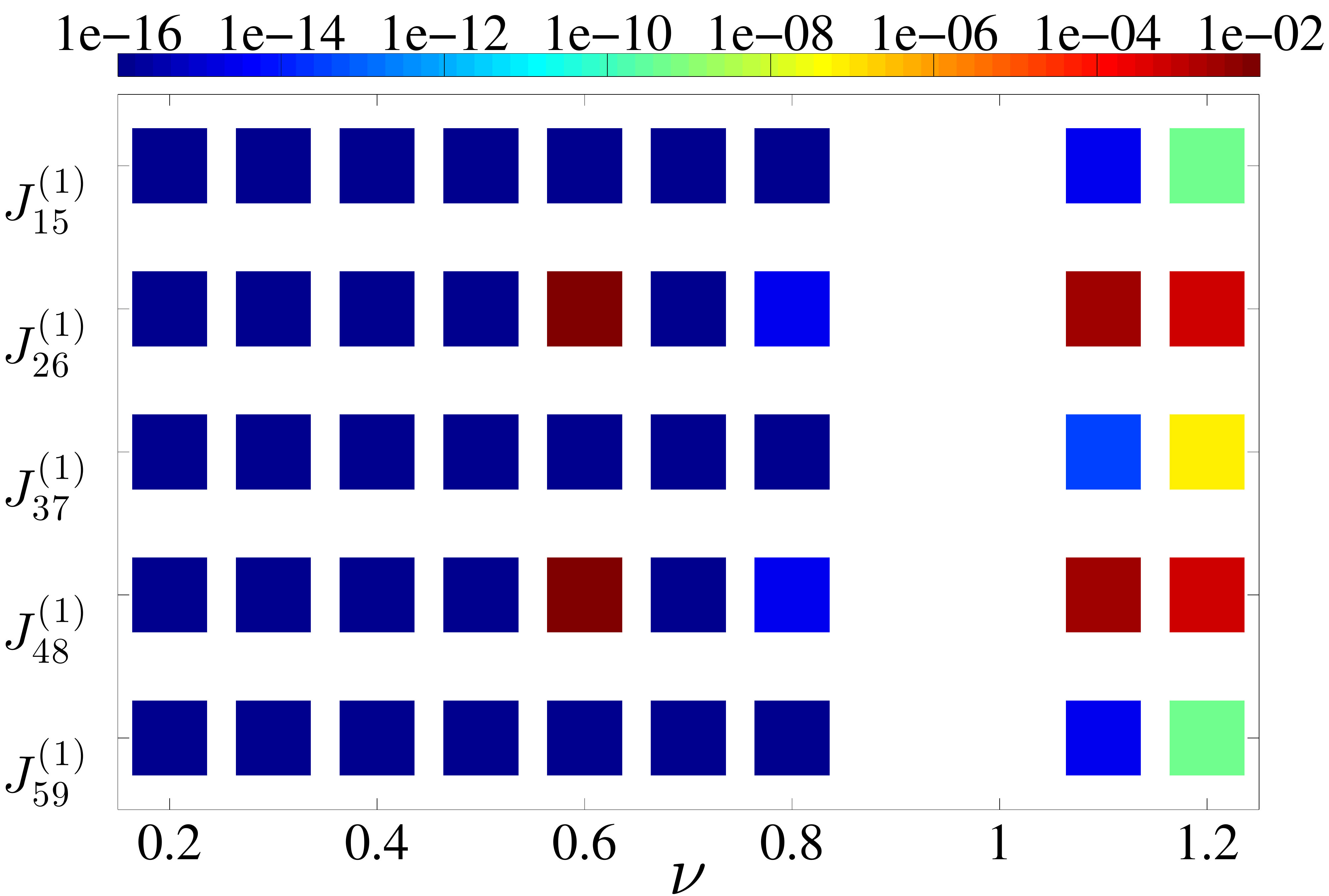}}\\
\subfloat[]{\includegraphics[width=0.49\textwidth]{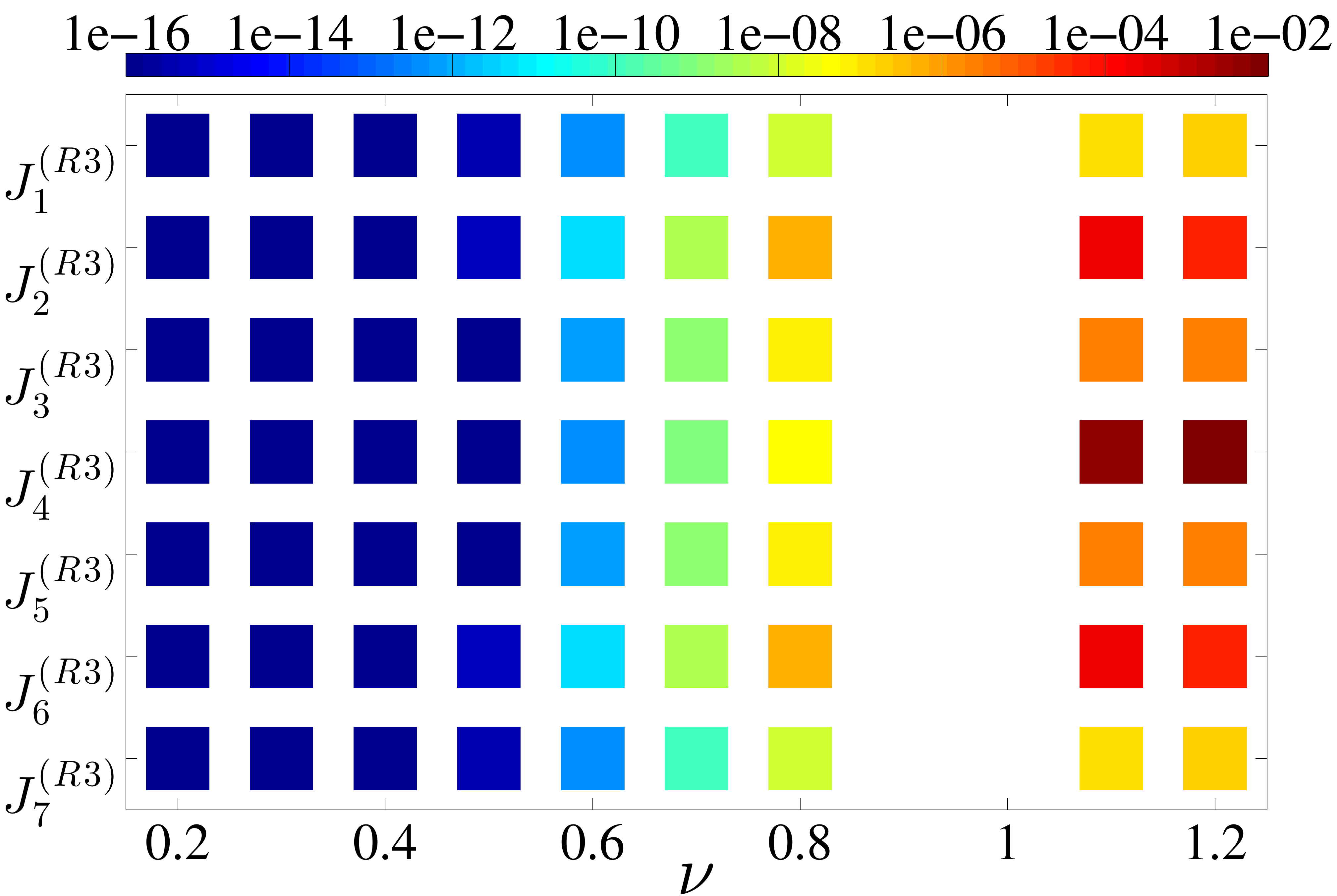}}
\subfloat[]{\includegraphics[width=0.49\textwidth]{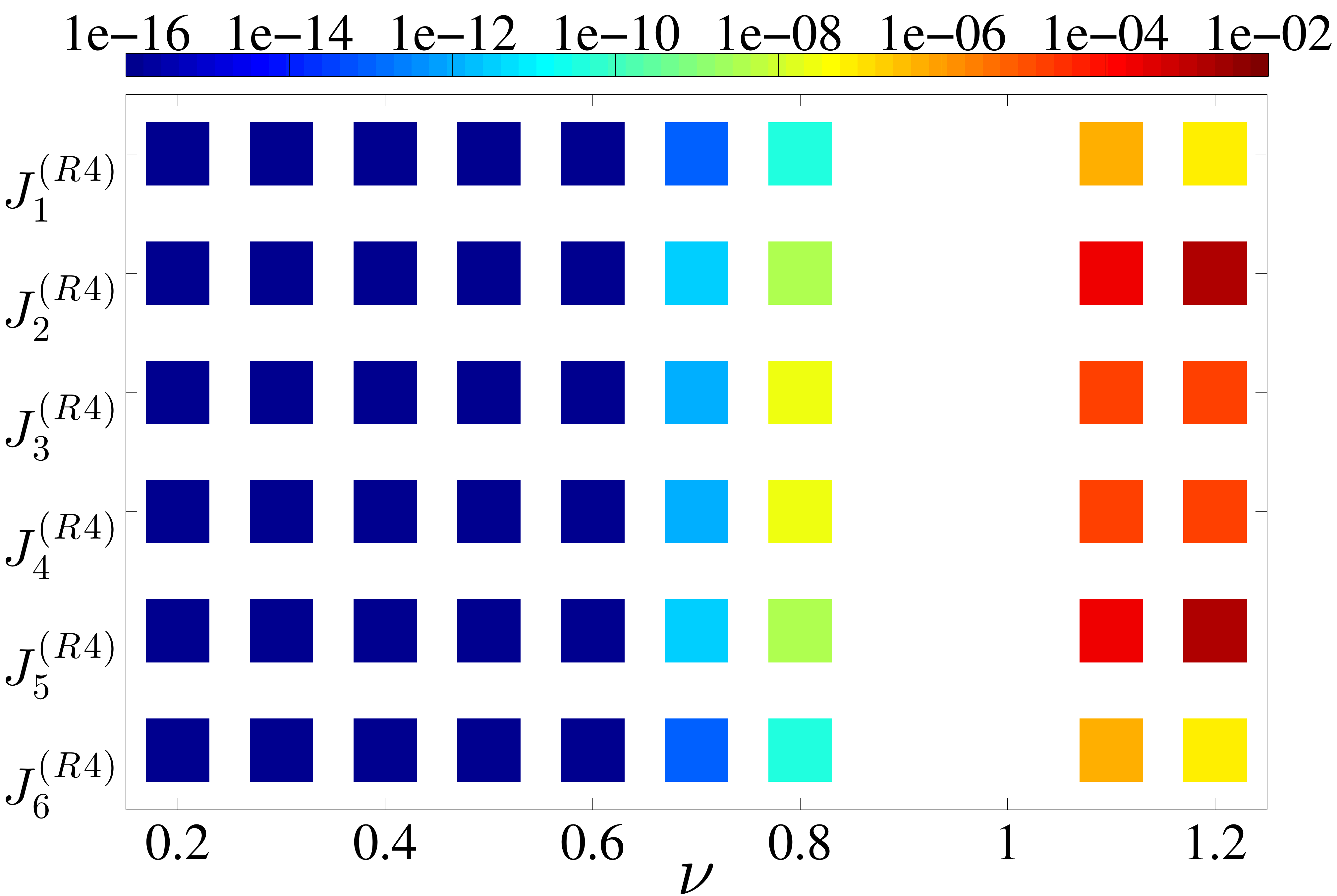}}\\
\subfloat[]{\includegraphics[width=0.49\textwidth]{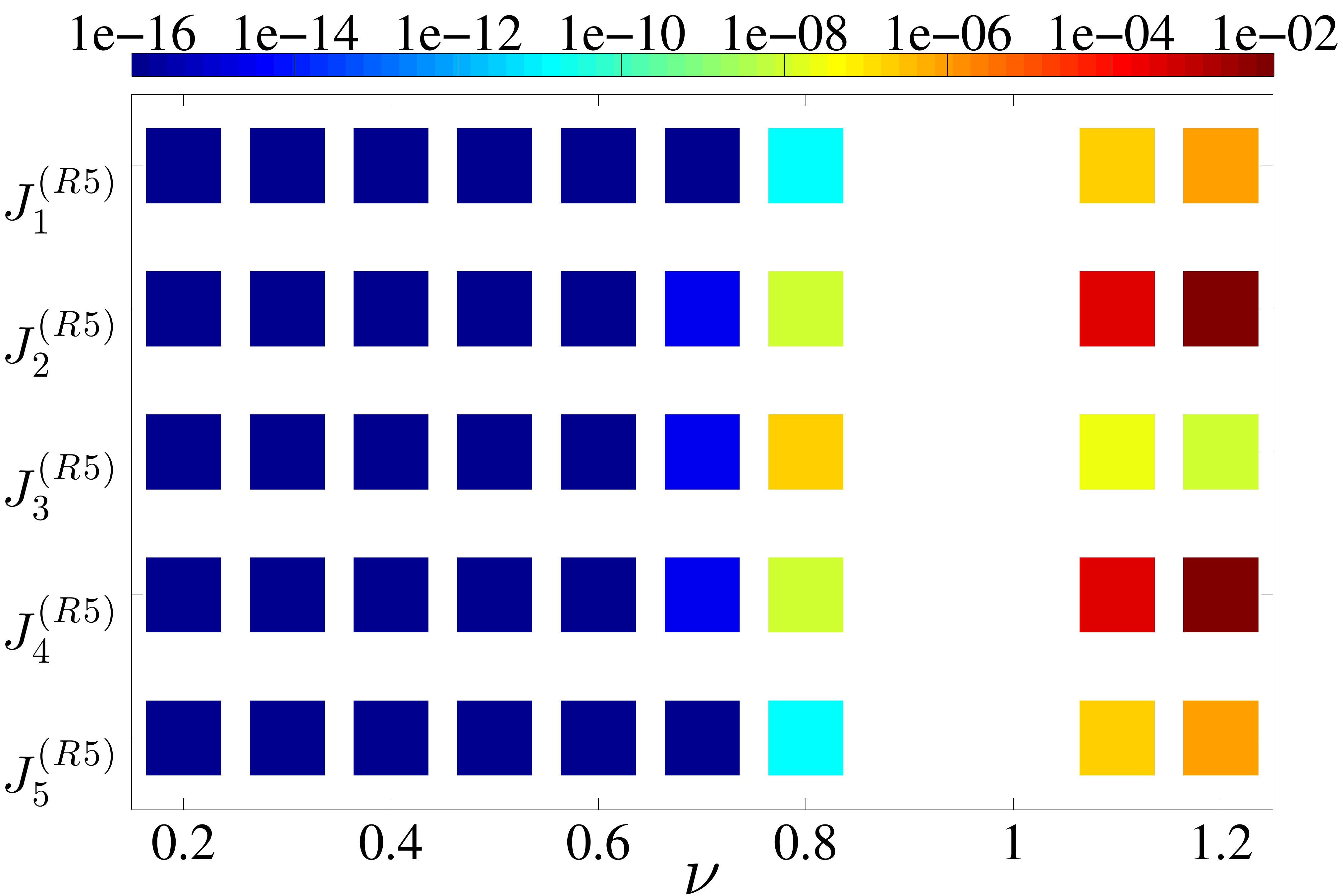}}
\subfloat[]{\includegraphics[width=0.49\textwidth]{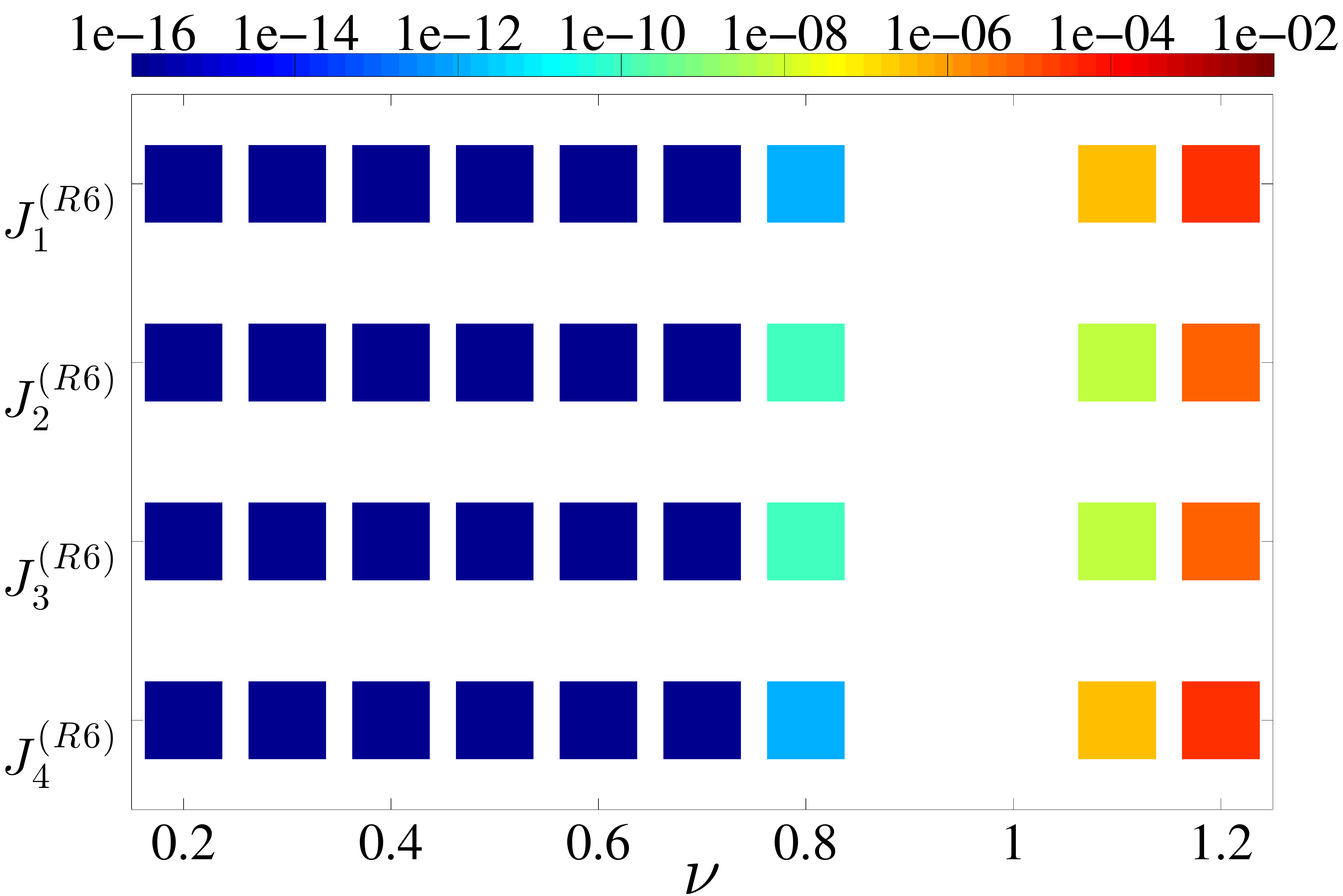}}
\caption{\label{fig:CouplingsN=9} $J_P$ for a Wigner crystal with 9 electrons, with $h_L = h_R = 0.1$: a) nearest-neighbour, b) 2$^{nd}$ neighbour, c) 3$^{rd}$ neighbour d) 4$^{th}$ neighbour, e) 3-body f) 4-body, g) 5-body and h) 6-body. $J_P$ is in units of meV. Note that there is a lower limit on the colour bar, which means that very small couplings are rounded up to $10^{-16}$.}
\end{center}
\end{figure}

\begin{figure}[h]
\begin{center}
\subfloat[]{\includegraphics[width=0.49\textwidth]{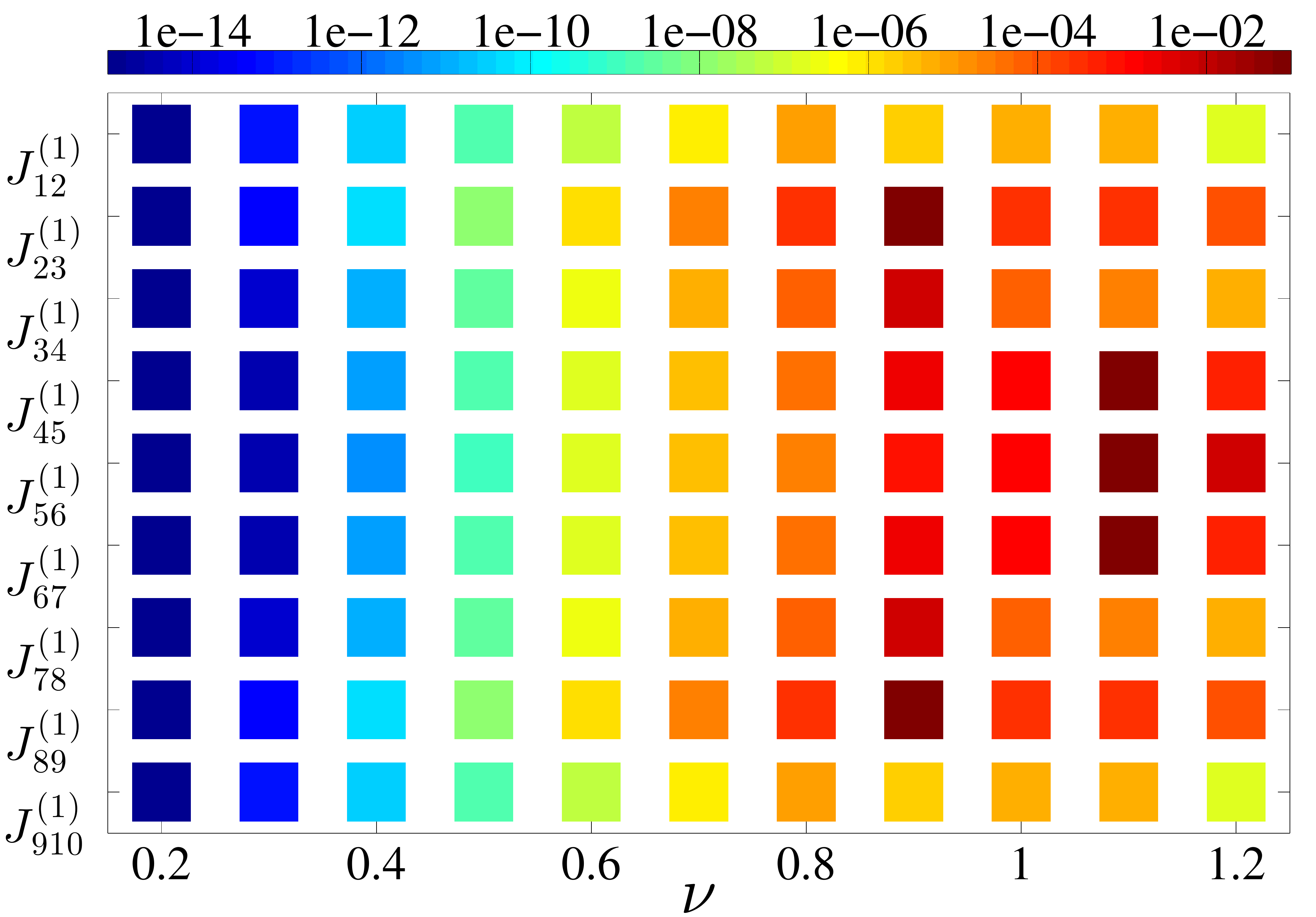}}
\subfloat[]{\includegraphics[width=0.49\textwidth]{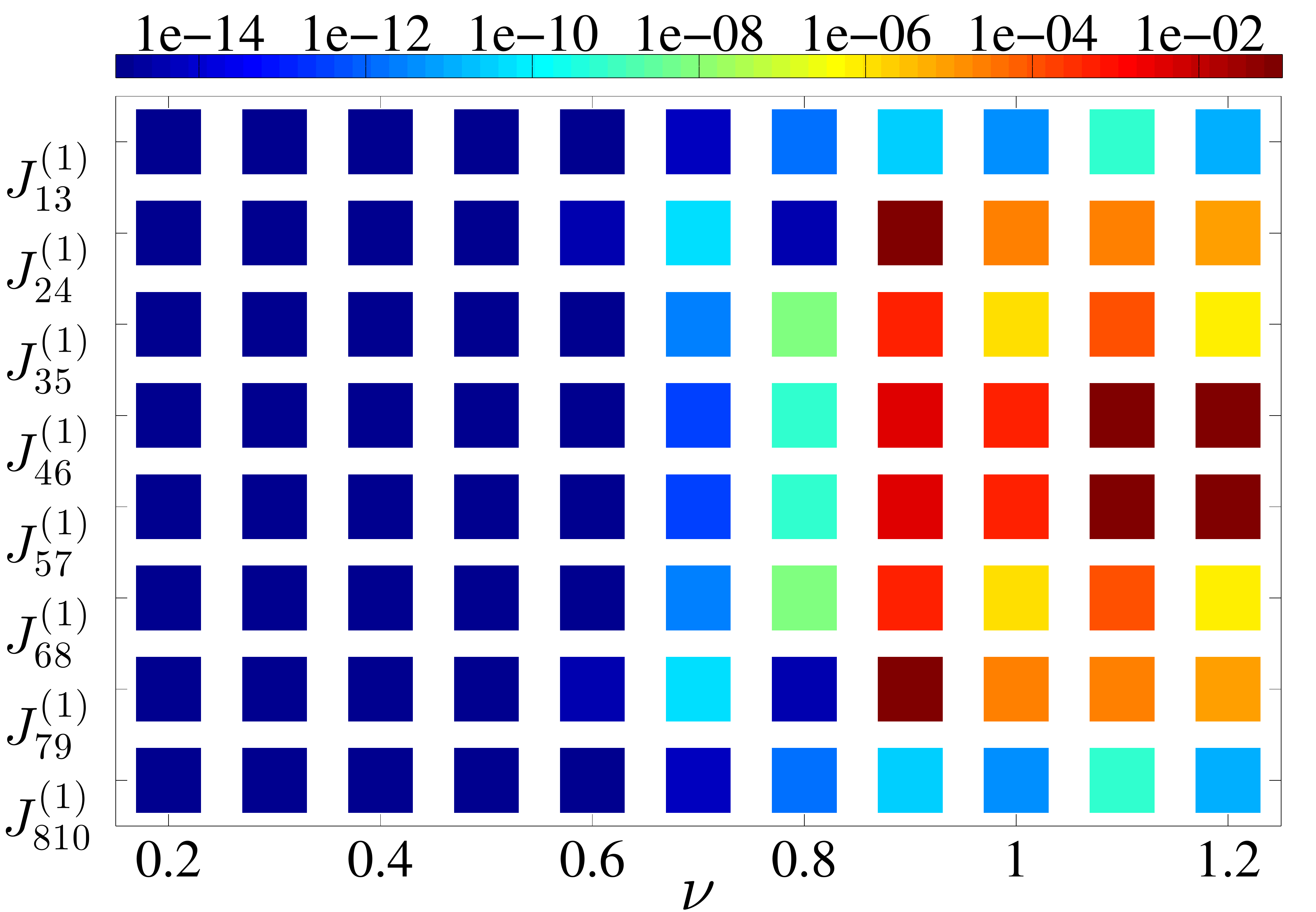}}\\
\subfloat[]{\includegraphics[width=0.49\textwidth]{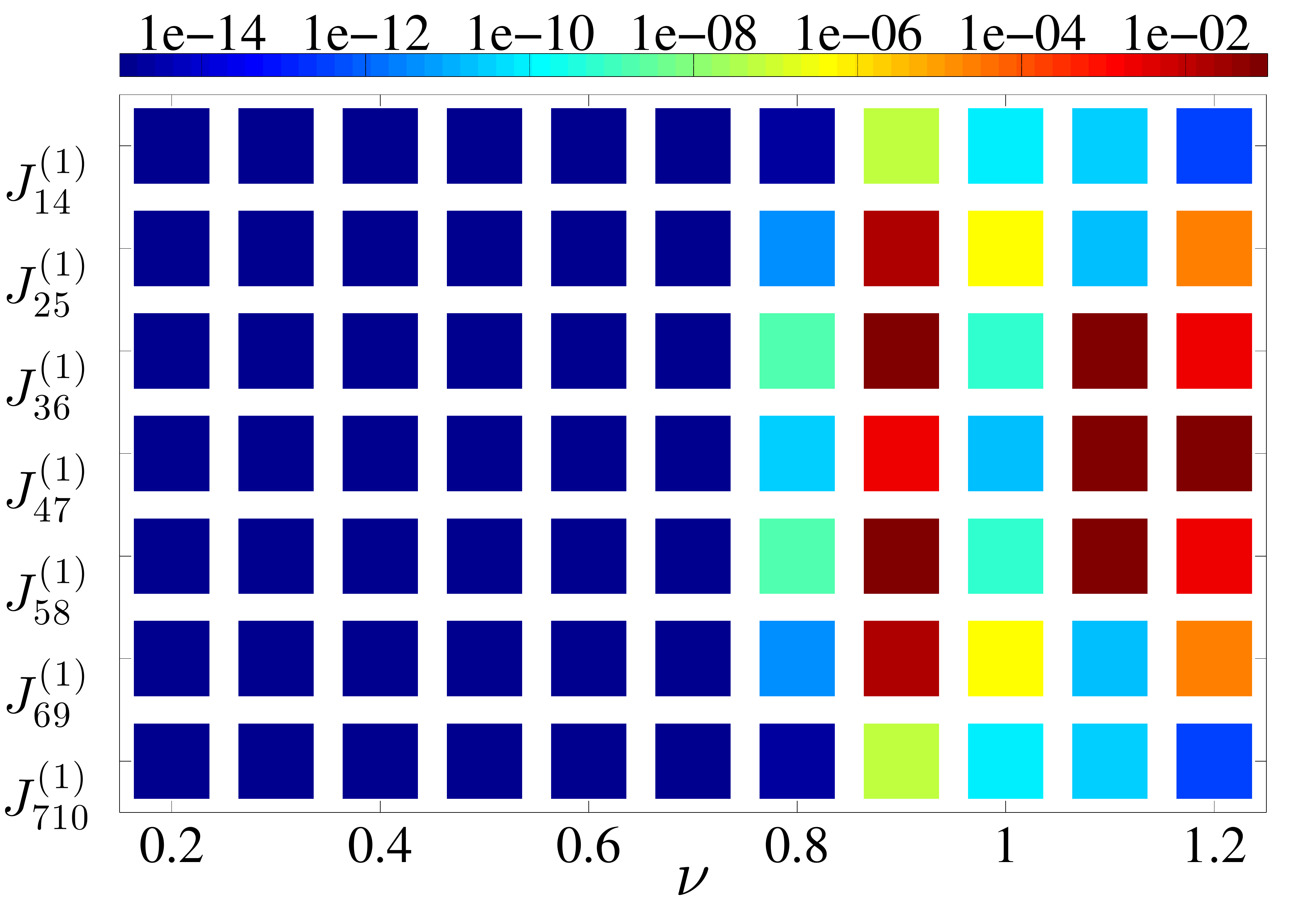}}
\subfloat[]{\includegraphics[width=0.49\textwidth]{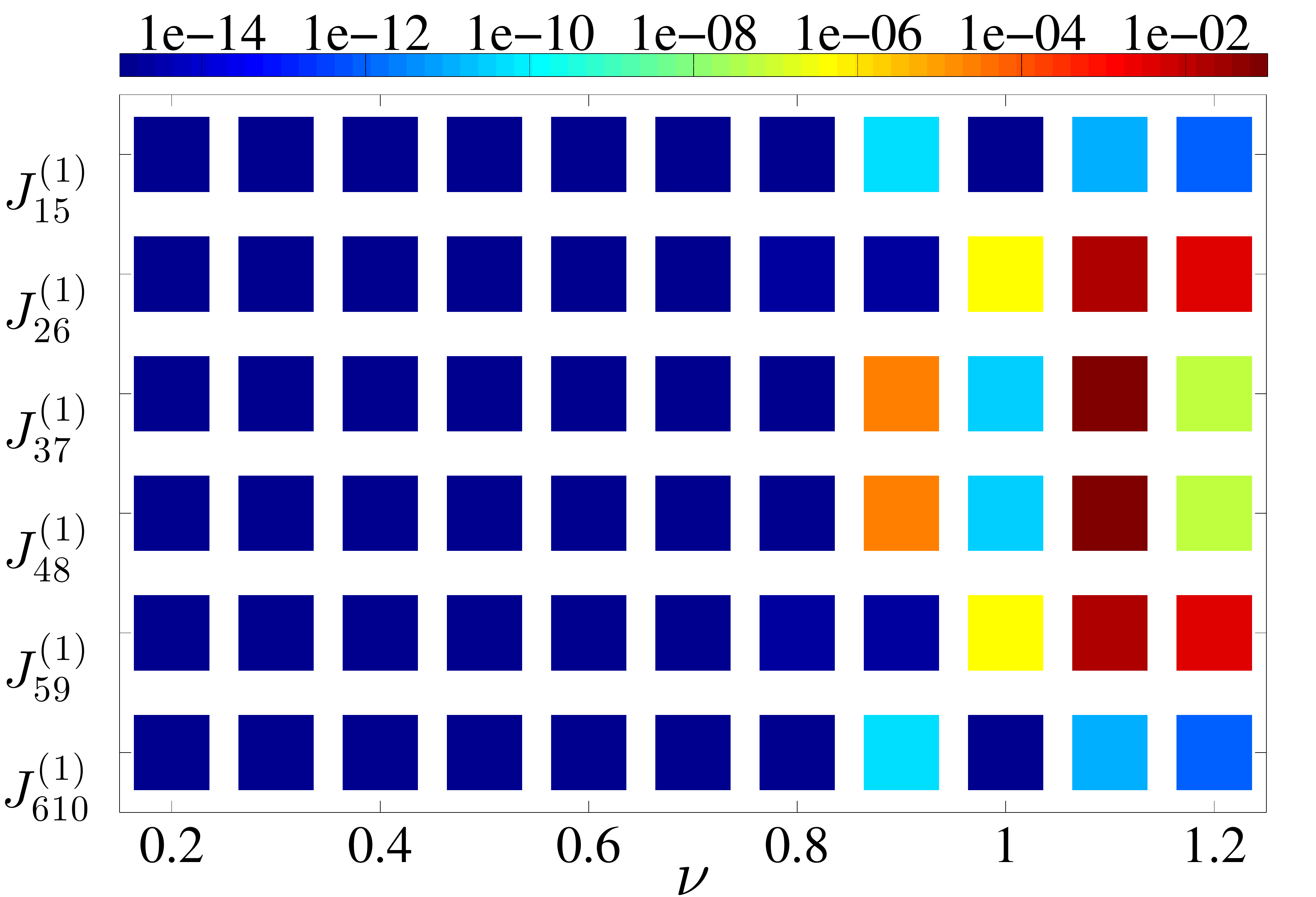}}\\
\subfloat[]{\includegraphics[width=0.49\textwidth]{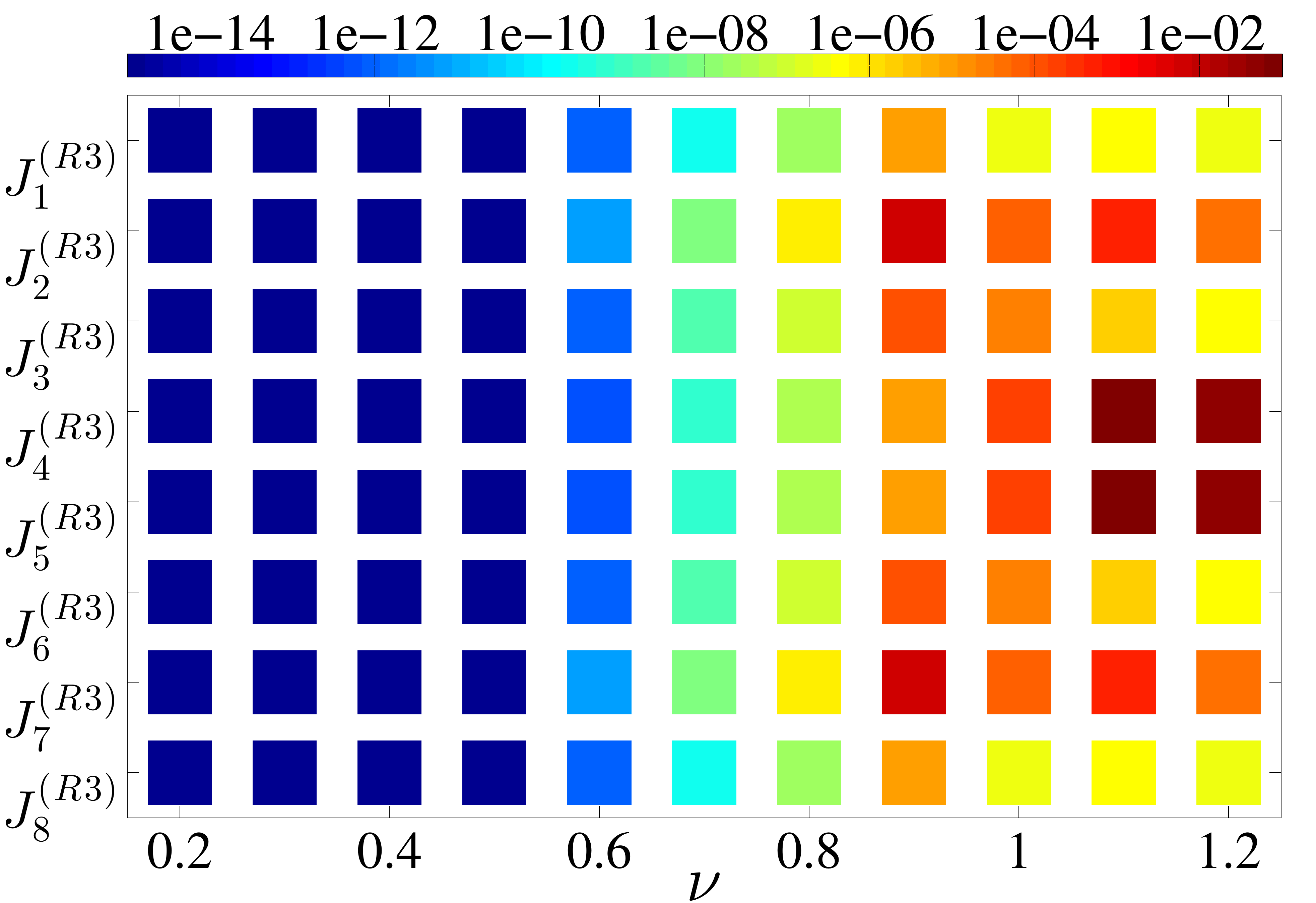}}
\subfloat[]{\includegraphics[width=0.49\textwidth]{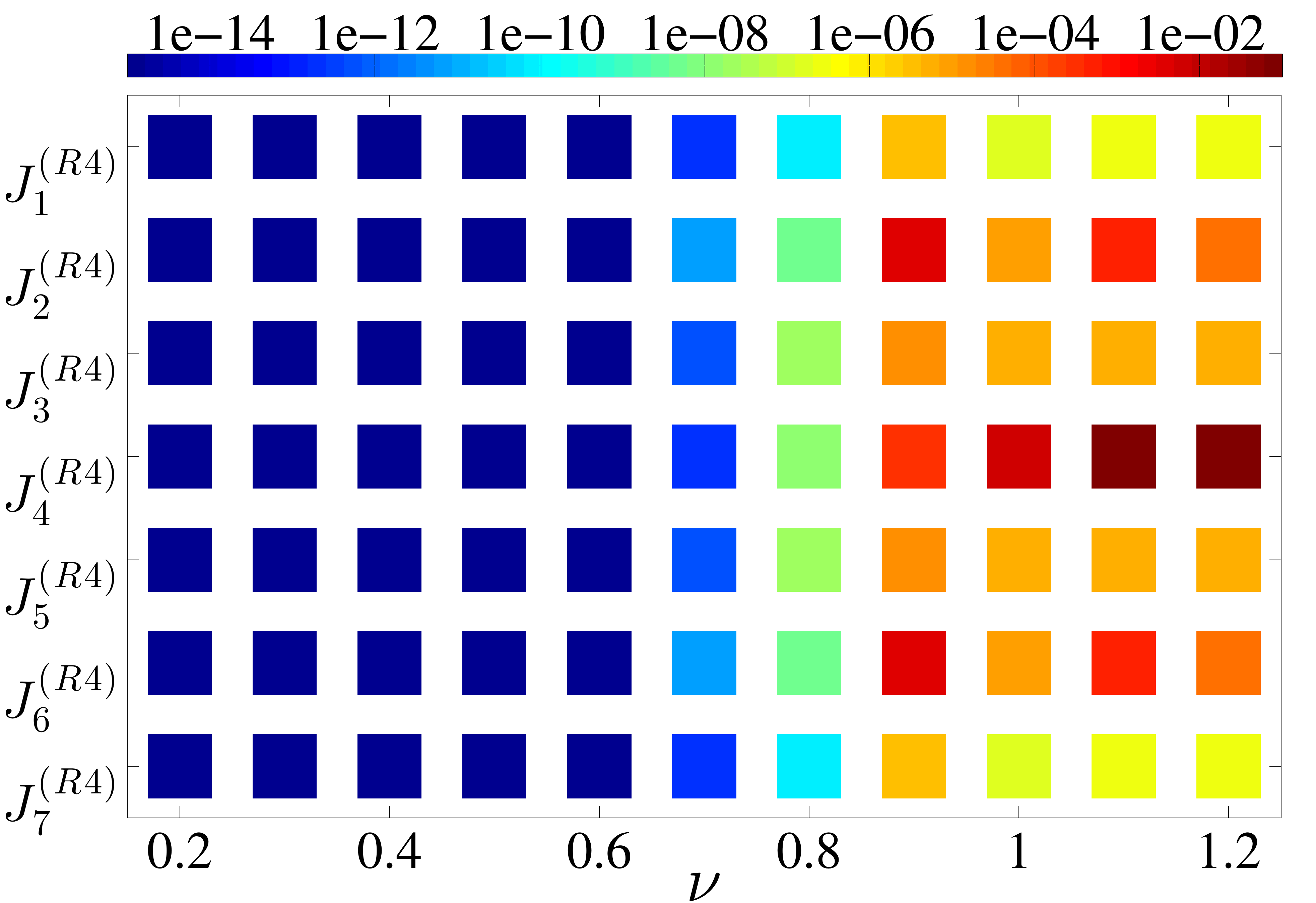}}\\
\subfloat[]{\includegraphics[width=0.49\textwidth]{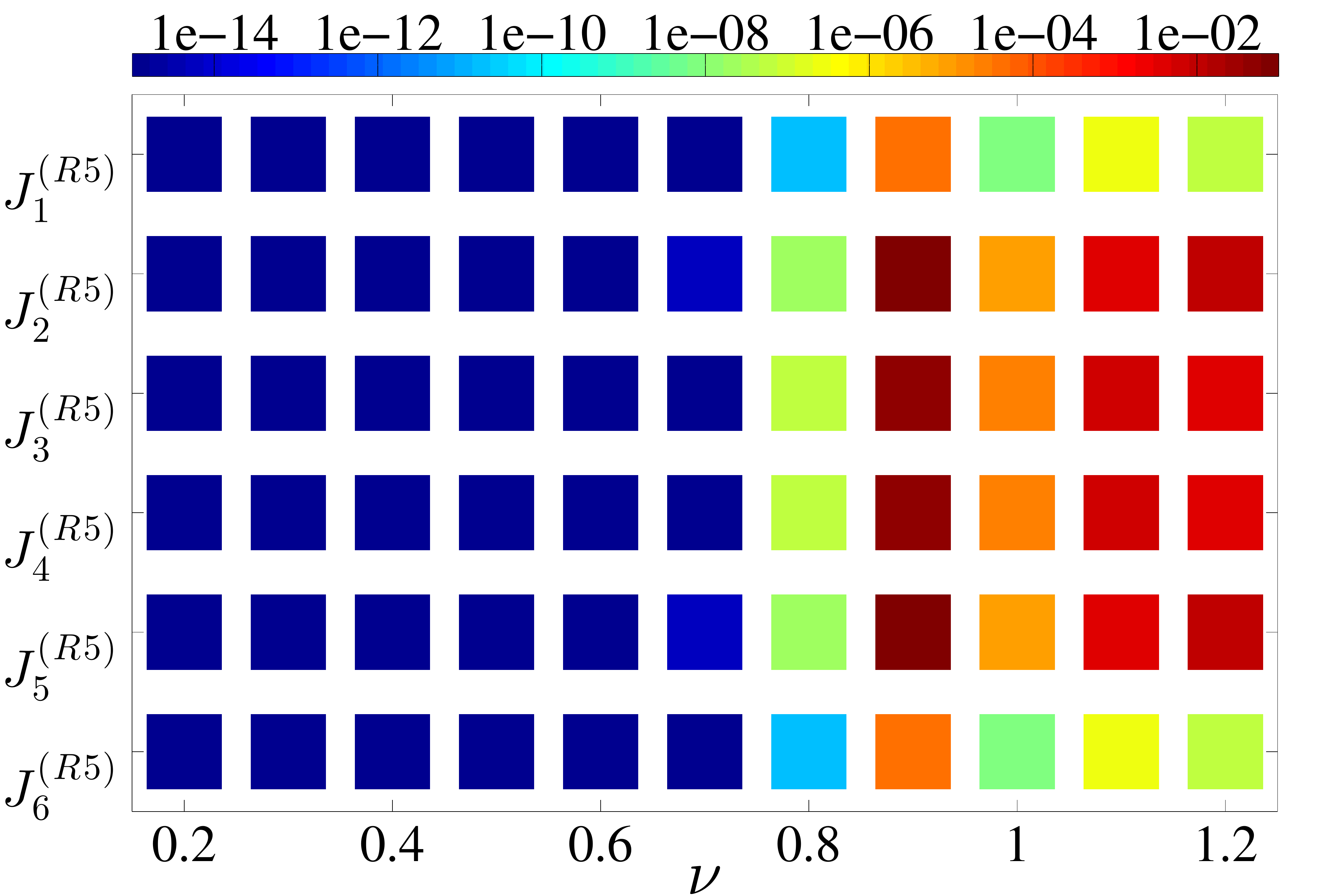}}
\subfloat[]{\includegraphics[width=0.49\textwidth]{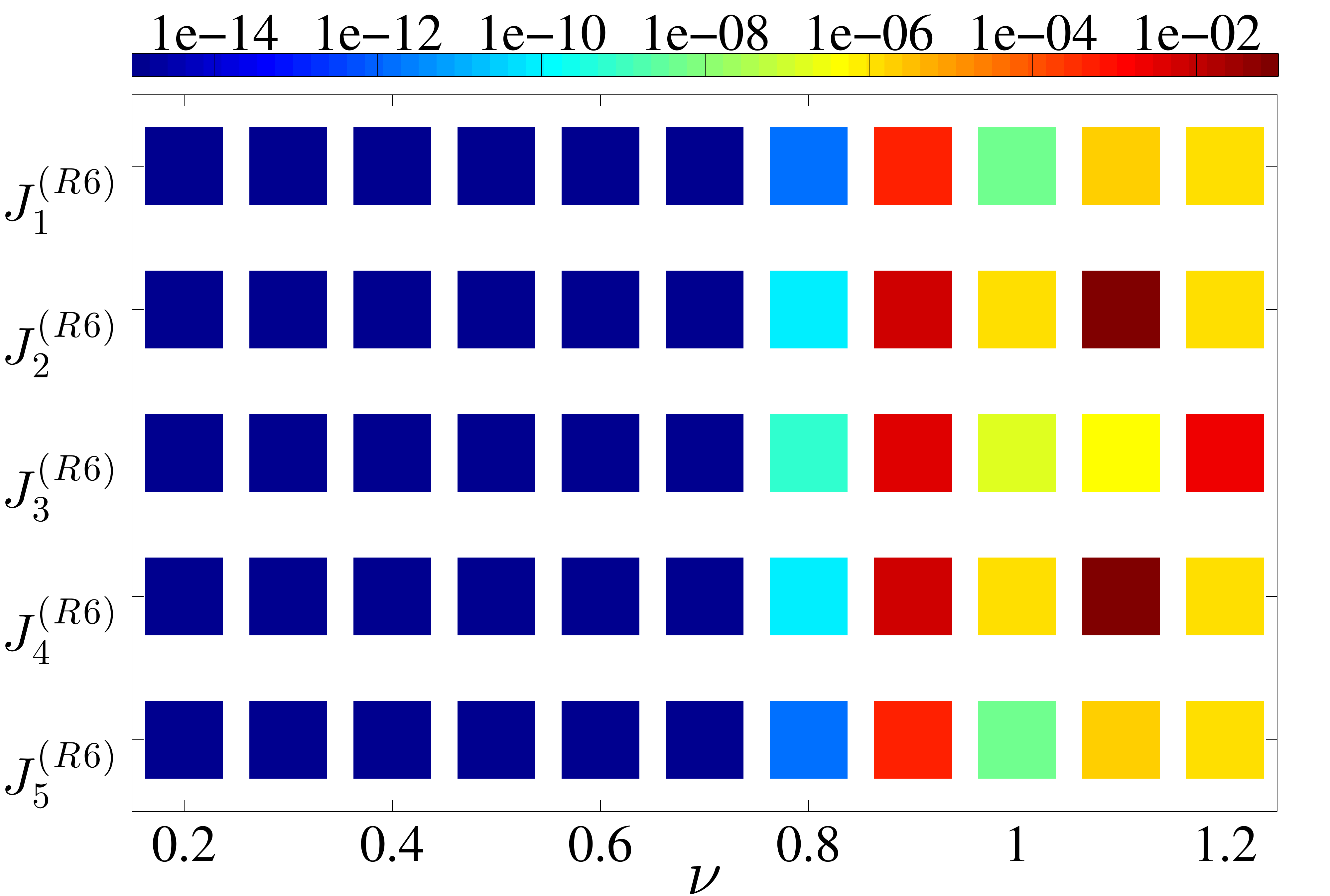}}
\caption{\label{fig:CouplingsN=10} $J_P$ for a Wigner crystal with 10 electrons, with $h_L = h_R = 0.1$:  a) nearest-neighbour, b) 2$^{nd}$ neighbour, c) 3$^{rd}$ neighbour d) 4$^{th}$ neighbour, e) 3-body f) 4-body, g) 5-body and h) 6-body. $J_P$ is in units of meV. Note that there is a lower limit on the colour bar, which means that very small couplings are rounded up to $10^{-16}$.}
\end{center}
\end{figure}

Results are shown for up to 4$^{th}$-nearest neighbour and 6-body exchange processes in Figs.~\ref{fig:CouplingsN=9} \& \ref{fig:CouplingsN=10}. The couplings are in meV and are labelled $J^{(1)}_{mn}$ for pairwise exchange between electrons $m$ and $n$, and $J^{(R3)}_n,J^{(R4)}_n,J^{(R5)}_n,J^{(R6)}_n$ to indicate 3-, 4- , 5- and 6-body ring exchanges starting from electron $n$.  The general features are low couplings at the boundary (due to the barrier between the dots and the chain) and a `U'-shaped coupling along the chain for low densities, becoming peaked at the middle for higher densities. We have excluded $\nu = 0.9$ and $\nu = 1$ for $N=9$, since for some couplings the algorithm did not converge to a minimum, which we attribute to the same instability seen for the quench in Fig.~\ref{fig:Quench9_2}.  

\begin{figure}[t]
\begin{center}
	\includegraphics[width=0.8\textwidth]{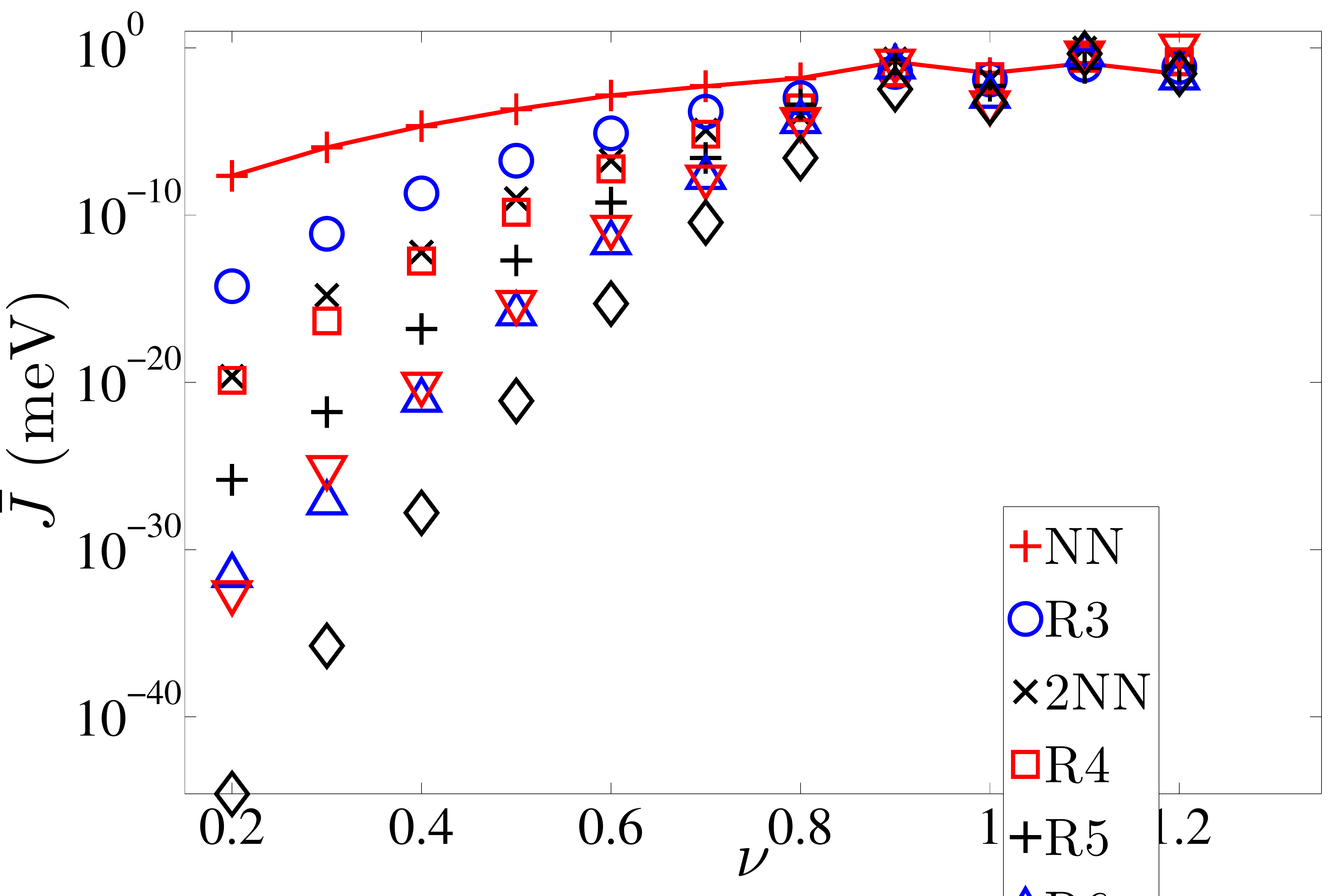}
	\caption{\label{fig:Javrg} $\bar{J}$ for a chain of 10 electrons, for the different exchange processes. The notation is such that 3NN indicates 3$^{rd}$-nearest neighbour etc.\ and R3 means 3-body ring exchange etc. The distribution of couplings for $N=9$ follows a similar trend.}
\end{center}
\end{figure}

Average couplings are shown in Fig.~\ref{fig:Javrg} for ten electrons, showing roughly which couplings dominate at different densities. Clearly the nearest-neighbour coupling dominates up to around $\nu = 0.8$, and the 3-body ring exchange is next highest, which is a qualitative agreement with other studies of Wigner crystals such as~\cite{Klironomos2007}. At around $\nu = 0.8$ the couplings all become of a similar magnitude, and there seems to be a plateau in coupling strength beyond this point. Although the average coupling shows some of the main behaviour, it is perhaps more informative to study the full coupling profile in Figs.~\ref{fig:CouplingsN=9} \& \ref{fig:CouplingsN=10}, from which it seems that including up to 2$^{nd}$ nearest neighbour and 5-body interactions captures the leading behaviour for $\nu \leq 0.8$. For $\nu > 0.8$, 6-body interactions and longer range pairwise interactions become significant. To be sure that our evolution is accurate, we therefore restrict ourselves to the regime $\nu \leq 0.8$ as it is not clear from our results whether higher order terms such as 7-body exchanges will also become important beyond this point. 

With this restriction, the Hamiltonian that we use to simulate the evolution is
\begin{align}\label{eqn:PermHam}
H =& \sum_{m \sim n } J_{mn}^{(2)}  \hat{P}_{mn}^\sigma  - \sum_{ \substack{ l,m,n \in \triangle \\ l < m < n} } J_{l}^{(R3)} \hat{P}_{lmn}^\sigma + \sum_{ \substack{ k,l,m,n \in \Box \\ k < l< n < m}}  J_{k}^{(R4)} \hat{P}_{klmn}^\sigma + \sum_{ \substack{ j,k,l,m,n \in \pentagon \\ j < k <n < l<m}} J_{j}^{(R5)} \hat{P}_{jklmn}^\sigma,
\end{align}
where e.g.\ $\hat{P}_{klmn}^{\sigma}$ means a cycle of length 4 which permutes the spins according to $(\sigma_k, \sigma_l,\sigma_m,\sigma_n) \to (\sigma_l, \sigma_m,\sigma_n,\sigma_k)$. We use the notation $m \sim n$ to mean $m$ and $n$ are up to 2$^{nd}$-nearest neighbours, $l,m,n \in \triangle$ means that $l,m,n$ are on the vertices of a triangle, $k,l,m,n \in \Box$ means that $k,l,m,n$ are on the vertices of a parallelogram and $j,k,l,m,n \in \pentagon$ means that $j,k,l,m,n$ are on the vertices of trapezium (see Fig.~\ref{fig:WigExch}). 

To convert this into a Hamiltonian, the spin permutation operators can be written in terms of pairwise spin exchange interactions, using the following identities~\cite{Klein1980}:
\begin{align}
&\hat{P}_{12}^\sigma = \frac{1}{2}(\idop + E_{12}), \\
&\hat{P}_{123}^\sigma + \hat{P}_{321}^\sigma = \hat{P}_{12}^\sigma + \hat{P}_{23}^\sigma + \hat{P}_{13}^\sigma - 1,  \\
&\hat{P}_{1234}^\sigma + \hat{P}_{4321}^\sigma = \hat{P}_{12}^\sigma \hat{P}_{34}^\sigma + \hat{P}_{14}^\sigma \hat{P}_{23}^\sigma - \hat{P}_{13}^\sigma \hat{P}_{24}^\sigma + \hat{P}_{13}^\sigma + \hat{P}_{24}^\sigma -1, \\
&\hat{P}_{12345}^\sigma + \hat{P}_{54321}^\sigma =  \frac{1}{2}(\hat{P}^\sigma_{12}\hat{P}^\sigma_{34} +  \hat{P}^\sigma_{14}\hat{P}^\sigma_{23} -  \hat{P}^\sigma_{13}\hat{P}^\sigma_{24})+ \frac{1}{2}(\hat{P}^\sigma_{15}\hat{P}^\sigma_{23} +  \hat{P}^\sigma_{12}\hat{P}^\sigma_{35} -  \hat{P}^\sigma_{13}\hat{P}^\sigma_{25} )\nonumber\\
&+\frac{1}{2}(\hat{P}^\sigma_{15}\hat{P}^\sigma_{24} +  \hat{P}^\sigma_{12}\hat{P}^\sigma_{45} -  \hat{P}^\sigma_{14}\hat{P}^\sigma_{25})+  \frac{1}{2}(\hat{P}^\sigma_{15}\hat{P}^\sigma_{34} +  \hat{P}^\sigma_{13}\hat{P}^\sigma_{45} -  \hat{P}^\sigma_{14}\hat{P}^\sigma_{35} ) +\frac{1}{2}(\hat{P}^\sigma_{25}\hat{P}^\sigma_{34} +  \hat{P}^\sigma_{24}\hat{P}^\sigma_{35} -  \hat{P}^\sigma_{25}\hat{P}^\sigma_{34}) \nonumber\\
&- \frac{1}{2}(\hat{P}^\sigma_{12}+\hat{P}^\sigma_{15}+\hat{P}^\sigma_{23}+\hat{P}^\sigma_{34}+\hat{P}^\sigma_{45}) +\frac{1}{2}(\hat{P}^\sigma_{13}+\hat{P}^\sigma_{35}+\hat{P}^\sigma_{25}+\hat{P}^\sigma_{24}+\hat{P}^\sigma_{14})- \frac{1}{2}.
\end{align}
where we have used $E_{mn} := X_m X_n + Y_m Y_n + Z_m Z_n$. 

Inserting these into eqn.\ (\ref{eqn:PermHam}), this gives the overall Hamiltonian, ignoring the factors of $\idop$ since they do not affect the dynamics:
\begin{align}
H &= \frac{1}{2} \sum_{m \sim n }J_{mn}^{(2)}  E_{mn} -  \frac{1}{2}\sum_{ \substack{ l,m,n \in \triangle \\ l < m < n} } J_{l}^{(R3)}  ( E_{lm} + E_{mn} + E_{ln})  + \frac{1}{4} \sum_{ \substack{ k,l,m,n \in \Box \\ k < l< n < m}}  J_{k}^{(R4)} \left[  E_{kl} + E_{km}  + E_{kn} + \right. \nonumber\\
&\left.  E_{lm} + E_{ln} + E_{mn} + \Upsilon_{klmn} \right]  -\frac{1}{8} \sum_{ \substack{ j,k,l,m,n \in \pentagon \\ j < k <n < l<m}} J_{j}^{(R5)} \left[ \Upsilon_{jklm} + \Upsilon_{jkln}  + \Upsilon_{jnkm} + \nonumber\right.\\
&\left. \Upsilon_{jnlm} + \Upsilon_{knlm} 
 + E_{jk}+E_{kl}+E_{lm}+E_{mn} +E_{jn}+E_{jl}+E_{km}+E_{ln}+E_{kn}+E_{jm} \right]
\end{align}
where $\Upsilon_{jklm} := E_{jk}E_{lm} +  E_{jm}E_{kl} -  E_{jl}E_{km}$. Note the minus sign in front of the $J^{(R3)}$ and $J^{(R5)}$ terms since they involve permutations of an odd number of particles, .

\begin{figure}[t]
\begin{center}
\subfloat[]{\includegraphics[width=0.6\textwidth]{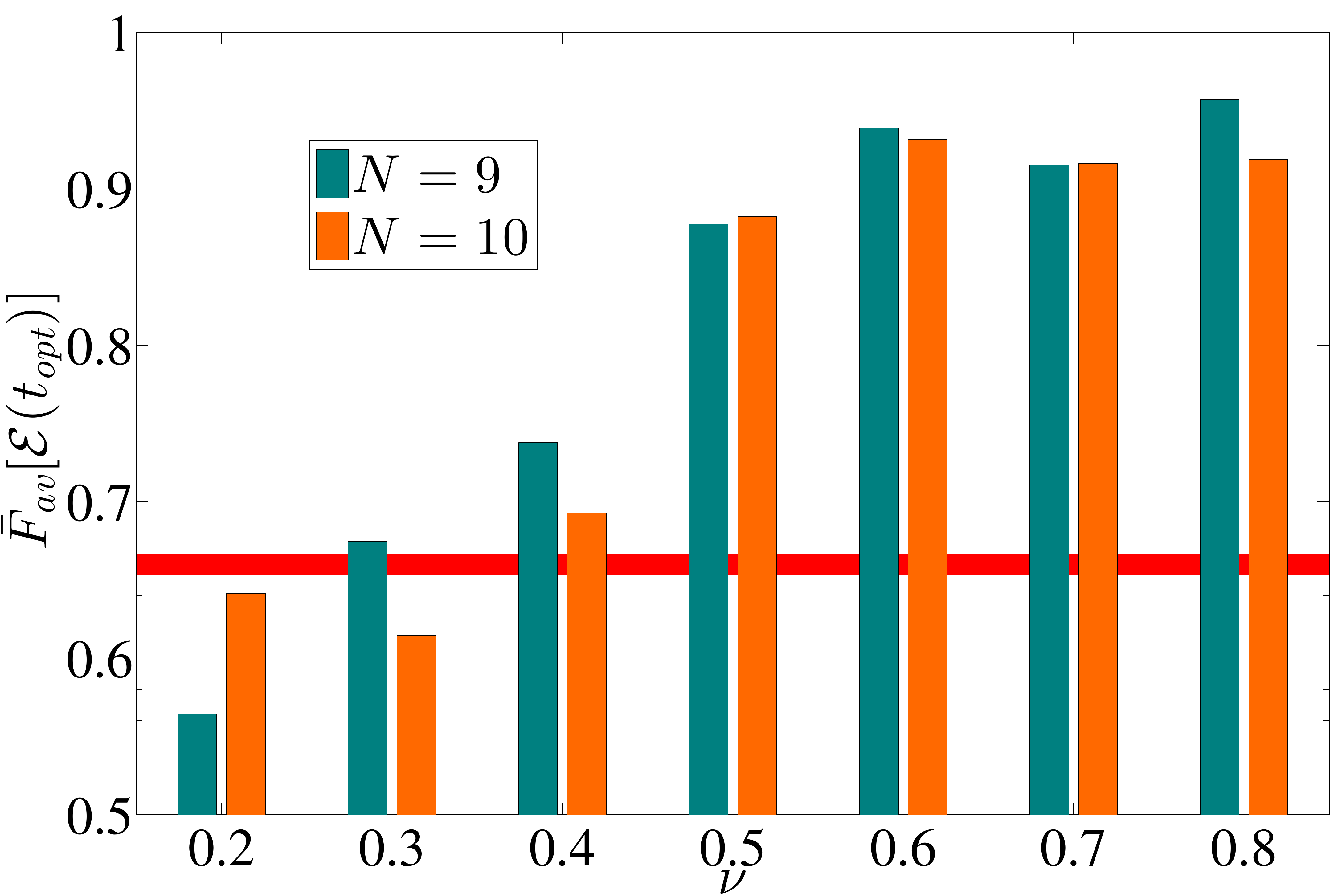}}\\
\subfloat[]{\includegraphics[width=0.6\textwidth]{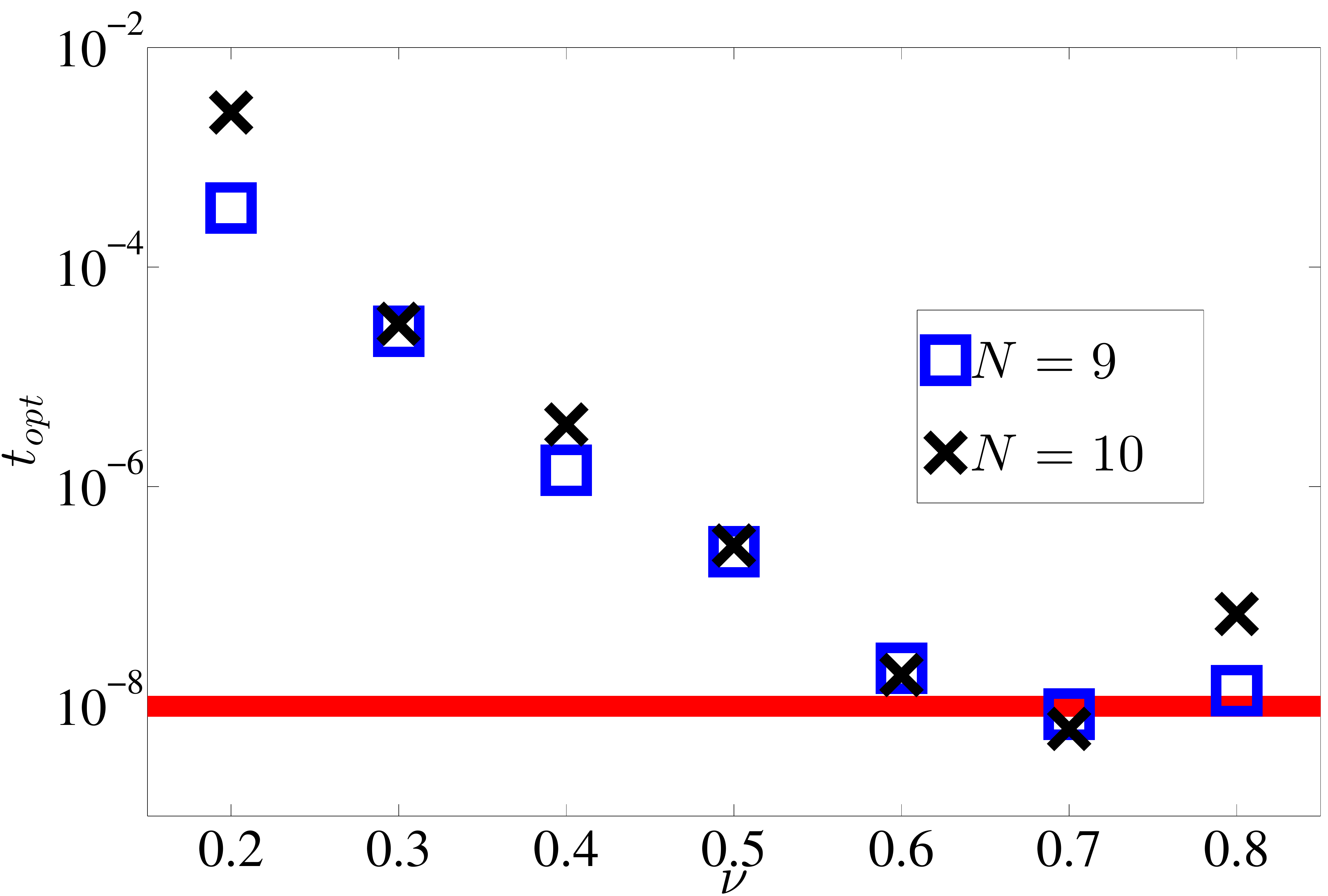}}
	\caption{\label{fig:Evo} a) Maximum value of the average fidelity $F_{av}[\mathcal{E}]$ with different densities for $N=9$ and $N=10$. The red line indicates the average fidelity achievable with a classical channel. b) $t_{opt}$, where $t_{opt}$ is the time at which maximum fidelity is reached, in seconds. The red line indicates the typical $T_2$ time in GaAs.}
\end{center}
\end{figure}

To measure the information transferring ability of this channel, the average fidelity was calculated using the formula in (\ref{eqn:FavrgHorod}) for qubits:
\begin{align}
F_{av}[\mathcal{E}(t)] = \frac{2 F_e[\mathcal{E}(t)] +1 }{3},
\end{align}
where $\mathcal{E}(t)[\rho] = \text{tr}_{\hat{N}} \left( e^{-iHt / \hbar} \rho e^{iHt / \hbar} \right)$. The average fidelity was calculated up to a timescale such that one peak had arrived at the other end, so does not necessarily represent the best achievable fidelity. The time at which the peak reaches the highest value of $F_{av}[\mathcal{E}(t)]$ is denoted $t_{opt}$. Results for the maximum achieved fidelity $F_{av}[\mathcal{E}(t_{opt})]$ are shown in Fig.~\ref{fig:Evo}, showing remarkably high fidelities generally increasing as $\nu$ increases. $t_{opt}$ ranges from around 0.1ns to 10$\mu$s; the decoherence timescale of an electron spin in GaAs is roughly 10ns at present~\cite{Petta2005} so this indicates that $\nu \simeq 0.6$ is the optimal density at which to create a Wigner spin chain in GaAs/AlGaAs. Clearly there is little difference in fidelity in using even or odd electrons in our protocol.

\subsection{Discussion and Conclusions}

We have calculated exchange couplings for a quasi-2D Wigner crystal using the semi-classical instanton approximation, and investigated the feasibility of using such a chain to transfer quantum information. Our results suggest that quantum information could be transmitted through a Wigner crystal with high fidelity (up to around 0.92 for chains of 9 or 10 electrons) using a quench, with transfer times that are below the typical decoherence times in GaAs, and using parameters typical of today's GaAs technology. This spin chain would be useful perhaps in conjunction with quantum dot qubits~\cite{Loss1998}, or could perhaps be extended to a full proposal of quantum computation, similar to the scheme in~\cite{Benjamin2003} with always-on interactions. Our scheme has many variables that can be altered without affecting the viability of the scheme, such as the barrier heights and the offset of the quantum dots. Magnetic fields are also known to alter the exchanges via the Aharonov-Bohm effect~\cite{Bernu2001,Zhu1995,Hirashima2001}. With some fine tuning of these parameters, we would expect even higher fidelities to be attainable. Using a quench may also not be the ideal protocol, and we would be interested to see whether other approaches such as adiabatically lowering the barriers is also viable.

Further work will need to be done to find a detailed experimental setup for realising this scheme, using either electrons or holes in GaAs/AlGaAs or other systems in which large $r_\Omega$ values can be achieved. The full effect of phonon excitations following the quench also needs to be taken into account to see how much error this leads to. There are also other effects that we have not taken into account, such as magnetic field fluctuations due to impurities, how the transfer is affected by a finite barrier dropping rather than an immediate quench, image charges produced by metal boundaries, dipole-dipole interactions and spin-orbit interactions that can produce anisotropic exchange terms in the Hamiltonian~\cite{Tserkovnyak2009}. It should also be noted that this scheme uses a semi-classical approximation, so is only accurate away from the Wigner crystal melting point ($r_\Omega \simeq 4 $ for 1D Wigner crystals). 

The best results seem to occur when there is a combination of nearest-neighbour, next-nearest neighbour and 3-body ring exchange present, which is perhaps because the 3-body exchange acts as an effective ferromagnetic interaction that works against the nearest- and next-nearest neighbour interactions and so makes the coupling strength more uniform throughout the chain. It would be interesting to investigate the roles of each of the exchange processes in more depth, to see if there are any processes which hinder or help information transfer, in order to search for the optimum regime and understand the underlying physical mechanisms. It would also be interesting to take this analysis to higher densities, including higher-order exchange processes to see if higher densities also allow high fidelity transfer.

\chapter{Information transfer with idealised spin chains}
\label{chap:NNN}

In the previous chapter, we explored a potential way to create a spin chain, through which information can be transferred. In this final chapter, we present three smaller studies, each exploring different protocols for sending information through an idealised spin chain, to see how these perform and how they can be optimised. It is hoped that each of these will provide a basis for more in-depth study in the future.

\section{Information transfer with next-neighbour interactions}

\label{sec:NNN}

In this section we investigate the effect of next-neighbour interactions on transfer through a uniform spin chain. The aim of this investigation is to test the robustness of existing nearest neighbour protocols to the addition of next-nearest neighbour (NNN) coupling, which may naturally be present, and to give new insights into what kind of systems are best for information transmission. The work follows a similar approach to the work by Bayat et.\ al.~\cite{Bayat10}, where the entanglement transferral across a uniform spin chain with Hamiltonian
\begin{eqnarray}
H=  \sum_{n=1}^{N-1} J (X_n X_{n+1}  +Y_n Y_{n+1}+ \Delta Z_n Z_{n+1}).
\end{eqnarray}
was investigated, for different values of $\Delta$. The protocol consists of attaching one half of a singlet to the end of  a spin chain in the ground state, and waiting for some time $t$ when the entanglement with the other end of the chain is maximised (see Section~\ref{sec:Prot} below). In this investigation they found that $\Delta = 1$ achieved the highest and fastest end-to-end entanglement

Following~\cite{Bayat10}, we investigate how next-nearest-neighbour (NNN) coupling in isotropic spin chains affects the quality of entanglement transferred using this protocol. The Hamiltonian we consider is the isotropic next-nearest-neighbour Hamiltonian (or `$J_1-J_2$' Hamiltonian):
\begin{eqnarray}
\label{eqn:HamNNN}
H_{\textsc{nnn}}^{(N)} = J_1\sum_{n=1}^{N-1} (X_n X_{n+1}  +Y_n Y_{n+1}+  Z_n Z_{n+1}) +J_2 \sum_{n=1}^{N-1} (X_n X_{n+2}  +Y_n Y_{n+2}+  Z_n Z_{n+2}),
\end{eqnarray}
where we have picked the isotropic model ($\Delta = 1$) since it already optimises the nearest-neighbour case, and the superscript $N$ indicates that this Hamiltonian acts on $N$ spins. Whilst the phase diagram of this $J_1-J_2$ model is qualitatively well known, it is still not exactly solvable except at certain points (see e.g.~\cite{SchollwockBook}). With $J_1,J_2 > 0$ the model is in the antiferromagnetic frustrated regime. Below a critical value of $J_2 < J_{2c} \approx 0.241J_1$, there is no energy gap between the ground state and excited states, and a gap opens up for $J_2 > J_{2c}$. At $J_2 = 0.5J_1$, the \emph{Majumdar-Ghosh} point, the ground state is analytically solvable~\cite{Majumdar69}. Here the ground state is completely dimerised, and is known as a \emph{dimer product state}:
\begin{eqnarray}
 |\phi_{\textsc{mg}} \rangle= |\psi^- \rangle_{1,2}  |\psi^- \rangle_{3,4}  ...  |\psi^- \rangle_{N-1,N},\: \mathrm{where} \quad |\psi^{-} \rangle =  \frac{1}{\sqrt{2}}(|\uparrow \downarrow \rangle - |\downarrow \uparrow \rangle ).
\end{eqnarray}
Otherwise, as $J_2$ grows and becomes larger than $J_1$ we effectively end up with two decoupled chains, so in this study we mainly focus on the region $0 < J_2 < 0.6 J_1$. In what follows we will set $J_1 = 1$ unless otherwise stated.

In the case of perfect state transfer, some investigations have already been done into the effects of different types of errors~\cite{Ronke11}, including small amounts of NNN interactions. Other work has also found that NNN coupling is detrimental to state transfer, for a protocol that transfers information by weakly coupling qubits to the two ends of a spin chain in the Kondo regime~\cite{Sodano10}. The differences between these studies and the work in this chapter is that we consider a spin chain with uniform interactions, and for larger NNN interactions.

\subsection{Protocol}
\label{sec:Prot}

We consider Hamiltonians of the form shown in (\ref{eqn:HamNNN}), where $J_1<0$ for the ferromagnetic case and $J_1>0$ for the antiferromagnetic case. We consider open chains (i.e.\ open boundary conditions) with uniform couplings, to minimise engineering requirements and to keep the number of variable parameters low.

Firstly we diagonalise $H_{\textsc{nnn}}^{(N)}$ on $N$ spins to find the ground state $\ket{\psi_G}$ (if there is degeneracy in the ground state a small magnetic field term $h\sum_n \sigma_n^z$ can be applied). We then couple a singlet state $ \ket{ \psi^{-} }_{0'0}$ of two qubits $0'$ and $0$ to one of the ends, so that the initial state is:
\begin{equation}
\ket{\psi(t=0)}= \ket{ \psi^{-} }_{0'0} \ket{\psi_G }.
\end{equation}
to do this, we imagine  a system in which there are barriers separating spins 0,0' and 1, so that the coupling can be selectively turned on or off. Spin $0'$ is completely uncoupled from the chain at all times, whilst spin $0$ interacts with the chain with the same couplings as in the chain (we have the choice of selecting weaker or stronger couplings to connect this singlet to the chain, however since faster transfer is likely to arise from stronger couplings, we choose the chain couplings to be the highest possible). Thus the Hamiltonian for the chain$+$singlet system is $H_{tot} = I_{0'} \otimes H_{\textsc{nnn}}^{(N+1)}$. From this we construct the evolution operator $U=e^{-iH_{tot}t}$, and use this to find the state after a time t, $|\psi(t) \rangle  =U(t) |\psi(t=0) \rangle$.
Since we have started with the singlet, which has two qubits maximally entangled, we might expect that after allowing the system to evolve there would be some entanglement between the $0'^{th}$ spin and the $N^{th}$ spin. To quantify this, first of all we trace out all of the qubits in the system except for the $0'^{th}$ and $N^{th}$ spins, leaving a density matrix:
\begin{equation}
\rho_{0'N}(t)= tr_{\widehat{0'N}} [|\psi(t) \rangle \langle \psi(t) | ],
\end{equation}
which we can then use to find the entanglement of formation $E_f$ between qubits $0'$ and $N$, where the entanglement of formation is defined in Section~\ref{sec:IntroSpinChain}. We then look for the optimal time where there is maximal entanglement between the 0' and $N^{th}$ spin, and find how this maximum entanglement varies with NNN coupling. 

To practically calculate the time-evolved state and the entanglement, we used exact diagonalisation in MATLAB, using a stepwise time evolution, where at each time step the evolution operator was applied to the state and then the entanglement at each time step could be calculated. Ideally we want to minimise the transmission time, as in a realistic system this would decrease the effects of decoherence. With this in mind, our results were limited to the first peak, which usually arrived on a timescale less than $N/2$, and we define $\max_1(E_f)$ as the entanglement of formation at the highest point of the first peak.

Exact diagonalisation works well for $N \leq 16$, but for more spins the computation time is very prohibitive, and so we use the technique of Density Matrix Renormalisation Group (DMRG) for larger systems (using the MATLAB code of Abolfazl Bayat, modified to include next-nearest neighbour coupling. See Appendix~\ref{sec:DMRG} for an overview of DMRG). Note that although $\sum_n Z_n$ commutes with Hamiltonians with the form that we use, so that evolution between states with different eigenvalues of $\sum_n Z_n$ is not allowed, this is not always incredibly useful here since the system is always initialised in the ground state, and so the major bottleneck is diagonalising the large matrices. 

\subsection{Results}
\label{sec:J1J2Results}

\begin{figure}
\begin{center}
\subfloat[]{\includegraphics[width=0.6\textwidth]{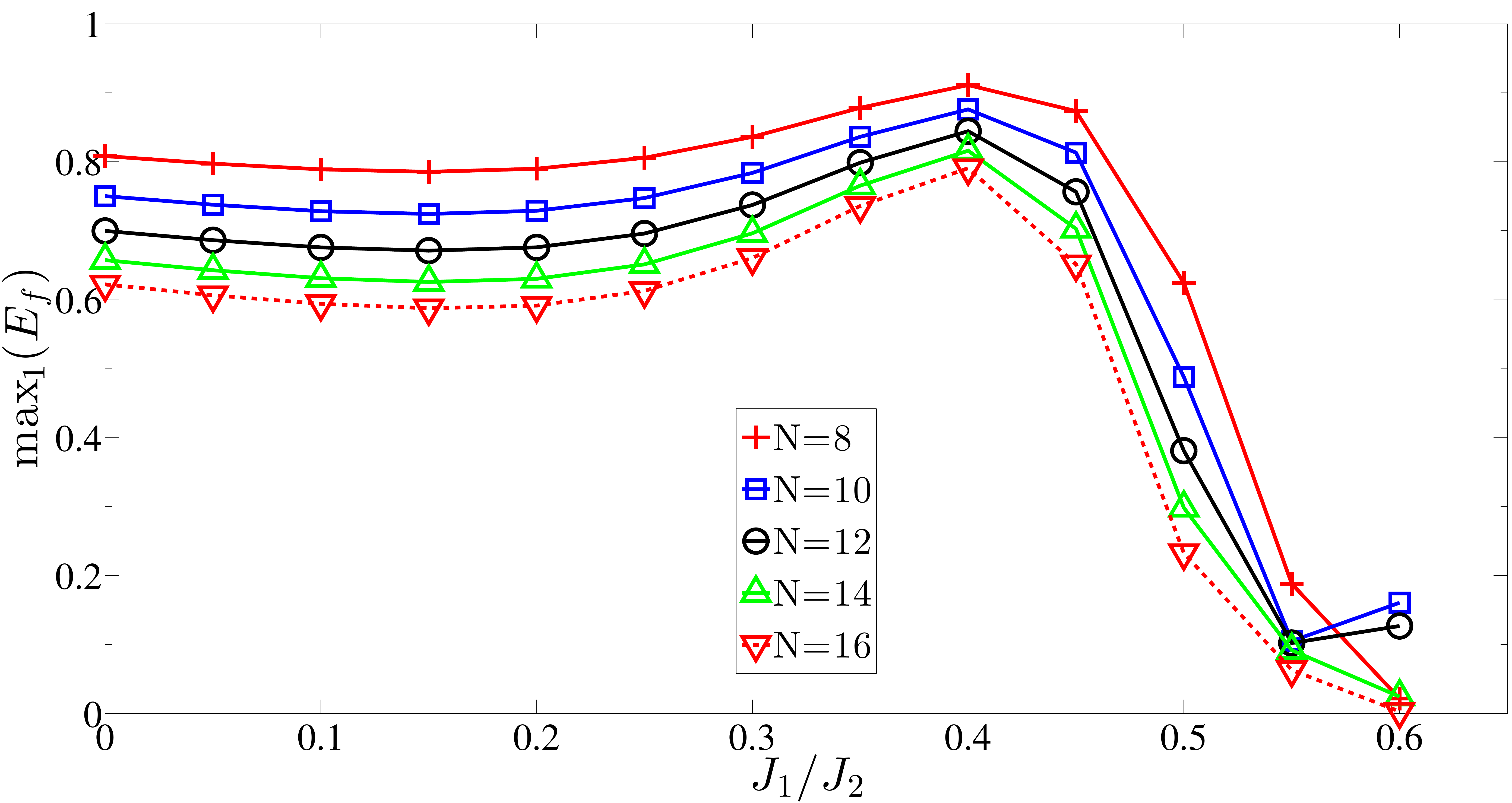} \label{fig:EvenNNN}} \\
\subfloat[]{\includegraphics[width=0.6\textwidth]{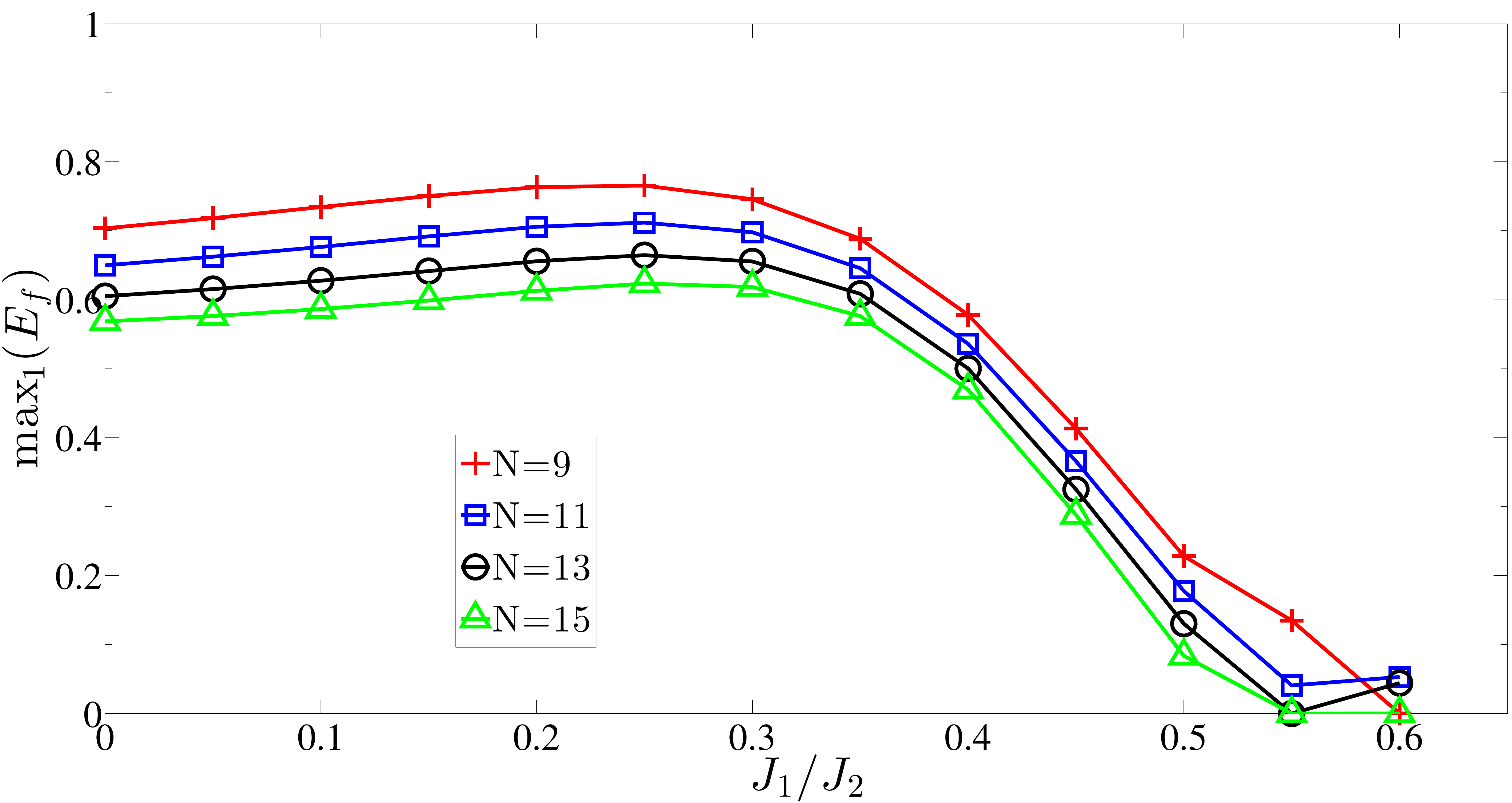} \label{fig:OddNNN}}
\end{center}
\caption{a) Maximum end-to-end entanglement $\max_1(E_f)$ vs. $J_2 / J_1$ for chains of even length $N=8$ to $N=16$ with a singlet added on. b) Maximum end-to-end entanglement $\max_1(E_f)$ vs. $J_2 / J_1$ for chains of odd length $N=9$ to $N=15$ with a singlet added on.}
\end{figure}

The results for maximum entanglement versus $J_2$ are shown in Figs.~\ref{fig:EvenNNN} \&~\ref{fig:OddNNN} for $N$ = 8 to 16. For even values of $N$, there is a slight decrease in entanglement up to $J_{2c}$,  followed by a significant rise until $J_2 \approx 0.4$, and then a sharp decline near the Majumdar-Ghosh point. For odd values of $N$, a rise is seen up to $J_{2c}$, and the peak around $J_2$ =0.4 is absent. 

\begin{figure}
\begin{center}
\subfloat[]{\includegraphics[trim=1.5cm 0cm 2.8cm 0cm, clip=true, width=0.5\textwidth]{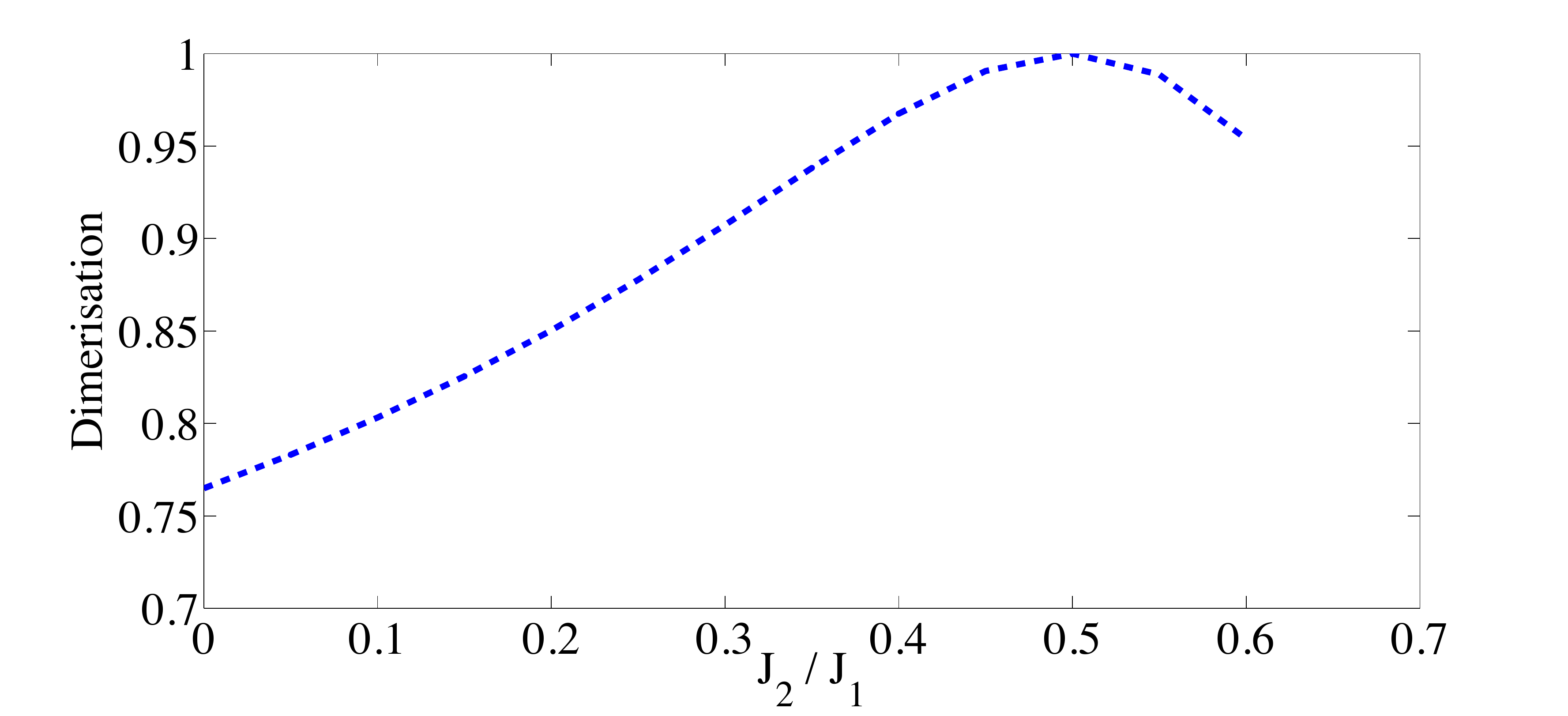} \label{fig:Dimers}}
\subfloat[]{\includegraphics[trim=1.5cm 0cm 2.8cm 0cm, clip=true, width=0.5\textwidth]{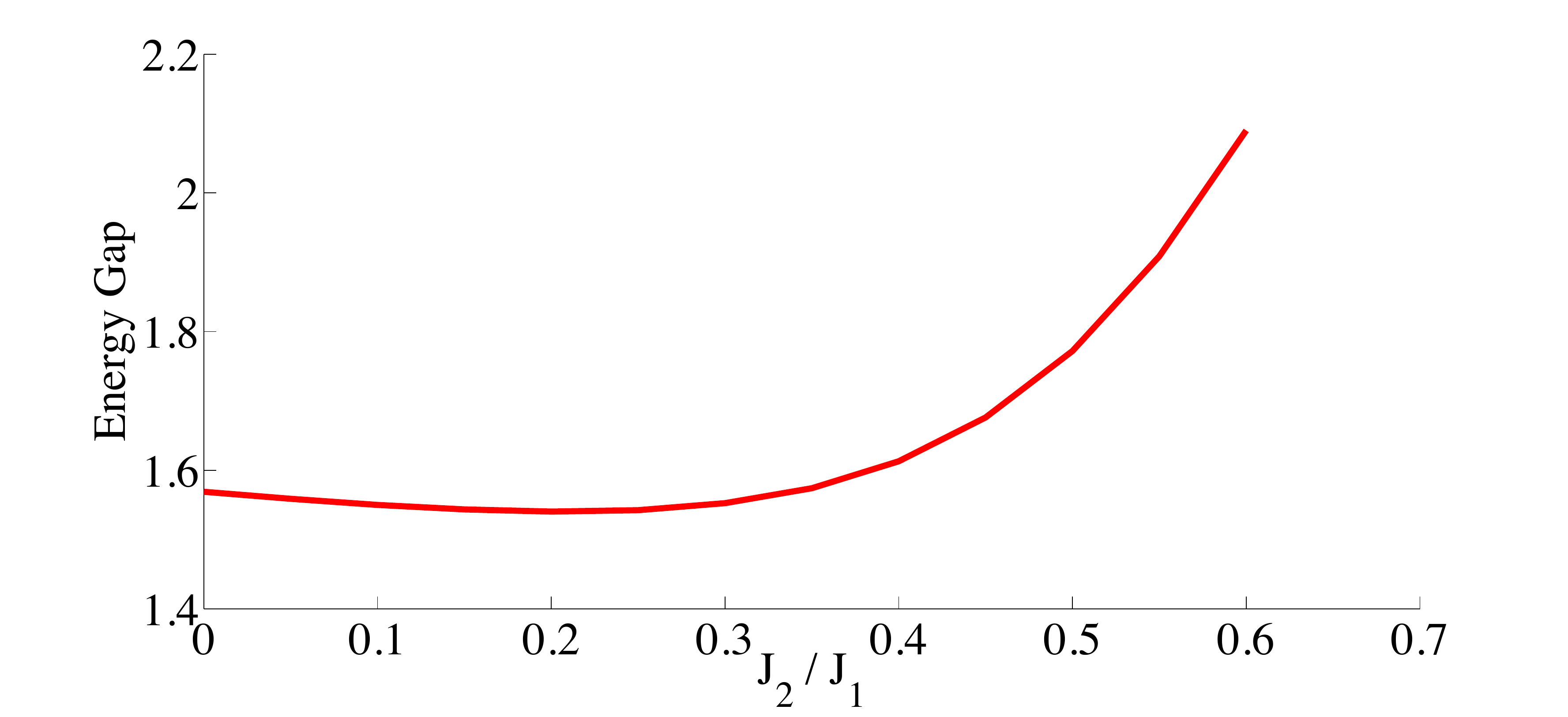}\label{fig:EgapJ2}}
\end{center}
\caption{a)Dimerisation, defined as the overlap between the chain ground state (before addition of a singlet) and the fully dimerised state (the Majumdar-Ghosh ground state). b) Energy gap as $J_2$ varies -  note that our system is not perfectly gapless as we are looking at finite chains, but we expect the chains to converge to a gapless system as the size increases.}
\end{figure}

We might expect that this gain in entanglement is to do with oscillations between degenerate states which have a single unpaired spin, i.e.\ we might expect there to be oscillations between degenerate states of the form:
\begin{eqnarray}
|\psi_1 \rangle=|\phi \rangle_1  |\psi_{chain} \rangle, \;|\psi_2 \rangle=|\psi_{chain} \rangle  |\phi \rangle_{N},
\end{eqnarray}
where $| \phi \rangle$ is an unpaired single qubit state. We refer to this as a `resonant' mechanism for transferring entanglement. This would only be possible for even $N$ values, for which the actual evolving part of the chain is odd. This is very similar to a result found in \cite{Gualdi2008}, where they found that perfect state transfer is achieved when there were two eigenstates of the Hamiltonian that had maximal overlap with the two ends of the chain but zero overlap with other sites in the chain (and is quite an intuitive result). We can try to look at the plausibility of this explanation in quite a rough way by looking at how the dimerisation and the energy gap vary with $J_2$, where the dimerisation is defined as the fidelity between the ground state of the chain and the Majumdar-Ghosh ground state $\ket{\psi^-}^{\otimes \frac{N}{2}}$. If this was the dominant effect we would expect high dimerisation and a high energy gap to give us maximal entanglement. Results are shown in Figs.~\ref{fig:Dimers} $\&$~\ref{fig:EgapJ2}. Using this model we would expect to see a peak in entanglement transfer at around $J_2=0.5$, which is not the case in our simulations. 

The difference between odd and even chains could perhaps be explained more or less using this idea; when $N$ is even, then after we have coupled the $0^{th}$ site to it our starting state has only one unpaired spin. Then as we go towards the Majumdar-Ghosh point where the ground state is dimerised, we would expect that the states with unpaired spins at either end might become more and more isolated, and so might oscillate only between themselves. Then for odd values of $N$, there would be more than two unpaired sites to begin with (since we add our $0^{th}$ spin to an odd chain) and so there may be more degeneracy and more states in which the evolution can disperse.

It is also in some ways analogous to the work done in \cite{Venuti06}. Here they looked at end-to-end entanglement of a chain with an explicitly dimerised Hamiltonian of the form
\begin{equation}
H = \sum\nolimits_{n=1}^{N} [1+(-1)^n \delta] (X_n X_{n+1}  +Y_n Y_{n+1}+  Z_n Z_{n+1}),
\end{equation}
which creates alternating strong and weak bonds along the chain, and has an energy gap $\Delta E \propto \delta^{2/3}$. They found that the end-to-end entanglement of this chain increased past a critical value of $\delta$. Then the behaviour we see may be of a similar type. Unfortunately due to the alternating weak and strong bonds in this Hamiltonian it is hard to create the same scenario as we have here, i.e.\ the two states:
\begin{eqnarray} \label{eq:OscStates}
 |\psi_1 \rangle=|\phi \rangle_1 \ |\psi^{-} \rangle |\psi^{-} \rangle ... |\psi^{-} \rangle, \; |\psi_2 \rangle=|\psi^{-} \rangle |\psi^{-} \rangle |\psi^{-} \rangle ...  |\phi \rangle_{N}
\end{eqnarray}
are not degenerate, because by changing the positions of the singlets we have put them over bonds of different strength, and so we have different energies. In addition this work is based around having weak coupling at either end (the idea being that stronger coupling would mean stronger entanglement to sites within the chain, and therefore less entanglement with the site we actually want to be entangled to, due to the monogamy of entanglement).

\begin{figure}
\subfloat[]{\includegraphics[trim=1.5cm 0cm 2.8cm 0cm, clip=true, width=0.475\textwidth]{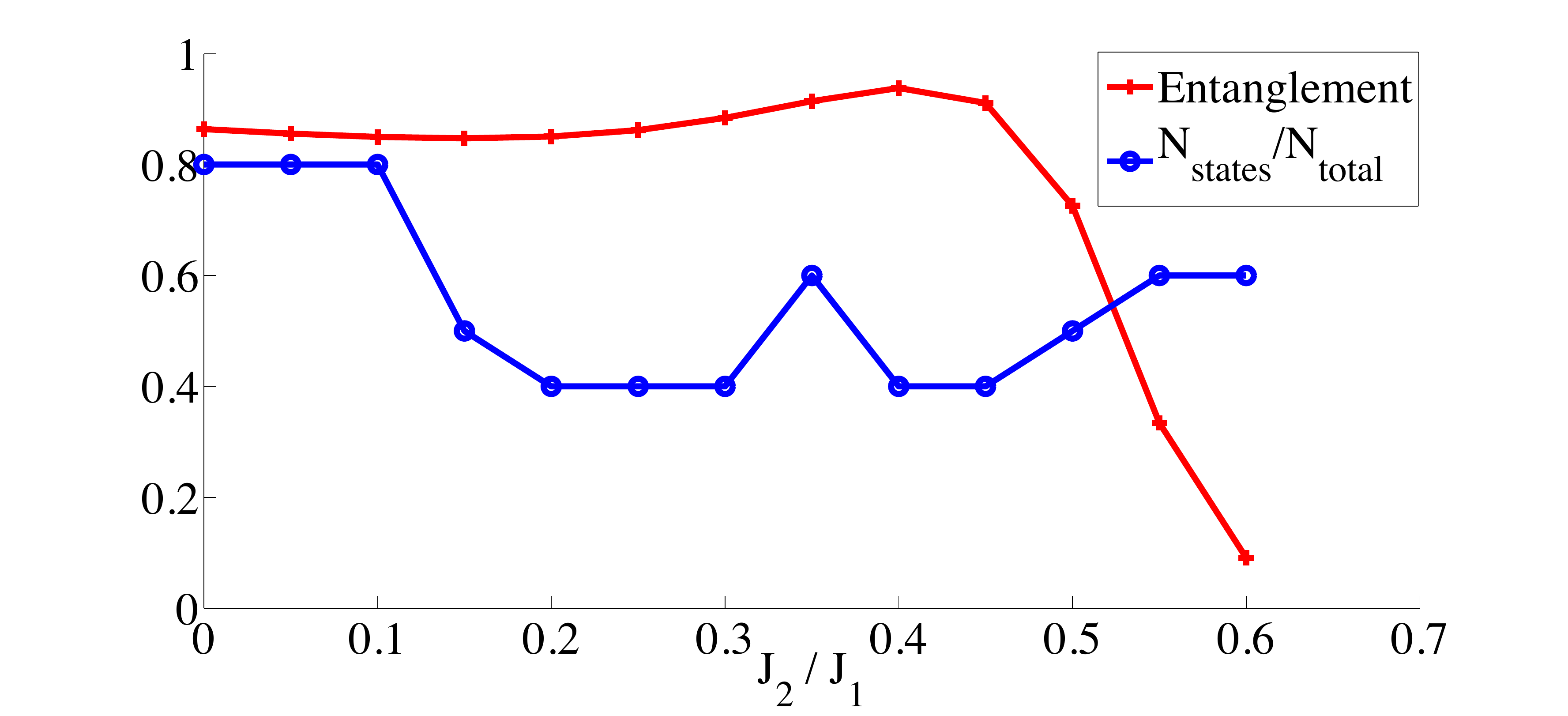} \label{fig:DoS8}}
\qquad
\subfloat[]{\includegraphics[trim=1.5cm 0cm 2.8cm 0cm, clip=true, width=0.475\textwidth]{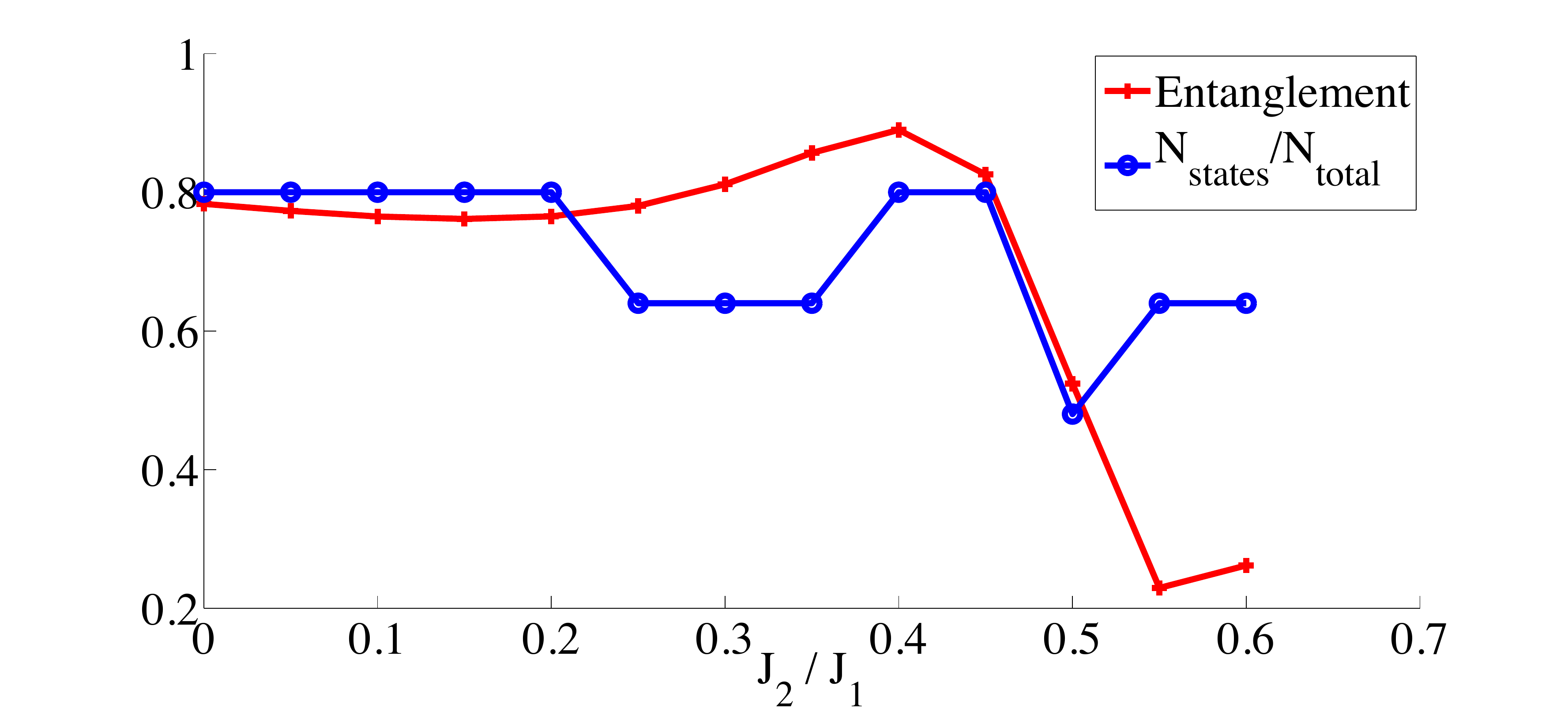}\label{fig:DoS12}}
\caption{Maximum end-to-end entanglement $\max_1(E_f)$, plotted alongside $N_{states}$ (the number of states within $\pm \Delta E $, rescaled to fit on this graph). a) $N = 8$, b) $N = 12$.}
\end{figure}

Similar ideas to the explanation presented above are looked at by Hartmann et.\ al in \cite{Hartmann06}. Here they argue that the more eigenstates that are involved in the evolution, the more dispersive the information transmission.  If we start with a state with energy expectation $\langle E \rangle$, then there is a variance in the energy of the state given by
\begin{equation}
\Delta E = \sqrt { \langle H^2 \rangle  - \langle H \rangle}.
\end{equation}
So we can try to quantify the number of eigenstates involved in the evolution by counting the number of eigenstates, $N_S$, within the energy range $\langle E \rangle - \Delta E \leq E_S \leq \langle E \rangle + \Delta E$. This energy range will be constant during the state's evolution, and we can see if this explains the behaviour seen above. The results are shown in Fig.~\ref{fig:DoS8} and~\ref{fig:DoS12} for $N = 8$ and  $N = 12$, respectively. Using this measure, there does appear to be some correlation, although not enough to justify using it as an explanation, and it appears to get worse as $N$ is increased. 

Since the effects we see do not seem to match up with the model of resonance being the only effect here, we hypothesise that there is information propagation through the chain as well as these resonating effects outlined above, and optimal entanglement transfer is for values of $J_2$ which allow some propagation and some resonance, in such a way that the resonance effects constructively add to the propagation through the chain. Then past this optimal point, propagation through the chain becomes more difficult (perhaps due in part to the argument above involving number of accessible eigenstates), and the resonating effects then become the only method of transferring entanglement. Currently this is as far as we have progressed in finding an explanation, which is not an entirely satisfactory one. To investigate this further we plan to look in more depth at the eigenstates involved in the evolution, to look for evidence that the states involved are really facilitating end-to-end resonance (which would most likely manifest itself as states with large overlap at the ends but not much in the middle, in a similar manner to \cite{Gualdi2008}).

\begin{figure}[h]
\begin{center}
\includegraphics[width=0.7\textwidth]{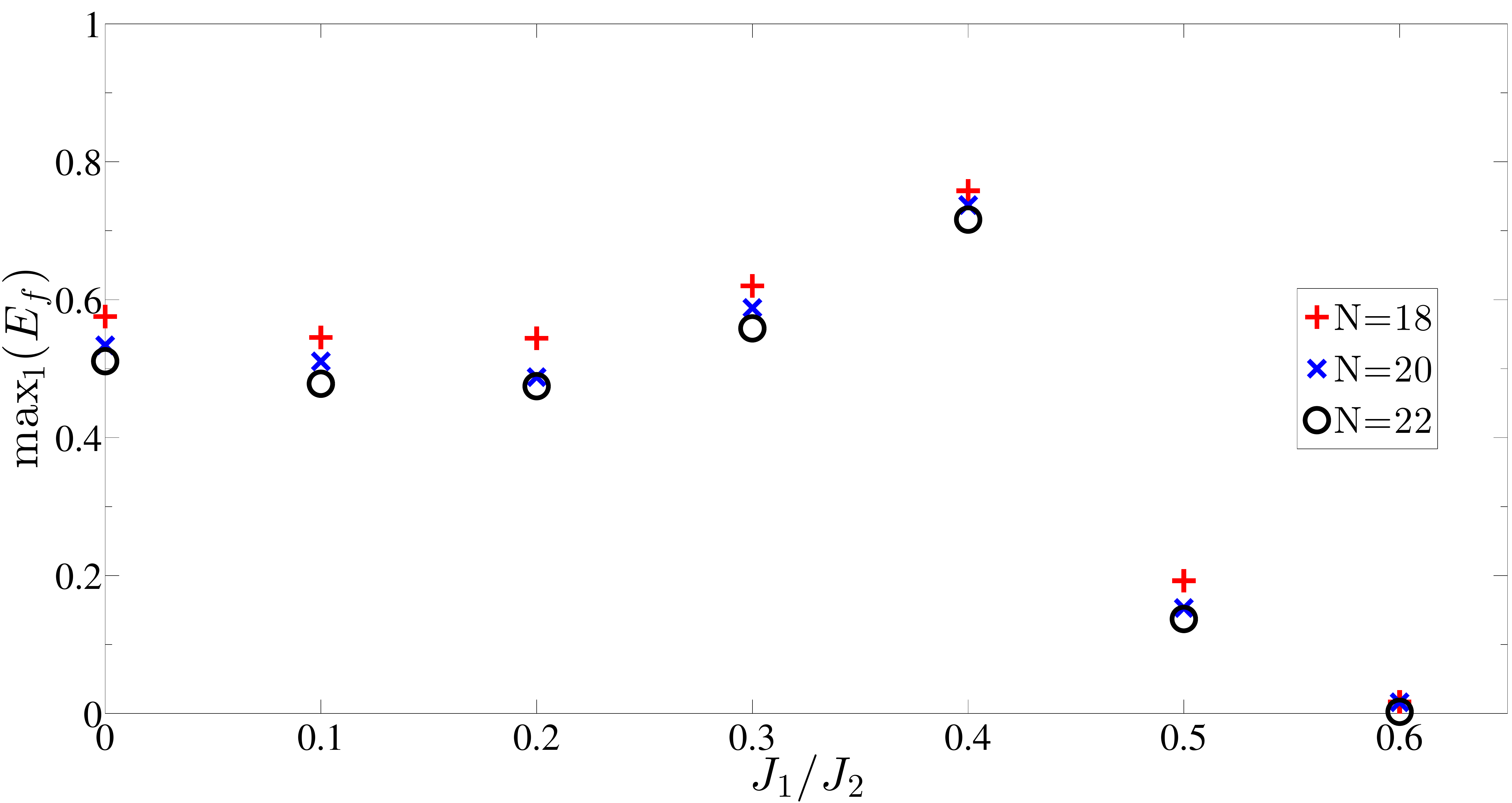}
\caption{Maximum end-to-end entanglement $\max_1(E_f)$, for different initial states, for $N = 18$, 20 and 22, found using DMRG.}
\label{fig:DMRG}
\end{center}
\end{figure}

\begin{figure}[h]
\begin{center}
\includegraphics[width=0.9\textwidth]{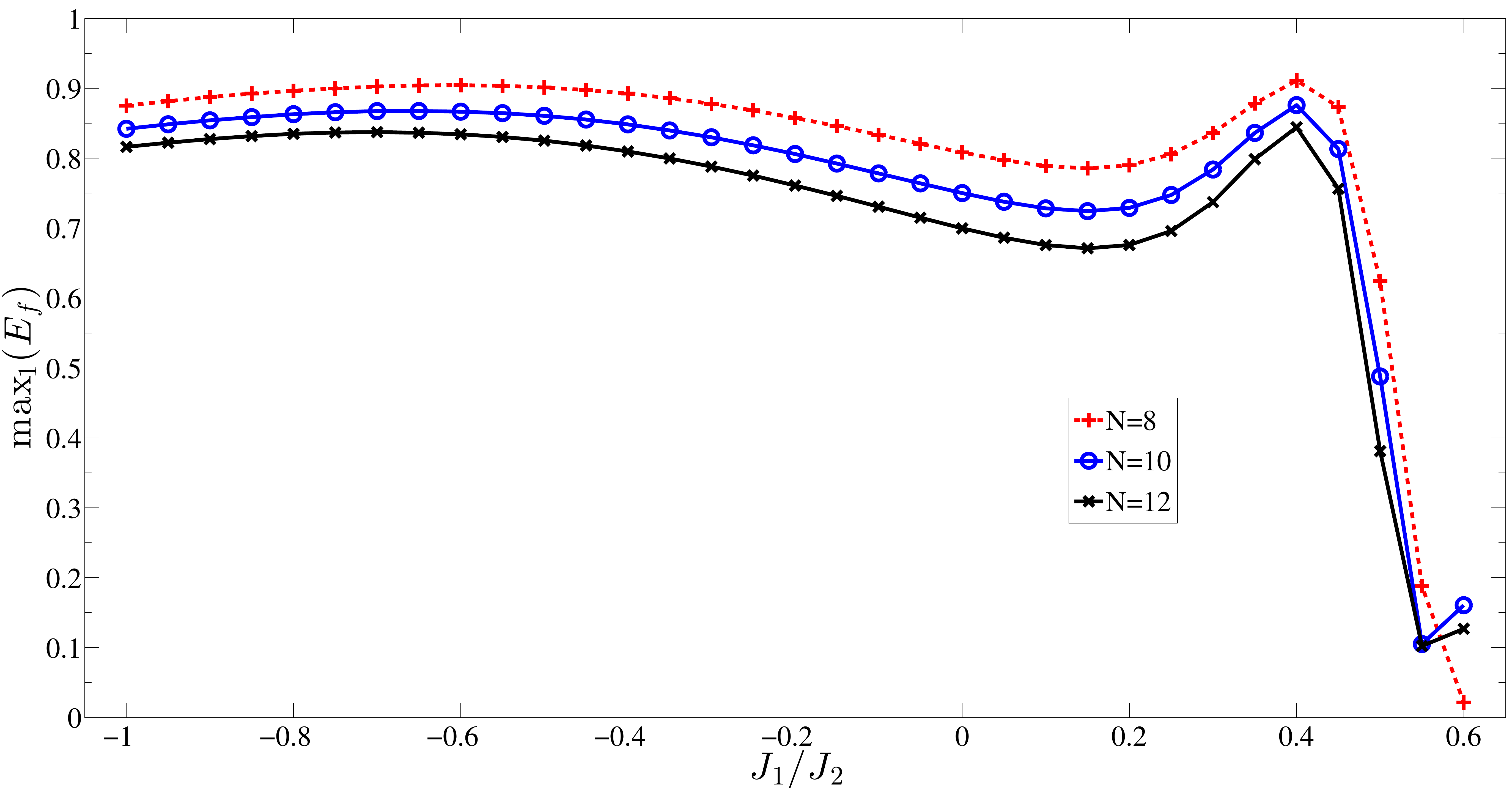}
\caption{Maximum end-to-end entanglement $\max_1(E_f)$ for positive and negative $J_2$, for chains of length $N = 8,10,12$}
\label{All}
\end{center}
\end{figure}

So far, all of the results have been found using exact diagonalisation of the Hamiltonian, however we can extend to larger $N$ by using the Density Matrix Renormalisation Group (DMRG) method (see Appendix~\ref{sec:DMRG}). Using this method, we extended the exact diagonalisation results up to values of $N=$ 18 and 24. From this we still see the characteristic rise after the critical value $J_{2c}$. Past chain lengths of around $N \sim 26$ the results appeared to diverge (i.e.\ there were sudden dramatic changes in the behaviour around these lengths, particularly for larger $J_2$ values). We attributed this to inaccuracies in the DMRG rather than any change in physical behaviour, since larger chains increase the time we have to wait to receive a signal at the other end, and it is likely that the simulation takes longer than the timescale over which the DMRG is accurate. This would also explain why the results start to diverge for larger values of $J_2$, since larger values of $J_2$ increase the time taken for the signal to reach the end.

\subsection{Extending to negative $J_2$}
\label{sec:FMtransfer}

When we extend this study to negative values of $J_2$ (whilst keeping $J_1=1$) we also see an increase (Fig.~\ref{All}). In this region increasing $J_2$ makes the chain less frustrated. In general then, there seem to be no large drops in entanglement transfer for a range of $-1<J_2<0.4$. This is reassuring, and suggests that generating end-to-end entanglement via this method is quite robust to the introduction of next nearest neighbour coupling.

\section{Edge locking}
\label{sec:EdgeLock}
We now investigate a protocol, based on the work in~\cite{Haque2010}, to transfer information, using certain states that are effectively stationary and so are said to be `locked'. We consider a nearest-neighbour XXZ Hamiltonian:
\begin{align}
H_{\textsc{xxz}} = \sum_n X_n X_{n+1} +Y_n Y_{n+1} + \Delta Z_n Z_{n+1}.
\end{align}
Provided the anisotropy $\Delta$ is large enough, this Hamiltonian allows two types of locked states; the first type is where the leftmost $n$ spins are up whilst the remaining spins are down, such as the following states
\begin{align}
\ket{\uparrow\downarrow \downarrow \downarrow ... \downarrow },\; \ket{\uparrow \uparrow \downarrow \downarrow ... \downarrow } ,\;\ket{\uparrow \uparrow \uparrow \downarrow... \downarrow  }.
\end{align}
These are locked since any movement of the spins introduces an extra $\downarrow \uparrow$ boundary (a `domain wall') which increases the energy (provided the $ZZ$ coupling is significantly larger than the $XY$ coupling). The second type of locked states are slightly less intuitive, having blocks of $\uparrow$ surrounded by $\downarrow$, which are sensitive to the number of spin-ups and where they are positioned. For example, the following states are edge-locked
\begin{align}
\ket{\downarrow \uparrow \uparrow \uparrow \downarrow... \downarrow  } ,\;  \ket{\downarrow \uparrow \uparrow \uparrow \uparrow\downarrow... \downarrow  } 
\end{align}
and in general, a state will be locked if it contains a block of $N_\uparrow$ spins, with the leftmost spin at position $k$ where $N_\uparrow \geq (2k-1)$ (and similarly for the right hand side of the chain). Following the notation in~\cite{Haque2010} we will label these states $\ket{L_{N_\uparrow,(k)}}$ ($\ket{R_{N_\uparrow,(k)}}$) for a block of $N_\uparrow$ spins with the leftmost (rightmost) spin at position $k$. The explanation in~\cite{Haque2010} for why these states are locked is that, provided $\Delta^{-1}$ is small but non-zero, performing degenerate perturbation theory on the Hamiltonian we find that terms in the Hamiltonian that connect states $\ket{L_{N_\uparrow,(k)}}$ to other states with $k'>k$ are of order $O(\Delta^{-N_\uparrow})$. Additionally one finds that the energy of the $\ket{L_{N_\uparrow,(k)}}$ states is modified by $O(\Delta^{-2(k-1)})$ relative to the other states with $k'>k$. Thus if $\Delta^{-N_\uparrow} > \Delta^{-2(k-1)}$ evolution of $\ket{L_{N_\uparrow,(k)}}$ to other states with $k' > k$ is dominant, and conversely if $\Delta^{-N_\uparrow} < \Delta^{-2(k-1)}$ the evolution is not allowed due to conservation of energy. So the states are approximately locked when $N_\uparrow \geq 2(k-1)$.

\subsection{Applications of edge locking}
This mechanism appears to be very useful as means to control quantum information in spin chains. Here we outline several possible applications, before exploring how robust the transfer is with more realistic conditions.

\subsubsection{Signal amplification}

Consider preparing an edge-locked state, e.g.\ $\ket{L_{3,(1)}}$, and attaching a spin $\ket{\phi}$ to the left edge:
\begin{equation}
\ket{\phi}\ket{\uparrow \uparrow  \uparrow\downarrow ... \downarrow  \downarrow}
\end{equation}
If $\phi = \uparrow$, then the chain is still locked. However if $\phi = \downarrow$, the chain becomes $\ket{L_{3,(2)}}$ and so is no longer locked, and the state is amplified by a ratio of roughly 2:1. Attaching a spin to the end like this is limited to ratios of 2:1, since if we tried this with larger number the resultant state would be edge-locked for both inputs. This can be extended to larger ratios if we can have spins to the left of our input spins, e.g.\ for a ratio of 6:1 we can start with:
\begin{equation}
\ket{\downarrow \downarrow} \ket{\phi} \ket{\uparrow \uparrow \uparrow \uparrow \uparrow \uparrow  \downarrow ... \downarrow  \downarrow},
\end{equation}
Such a protocol could also perhaps be used as a method to detect a spin current. Note that the higher the ratio of input to output, the smaller the energy gap, and the more sensitive the scheme will be to fluctuations in energy (e.g.\ there may be additional energy added to the system in the process of coupling the state $\ket{\phi}$ to the end of the chain).

\subsubsection{\textsc{nand} gates}
A classical \textsc{nand} gate ($\textsc{nand}(a,b) = 1 \oplus ab$) can be constructed using a chain with an edge-locked state at each end, e.g.:
\begin{equation}
\ket{\phi_1}  \ket{\uparrow \downarrow  \downarrow ...  \downarrow \downarrow  \uparrow} \ket{\phi},
\end{equation}
where $\phi_1$ and $\phi_2$ are our input states. Then for the cases where $\ket{\phi_1\phi_2} = \ket{\downarrow \downarrow},\ket{\uparrow \downarrow}$ and $ \ket{\downarrow \uparrow}$, we will measure a signal travelling through the middle of the chain. If we can measure the presence or absence of this signal, and choose $\ket{0} = \ket{\downarrow},\ket{1} = \ket{\uparrow}$, this performs a \textsc{nand} gate.

\subsubsection{Heralded entanglement}
With the same setup as for the \textsc{nand} gate, we can create heralded entanglement: Firstly, we would expect that the signal travelling through the chain when $\ket{\phi_1\phi_2} = \ket{\uparrow \downarrow}$ to be the same as the case where $\ket{\phi_1\phi_2} =\ket{\downarrow \uparrow}$, and we would expect both of these signals to be different to the case where $\ket{\phi_1\phi_2} = \ket{\downarrow \downarrow}$. If so, then if it is possible to find a measurement that differentiates between the two types of signal, we could distinguish between the $(\ket{\uparrow \downarrow}, \ket{\downarrow \uparrow})$ cases and the $\ket{\downarrow \downarrow}$ case. If the end states are then initialised as $ \ket{\phi_1} = \ket{\phi_2} = \ket{+}$, this creates a superposition of different signals, and if a measurement is made that distinguishes between the $(\ket{\uparrow \downarrow}, \ket{\downarrow \uparrow})$ cases and the $\ket{\downarrow \downarrow}$ case, we know we have projected the end qubits into the superposition $\frac{1}{\sqrt{2}}(\ket{\uparrow \downarrow}+ \ket{\downarrow \uparrow})$.

\subsubsection{Charge qubit to spin qubit transfer}

We could also use edge locking as a means to interface between charge and spin qubits. An example of a charge qubit is an electron in one of two quantum wells, with presence in the left well indicating $\ket{0}$ and the right well indicating $\ket{1}$. If the spin state of this electron is set to $\ket{\downarrow}$, then depending on if it is on the left or the right well, the spin chain will be unlocked or remain locked, and thus the state of the charge qubit can be transferred into spin degrees of freedom. 

\subsection{Release under more realistic conditions}

We have just seen several examples of how to use this edge locking state, must of which involve coupling a state to one edge of the chain, and measuring at the middle or the ends. In a perfect world, these operations could be done instantaneously, but of course it is more realistic to have operations that take a finite amount of time. In this subsection we first of all explore the effect of having less instantaneous operations when releasing an edge locked state (for example when performing the signal amplification shown above). We then explore how well a signal can be captured at the opposite end of the chain in the next subsection, by using the reverse strategy. Our motivation is to explore how realistic the applications of the edge locking could be, as well as seeing if there is some optimisation that can be done.

We use an XXZ Hamiltonian with anisotropy $\Delta$:
\begin{eqnarray}\label{XXZ}
H_{\textsc{xxz}} = J \sum_{n=1}^{N-1} (X_n X_{n+1} + Y_n Y_{n+1}) + \Delta Z_n Z_{n+1}
\end{eqnarray}
where $X_n$, $Y_n$ and $Z_n$ are the Pauli matrices acting on site n. To make it easier to sweep the whole range of $0 < \Delta \rightarrow \infty$, we parameterise it as $\Delta = \tan \theta_{\Delta}$, so $H_{XXZ}$ becomes:
\begin{eqnarray}
H_{\textsc{xxz}} &=& J \sum_{n=1}^{N-1} \sin \theta_{\Delta}(X_n X_{n+1} + Y_n Y_{n+1}) + J\cos \theta_{\Delta} Z_n Z_{n+1}.
\end{eqnarray}
To perform the release, we start with the edge locked state:
\begin{equation}\label{eqn:Init}
\ket{\phi_0} \equiv \ket{\uparrow \uparrow \downarrow ... \downarrow  \downarrow}.
\end{equation}
Ideally, we would be able to switch off all other interactions and perform a perfect $X$ rotation on the first spin, to give us the state $\ket{L_{1,(1)}} \equiv \ket{\downarrow \uparrow \downarrow ... \downarrow}$ which is no longer locked and so will travel along the chain. However, this may not be realistic, so we model the release as having a mixture of a rotation on the first site (magnetic field in the $X$ direction) along with $H_{XXZ}$, with the relative strengths of these two operators given by $\theta_r$, giving the following Hamiltonian:
\begin{eqnarray}
H_{r}(\theta_r) &=& J \cos \theta_r X_1 + J\sin \theta_r H_{\textsc{xxz}} \nonumber\\
&=& J \cos \theta_r X_1 + J \sin \theta_r \left[ \sum_{n=1}^{N-1} \sin \theta_{\Delta}(X_n X_{n+1} + Y_n Y_{n+1})  +  \cos \theta_{\Delta} Z_n Z_{n+1} \right]
\end{eqnarray}
This means we have some ratio $\tan \theta_r$ between the applied magnetic field and the couplings in the chain, with $\theta_r=0$ corresponding to a pure magnetic field (the ideal case) and $\theta_r = \pi/2$ corresponding to no magnetic field (no release). To perform the flip we would then evolve the state in eqn.\ (\ref{eqn:Init}) via:
\begin{eqnarray}
\ket{\phi_f} &=& e^{-iH_{r} t_{r} } \ket{\phi_0} = e^{-iH_{r} \pi / 2 } \ket{\phi_0}
\end{eqnarray}
where we have set $t_{r} =  \pi / (2J\cos \theta_r)$, since the maximum overlap of $\ket{\phi_f}$ with the ideal state $\ket{L_{1,(2)}}$ occurs at or very close to this time. Unless otherwise noted, we will express time in units of $\hbar/J$.

We then evolve the state with $H_{\textsc{xxz}}$, so that the time-evolved state is:
\begin{eqnarray}
\ket{\phi(t)} &=& e^{-iH_{\textsc{xxz}} t }\ket{\phi_f} =  e^{-iH_{\textsc{xxz}} t } e^{-iH_{r} \pi / 2 } \ket{\phi_0}
\end{eqnarray}
Following this evolution, we calculate the fidelity of $\ket{\phi(t)}$ with state $\ket{R_{1,(2)}} = \ket{\downarrow \downarrow...\downarrow  \uparrow  \downarrow}$, giving us the `release fidelity', $F_r$:
\begin{equation}
F_{r}(t) = |\sprod{{R_{1,(N-1)}}}{\phi(t)}|^2.
\end{equation}

\begin{figure}[!htb]
\begin{center}
\includegraphics[width=0.7\textwidth]{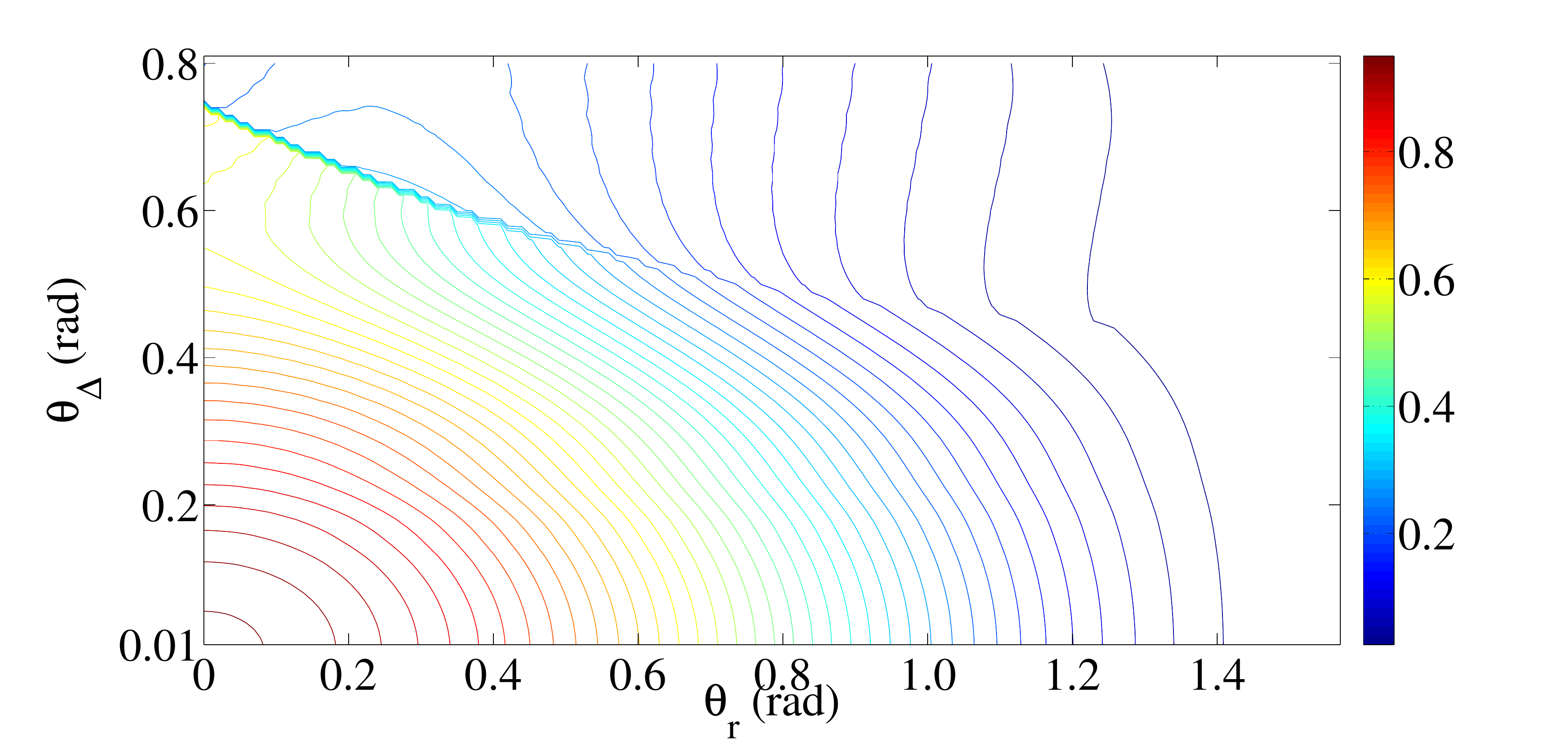}
\caption{Contour plot showing the maximum of $F_r$ as a function of $\theta_{\Delta}$ and $\theta_1$, with $N=6$, and using exact evolution. Note that the minimum value of $\theta_\Delta$ is 0.01}
\label{Fid1}
\end{center}
\end{figure}

\begin{figure}[!htb]
\begin{center}
\includegraphics[width=0.7\textwidth]{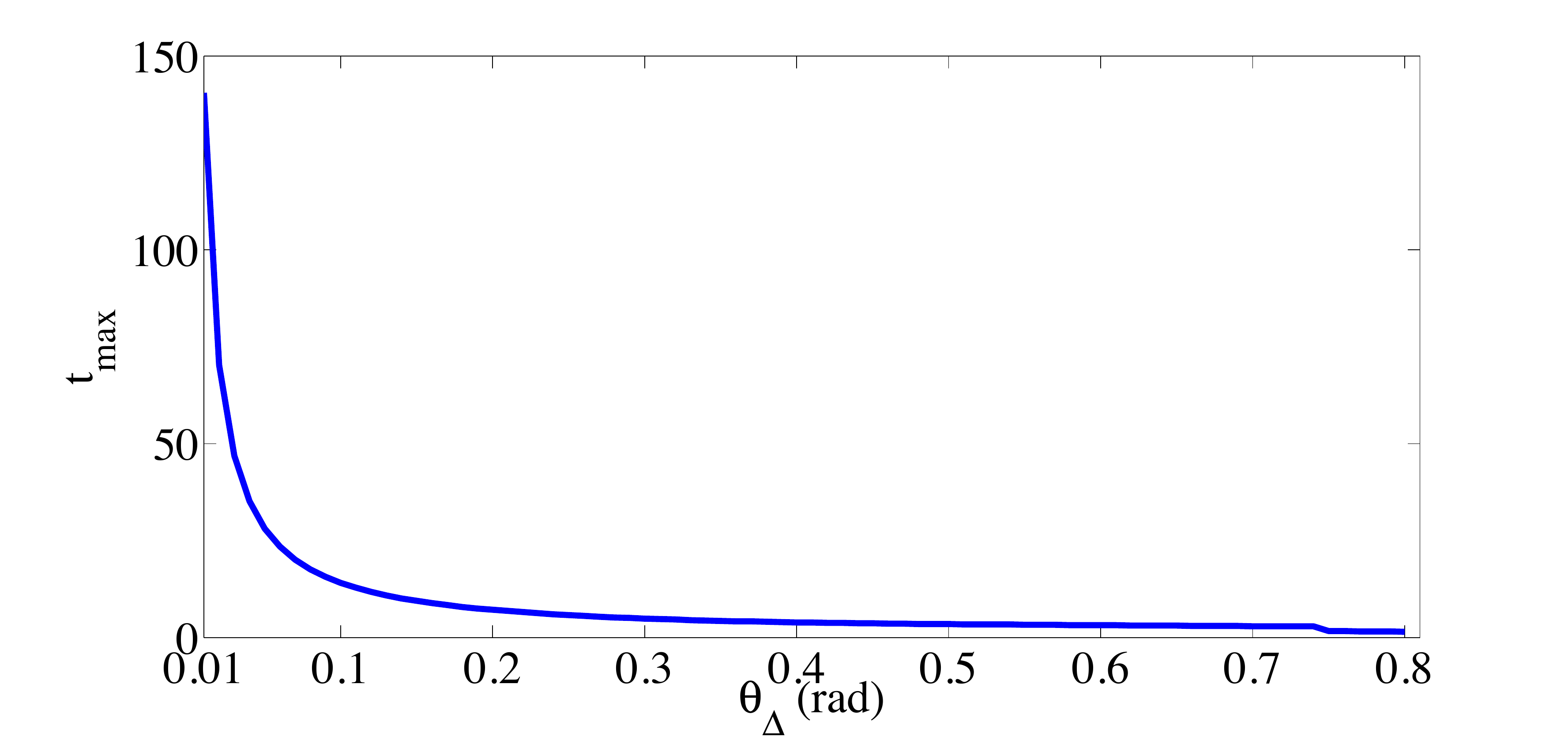}
\caption{Time taken to reach the first peak in $F_r$ after the release (in units of $\hbar/J$), for $\theta_r=0$. A similar variation is seen for all other values of $\theta_r$}
\label{tmaxDelta}
\end{center}
\end{figure}

To begin with, we looked at evolving the state until $F_r$ reaches its first peak, since in an application it would probably be best to wait for the shortest time possible, to limit the effects of decoherence. We thus define $t_{max}$ as the time taken for $F_r$ to reach the first peak. Firstly we looked at the variation of the size of $F_r(t_{max})$ with $\theta_{\Delta}$ and $\theta_{r}$ (see Fig.~\ref{Fid1}); evidently the best signal is for $\theta_{\Delta}$ and $\theta_r$ close to 0, as we might expect. Notice the rapid increase in time taken to reach the first maximum (see Fig.~\ref{tmaxDelta}), which limits how close to zero we can get as the simulations take too long (the minimum value of $\theta_{\Delta}$ in Figs.~\ref{Fid1} and \ref{tmaxDelta} is 0.01 radians, which means the $Z$ terms in $H_{XXZ}$ are roughly 100 times stronger than the $X$ or $Y$ terms). This suggests that there will be an optimal value of $\theta_{\Delta}$, depending on the particular experimental parameters such as decoherence time. There are also some strange discontinuities in Fig.~\ref{Fid1} for values of $\theta_{\Delta} > 0.4$, which we are currently unable to explain.

\begin{figure}[!htb]
\begin{center}
\includegraphics[width=0.7\textwidth]{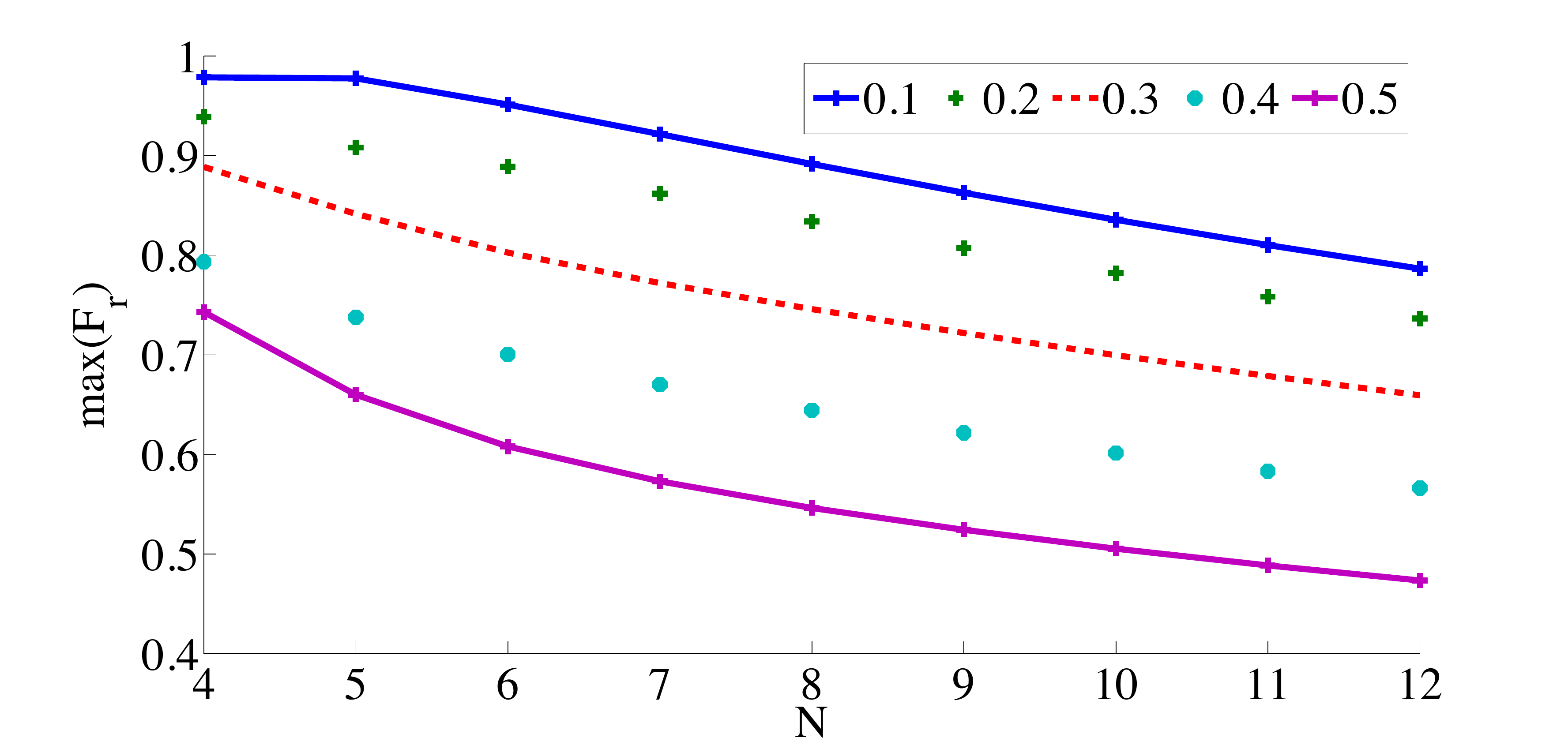}
\caption{Change in $F_r$ with $N$, with $\theta_r = 0$, where $F_r$ is taken at the first peak. The different coloured lines are for different values of $\theta_{\Delta}$ (in radians)}
\label{VaryingN}
\end{center}
\end{figure}

Next we look at how $F_r$ varies with the length of the chain $N$, and $\theta_{\Delta}$. This was done similarly to above by finding the height of the first peak, which shows a steady decrease (see Fig.~\ref{VaryingN}) that appears to decay as $\sim 1/N$ for all $\theta_\Delta$, as we would intuitively expect.

\subsection{Capture under more realistic conditions}

We now explore flipping the $N^{th}$ spin at some point in the evolution, with the hope of catching the pulse and keeping it locked in place. In a similar manner to the release protocol above, we capture the state by evolving with the Hamiltonian:
\begin{eqnarray}
H_{c}(\theta_c) = J\cos \theta_c X_N + \sin \theta_c H_{\textsc{xxz}}.
\end{eqnarray}
Then, after the release protocol above has been followed and the state has been evolved by $H_{\textsc{xxz}}$ for a time $\tau$, we switch on $H_{c}$ for a time $t_c$ and find the fidelity between this state and the ideal captured state, $\ket{R_{2,(1)}} = \ket{\downarrow \downarrow ... \downarrow  \uparrow   \uparrow}$. We define this as the capture fidelity, $F_c$:
\begin{equation}
F_c =   \bra{R_{2,(1)}} e^{-i H_c } \ket{\phi(t)}, \; t > \tau .
\end{equation}
For smaller values of $\theta_c$, $t_c$ could be set to $\pi/2$ and this would contain the maximum value of $F_c$, however for larger $\theta_c$ the evolution was more chaotic and so longer times had to be used. 

\begin{figure}[!htb]
\begin{center}
\includegraphics[width=0.7\textwidth]{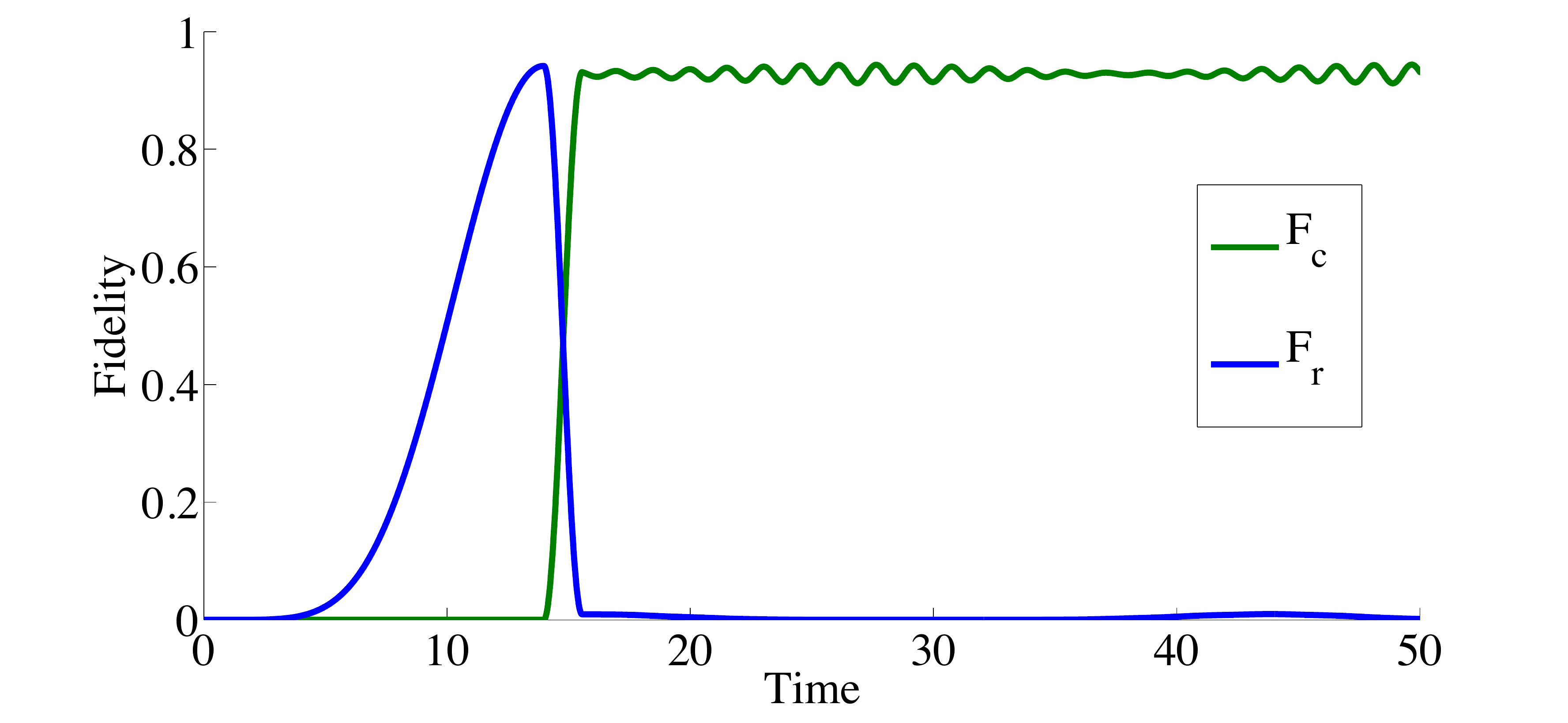}
\caption{An example of the capturing protocol, with $N=6$, $\theta_{\Delta} = 0.1$, $\theta_{r} = 0.1$, $\theta_{c} = 0.1$.}
\label{Capture1}
\end{center}
\end{figure}

\begin{figure}[!htb]
\begin{center}
\includegraphics[width=0.7\textwidth]{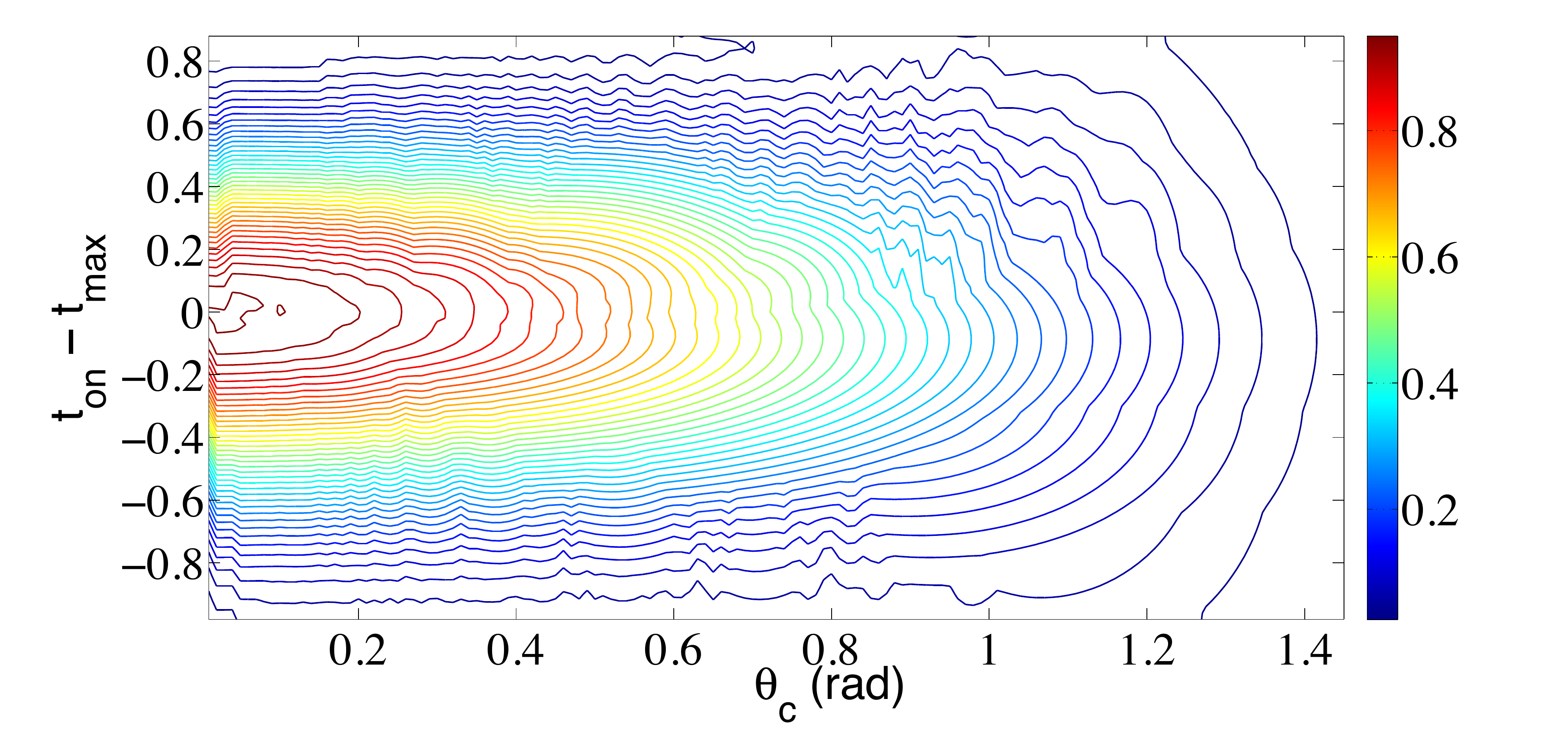}
\caption{The maximum achievable capture fidelity $\max (F_c)$ as a function of $t_{on} - t_{max}$ and $\theta_c$, with $\theta_{\Delta} = 0.1$.}
\label{Offset}
\end{center}
\end{figure}

An example of release and capture is shown in Fig.~\ref{Capture1}, for which the capturing magnetic field is turned on at $t_{on} =t_{max}$ (the point where $F_r$ reaches a maximum) and then is turned off again when $F_c$ reaches a maximum. We see that capture is possible, with $F_c$ bounded by the maximum value of $F_r$, as we would expect. We also tried varying the timing of this magnetic field; results are shown in Fig.~\ref{Offset}, showing the variation $F_c$ as $t_{on}$, is varied relative to $t_{max}$, and also the variation with $\theta_c$. Clearly the best results happen where $t_{on} = t_{max}$, and with $\theta_c = 0$.

\section{Ground state transfer}
\label{sec:MagFieldTransfer}
In this section we outline a potential information transferring protocol, based on the work in~\cite{Khajetoorians2012}. where classical information was shown to be transferred through an Ising chain. The experiment uses an even or odd chain of Fe atoms adsorbed to a copper surface, and through changing the local magnetic field experienced by one end of the chain and then preparing the system in the ground state, the spin of the atoms at opposite ends of the chain are aligned (anti-aligned) for odd (even) chains.

The studies in~\cite{Khajetoorians2012} are performed with Ising interactions. Here we investigate the transferring ability of chains with different Hamiltonians (but not necessarily those which are achievable using chains of Fe atoms). We will use a Hamiltonian of the form
\begin{align}
H_{\pm} &= \pm B_1 Z_1 +J_1\sum_{n=1}^{N-1}   (\sin \theta_\Delta (X_n X_{n+1} + Y_n Y_{n+1} )+ \cos\theta_\Delta  Z_n Z_{n+1} ) \nonumber \\
&+ J_2  \sum_{n=1}^{N-2}  ( \sin \theta_\Delta( X_n X_{n+2} + Y_n Y_{n+2} )+ \cos\theta_\Delta  Z_n Z_{n+2} )
\end{align}
To perform the protocol, we select either $H_+$ or $H_-$, and prepare the system in the ground state, then trace out all but the leftmost $n_{out}$ qubits, which acts as the output. These $n_{out}$ output qubits will in general be in a mixed state $\rho_N^{\pm}$. If the probability distribution of the inputs is such that $H_+,H_-$ are chosen with probabilities $p_+,p_-$, the state of the output qubits is $\sum_{x=\pm}  p_x \rho_N^{x}$. To assess the ability of this chain to send information, we require a measure of how distinguishable the output qubits are for the ground states $H_+$ and $H_-$. An appropriate way to measure this is to use the Holevo information $\chi$, which for a system prepared in one of the states $\{ \rho_1,\rho_2,...,\rho_n\}$ with probabilities $\{p_1,p_2,...,p_n\}$, is
\begin{align}
\chi_{H} : = S(\rho) - \sum_{x} p_x  S(\rho_x)
\end{align}
where $\rho = \sum_x p_x \rho_x$. From Holevo's theorem, this is an upper bound on the amount of information that can be extracted from the output qubits using any possible measurement~\cite{NielsenChuang}.

\begin{figure}[h]
\begin{center}
\subfloat[]{\includegraphics[width=0.49\textwidth]{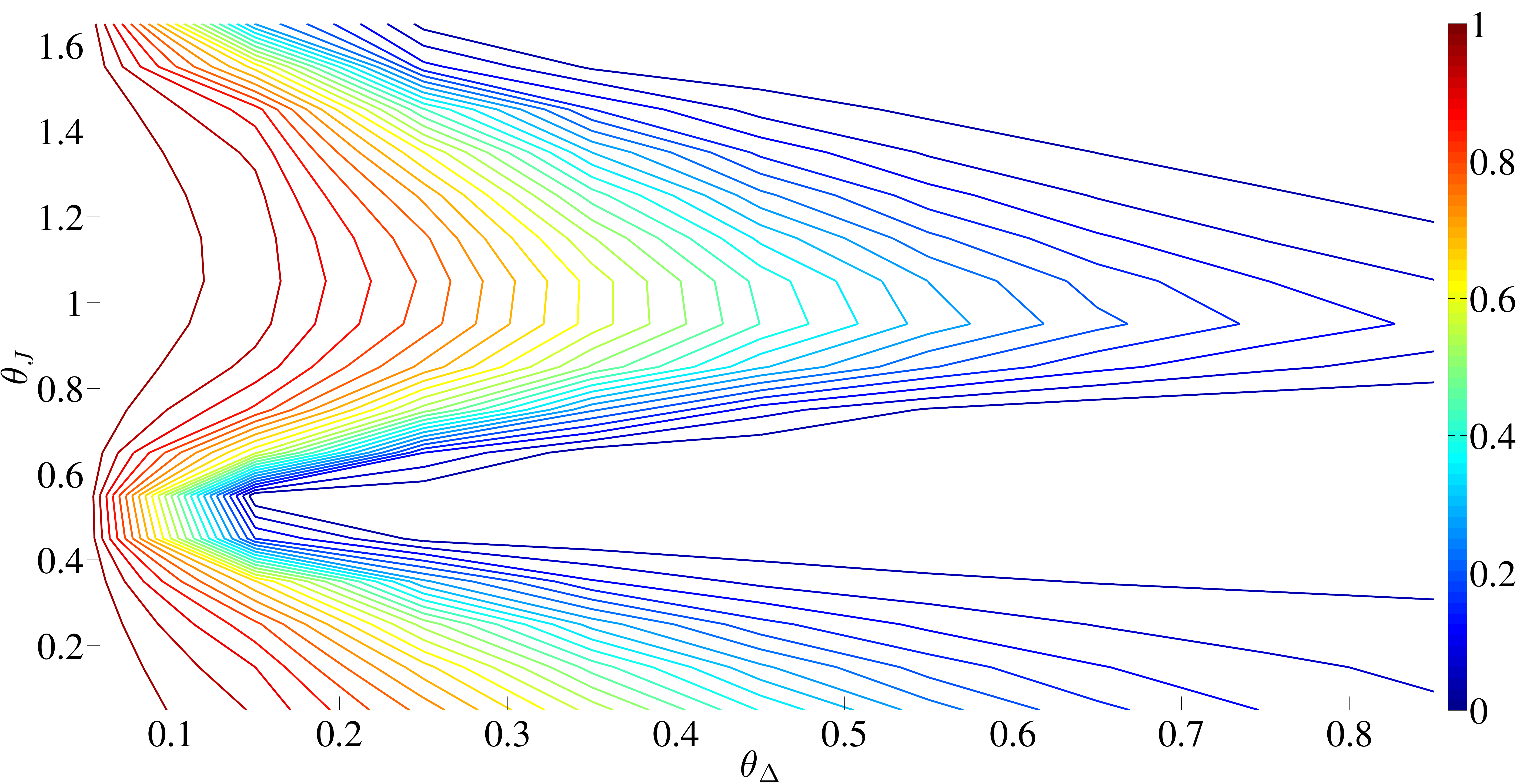}}
\subfloat[]{\includegraphics[width=0.49\textwidth]{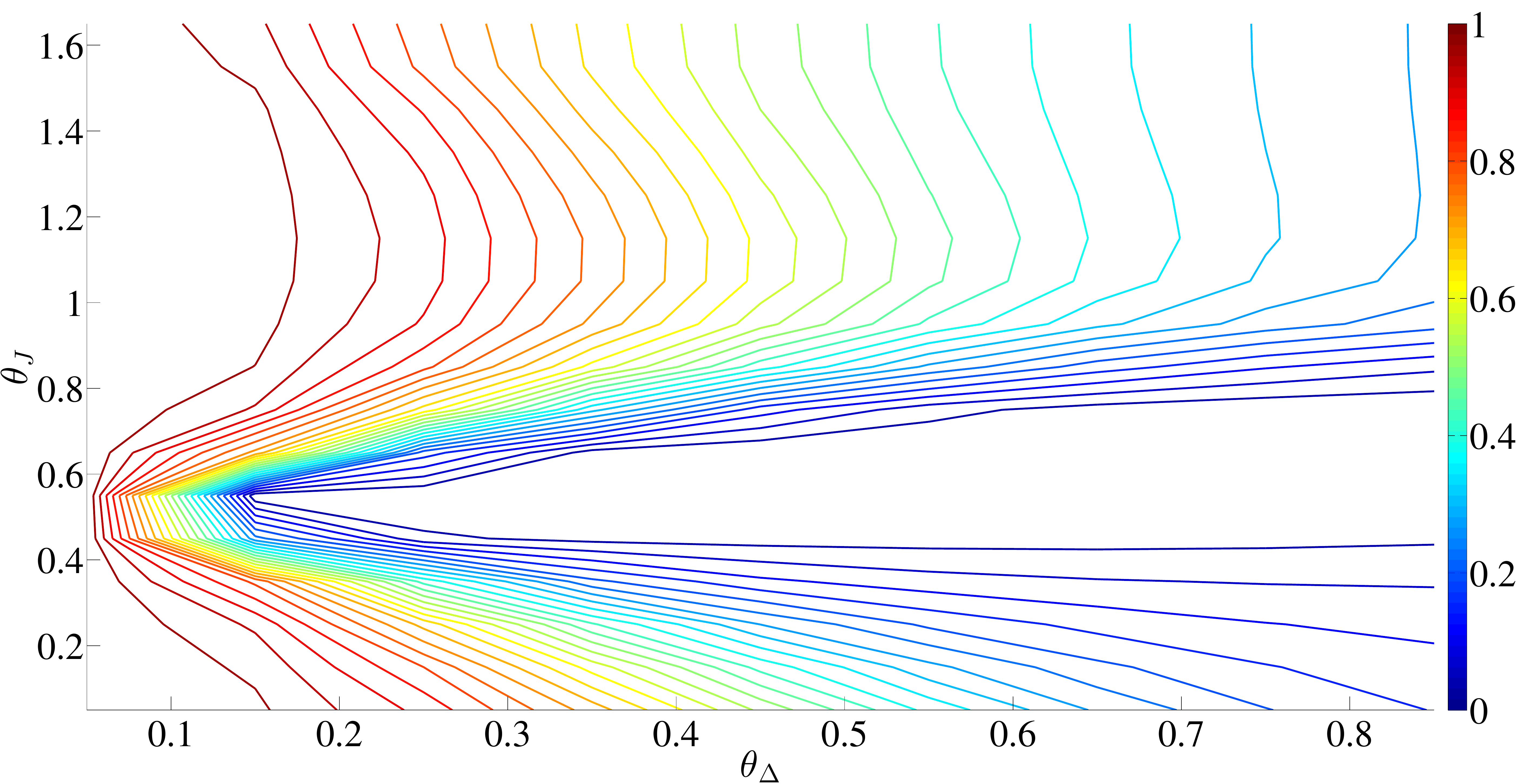}}
\caption{\label{fig:MagFieldTrans_Even}. Information transferring ability using the ground state of a chain with $N=8$ qubits. a) Reading out the end qubit ($n_{out} = 1$) b) reading out from the two end qubits ($n_{out} = 2$).}
\end{center}
\end{figure}

We parameterise $J_1 = \sin\theta_M \cos \theta_J$, $J_2 = \sin\theta_M \sin \theta_J$, $B = \cos \theta_M$, so that the full range of $J_1 / J_2$ and $J_1 / B$ can be covered ($\theta_M = 0$ corresponds to a large magnetic field and weak coupling, whilst $\theta_J=0$ corresponds to $J_1 = 1,J_2 = 0$ and $\theta_J = \frac{\pi}{2}$ corresponds to $J_1 = 0,J_2 = 1$. The Holevo information was then found over a range of parameters by finding the ground states of $H_{\pm}$. Results are shown in Fig.~\ref{fig:MagFieldTrans_Even} a) for a chain of length $N=8$ qubits, with $\theta_M = 0.1$ and for $\theta_\Delta \lesssim \pi /4$  (for larger $\theta_\Delta$, $\chi_{H}$ continued to decay to even smaller values). Fig.~\ref{fig:MagFieldTrans_Even}b) shows the same set-up but reading out the last $n_{out} = 2$ qubits.

The main features are high transferring ability for small $\theta_J$ or $\theta_J \simeq \frac{3 \pi}{8}$, which information transferral decreasing rapidly with increasing $\theta_\Delta$. There is also very poor information transferral around $\theta_J = 0.5$, which is most likely to be a result of the Majumdar-Ghosh point that occurs at $\theta_J = \pi/4, \theta_\Delta = \pi/4$, at which point the ground state is fully dimerised and so there is no long range order. Information is also poorly transferred for $\theta_J \simeq \frac{\pi}{2}$, which is likely to be because at this point there is pure next-neighbour coupling, which for an even chain means no correlation between inputs and outputs. 

Expanding the output to include more than one qubit predictably leads to better transfer of information. In future work, we would like to understand the theoretical basis for this, and perhaps derive a formula for the expected Holevo information as a function of the number of end qubits used.

\section{Conclusions}

At the end of this section, we are left with more questions than answers; we have seen that state transfer protocols through anisotropic spin chains can be robust to next-neighbour interactions, and there is a peak near to the Majumdar-Ghosh point where nearest-neighbour and next-nearest-neighbour interactions are equal, although the mechanism by which this peak occurs is not clear. 

We have also seen protocols for sending information using edge-locking in XXZ Hamiltonians, for which there do not seem to be any other parameter regimes except the ideal cases (where the spins are flipped instantaneously, and the Hamiltonian is very close to an Ising Hamiltonian) for which edge locking could be used transfer states at high fidelity. Given the long time for the evolution close to $\Delta^{-1} = 0$, the value of $\Delta$ will be limited by the decoherence rates in the system of interest, and future work could involve finding the optimum regime for a particular experimental set up. This study is limited in that we only use square pulses: more sophisticated pulses could also be tried, with which it is likely that near perfect transfer could be achieved with realistic parameters. Non-uniform spin chains may also give an advantage.

Finally we investigate and the natural alignment of the ground state of a XXZ with a boundary magnetic field, and the use of this to transfer information. The information transfer is somewhat robust to the presence of XX and YY interactions, and next-nearest neighbour interactions. Better fidelity can be achieved by measuring more of the end qubits. There are many avenues that these preliminary results could eventually go down, and our hope is that the small results presented here will lay the foundations for future work.

\chapter{Summary and Outlook}
In this thesis, we have introduced novel results aimed at simplifying and exploring the possibilities of quantum computation. A scheme to implement two-qubit quantum gates on decoherence-free qubits has been demonstrated, giving high fidelities above known error thresholds even with reasonable levels of noise. This scheme makes use of many-body interactions that are predicted to occur in certain systems, and requires far fewer pulses than existing gates, so could be useful in implementations where there is limited control, or where control errors are an issue. Further work could be done to explore encoded qubits on more than 4 qubits, and include effects due to Aharanov-Bohm phases from magnetic fields, either as a way to tune the interactions or as another source of error.

Three schemes to realise 3-qubit gates on arrays of qubits have also been proposed, using Hamiltonians that are time-independent during the gate operation, and using interactions that are as uniform as possible. The viability of implementing these using ion traps or bismuth donors in silicon has also been explored, finding that a fault-tolerant Toffoli gate could be realised using bismuth donors with today's technology, and it is hoped that experimental realisation of these gates could happen in the near future. These gates could be useful as primitives for quantum computation, or could prove useful as classical gates, making use of the many advances in quantum technologies before a quantum computer is within our reach, potentially allowing smaller or faster classical gates.More could be done to explore how much dissipation we would expect in a practical implementation of these gates including e.g.\ the effects of heating from laser fields, and it would also be interesting to explore the possibility of creating larger time-independent automated structures that could perform more than just a single gate. 

An extension of the adiabatic cluster-state model of computation has also been presented, using the tools of measurement-based quantum computation (MBQC) to transform adaptive measurements into adiabatically driven holonomic transformations. This translation has allowed a study of how properties of MBQC are translated into a holonomic setting, showing interesting trade-offs between properties of the Hamiltonian and the number of adiabatic steps. The allowed ordering of operations in this model was explored, finding that that for some approaches the allowed ordering is very limited compared to MBQC. It is hoped that this work opens up new channels of communication between the adiabatic and measurement-based schools of thought, providing a fruitful comparison and sharing of concepts. This line of thought could also possibly be extended to other MBQC patterns for which the tools used here are not valid. It would also be useful to explore practical implementations of this computational method, to see whether or not there is a significant practical advantage in using adiabatic processes rather than measurements.

The practicality of using a Wigner crystal as a quantum communication channel was explored using the semiclassical instanton approximation, yielding average fidelities of up to 0.92 with nine or ten electrons, with a transfer time significantly below the typical decoherence time of electrons in GaAs quantum dots. Whilst the parameters we use are for electrons in GaAs, the results are quite general and are applicable to any system in which a Wigner crystal can be created. We hope that this study inspires future technological applications using Wigner crystals, and hope that it might be developed into an experimentally realisable scheme in the near future. There are many parameters and additional effects such as noise, magnetic fields and image charges that we have not included in this analysis, so future work will include these effects and look for better ways of transferring information. There is also scope to go further, to use a Wigner crystal to perform gates between qubits, or even to use a Wigner crystal to perform universal quantum computation. 

Finally, methods of communicating using more idealised spin chains were also studied, extending a previous study to include next-nearest-neighbour interactions, finding an interesting although as yet not fully explained peak close to the Majumdar-Ghosh point. This demonstrates that previous spin chain communication protocols are robust to next-neighbour interactions, and provides interesting questions for future work. Novel schemes to transfer information using edge-locking and ground state transfer effects were also briefly studied. These schemes lay the foundation for further analysis exploring how these effects can best be put to use.

\bibliography{Thesis_arXiv.bbl}

\appendix

\chapter{Eigenvalues of commuting matrices}
\label{sec:LinAlg}

Here we prove a useful result for Chapter~\ref{chap:AGQC}, on the eigenvalues of commuting operators. We begin by presenting the proof found in Michael Nielsen's blog\footnote{\url{http://michaelnielsen.org/blog/archive/notes/fermions_and_jordan_wigner.pdf}}, and which was used in~\cite{Antonio2014} to calculate the minimum energy gap. However, there is a flaw in our interpretation of this proof (thanks to Daniel Burgarth for pointing this out). To amend this, we present another proof of the form of the minimum energy gap for the Hamiltonians considered in~\ref{chap:AGQC}, but the first misinterpreted proof is included since it may be of interest to other researchers in the field.

The spectral decomposition theorem states that normal matrices (of which Hermitian matrices are a special case) can be diagonalised, and therefore a Hermitian matrix $H$ can be written as
\begin{eqnarray}
H = \sum_j E_j P_j,
\end{eqnarray}
where $E_j$ are the eigenvalues of $H$ (real numbers if $H$ is Hermitian), and $P_j \neq \idop $ are operators projecting into the subspace with energy $E_j$, satisfying
\begin{eqnarray}\label{eqn:OrthBasis}
P_jP_k = \delta_{jk} P_j, \; \sum_j P_j = \idop, \; P_j^2 = P_j.
\end{eqnarray}
The projectors $P_j$ are said to form a complete orthonormal basis if they satisfy these properties. 

If two normal matrices $H_1$ and $H_2$ commute, then they can be diagonalised in the same basis, since if $\{ |E^{(1)}_n \rangle \}$ is a basis in which $H_1$ is diagonal, then
\begin{eqnarray}
H_2H_1 | E^{(1)}_n \rangle = E_n^{(1)} \left( H_2  | E^{(1)}_n \rangle \right)= H_1 \left( H_2 | E^{(1)}_n \rangle \right).
\end{eqnarray}
So $H_2 | E^{(1)}_n \rangle = E_n^{(2)} | E^{(1)}_n \rangle$, and so $H_j$ and $H_k$ have common eigenstates. Therefore the eigenvalues of $H_1 + H_2$ will be $E_n^{(1)} + E_n^{(2)}$ for certain combinations of $m$ and $n$; which combinations will depend on the structure of the matrices involved. 

Now consider a set of commuting normal matrices $\{ H_j\}$. We can spectrally decompose each one of these matrices:
\begin{eqnarray}
H_j = \sum_k E_{jk} P_{jk}
\end{eqnarray}
Since all of the $H_j$ commute, $[P_{jk},P_{lm}] = 0$ for any $j,k,l,m$. This follows since if two commuting matrices $A$ and $B$ have spectral decompositions $A = \sum_k a_{k} P_{k}^{(A)}$ and $B = \sum_k b_{k} P_{k}^{(B)}$, then the projectors can be written $P_{k}^{(A)} = poly(A)$, $P_{k}^{(B)} = poly(B)$. This can be seen since $A^n = \sum_k a^n_k P_k^{(A)} = M \mathbf{P}$ where $M_{mn} := a^m_n$ and $\mathbf{P} : = (P_1^{(A)},P_2^{(A)},...,P_D^{(A)})$ where $D$ is the dimension of $A$. The determinant of $M$ is then related to the determinant of a Vandermonde matrix (assuming that none of the $a_k$ are 0, which can be made true by excluding any projectors for which this is the case), which is non-zero, and so $M$ can be inverted so that $\mathbf{P} = M^{-1} \mathbf{A}$, where $\mathbf{A} := ( A,A^2,...,A^D)$. In other words, $P_k^{(A)} = poly(A)$ (this part of proof courtesy of Daniel Burgarth).

This means we can construct a complete orthonormal basis $P_{\vec{k}}$ where
\begin{eqnarray}\label{eqn:Pk}
P_{\vec{k}} = P_{1k_1} P_{2 k_2} ... P_{mk_m}
\end{eqnarray}
To prove it is a complete orthonormal basis, we must show that it satisfies the same properties in eqn.(\ref{eqn:OrthBasis}). Using the fact that $[P_{jk},P_{lm}] = 0$, 
\begin{align}
P_{\vec{k}} P_{\vec{k'}} & = \left( P_{1k_1} P_{2 k_2} ... P_{mk_m} \right) \left(P_{1k'_1} P_{2 k'_2} ... P_{mk'_m} \right) = \left( P_{1k_1} P_{1 k'_1} \right)\left( P_{2k_2} P_{2 k'_2} \right)...\left( P_{mk_m} P_{m k'_m} \right) \nonumber\\
&=\left( P_{1k_1} \delta_{k_1 k'_1} \right)\left( P_{2k_2} \delta_{k_2 k'_2}  \right)...\left( P_{mk_m} \delta_{k_m k'_m} \right) = P_{\vec{k}} \delta_{\vec{k} \vec{k'}}. \\
 \sum_{\vec{k}} P_{\vec{k}}  &= \sum_{k_1} \sum_{k_2}...\sum_{k_m} P_{\vec{k}} = \left( \sum_{k_1} P_{1k_1} \right)\left( \sum_{k_2} P_{2k_2} \right)...\left( \sum_{k_m} P_{mk_m} \right) = \idop.
\end{align}
So the $P_{\vec{k}}$ form an orthonormal basis, and the eigenvalues of $\sum_j H_j$ are formed from adding the eigenvalues of each of the $H_j$ together. Since each $P_{\vec{k}}$ projects into a different combination of eigenvalues of the individual $H_j$ matrices, we (incorrectly) interpreted this as equivalent to proving that the eigenvalues of $\sum_jH_j$ of a set of commuting matrices $\{H_1,H_2,...,H_m\}$ are $\{ \sum_{\vec{k}} E^{(1)}_{k_1} + E^{(2)}_{k_2} + ... + E^{(m)}_{k_m} \}$, which contains all possible combinations of the individual eigenvalues of the $H_j$. Clearly this cannot be true since, if the vector space has dimension $d$, this implies there are $d^m$ eigenvalues. The resolution to this is to note that if any adjacent pair of the projectors in $P_{\vec{k}} $ are orthogonal, then $P_{\vec{k}} $ becomes 0. Thus to find the eigenvalues of a sum of commuting Hermitian matrices $H = H_1 + H_2 + ... H_n$, we need to find the combinations of projectors in $P_{\vec{k}} $ such that no two projectors are orthogonal.

%
%

We now present a proof that, for Hamiltonians of the form in (\ref{eqn:HamSplit}), the eigenvalues are given by all combinations of the eigenvalues of the commuting terms. This proof is based on the explanation given in~\cite{PreskillNotes} for why stabilisers each halve the number of encoded qubits in a stabiliser code. Say we have a set of Hermitian operators $\mathcal{M} = \{ M_n \}$ which have eigenvalues $\{ \pm \lambda_n\}$ and eigenvectors $\{ \ket{\pm \lambda_n}\}$. Consider that we also have a set of unitary operators $\mathcal{Q} = \{ Q_n \} $ such that $\{ Q_n , M_n \} = 0$, $[Q_n,M_m]_{n \ne m} = 0$ for all $n,m$. Then $Q_n\ket{+\lambda_n}$ is an eigenstate of $M_n$ with eigenvalue $-\lambda_n$. Since $Q_n$ is a unitary operator, it is a one-to-one mapping, so that there must be an equal number of eigenstates of $M_n$ with eigenvalue $+\lambda_n$ or $-\lambda_n$ (i.e.\ they have the same degeneracy).

Now extending this, consider a state $\ket{\psi}$ which is a simultaneous eigenstate of all the $M_n$ operators. Then
\begin{eqnarray}
\left( \sum_{n=1}^{| \mathcal{M} |} M_n \right)  Q_m \ket{\psi} =\left( \sum_{n=1}^{| \mathcal{M} |}  \lambda_n(-1)^{\delta_{nm}} \right) Q_m \ket{\psi}.
\end{eqnarray}
So $Q_m \ket{\psi}$ is an eigenstate of $\left( \sum_{n=1}^{|\mathcal{M}|} M_n \right)$, with eigenvalue $\sum_{n=1}^{|\mathcal{M}|}  \lambda_n(-1)^{\delta_{nm}} $, and with the same degeneracy as $\ket{\psi}$. Therefore if there exists a unitary operator $Q_n$ for every operator $M_n$ such that $\{ Q_n , M_n \} = 0$, $[Q_n,M_m]_{n \ne m} = 0$ for all $m$, and if each $M_n$ has eigenvalues $\pm \lambda_{n}$, the spectrum is $\pm \lambda_{1} \pm \lambda_{2} \pm... \pm\lambda_{N}$. If the dimension of the space on which these operators act is $D$, then the degeneracy of each of these eigenvalues will be $\frac{D}{ 2^{|\mathcal{M}|}}$.

We now apply this argument to the Hamiltonian in (\ref{eqn:HamSplit}):
\begin{align}
H_{L_k}(s) =  - \gamma \left( \sum_{v <  L_k}X_v + \sum_{v>L_k}T_v \right) - \gamma \left(\sum_{v\in L_k} s X_v + (1-s)T_v \right).
\end{align}
where the operators $\{ T_v \}$ are stabilisers of the form in (\ref{eqn:GflowStab}) for a graph with \emph{gflow}. Each of the individual summands has two eigenvalues of equal magnitude and opposite sign, so the analysis above can be applied. For the summands in the $\sum\nolimits_{v <  L_k} X_v$ term, operators  that satisfy the conditions above are $Q_v = Y_v$. For any summand in the $\sum\nolimits_{v >  L_k} T_v$ term, we can select $Q_v = X_v^{\theta_v}$ (chosen so that it will commute with any $T_w$ such that $v \in g(w)$). Finally, for any of the terms in the $\sum_{v\in L_k} [(1-s) T_v +  s X_v]$ sum, we can select $Q_v = Y_v$. This follows since $Y_v$ anticommutes with $[(1-s) T_v +  s X_v]$ and commutes with $[(1-s) T_u +  s X_u]_{u \neq v}$, since the stabiliser $T_u$ has an identity operator acting on all vertices in the same layer as $u$ (from the rules of \emph{gflow}, there is an even connection between $g(u)$ and $v$ if $u$ and $v$ are in the same layer. Thus if $u$ and $v$ are in the same layer and $u \neq v$, then $[(1-s) T_u +  s X_u , Y_v] = 0$). Similarly $[Y_v,\sum\nolimits_{u >  L_k} T_u]=0$ since each $T_u$ operator has identity acting on vertex $v$, and $[Y_v,\sum\nolimits_{u <  L_k} X_u] = 0$.

%
%
%
%
%
%
%
%
%
Thus, given that it is possible to find a set of $\{ Q_n \}$ operators for each of the individual commuting summands $M_n$ in $H_{L_k}(s)$ that satisfy the commutation rules $\{ Q_n , M_n \} = 0$, $[Q_n,M_m]_{n \ne m} = 0$, the eigenvalues of $H_{L_k}(s)$ are all combinations of the eigenvalues of the commuting parts. The eigenvalues of the individual $[(1-s) T_u +  s X_u]$ terms are $\pm \sqrt{ (1-s)^2 +s^2 } := \pm \eta$, so the eigenvalues of $H_{L_k}(s)$ are $-|L_k|\gamma\eta,-(|L_k|-2)\gamma\eta,...,(|L_k|-2)\gamma\eta,|L_k|\gamma\eta$ plus some integer multiples of $\pm \gamma$. The total number of summands in $H_{L_k}$ is equal to the number of qubits encoded into the graph $N_q$, which means that the degeneracies of each eigenvalue are the same and equal to $\frac{2^N}{2^{N-N_q}} = 2^{N_q}$.

Similar reasoning can be applied to the time derivative $\dot{H}_{L_k}(s)$, with the result that the eigenvalues of $\dot{H}_{L_k}(s)$ are $\{ -|L_k|\gamma,-(|L_k|-2)\gamma,...,(|L_k|-2)\gamma,|L_k|\gamma \}$. The spectral norm of $\dot{H}_{L_k}(s)$ is therefore $\| \dot{H}_{L_k}(s) \| = |L_k| \gamma$.

A similar reasoning can also be applied to the Hamiltonian in eqn.~\ref{eqn:AGQCallatonce}:
\begin{align}
H(s) = -\gamma\sum_{v \in V \setminus O} (1-s)\tilde{T}_v + s X_v
\end{align}
where $[\tilde{T}_v, X_w] = 0\; \mbox{for all} \; v \ne w \; \mbox{in} \; V \setminus O$. In this case, for a summand $ (1-s)\tilde{T}_v + s X_v$ we can choose $Q_v = Y_v$. Since we are not just replacing stabilisers in the same layer as in the case above, other stabilisers may either have an identity operator or $X$ operator located at site $v$, and so there may be other stabilisers $\tilde{T}_u$ that anticommute with $Y_v$. This can be remedied by multiplying $Q_v$ by $X_u$ for any stabilisers $\tilde{T}_u$ that anticommute with $Y_v$, since $[(1-s)\tilde{T}_u + s X_u,Y_vX_u] = 0$ if $\{ T_u,Y_v\} = 0$. The eigenvalues of each summand will still be $\pm \eta$, which we can see since squaring a summand gives $\gamma^2[(1-s)\tilde{T}_v + s X_v]^2 = \gamma^2[(1-s)^2 + s^2]\idop = \gamma^2 \eta^2 \idop$. Thus the eigenvalues of the Hamiltonian in eqn.~\ref{eqn:AGQCallatonce} are multiples of $\pm\gamma \eta$, and the energy gap is $2 \gamma \eta$. Following similar reasoning, the maximum eigenvalue of $\| \dot{H} \|$ scales with $N \gamma$.

Finally we show how to arrive at the energy gap of the Hamiltonian given in equation (\ref{eqn:HamReorder}). The Hamiltonian has the form:
\begin{align}
H(s) = -\sum_{j < n -2} T_j  - \left( T_{n-2} + T_{n-1} + (1-s) T_n + sX_n^{\theta_n} \right) - \sum_{k > n} T_k.
\end{align}
Using the identity $\textsc{cz}_{m,n}^\dagger X_n Z_m \textsc{cz}_{m,n} = X_n$, then conjugating with \textsc{cz} operators acting between sites $(n-2,n-1)$ and $(n+1,n+2)$, gives
\begin{align}
H'(s) &= \textsc{cz}_{n-2,n-1}^\dagger\textsc{cz}_{n+1,n+2}^\dagger H(s) \textsc{cz}_{n+1,n+2}\textsc{cz}_{n-2,n-1} \nonumber\\
&= -\left( \sum_{j < n -3} T_j + Z_{n-3}X_{n-2} \right) - \left( X_{n-1} Z_{n} + T_{n-1} + (1-s) Z_n X_{n+1} + sX_n^{\theta_n} \right) \nonumber\\
&- \left( X_{n+2} Z_{n+3} + \sum_{k > n} T_k  \right).
\end{align}
$H'(s)$ has the same spectrum as $H(s)$, but is in a particularly useful form, as it is a \emph{Kronecker sum}. Given two square matrices $A$ and $B$ acting on vector spaces of dimension $d_A$ and $d_B$ respectively, we define a Kronecker sum as $A \oplus B : = A \otimes \idop_{d_B} + \idop_{d_A} \otimes B$. Thus $H'(s)$ can be written:
\begin{align}\label{eqn:KronSum}
-H'(s) &= \left( \sum_{j < n -3} T_j + Z_{n-3}X_{n-2} \right)  \oplus \left( X_{n-1} Z_{n} + T_{n-1} + (1-s) Z_n X_{n+1} + sX_n^{\theta_n} \right) \nonumber \\
&\oplus \left( X_{n+2} Z_{n+3} + \sum_{k > n} T_k  \right).
\end{align}
A particularly useful property of Kronecker sums is that the eigenvalues of $A \oplus B$ are all combinations of the eigenvalues of $A$ and $B$~\cite{HornJohnson2}. Thus the eigenstates of $H'(s)$ are all combinations of the eigenvalues of the individual terms in the Kronecker sum. The terms on the left and right in (\ref{eqn:KronSum}) have integer eigenvalues, whilst the middle term has energy gap $\leq 1$, and so the energy gap is given by the energy gap of the middle term.

\chapter{Interaction picture and the rotating wave approximation}
\label{app:IntPic}

In quantum mechanics, typically we consider states $\ket{\psi(t)}$ that evolve according to the Schr\"{o}dinger equation $i\hbar | \dot{\psi}(t) \rangle= H\ket{\psi(t)}$, so that the state vector carries the information of how the system changes, whilst operations are performed. This picture of quantum mechanics is usually called the Schr\"{o}dinger picture. For some applications, it is often more informative to use different ways of depicting the evolution other than the Schr\"{o}dinger picture. One alternative is the \emph{Heisenberg picture}, in which states are stationary but the operators carrying the time dependence (see e.g.\ Chapter~\ref{chap:AGQC} for an example of using the Heisenberg picture), for which we can express the Schr\"{o}dinger equation as an equation involving operators $O(t)$ rather than states:
\begin{align}
i\hbar \frac{d}{dt} O(t) = H(t) O(t)
\end{align}
In cases where we have a Hamiltonian $H(t) = H_0(t) + V(t)$ which can be split up into a part which is easily solvable $H_0$, and an additional part $V$, it is often convenient to use the \emph{intermediate} or \emph{Dirac} picture to follow the dynamics, which is effectively halfway between the Schr\"{o}dinger and Heisenberg pictures. To transform into the intermediate picture, we represent operators and states relative to the evolution due to $H_0(t)$, which can be done by operating with $A(t) := e^{-iH_0t/\hbar}$. States in this picture then become $\ket{\psi_I(t)} := A^{\dagger}(t) \ket{\psi(t)}$, and operators are $O_I(t) := A^{\dagger}(t) O(t) A(t)$. If we insert these into the time-dependent Schr\"{o}dinger equation, we see that:
\begin{align}
\frac{d}{dt} \ket{\psi_I(t)}  &= \frac{d}{dt} [ A(t) \ket{\psi_I(t)}] = \frac{d}{dt} [ e^{iH_0t/\hbar} \ket{\psi_I(t)}] \nonumber\\
&=\frac{i}{\hbar}H_0 \ket{\psi_I(t)} + e^{iH_0t/\hbar} \left[ \frac{1}{i\hbar} H(t) \ket{\psi(t)} \right]= - \frac{ie^{iH_0t/\hbar} }{\hbar} V \ket{\psi(t)}=- \frac{i}{\hbar} V_I \ket{\psi_I(t)}
\end{align}
So that
\begin{align}
i\hbar \frac{d}{dt} \ket{\psi_I(t)} = V_I(t)  \ket{\psi_I(t)}
\end{align}
which can be a useful way to frame a particular problem.

A useful tool when using the interaction picture is the rotating wave approximation, in which terms in the interaction picture with a high frequency time dependence can be ignored. To see this, start with the Dyson series for a time-dependent Hamiltonian $H_I$ (see e.g.~\cite{DalessandroBook}) 
\begin{align}
U(t,0) & = \mathcal{T} e^{-\frac{i}{\hbar} \int_0^t H_I(t_1) dt_1} =\idop - \frac{i}{\hbar}\int_0^t H_I(t_1) dt_1  + \frac{1}{2\hbar^2}\int_0^t \int_0^{t}  \mathcal{T}[H_I(t_1) H_I(t_2)]  dt_1 dt_2  + ... 
\end{align}
where $\mathcal{T}$ is the time-ordering operator, $\mathcal{T}[x(t_1)x(t_2)] = \Theta(t_1-t_2)x(t_1) x(t_2) + \Theta(t_2 - t_1) x(t_2)x(t_1)$, and $\Theta(t)$ is the Heaviside step function. Consider a Hamiltonian of the form $H_I = V \cos(\omega t)$ where $V$ is time-independent and Hermitian. If it is the case that $\omega$ is a high frequency, this means that $V \cos(\omega t)$ is a fast oscillating term which may have negligible effect on the dynamics. Using the identity $\int_0^t \cos (\omega t') dt' = \frac{\sin (\omega t')}{ \omega} $, the evolution operator can be written
\begin{align}
U(t,0) &= \idop - \frac{i}{\hbar}\int_0^t V \cos(\omega t_1) dt_1  + \frac{1}{2\hbar^2}\int_0^t \int_0^{t}  \mathcal{T}[V \cos(\omega t_1) V \cos(\omega t_2)]  dt_1 dt_2  + ... \nonumber \\
& = \idop + O \left( \frac{\Vert V \Vert}{\hbar \omega} \right) 
\end{align}
provided that $\Vert V \Vert < \hbar\omega$. This means that, whenever we have terms in a Hamiltonian which have a high frequency time dependence, these can be ignored with an error of order $\Vert V \Vert / \hbar \omega$. 
We now derive some identities that will be useful in the next subsections:

Consider a Hamiltonian term of the form $C\Omega e^{-i\omega Z t} \cos (\omega_x t) X$. Setting $\omega_x = \omega$ gives
\begin{align}\label{eqn:Iden1}
& C e^{-i\omega Z t} \cos (\omega t) X =\frac{C}{2} e^{-i\omega Z t} (e^{i\omega t} +  e^{-i\omega t})X \nonumber\\
&=\frac{C}{2}(e^{i\omega t} +  e^{-i\omega t})(e^{-i\omega t } \sigma^+ + e^{i\omega t } \sigma^- )= \frac{C}{2}\left(  \sigma^+ + e^{ 2i\omega t } \sigma^- +  e^{-2i\omega t} \sigma^+ +  \sigma^- \right)\nonumber\\
& = \frac{C}{2} X + O \left( \frac{C}{\hbar\omega} \right)
\end{align}
provided that $C <  \hbar \omega$. Strictly speaking the last line is incorrect: the Hamiltonian itself does not have terms of order $C / \hbar \omega$, but the evolution due to the Hamiltonian does. We will use this type of notation in Chapter~\ref{chap:3qubit}.

For the second identity, consider the following Hamiltonian:
\begin{align}\label{eqn:Iden2}
&e^{-it [\omega_1 Z_1 +  \omega_2 Z_2]}  J ( X_1X_2 + Y_1 Y_2 + Z_1 Z_2)e^{it [\omega_1 Z_1 +  \omega_2 Z_2]} =2J e^{-2 it [\omega_1 Z_1 +  \omega_2 Z_2]} (\sigma^+_1 \sigma^-_2 + \sigma^-_1 \sigma^+_2) + JZ_1 Z_2 \nonumber\\
&=2J\sigma^+_1 \sigma^-_2 e^{2it[\omega_1 - \omega_2]}  + 2J\sigma^-_1 \sigma^+_2 e^{-2it[\omega_1 - \omega_2]} + JZ_1 Z_2=  JZ_1 Z_2 + O \left( \frac{2J}{\hbar( \omega_1 - \omega_2)} \right)
\end{align}
provided $2 J < \hbar(\omega_1 - \omega_2)$.

\chapter{Density matrix renormalisation group}
\label{sec:DMRG}

The Hilbert space of a spin chain grows exponentially with $N$, and so for even a modest number of spins, exactly diagonalising a Hamiltonian can take a very long time. Fortunately, several tools have been developed to approximate the behaviour of a system, by ignoring the states in a system which have the smallest effect. This is the basis of the density matrix renormalisation group (DMRG) approach~\cite{White92}. 

\subsubsection{Time-independent DMRG}
This algorithm can be split into two parts: the `infinite' algorithm and the `finite' algorithm, which we outline here for a simple Hamiltonian with nearest-neighbour interactions. The overall process is that we build up the ground state of a chain, starting off with two sites, and sequentially adding on 2 sites at a time, choosing a truncated basis in which to represent this ground state (if necessary). Say we start with a chain which we divide into two `blocks', left and right, which both have dimension M, and which have individual Hamiltonians $H_L$ and $H_R$ (i.e.\ for the moment we assume the two halves of the chain are non-interacting)
\begin{equation}
|\psi \rangle = \sum_{L,R} A_{L,R} | L \rangle | R \rangle.
\end{equation}
Now we add sites of dimension d to the left and the right sides to get
\begin{eqnarray}
|\psi \rangle &=& \sum_{L,R,l,r} A_{L,R,l,r} | L \rangle | l \rangle | r \rangle |R \rangle := \sum_{i,j} B_{i,j} | i \rangle |j \rangle,
\end{eqnarray}
so that the left and right sides now have dimensions $Md$, and we have a new `extended' Hamiltonian for both sides, $H_{eL} $ and $H_{eR}$, and an overall Hamiltonian $H_S$:
\begin{eqnarray}
H_{eL} &=&  H_{L} \otimes I_d +  J \vec{\sigma}_{N_L} \cdot \vec{\sigma_l} \nonumber\\
H_{eR} &=&  I_d \otimes H_{R}  + J \vec{\sigma}_r \cdot \vec{\sigma}_{N_R} \nonumber\\
H_S &=& H_{L} \otimes I_{Md} + I_{Md} \otimes H_{R} +  I_M \otimes (J \vec{\sigma}_l \cdot \vec{\sigma}_r) \otimes I_M,
\end{eqnarray}
where $\vec{\sigma}_{N_L}$ and $\vec{\sigma}_{N_R}$ indicate, respectively, the Pauli spin operator of the right or leftmost site of L or R (expressed in the basis of L or R), and  $\vec{\sigma}_{l}$, $\vec{\sigma}_{r}$ are the Pauli spin operators acting on the added sites. We now ask: what is the best way to express the states $|i \rangle$ and $| j \rangle$ in a basis of dimension M? The answer is that if we find the ground state $| \psi_G \rangle$ of $H_S$, and then find the partial trace of the this wavefunction for the right and left side
\begin{eqnarray}
\rho_{eL} = tr_R(|\psi_G \rangle \langle \psi_G |), \; \rho_{eR} = tr_L(|\psi_G \rangle \langle \psi_G |),
\end{eqnarray}
and keep only the eigenstates corresponding to the M largest eigenvalues, then this will give the  state with least error \cite{SchollwockThesis}. Then after we have discarded the lowest eigenvectors, we have a reduced basis in which to express the right and left sides of the chain, and we transform all the Hamiltonians and operators relevant to our calculations into this basis.To do this we construct operators $T_L$ and $T_R$ made up of the first m eigenvectors of $\rho_{eL}$ and $\rho_{eR}$, and express the extended Hamiltonians in terms of this basis, to form new block Hamiltonians, $H_{L}'$ and $H_{R}'$, which are truncated into the $M \times M$ Hilbert space.
\begin{eqnarray}
H_{L}' = T_L^{\dag} H_{eL} T_L , \; H_{R}' = T_R^{\dag} H_{eR} T_R.
\end{eqnarray}
We are now back to where we started, with a left and right block, and we can repeat the algorithm, adding sites onto these blocks, and repeating until the desired system size is reached, but avoiding the exponential growth in Hilbert space (and therefore the growth in computing resources needed). By truncating the Hilbert space we will introduce an error $\epsilon$, which is just given by the sum of eigenvalues of the eigenstates we have chosen to exclude
\begin{equation}
\epsilon =  \sum_{j \not \in m} \lambda_j = 1 - \sum_{j  \in m} \lambda_j,
\end{equation}
where the $\lambda_j$ are the eigenvalues of $\rho_{eL}$ or $\rho_{eR}$. This is the end of the `infinite' DMRG algorithm; however to achieve a really accurate ground state we need to use a second algorithm, called the `finite' DMRG algorithm, where the chain length is kept fixed. The first part is the same as for the infinite case. We find $\rho_{eL}$ and find $H_{L}'$ in terms of the eigenstates of $\rho_{eL}$, but we leave $H_{R}$ unchanged. In the next step we use $H_{L}'$ as our new left block Hamiltonian, which means we have increased the size of the left part by one site, and so we use a right block Hamiltonian which is smaller by one site (for this reason we have to keep in memory all of the transformation matrices $T_L$, $T_R$ that we found in the infinite algorithm). Then the chain length is unchanged, but we have advanced towards the right by one site (but we have only made changes to the left hand side), and we can repeat the process again until we reach the right hand end of the chain
\begin{equation}
|\psi_{End} \rangle = \sum_{L,l,r} A_{L,l,r} | L \rangle | l \rangle | r \rangle.
\end{equation}
The process is reversed so that the right blocks are updated, and so on back and forth until there is some convergence (e.g. convergence of the ground state energy).

\subsubsection{DMRG with time evolution}

The next step is to evolve the system in time. For systems with nearest-neighbour interactions, we would normally use the Suzuki-Trotter time-step method \cite{Vidal2003} \cite{White2004}, whereby the evolution operator can be expressed as
\begin{eqnarray}
e^{-i H \tau} \approx e^{-i H_{12} \tau/2} e^{-i H_{23} \tau/2}  ... e^{-i H_{23} \tau/2} e^{-i H_{12} \tau/2} + \mathcal{O}(\tau^3),
\end{eqnarray}
where $H_{ij}$ is the Hamiltonian acting between sites i and j only, and we operate on each pair up to $(N-1,N)$ and back. However, since we are using next-nearest neighbour interactions, this pairwise method is not suitable here, and instead we use the time-step targeting method, introduced in \cite{Luo2003} and developed further in \cite{Feiguin2005}, which is generally less accurate than using the Suzuki-Trotter time-step method but more suitable for our calculations.

To do this, we start at one end of the chain (i.e. after the finite algorithm has reached one end). Then we sweep back to the other end, updating the basis of the block+site at each step as before, but this time updating them with respect to some target states \cite{Feiguin2005}, which are given by the fourth order Runge-Kutta method, and will have some dependence on time. To perform this we construct a set of four vectors:
\begin{eqnarray}
|k_1 \rangle &=& -i \tau \tilde{H}(t) | \psi (t) \rangle  \nonumber\\
|k_2 \rangle &=& -i \tau \tilde{H}(t+\tau/2) [| \psi (t) \rangle + \frac{1}{2} | k_1 \rangle ] \nonumber\\
|k_3 \rangle &=& -i \tau \tilde{H}(t+\tau/2) [| \psi (t) \rangle + \frac{1}{2} | k_2 \rangle] \nonumber\\
|k_4 \rangle &=& -i \tau \tilde{H}(t+\tau) [| \psi (t) \rangle + | k_3 \rangle],
\end{eqnarray}
Where $\tilde{H}(t) = H(t) - E_0$, and in our case the Hamiltonian is time invariant. We then find expressions for the state at times $(t+\frac{\tau}{3})$, $(t+\frac{2\tau}{3})$ and $(t+\tau)$
\begin{eqnarray}
|\psi(t+\frac{\tau}{3}) \rangle &\approx& |\psi(t) \rangle + \frac{1}{162}[ 31|k_1 \rangle + 14|k_2\rangle + 14|k_3\rangle -5|k_4 \rangle] \nonumber \\
|\psi(t+\frac{2\tau}{3}) \rangle &\approx& |\psi(t) \rangle + \frac{1}{81}[ 16 |k_1 \rangle + 20 |k_2\rangle + 20|k_3\rangle -2 |k_4 \rangle] \nonumber \\
|\psi(t + \tau) \rangle &\approx& |\psi(t) \rangle + \frac{1}{6}[ |k_1 \rangle + 2|k_2\rangle + 2|k_3\rangle + |k_4 \rangle].
\end{eqnarray}
We then construct a reduced density matrix using a weighted sum over the states at these four times e.g. for the left block:
\begin{equation}
\rho_L = \sum_{n=0}^3 w_n tr_R(| \psi(t + n \tau / 3) \rangle \langle \psi(t + n \tau / 3) |),
\end{equation}
where the optimal weightings were determined in \cite{Feiguin2005} to be $w_0=1/3, w_1=1/6, w_2=1/6, w_3=1/3$. And as before we can diagonalise this and discard a certain number of states to find a new more efficient basis in which to do our time evolution (without doing this kind of basis transformation, we would quickly run into errors \cite{GarciaRipoll2006}). Then we repeat this process along the chain, updating the basis at each step (and importantly, updating any operators, Hamiltonians, or transformation matrices that we may need to use later on) but without advancing in time. Then at the end of this half-sweep, when we have updated each basis and thus found the optimal basis in which to perform our time evolution, we evolve the state by a single time step (and since the computational time required to do this last part is small, we can use a more accurate procedure than that above, so instead we use 10 Runge-Kutta iterations each with time step $\tau / 10$). Then we repeat the process, making sweeps along the chain, advancing in time only at the end of each half sweep, until we have evolved the state for the required length of time.

\end{document}